\documentclass[12pt, openright,	oneside, a4paper, sumario=tradicional, english, brazil]{abntex2}

\usepackage{lmodern}			
\usepackage[T1]{fontenc}
\usepackage{pstricks-add}
\usepackage[utf8]{inputenc}		
\usepackage{lastpage}			
\usepackage{indentfirst}		
\usepackage{color}				
\usepackage{graphicx}			
\usepackage[font=small]{caption}
\usepackage{subcaption}
\usepackage{microtype} 			
\usepackage{cite}
\usepackage{textcomp}
\usepackage{calc}
\usepackage{epstopdf} 
\usepackage{tikz}
\usepackage{rotating}
\usetikzlibrary{arrows,shapes,trees}
\usepackage{{amsmath,amsfonts,amssymb,amsthm}}
\usepackage[mathscr]{euscript}

\usepackage{ifthen}
\usepackage{textcase}
\usepackage{hyperref}
\usepackage{framed}
\usepackage{setspace}
\usepackage{amsmath}
\usepackage{wrapfig}
\usepackage{url}
\usepackage{mathptmx}
\usepackage{multirow}
\usepackage{xfrac}
\usepackage{icomma}

\usepackage{multirow}
\usepackage{float}
\usepackage{pdfpages} 
\usepackage{lipsum}				
\usepackage[alf]{abntex2cite}	

\usepackage{titlesec/titlesec}

\usepackage{tocloft}
\cftsetindents{chapter}{0in}{0.5in}
\cftsetindents{section}{0in}{0.5in}
\cftsetindents{subsection}{0in}{0.5in}
\usepackage{titletoc}

\renewenvironment{agradecimentos}
{\pdfbookmark[0]{Agradecimentos}{ag}\begin{center}
\textbf{AGRADECIMENTOS}
\end{center}}{\pagebreak}

\renewenvironment{resumo}
{\pdfbookmark[0]{Resumo}{re}\begin{center}
\textbf{RESUMO}
\end{center}}{\pagebreak}

\renewenvironment{abstract}
{\pdfbookmark[0]{Abstract}{abs}\begin{center}
\textbf{ABSTRACT}
\end{center}}{\pagebreak}

\titleformat{\chapter}{\normalfont\bfseries}{\thechapter}{1em}{}
\titlespacing*{\chapter}{0pt}{-30pt}{40pt}
\titleformat{\section}{\normalfont}{\thesection}{1em}{}
\titleformat{\subsection}{\normalfont\itshape}{\thesubsection}{1em}{}

\titulo{Interactions between topological defects in (1+1) dimensions}
\autor{João Guilherme Ferreira Campos}
\local{Recife}
\data{2022}
\orientador{Profa. Dra. Azadeh Mohammadi}
\instituicao{%
  Universidade Federal de Pernambuco
  \par
  Centro de Ciências Exatas e da Natureza
  \par
  Departamento de Física}
\tipotrabalho{Tese de Doutorado}
\preambulo{\textit{Tese apresentada ao Programa de Pós-Graduação em Física da Universidade Federal de Pernambuco,
como requisito parcial para obtenção do título de Doutor em Física.}}

\definecolor{blue}{RGB}{41,5,195}

\makeatletter
\hypersetup{
		pdftitle={\@title}, 
		pdfauthor={\@author},
    	pdfsubject={\imprimirpreambulo},
	    pdfcreator={LaTeX with abnTeX2},
		pdfkeywords={abnt}{latex}{abntex}{abntex2}{trabalho acadêmico}, 
		colorlinks=true,       		
    	linkcolor=black,          	
    	citecolor=black,        		
    	filecolor=magenta,      		
		urlcolor=black,
		bookmarksdepth=4
}
\makeatother

\makeindex
\nobibintoc
\begin{document}
\selectlanguage{english}
\frenchspacing

\includepdf{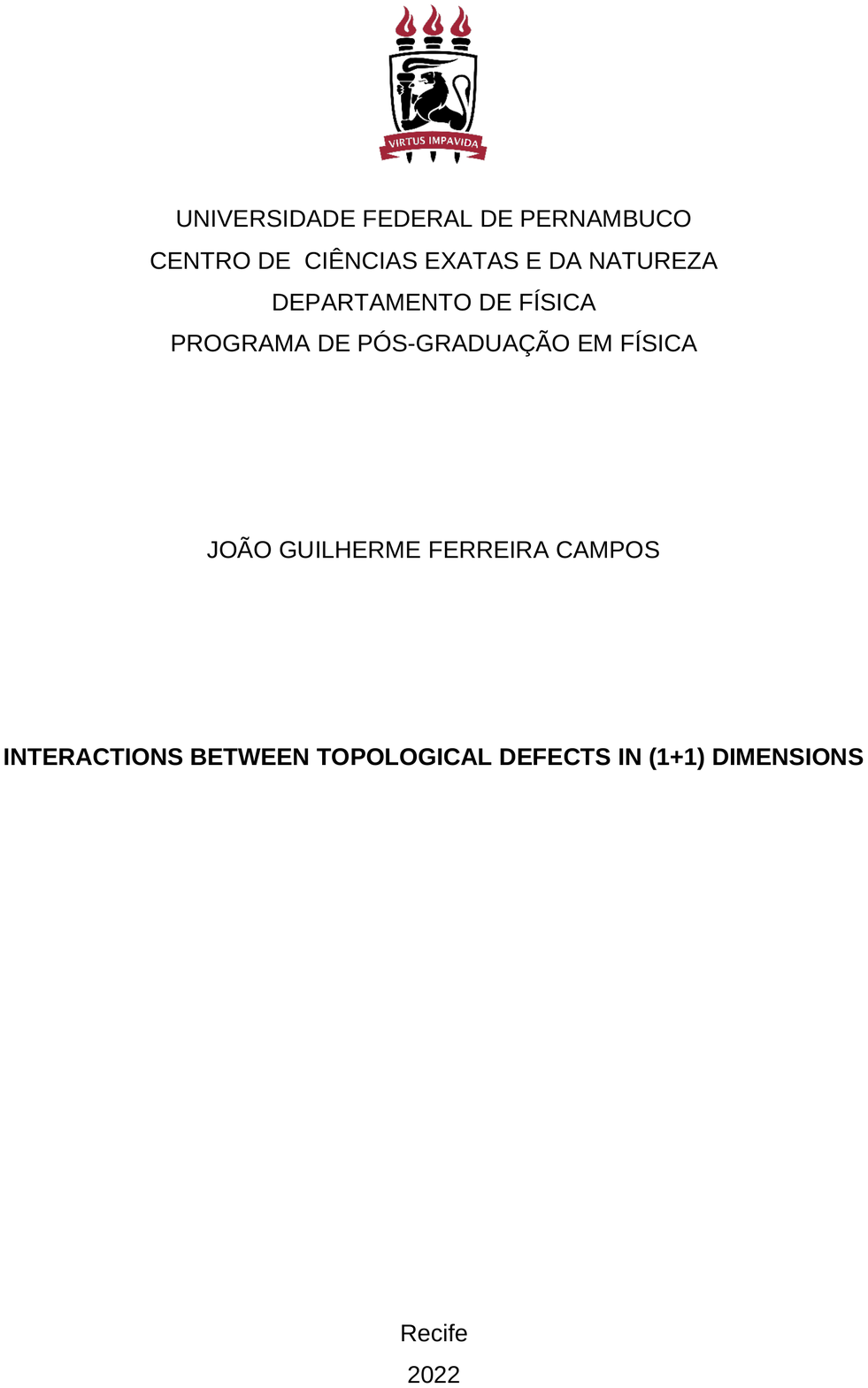}
\pagenumbering{arabic}
\includepdf{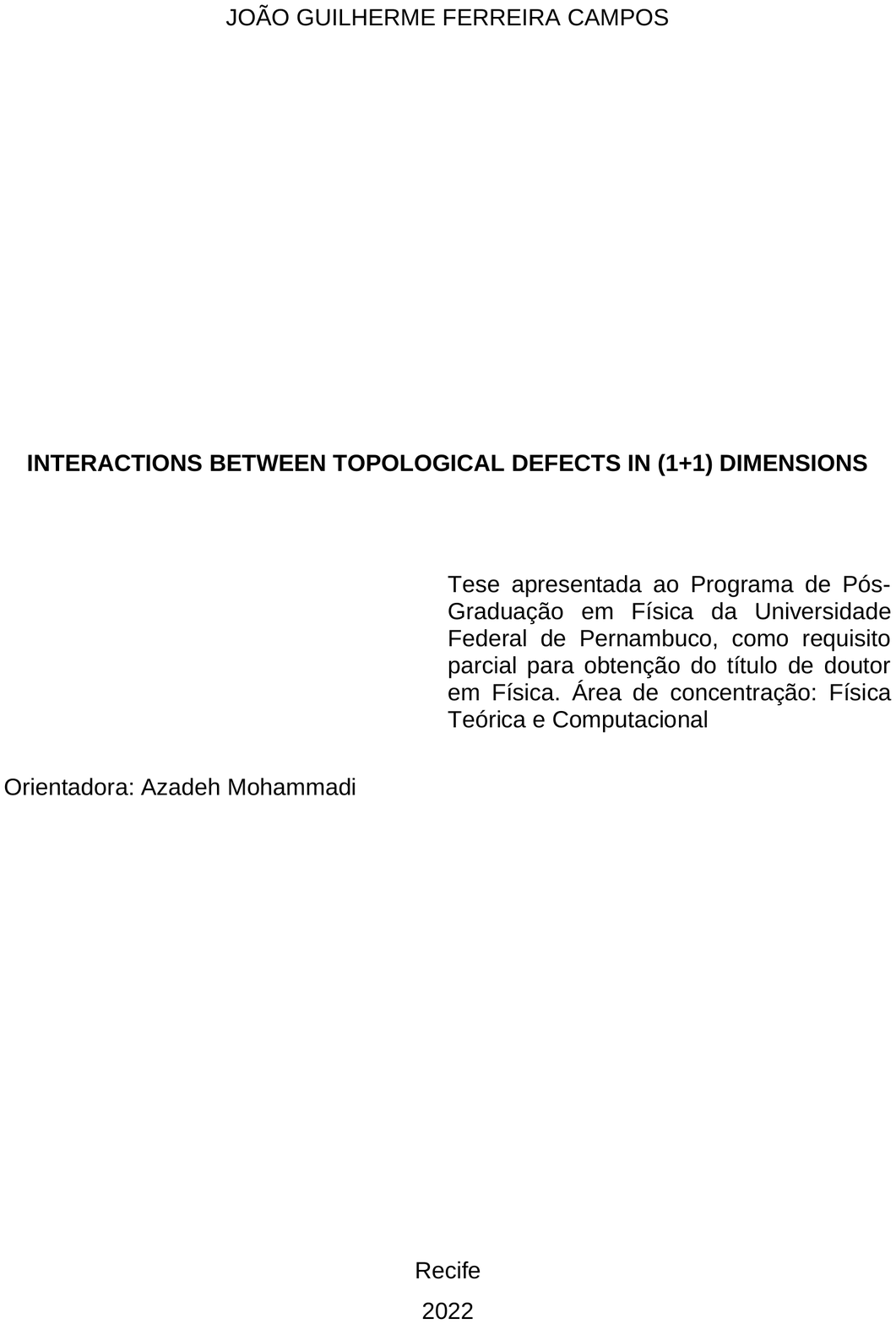}
%
%
\includepdf{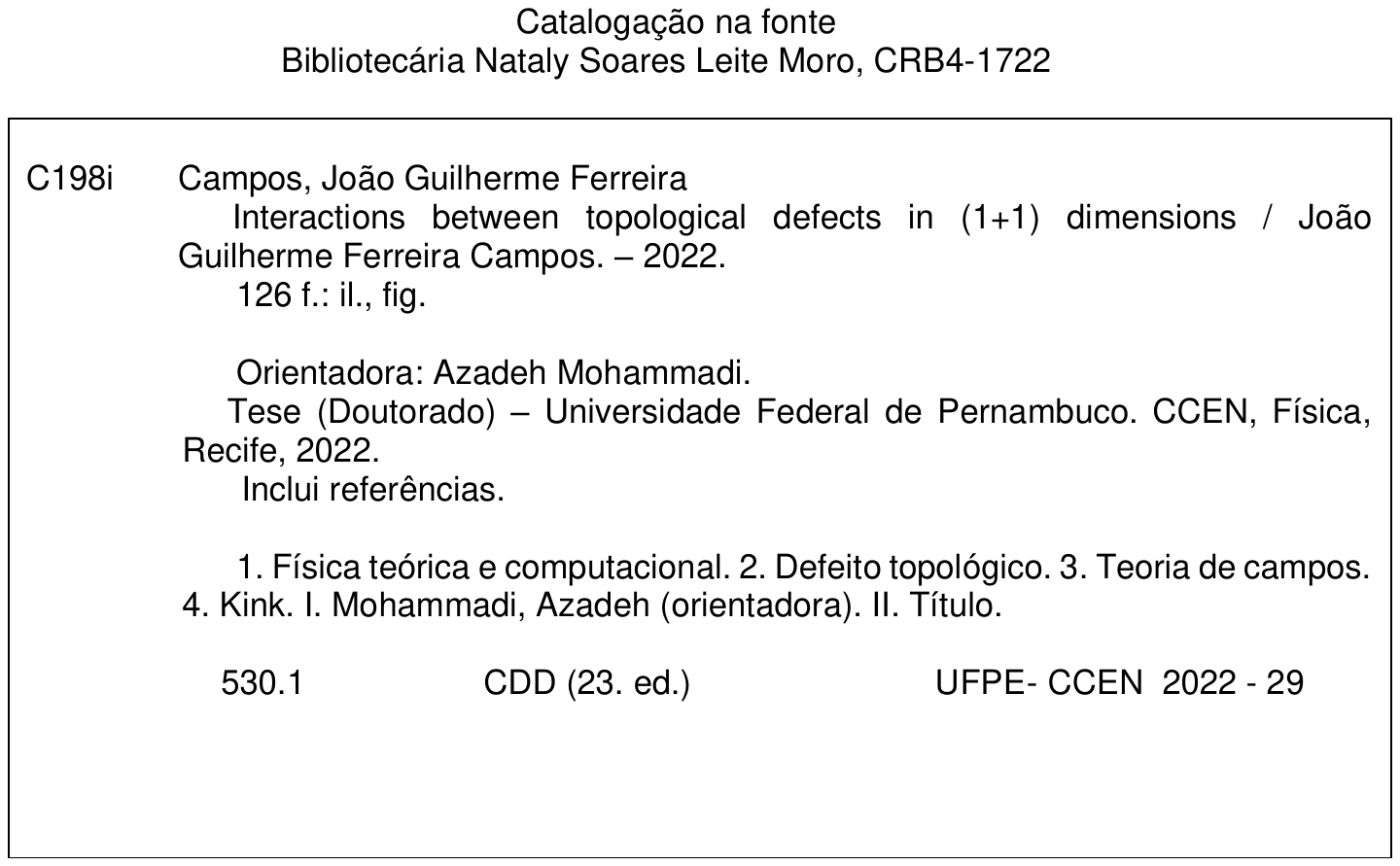}
\includepdf{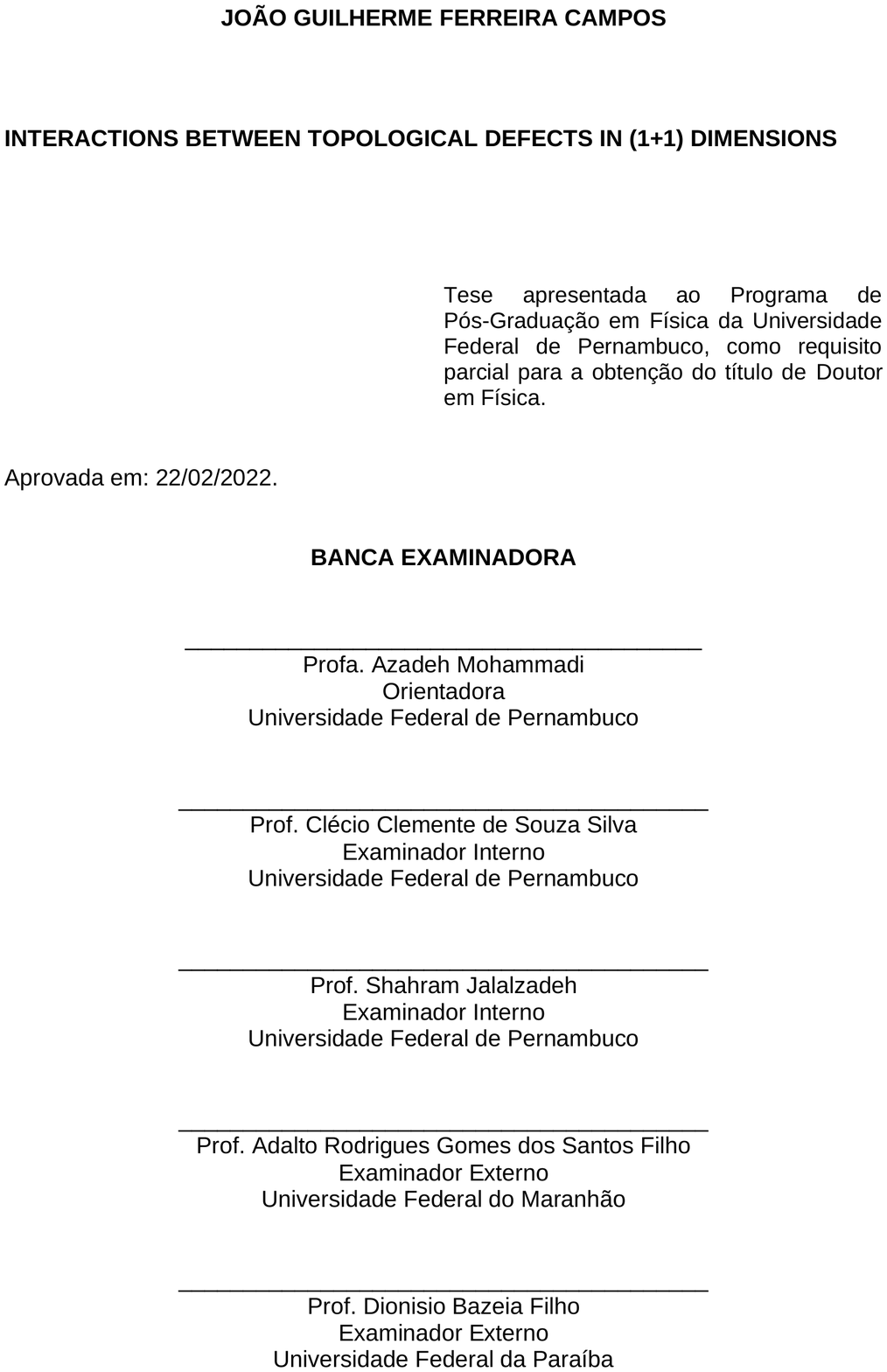}

\begin{dedicatoria}
   Aos meus pais, Itaciana Maria de Souza Ferreira e Hélio Araújo Cavalcanti Campos Filho e à minha irmã Mariana Lídia Ferreira Campos.
\end{dedicatoria}

\begin{agradecimentos}

Gostaria de agradecer ao departamento de física da UFPE, por ter proporcionado o ambiente para a realização desse trabalho e à professora Azadeh Mohammadi, pela leveza em sua condução durante o árduo processo que é o doutorado. Além disso, gostaria de agradecer aos meus colegas do grupo de física de altas energia pelas discussões e ao professor Mauro Copelli por permitir o uso do cluster de seu laboratório, sem o qual essa pesquisa não seria possível. Gostaria de agradecer também ao CNPq, à CAPES,  e à UFPE pelo apoio financeiro.

\end{agradecimentos}


\setlength{\absparsep}{18pt} 
\begin{abstract}

\noindent 
In this thesis, we study interactions between topological defects in two-dimensional spacetimes. These defects are called kinks. They are solutions of scalar field theories with localized energy which propagate without losing its shape. In order to understand the resonance phenomenon exhibited by those models, we built a toy model where the kink's vibrational mode turns into a quasinormal mode. This causes the suppression of resonance windows and, consequently, its fractal structure is lost. Considering a higher order polynomial as the scalar field potential, we find kinks with long-range tails, which decay as a power law. We developed a numerical method to correctly initialize this systems and applied it to a scalar field model containing kinks with long-range tails in both sides. After the collision, the kink-antikink pair is annihilated for velocities below an ultra-relativistic critical velocity without bion formation. We also investigated a collision between wobbling kinks of the double sine-Gordon model. When the kinks are already wobbling before colliding, there appears resonance windows with separation after a single bounce. On the second half of the thesis, we focused on fermion-kink interactions. We studied what happens when a fermion binds to a wobbling kink. The result is that the fermion escapes from the kink as radiation and at a constant rate. This occurs if the energy gap between the initial state and the continuum threshold is not too large. Lastly, we investigated the interaction of a fermion with a background scalar field with an impurity that preserves half of the Bogomol'nyi–Prasad–Sommerfield (BPS) property. We found an adiabatic evolution near the BPS regime, which means that the system is at a static BPS solution at every moment.
 
   \vspace{\onelineskip}
 
\end{abstract}

\begin{resumo}

\noindent
Nesta tese, estudamos interações entre defeitos topológicos em uma dimensão espacial e uma temporal. Esses defeitos são chamados de kinks. Eles são soluções de teorias de campos escalares que possuem energia localizada e se propagam sem perder sua forma. Para entender melhor o fenômeno de ressonância exibido nesses modelos, construímos um modelo simplificado onde o modo de vibração do kink se torna um modo quasinormal. Isso acarreta na supressão das janelas de ressonância e, consequentemente, na perda da estrutura fractal formada pelas mesmas. Já no caso em que o pontencial do campo escalar é um polinômio de ordem alta, a cauda do kink pode ser de longo alcance, por decair como uma lei de potência. Nós desenvolvemos um método numérico para inicializar esses sistemas corretamente e aplicamos a um modelo de kinks com caudas longas em ambos os lados. Após a colisão, o sistema se aniquila para velocidades abaixo de uma velocidade crítica ultra-relativística e não forma bions. Também investigamos uma colisão entre kinks vibrantes dentro do modelo sine-Gordon duplo. Um dos efeitos de excitar o modo de vibração antes da colisão é a formação de janelas de ressonância onde ocorre apenas um contato entre os kinks. Já na segunda metade da tese, focamos em interações entre kinks e férmions. Estudamos o que acontece com o férmion quando ele se liga a um kink que está vibrando. O férmion escapa do kink na forma de radiação a uma taxa contínua se a diferença de energia entre o estado inicial e o contínuo não for muito grande. Por último, nós investigamos a interação de um férmion na presença de um campo escalar com uma impureza que preserva metade da propriedade Bogomol'nyi–Prasad–Sommerfield (BPS) do sistema. O resultado desse processo é que perto do regime BPS a evolução do sistema é adiabática, pois sempre corresponde a uma solução BPS estática.

 \vspace{\onelineskip}
 
 \noindent
 \textbf{Palavras-chave}: defeito topológico; teoria de campos; kink; colisão; modos normais e quasinormais; férmion.
\end{resumo}

\pdfbookmark[0]{\contentsname}{toc}
\renewcommand{\contentsname}{\textbf{\normalsize CONTENTS}}
\tableofcontents*
\cleardoublepage

\textual

\chapter{INTRODUCTION}

\section{Solitons, solitary waves and topological defects}

We will start this thesis by defining our objects of study. These are solitons, solitary waves and topological defects. The definition of solitons can be found, for instance, in the excellent book by Rajaraman \cite{rajaraman1982solitons}. A soliton is a nontrivial solution of a wave field theory with three properties
\begin{itemize}
\item It has localized and finite energy;
\item It propagates without losing the shape;
\item When two or more solitons interact, they emerge with the same velocities as the initial ones. The only effect of the interaction is a displacement from their original trajectories.
\end{itemize}
The first two properties define a solitary wave. When the third property is included, the solution is called a soliton. 

In many cases, these solutions are called topological defects. We use this term when the wave properties and stability come from topological reasons. To understand that, let us first recall what topology is. Topology is the study of properties that are preserved when a geometric object is continuously deformed. A topological defect can be defined as a solution that cannot be continuously deformed to the trivial vacuum without reaching an infinite energy configuration. Therefore, the topology of the solution guarantees its stability. We will give a concrete example of how this works when we analyze the $\phi^4$ model.

\section{A little bit of history}

This section will discuss a few historical discoveries regarding solitons and solitary waves. We will give special attention to solutions in (1+1) dimensions, one time and one spatial dimension.  These types of solutions are the subject of this thesis.

\subsection{Integrable models}

A model is called integrable when it possesses an infinite number of conserved quantities. In general, models that exhibit solitons are integrable because the conserved quantities allows the initial configuration to be recovered after a collision, making it ellastic. The study of those systems was very important for the understanding of topological defects in field theories. For this reason, let us review a few key results. The first documented experimental observation of solitons dates back to the observation of John Scott Russell in 1834. He observed a mass of water that propagated in a narrow channel. He followed the mass for one or two miles and observed that it retained its shape during the whole path. In 1895, Kortweg and de Vries (KdV) found an equation describing waves' propagation in a narrow channel with shallow water. This equation was able to describe the phenomenon observed by John Scott Russel. An important method used to solve the KdV equation analytically is the inverse scattering method. It was discovered 1967 \cite{gardner1967method}. The history of the inverse scattering method is summarized, for instance in \cite{novikov1984theory}. This method is very powerful and allows one to find all the exact solutions of the KdV equation. With this formalism, it was proven that the KdV equation is integrable and exhibits solitons. A few years later, other equations such as the nonlinear Schr\"{o}dinger equations and the sine-Gordon model could also be solved by the inverse scattering method. This also means that they are integrable and exhibit solitons.

\subsection{Interactions between non-integrable kinks}

In integrable theories, we obtain soliton solutions. Their interaction is elastic and can be computed analytically. On the other hand, interactions between topological defects in non-integrable models are much more complex. Their investigation started later because it had to wait for the development of modern computers. These defects do not collide elastically due to the non-integrable behavior and are actually solitary waves. 

In (1+1) dimensions, topological defects are usually called kinks. In general, the kink solution possesses a reverse solution, the antikink. The sine-Gordon and $\phi^4$ models are examples of models containing kinks. The former is integrable, while the latter is non-integrable. One of the earliest investigations of interaction in the $\phi^4$ model is the work by Sugiyama \cite{sugiyama1979kink}. There, the author analyzed the collision between a kink and an antikink. He showed that after a critical velocity, the two reflect and, below that velocity, they form a long-lived bound state. Nowadays, this bound state is called a bion. Moreover, Sugiyama also proposed an effective model for the kink-antikink system. This effective model is called the collective coordinates method and consists of reducing the infinite degrees of freedom of a kink-antikink system to only two: the relative position and the vibration amplitude. However, the equations had a typo that was only corrected many years later \cite{kevrekidis2019four, takyi2016collective}. 

A few years after the work of Sugiyama, Campbell et al. published a triplet of seminal papers \cite{campbell1983resonance, peyrard1983kink, campbell1986kink}. There, the authors also observed that,  in a collision, the kink and the antikink reflect after a critical velocity. However, below the critical velocity, there are also regions where the kinks\footnote{Sometimes, we will abbreviate kink and antikink by the term kinks.} reflect, but only after multiple bounces. These regions alternate between regions where the kinks annihilate and are called resonance windows. They argued, using simplified models, that the resonances occur due to a resonant energy exchange mechanism between the translational and vibrational modes of the kink. This mechanism states that, in the first bounce, the translational energy of the kink is transformed into vibrational energy. If the translational energy is not large enough afterward, the kinks will not be able to separate and will collide again due to the mutual attraction. However, the vibrational energy can be converted back into translational energy at subsequent bounces if the timing is right. This would allow the kinks to separate. Moreover, the authors found that at the border of two-bounce resonance windows, there is a nested structure of three-bounce windows, while, at the border of three-bounce resonance windows, there is a nested structure of four-bounce resonance windows, and so on. 

A few years later, Anninos et al. showed numerically that the resonance windows exhibit a fractal structure \cite{anninos1991fractal}. This result was backed up by an analysis of the reduced Sugiyama model. At the time, the authors were not aware of the typo, and surprisingly, the reduced model worked exceptionally well. This analysis was considered as a quantitative explanation of the resonant energy exchange mechanism. Many years later, the typo was corrected \cite{takyi2016collective}, and the resemblance of the reduced model with the full one was lost. Even worse, the reduced model was singular for vanishing separation. More recently, Manton et al. showed that the collective coordinates model had a poor choice of coordinates \cite{manton2021kink, manton2021collective}. After finding a more suitable set of coordinates, they were able to reproduce the behavior of the full system very well. Thus, the resonant energy exchange mechanism finally acquired a correct quantitative explanation.

\section{Some relevant works}

There are many excellent books about solitons, where the essential theoretical aspects are discussed. To cite a few, see \cite{vilenkin2000cosmic, manton2004topological, vachaspati2006kinks, rajaraman1982solitons}. 

The area of interactions between kinks is a very active field of research with many theoretical developments. Let us briefly mention some relevant works in the field. The reduced Sugiyama model was discussed in detail in \cite{goodman2005kink, goodman2007chaotic}. The interaction of kinks with impurities was discussed in \cite{kivshar1991resonant, fei1992resonant, fei1992resonant2, malomed1992perturbative, goodman2004interaction}. Interestingly, when kinks interact with impurities, resonance windows also appear due to a vibrational mode of the impurity. 

The vibrational mode of a single kink was studied in \cite{manton1997kinks, barashenkov2009wobbling, oxtoby2009resonantly}. The important property is that the vibration amplitude decays slowly in time due to the coupling with the radiation modes. This property was also observed in oscillons \cite{romanczukiewicz2018oscillons}.

The double sine-Gordon model, which is a modification of the integrable sine-Gordon, was studied perturbatively in \cite{malomed1989dynamics, kivshar1989dynamics, kivshar1987radiative} using the inverse scattering method. This is an interesting modification of the sine-Gordon model, which is not integrable. A significant result is that there is still no energy and momentum transfer between the kinks in a collision in first-order perturbation theory.
Then, this model was studied numerically in \cite{campbell1986kink, gani1999kink, gani2018scattering, gani2019multi, simas2020solitary}. It contains kink solutions that are composed of two sine-Gordon subkinks. This means that the kinks have an inner structure. This property is also found in other models such as \cite{zhong2020collision, dorey2021resonance}. The double sine-Gordon model possesses resonance windows due to the presence of a vibrational mode. However, we will see in section \ref{chap_wob} that near the integrable regimes, the resonance windows are hidden. The work we developed on collisions between wobbling kinks in this model was published in \cite{campos2021wobbling}, and will be discussed in detail in the aforementioned section.

Polynomial potentials are very important due to their simplicity. They are usually called $\phi^n$, where $n$ is the polynomial order. Modification of the $\phi^4$ model have been discussed in \cite{bazeia2018scattering, gomes2018false, simas2016suppression, yan2020kink}, the $\phi^6$ model has been studied in \cite{dorey2011kink,gani2014kink}. Modification of the $\phi^6$ model has been studied in \cite{demirkaya2017kink} and a hybrid model between $\phi^4$ and $\phi^6$ has been studied in \cite{bazeia2019kink}. The $\phi^6$ model is an interesting exception to the resonant energy exchange mechanism because it exhibits resonance windows even though the kink does not have a vibrational mode. In \cite{dorey2011kink}, the authors showed that the resonance appears due to the exchange of the translational energy with a vibrational mode of the kink-antikink pair.

Scalar field theories $\phi^n$ containing higher-order polynomials with $n\geq 8$ were studied in \cite{gomes2012highly, khare2014successive, gani2015kink, bazeia2018analytical, belendryasova2019scattering, manton2019forces, khare2019family, christov2019long, christov2019kink, christov2021kink}. Interestingly, in these theories, the kinks may contain one or two long-range tails, which decay as a power-law. The simulation of these systems requires specialized methods, such as the method we developed in \cite{campos2021interaction}. It will be discussed in section \ref{chap_long}. The opposing limit, where the kink possesses short-range tails, was discussed in \cite{bazeia2021semi}.

An interesting question is whether the resonance phenomenon still exists when the vibrational mode of the kink is turned into a quasinormal mode. The reason is that it tests other regimes where the resonant energy exchange mechanism may be valid. A quasinormal mode is a solution of the kink stability equation that obeys purely outgoing boundary conditions. In some cases, including the one we are interested, it signals that the kink had a normal mode that became unstable and can slowly decay. The aforementioned question was studied in \cite{dorey2018resonant}. The authors found that the resonance structure is gradually lost as the decay rate of the quasinormal mode increases. We also studied this effect for a toy model in \cite{campos2020quasinormal}. It will be discussed in section \ref{chap2}. In other models, the potential approaches a vacuumless configuration, which has a local maximum but no local minimum. In this limit, the vibrational mode of the kink ceases to exist. Thus, the resonance windows also disappear \cite{bazeia2017sine, simas2017degenerate}.

One can also study a scalar field in the half-line subject to a boundary condition at the origin \cite{arthur2016breaking, dorey2017boundary, lima2019boundary}. In this case, a great variety of processes can occur, and in many cases, the system exhibits resonance. Another type of study with a great variety of outcomes is multi-kink collisions \cite{marjaneh2017multi, marjaneh2017high, marjaneh2018extreme}. In those works, the authors studied the interaction of three and four kinks colliding at a single point. One can also study scalar fields with two components \cite{halavanau2012resonance, alonso2018reflection, alonso2020non, ALONSOIZQUIERDO2021106011}. This system also exhibits a wide variety of behaviors due to the increasing complexity.

In scalar field models, the field is also interacting with radiation. Interestingly, this interaction may create kink-antikink pairs \cite{romanczukiewicz2006creation, dutta2008creating, romanczukiewicz2010oscillon}. Also radiation incident on a kink generates negative radiation pressure \cite{forgacs2008negative}. This could explain the absence of topological defects in a cosmological scale \cite{romanczukiewicz2017could}. 

Another interesting question is how the kink-antikink collision is modified if the vibrational mode is excited beforehand. This was studied in \cite{izquierdo2021scattering}. The authors argued that this investigation is relevant to understanding the resonance phenomenon because the multiple bounces in resonance can be viewed as an iteration of wobbling kink collisions.

Finally, we would like to mention a series of papers about scalar field models with an impurity that preserves half of the Bogomol'nyi–Prasad–Sommerfield (BPS) property \cite{manton2019iterated, adam2019spectral, adam2019solvable, adam2019phi, adam2019bps, adam2020kink}. Solutions with BPS property are static and stable, meaning that they have vanishing force. Interestingly, the BPS sector in these models has nontrivial dynamics. The authors were able to show a phenomenon called a spectral wall. A spectral wall appears when a vibrational mode of the BPS solution enters a continuum. An excited defect passing through a spectral wall experiences a force that may cause it to be reflected by the wall. This phenomenon can be isolated in those models because the inter-defect force vanishes.
Interestingly, this phenomenon was also observed in models with two scalar fields without the need for an impurity \cite{adam2021spectral}. Moreover, the authors found a model which displays another exception to the resonant energy exchange mechanism. They found that the resonant structure appears due to the exchange between the translational energy with the vibrational mode of a sphaleron, an unstable static solution \cite{adam2021sphalerons}.

\section{Applications of kink models}

Many systems in physics are described by fields. Moreover, if they are effectively one-dimensional in space and have degenerate lowest energy states, they probably will exhibit kinks. Examples of such models are deformation in polyacetylene \cite{su1979solitons} and in graphene \cite{yamaletdinov2017kinks}, the potential in Josephson junctions \cite{ustinov1998solitons}, vortices in superfluid Helium-3 \cite{volovik2003universe}, properties of DNA \cite{yakushevich2006nonlinear}, signals in optical fibers \cite{mollenauer2006solitons} and in surface displacement in a long channel of shallow liquid \cite{denardo1990observation}. Moreover, kinks can be used to describe domain walls in the direction perpendicular to the wall. Domain walls appear in magnetic materials \cite{kardar2007statistical, koyama2011observation, parkin2008magnetic}. Moreover, in some cosmological theories, the universe is described by a domain wall \cite{rubakov1983we} and the Big Bang by domain wall collisions \cite{khoury2001ekpyrotic}. 

Polyethylene is another effectively one-dimensional system that exhibit topological defects. These are called twistons. They are related to 180$^{\circ}$ degrees twists in the polymeric chain and can be described by coupled scalar fields \cite{bazeia1999topological}. We should also mention that one-dimensional solitons also appear in optics \cite{radhakrishnan1995bright, abdullaev2014optical}, but these solitons are not kinks. The same is true for systems described by the KdV equation, such as the soliton observed by John Scott Russell and deformations in plasma \cite{chen2016introduction}.

As one of the simplest topological defects available, kinks serve as a guide to study more complicated defects in higher dimensions. It is more difficult to study these structures in higher dimensions because they are much harder to simulate numerically. Hence, a better understanding of kinks can elucidate the behavior of higher dimensional topological defects. Examples of such defects are vortices in superconductors \cite{abrikosov2004nobel, auslaender2009mechanics, fetter2018theory}, knots in fluids \cite{kleckner2013creation}, instantons in quantum field theory \cite{polyakov2018nucleon, schaefer2002instanton, schneider2018discrete}, magnetic skyrmions \cite{fert2017magnetic} and skyrmions in nuclear physics \cite{skyrme1962unified}. Other topological defects that are relevant theoretically are cosmic strings \cite{hindmarsh2016new} and monopoles \cite{t1974magnetic, csaki2018kaluza}. All these structures appear in either in real physical systems or effective models. There are some contexts where they have not been observed yet, but it is possible that they will be observed in the future.

\chapter{INTRODUCTION TO KINK-ANTIKINK COLLISIONS}

In this main section, we will discuss two prototypical models that exhibit kink solutions. The models are the sine-Gordon, which is integrable, and the $\phi^4$, non-integrable. Then we will discuss collisions in these models and discuss how they differ. None of the results in this main section is original. They can be found in \cite{rajaraman1982solitons, manton2004topological, vachaspati2006kinks} for instance. However, the section serves as an introduction to the more complicated cases discussed in the following main sections.

\section{The $\phi^4$ model}

\begin{figure}[tbp]
\centering
  \caption{(left) Potential and (right) kink solution for the $\phi^4$ model.}
  \includegraphics[width=0.9\columnwidth]{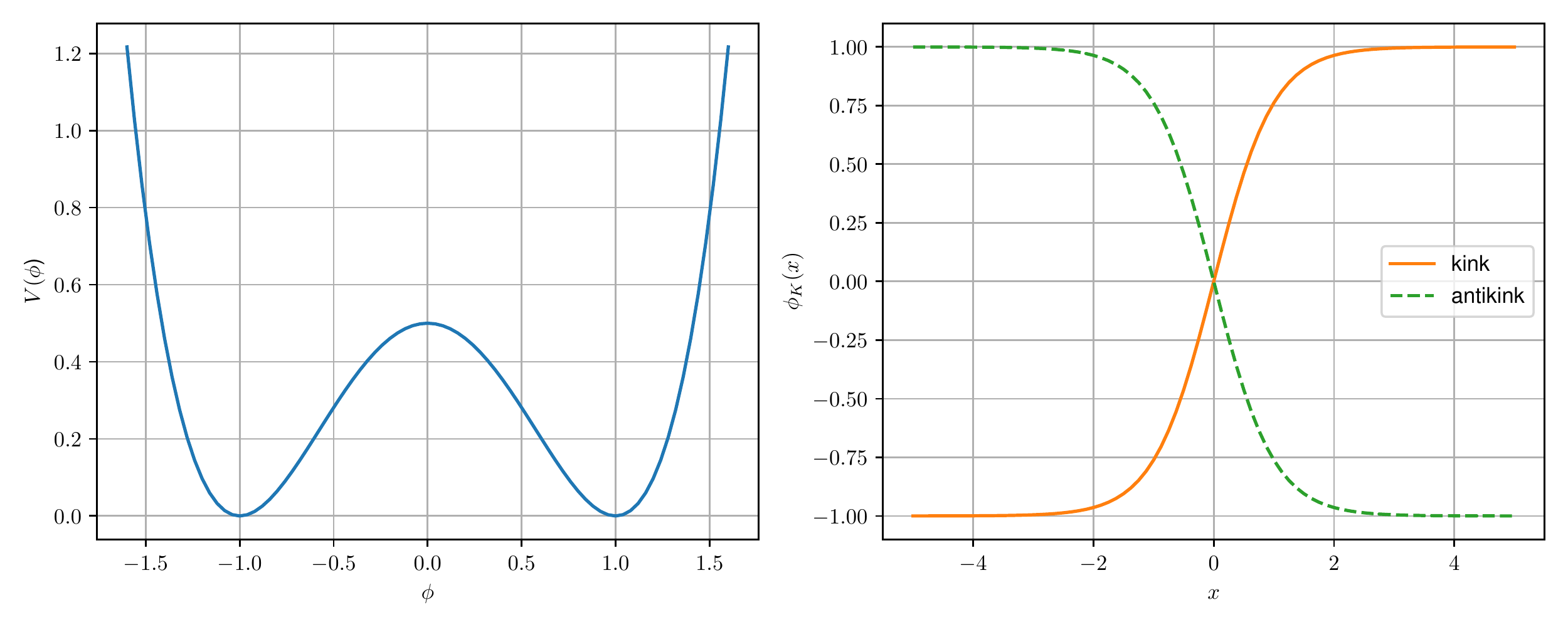}
  \caption*{Source: The author (2022).}
  \label{fig_p4}
\end{figure}

One of the simplest examples that exhibits kinks is the $\phi^4$. It consists of a single scalar field theory in $(1+1)$ dimensions described by the following Lagrangian
\begin{equation}
\mathcal{L}=\frac{1}{2}\partial_\mu\phi\partial^\mu\phi-V(\phi).
\end{equation}
We are considering the metric $g_{00}=-g_{11}=1$ and $g_{01}=g_{10}=0$. This Lagrangian describes relativistic theories in units where $c=\hbar=1$ as well as non-relativistic wave equations in units where the wave velocity has been set to unit. It leads to the following equation of motion
\begin{equation}
\label{kinkeom}
\partial_t^2\phi-\partial_x^2\phi+V^\prime(\phi)=0.
\end{equation}
So far, this is very general. However, for the $\phi^4$ model we set $V(\phi)=\frac{\lambda}{2}(\phi^2-\eta^2)^2$. The constant $\lambda$ has units of mass squared, and the constant $\eta$ is dimensionless. They will depend on the system that we are modeling. However, we can remove these constants by working with adimensional units. For the $\phi^4$, this is done by setting $\phi\to\eta\phi$ and $x^\mu\to x^\mu/(\sqrt{\lambda}\eta)$. The rescaled system possesses the following potential $V(\phi)=\frac{1}{2}(\phi^2-1)^2$. In the following sections, we will always work with adimensional units. The rescaled potential contains two symmetric vacua at $\phi=\pm 1$ as shown in Fig.~\ref{fig_p4}. In this case the $Z_2$, $\phi \to -\phi$, symmetry is spontaneously broken. 

The equation of motion for the $\phi^4$ model becomes
\begin{equation}
\label{kinkeom2}
\partial_t^2\phi-\partial_x^2\phi+2\phi(\phi^2-1)=0.
\end{equation}
We see that the trivial vacua $\phi=\pm 1$ are solutions to this equation. We are looking for static solutions which are nontrivial with finite energy. First, let us take a look at the system's total energy. It can be easily found from the Lagrangian. The result is
\begin{equation}
\label{eq_lagr}
E=\int dx\left[\frac{1}{2}(\partial_t\phi)^2+\frac{1}{2}(\partial_x\phi)^2+V(\phi)\right].
\end{equation}
From this expression, we see that a static solution with finite energy must reach a vacuum configuration for the boundaries $x\to\pm\infty$, that is, $V(\phi(x\to\pm\infty))=0$. This is necessary for the integral to converge. For the $\phi^4$ model, the vacuum is degenerate. It can be either $\phi=\pm 1$. If the vacua at the limits $x\to\pm\infty$ are identical, the static solution is the trivial vacuum, which has zero total energy. If the vacuum at the limits $x\to\pm\infty$ is different, we may have nontrivial, static, and finite energy solutions\footnote{A solution is nontrivial when it depends on the position $x$. These three properties only appear if the topology is also nontrivial, which means that the value of the field at $\pm\infty$ are different. It is also possible to have nontrivial solution (depending on $x$) with trivial topology, but these are not static. This can be proven using the static equation of motion, which will be derived below.}.

Now, we will derive these solutions with nontrivial boundary conditions. The static kink solutions can be found from eq.~(\ref{kinkeom}) after setting the time derivatives to vanish. We find
\begin{equation}
\label{eq12_static}
\partial_x^2\phi=V^\prime(\phi).
\end{equation}
Multiplying this equation by $\partial_x\phi$ on both sides and integrating, we obtain
\begin{equation}
\label{eq1_BPS}
\frac{1}{2}(\partial_x\phi)^2=V(\phi).
\end{equation}
The constant of integration is zero due to the boundary conditions obtained by the finite energy condition. Substituting the $\phi^4$ potential, this equation is easily integrated and has two nontrivial solutions. The first one is the kink
\begin{equation}
\phi_{K}(x)=\tanh(x-x_0).
\end{equation}
It interpolates between $-1$ and $1$. The integration constant $x_0$ is the center of the kink. The freedom to set the center at any point is related to the translation symmetry of the system. There is also the antikink solution, which interpolates between $1$ and $-1$
\begin{equation}
\phi_{\bar{K}}(x)=-\tanh(x-x_0).
\end{equation}
These solutions are shown in Fig.~\ref{fig_p4}. It is easy to see that the energy density is $\text{sech}^4(x-x_0)$ and total energy is $E=4/3$. As the model obeys Lorentz symmetry, to find the kink solutions in a boosted frame, one only needs to Lorentz transform the coordinates. The boosted kink solution is
\begin{equation}
\phi_K(x,t)=\tanh(\gamma(x-x_0\pm vt)),
\end{equation}
where $v$ is the kink's velocity and $\gamma\equiv1/\sqrt{1-v^2}$.

Due to the boundary conditions, the kink and antikink solutions are stable and cannot decay into the trivial vacuum. To prove this property, let us define a topological current
\begin{equation}
j^\mu=\frac{1}{2}\epsilon^{\mu\nu}\partial_\nu\phi.
\end{equation}
This current is conserved by construction due to the presence of the antisymmetric tensor $\epsilon^{\mu\nu}$. The conserved topological charge of this current, or winding number, is
\begin{equation}
Q=\int dx\,j^0=\frac{1}{2}[\phi(x=+\infty)-\phi(x=-\infty)].
\end{equation}
Therefore, the kink cannot decay to the trivial vacuum because they have different winding numbers. Of course, if we consider finite boundaries, the winding number is not conserved anymore.

Now, we would like to study what happens to the field when the kink is perturbed. The stability equation for fluctuations around the kink solution can be obtained by considering the behavior of solutions close to the kink solution. We write $\phi=\phi_K+\psi\cos(\omega t)$, where $\psi$ is a small perturbation. Substituting in the equations of motion, it yields
\begin{equation}
H\psi\equiv-\partial_x^2\psi+V^{\prime\prime}(\phi_K)\psi=\omega^2\psi.
\end{equation}
It can be shown that the eigenvalues of this equation are non-negative and, therefore the kink is stable. For the $\phi^4$ model, it reads
\begin{equation}
-\partial_x^2\psi+(6\tanh^2(x)-2)\psi=\omega^2\psi,
\end{equation}
which is a P\"{o}schl-Teller potential with discrete solutions $\omega^2=0,3$. The existence of a mode with $\omega=0$, called zero mode, is expected due to the translation invariance of the system. The mode with $\omega=\sqrt{3}$ is the vibrational mode of the kink, also known as shape mode. Its solution is given by $\psi_S=\text{sech}(x)\tanh(x)$.

Another way to obtain the equation for a static kink is to complete squares in the energy functional. This is done as follows
\begin{align}
E&=\int dx\left[\frac{1}{2}(\partial_t\phi)^2+\frac{1}{2}(\partial_x\phi)^2+V(\phi)\right]\\
&=\int dx\left[\frac{1}{2}\left(\partial_t\phi\right)^2+\frac{1}{2}\left(\partial_x\phi\mp\sqrt{2V(\phi)}\right)^2\pm \sqrt{2V(\phi)}\partial_x\phi\right]\\
&\geq\int^{\phi(\infty)}_{\phi(-\infty)} d\phi^\prime\sqrt{2V(\phi^\prime)}
\end{align}
Therefore, the energy is minimized, for fixed boundary conditions, when the equality is satisfied. Consequently, the action will be extremized. The equality occurs for static fields obeying
\begin{equation}
\partial_x\phi=\pm\sqrt{2V(\phi)}.
\label{eq12_BPS2}
\end{equation}
This equation is know as the Bogomol'nyi–Prasad–Sommerfield (BPS) \cite{prasad1975exact, bogomol1976stability}. It is equivalent to eq.~(\ref{eq1_BPS}). This construction from the energy functional is important because it is also used to simplify the equations of motion in more complicated models.

\section{Sine-Gordon model}

\begin{figure}[tbp]
\centering
  \caption{(left) Potential and (right) kink solutions for the sine-Gordon model.}
  \includegraphics[width=0.9\columnwidth]{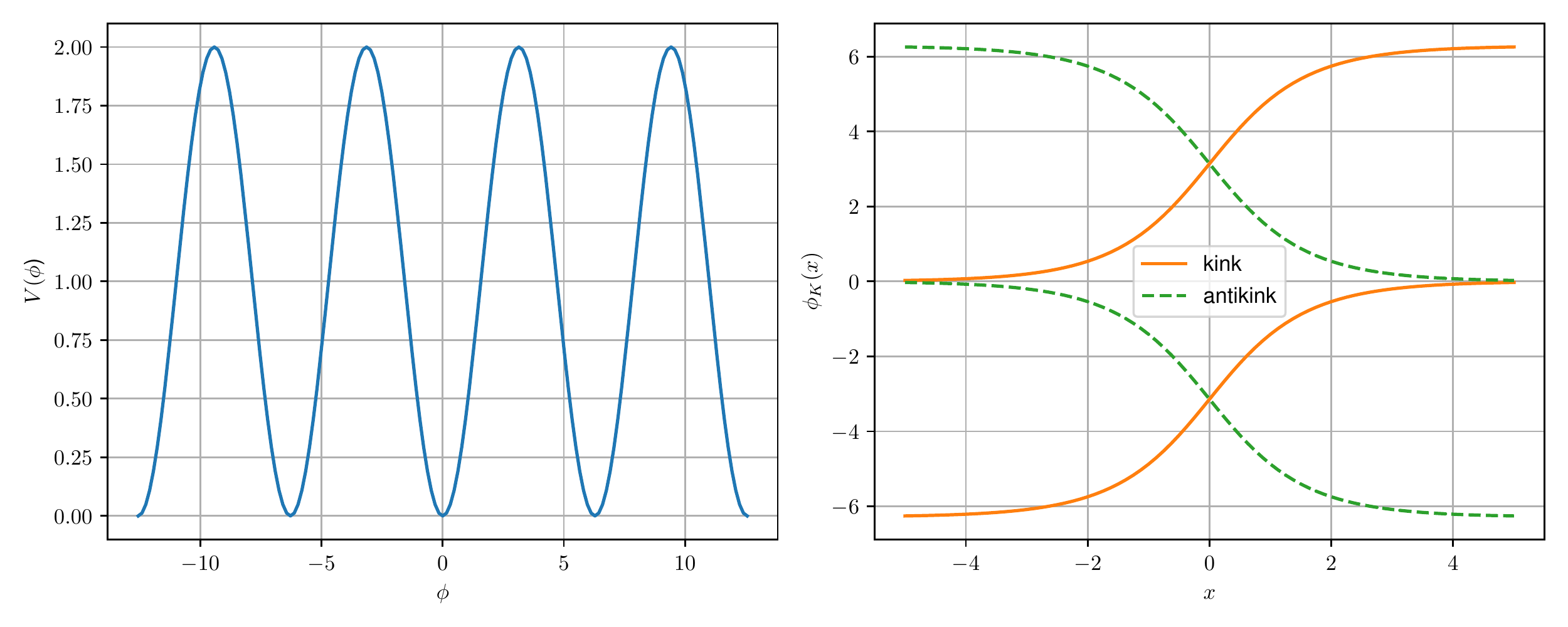}
  \caption*{Source: The author (2022).}  
  \label{fig_SG}
\end{figure}

The most familiar example of a soliton is the sine-Gordon model. It is a soliton because the model is integrable with infinite conserved quantities. Hence, the solitons emerge after a collision with the same velocity as before and the only effect of the interaction is a time delay in the original trajectory. The Lagrangian for this model is given by
\begin{equation}
\mathcal{L}=\frac{1}{2}(\partial_\mu\phi)^2+(\cos\phi-1)
\end{equation}
The BPS equation can be easily solved. It yields the following kink solution
\begin{equation}
\phi_K(x)=4\tan^{-1}\left(e^{x-x_0}\right).
\end{equation}
The solution and potential are shown in Fig.~\ref{fig_SG}. Due to the periodicity of the potential, one can obtain a different kink solution for any integer $n$ by adding a constant $2\pi n$ to the field. Moreover, one can obtain an antikink solution reversing the field. It is simple to show that these solutions have energy density $4\text{sech}^2(x-x_0)$ and total energy $E=8$.

\begin{figure}[tbp]
\centering
  \caption{Evolution in spacetime of (left) a breather solution and (right) a kink-antikink collision. The model considered is the sine-Gordon and we set $v=0.2$.}
  \includegraphics[width=0.9\columnwidth]{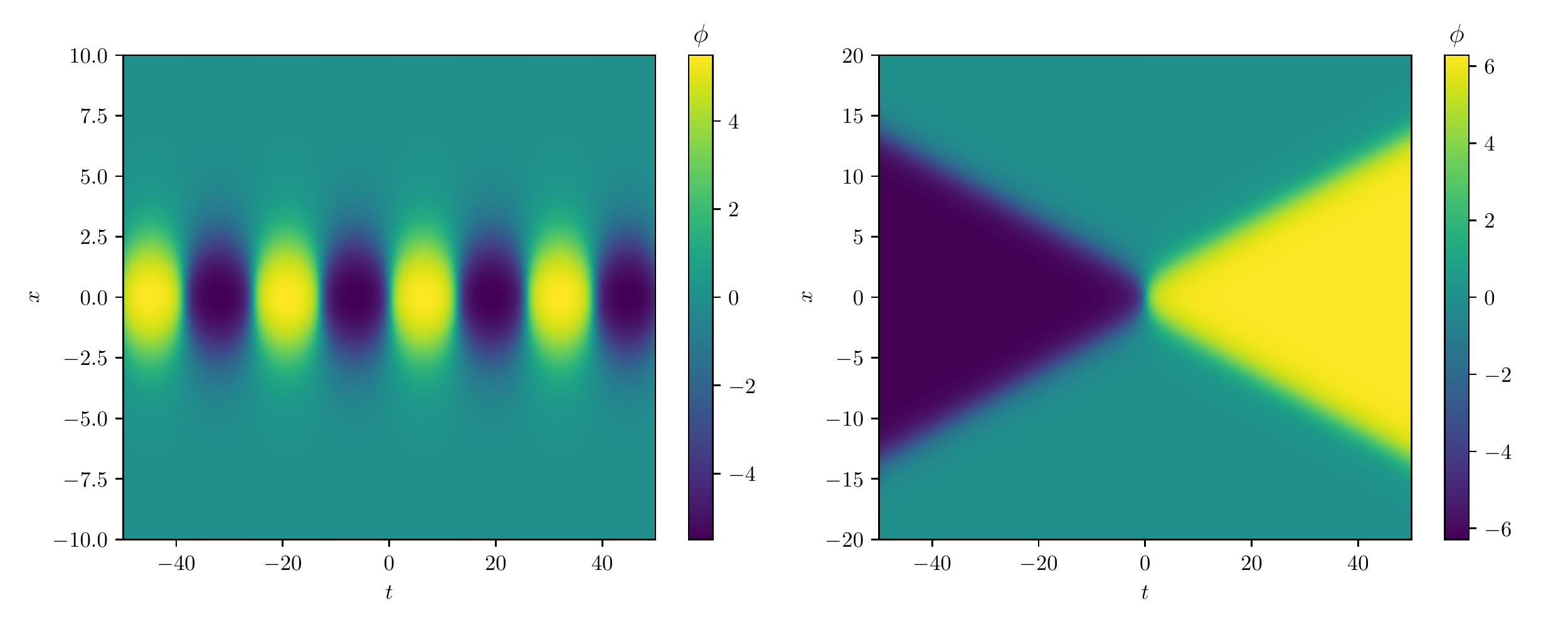}
        \caption*{Source: The author (2022).}
       \label{fig_fieldSG}
\end{figure}

Due to the integrability of the model, there are exact solutions for soliton collision and oscillating bound states, which are called breathers. They can be obtained using specialized methods like the B\"{a}cklund transformation or the inverse scattering method. We will not derive these solutions but merely state them. A breather is described by
\begin{equation}
\phi_v(x,t)=4\tan^{-1}\left[\frac{\sin(\gamma vt)}{v\cosh(\gamma x)}\right],
\end{equation}
where $v$ is a parameter related to the frequency of the breather and $\gamma=1/\sqrt{1-v^2}$. The evolution of this field configuration in spacetime is shown in Fig.~\ref{fig_fieldSG}.

Another example is the kink-antikink collision
\begin{equation}
\phi_{KA}(x,t)=4\tan^{-1}\left(\frac{\sinh(\gamma vt)}{v\cosh(\gamma x)}\right).
\end{equation}
One can show that the asymptotic behavior of this solution for $t\to-\infty$ describes widely separated kink and antikink approaching each other. Similarly, for $t\to\infty$, it consists of widely separated kink and antikink moving away from each other with the same velocity but with a time delay. This is what one expects from an interaction between solitons. The evolution of this field configuration in spacetime is shown in Fig.~\ref{fig_fieldSG}.

\section{kink-antikink collision in the $\phi^4$ model}

Now, let us consider a kink-antikink configuration. In this case, there is no analytical expression, but we can find good approximations. One widely used approximation is the additive ansatz
\begin{equation}
\phi(x,t)=\phi_K(x+x_0)+\phi_{\bar{K}}(x-x_0)-1.
\end{equation}
If $x_0$ is large, the kinks are far away from each other and can be linearly superposed. This field configuration approximates a kink and an antikink located at $-x_0$ and $+x_0$, respectively. To understand the need for the constant $-1$, we can examine the ansatz for $x>0$. In this region, we are far from the kink and, thus, the first term is approximately $1$. Therefore, it is canceled by the third term, and only the second term is left, as desired.

The kink-antikink configuration does not solve the BPS equation (\ref{eq12_BPS2}). In this model, the BPS equation is equivalent to the static equation of motion (\ref{eq12_static}) for configurations with finite energy. Therefore, the kink-antikink configuration is not static. In fact, there is an attractive force between the two. One can show that, for a large separation $R=2x_0$, the force is given by $F=32e^{-2R}$. 

\begin{figure}[tbp]
\centering
  \caption{Evolution in spacetime of kink-antikink collisions in the $\phi^4$ model. The different scenarios are (left) annihilation, (center) resonance and (right) reflection.}
  \includegraphics[width=0.9\columnwidth]{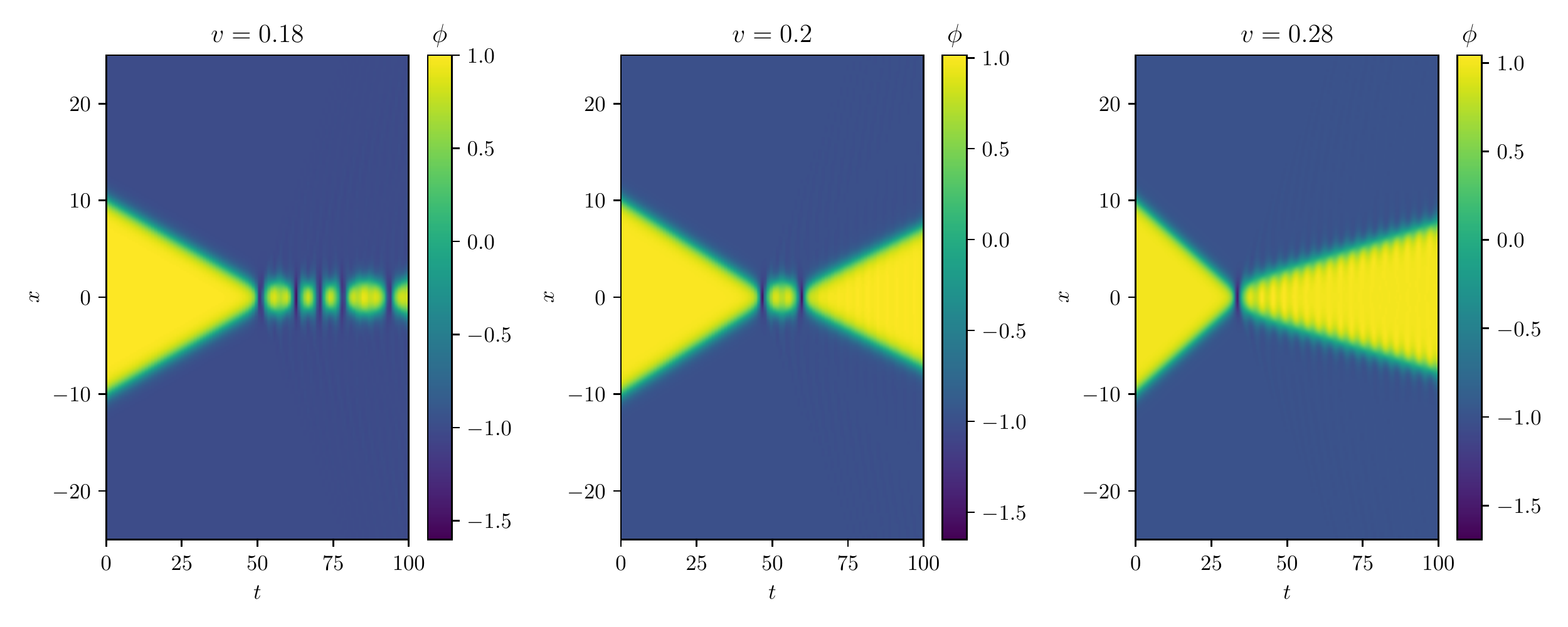}
        \caption*{Source: The author (2022).}
       \label{fig_fieldp4}
\end{figure}

The additive ansatz can be used to study the interaction of boosted kink-antikink configurations. In this case, the initial condition for the field equations is given by the following expression
\begin{equation}
\phi(x,t)=\phi_K(\gamma(x+x_0-vt))+\phi_{\bar{K}}(\gamma(x-x_0+vt))-1.
\end{equation}
The evolution of this solution can be obtained by integrating the equations of motion numerically. By virtue of the non-integrability of the model, there are more possible outcomes to this interaction. If the velocity is small, the kinks annihilate, as shown in the left plot of Fig.~\ref{fig_fieldp4}. Before decaying into the trivial vacuum, they form an oscillating bound state known as bion. This bound state is long-lived but slowly decays by emitting radiation at each bounce. Another possibility is shown in the right plot of Fig.~\ref{fig_fieldp4}. In this case, the kinks bounce, reflect and separate to infinity. Surprisingly, there is a third possibility called resonance. In this scenario, the kinks also separate, but only after multiple bounces. A two-bounce scenario is shown in the center plot of Fig.~\ref{fig_fieldp4}.

\begin{figure}[tbp]
\centering
  \caption{Final velocity $v_f$ as a function of the initial velocity $v_i$ for the $\phi^4$ model. The color code indicates the number of bounces before separation.}
  \includegraphics[width=0.9\columnwidth]{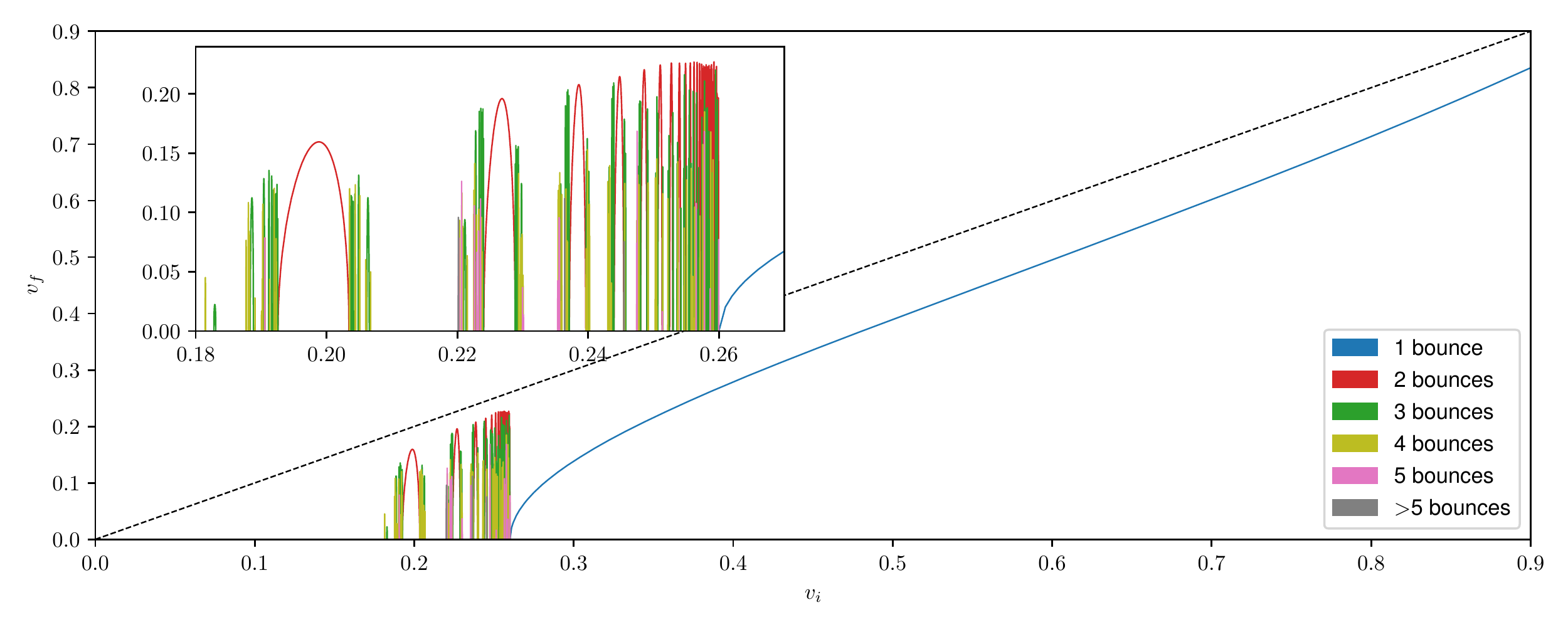}
        \caption*{Source: Reproduction made by the author of results obtained in \cite{izquierdo2021scattering}.}
       \label{fig_resonancep4}
\end{figure}

The regions of initial velocity $v_i$ where each behavior occurs are summarized in Fig.~\ref{fig_resonancep4}. In this figure, we show the final velocity $v_f$ as a function of the initial one. The final velocity is computed as follows. While we integrate the equations of motions, we keep track of the kink's position. In order to do so, we look for the position where $\phi=\phi_K(x=0)$. Then, we wait until the distance between the kink and the origin is larger than $\Delta x\in[10.0,15.0]$. If this separation is never reached, we set $v_f$ to zero, meaning that there is annihilation for the corresponding value of $v_i$. If the kinks separate, we start using the position measurements to compute the final velocity. We perform a linear regression using 80 position values separated by $\Delta t=0.5$. Therefore, we obtain a nonzero $v_f$ when there is separation after one or multiple bounces. The region where there is only a single bounce is drawn in blue. The other colors correspond to multiple bounces. The regions where multiple bounces occur are called resonance windows. The term $n$-bounce window can be used to describe separation after $n$ bounces. Interestingly, there is a sequence of $(n+1)$-bounce windows at the edge of $n$-bounce windows, forming a fractal structure.   

The resonance phenomenon was explained in \cite{campbell1983resonance}. The mechanism behind its appearance is called the resonant energy exchange mechanism. According to this mechanism, there is an exchange between the translational and vibrational energy of the kink at each bounce. At the first bounce, part of the translational energy is lost and is stored as vibrational energy. Due to the mutual attraction, if the translational energy is too small afterward, the kinks will be forced to bounce one more time. The vibrational energy can be converted back into the translational energy at the subsequent bounces, and the kinks may separate. Whether it happens or not should be sensitive to the state of vibrational mode at the moment of the subsequent bounces.

\begin{figure}[tbp]
\centering
  \caption{Evolution of the field at the origin for the first six two-bounce resonance windows. We are considering the $\phi^4$ model.}
  \includegraphics[width=0.9\columnwidth]{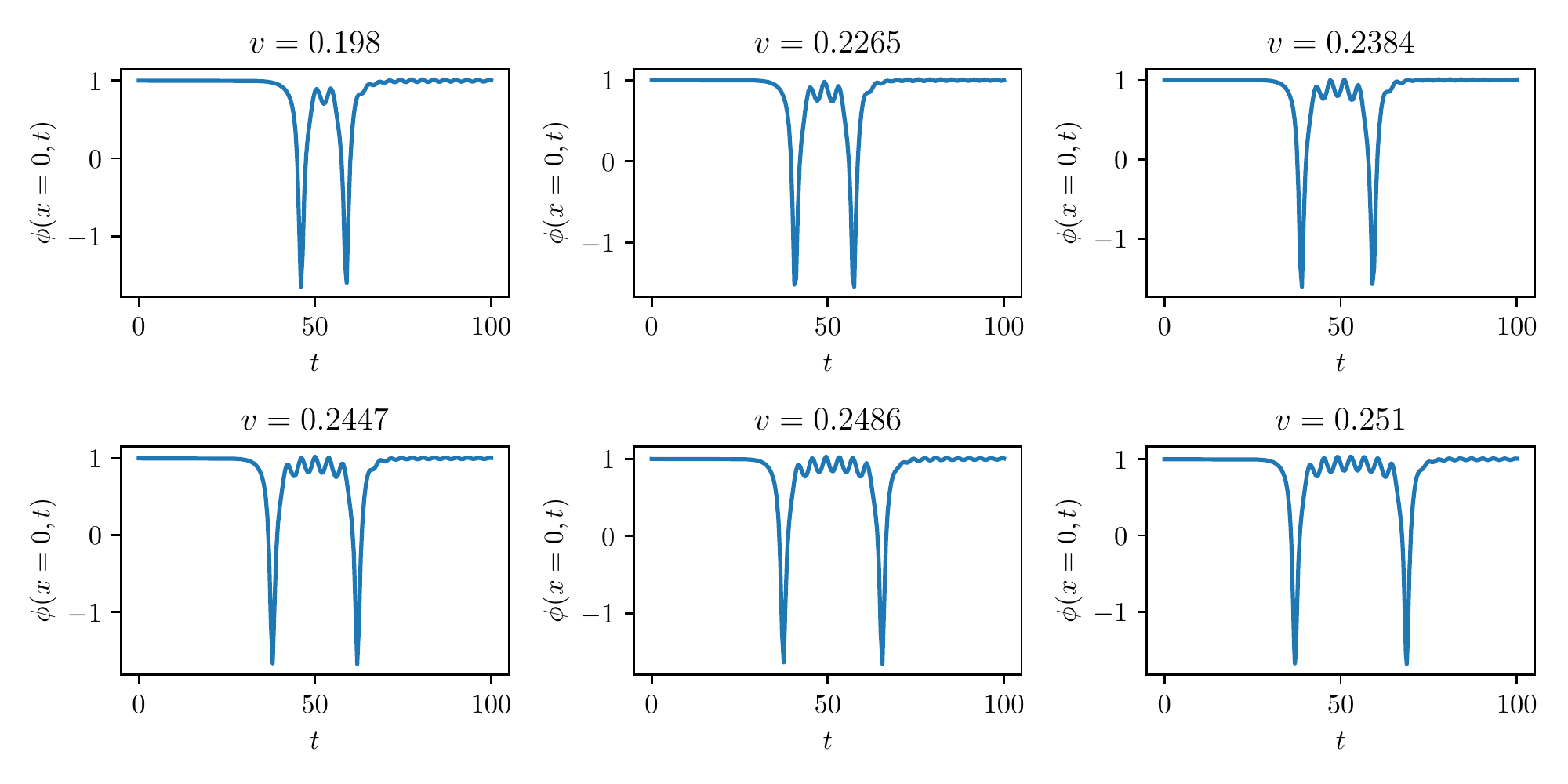}
        \caption*{Source: Reproduction made by the author of results obtained in \cite{campbell1983resonance}.}
       \label{fig_centerp4}
\end{figure}

The authors plotted the field's evolution at the origin of several two-bounce resonance windows to prove this conjecture. The result for the first six two-bounce windows is shown in Fig.~\ref{fig_centerp4}. The figure shows that the field starts at $1$ because the kinks are separated. Then, there is a sharp inverted peak corresponding to the first bounce. After that, there is an important oscillation of the field at the center and, after a second bounce, the kinks separate. If it is hard to interpret the figure, try following the field at the origin in the center plot of Fig.~\ref{fig_fieldp4}. The crucial point is that, as we move from one resonance window to the next, the number of oscillations between bounces increases by one. If $\omega$ is the angular frequency of oscillation and $T_n$ is the time between bounces for the $n$-th two-bounce resonance window, this can be written mathematically as
\begin{equation}
\omega T_n=\delta+2\pi n,
\end{equation}
where $\delta$ is a constant. This means that whether the system separates depends on the state of the vibration appearing between the sharp inverted peaks in Fig.~\ref{fig_centerp4}. To confirm the resonant energy exchange mechanism, the authors had to show that this oscillation at the center corresponds to the kink's vibration measured at the origin. Comparing the numerical value of $\omega$ with the theoretical value of the vibrational mode's frequency $\omega=\sqrt{3}$, they found an excellent agreement, confirming the exchange mechanism.

One important consequence of the resonant energy exchange mechanism is that resonance windows are absent when the kink does not have a vibrational mode. As explained in the introduction, there are exceptions to this rule, but it is generally valid. In the three subsequent main sections, we will study kink-antikink collisions in three different models. These consist of a toy model with quasinormal modes, kinks with double long-range tails, and double sine-Gordon, respectively. The second half of the thesis will be dedicated to studying fermion-kink systems.

\chapter{KINK INTERACTIONS WITH QUASINORMAL MODES}
\label{chap2}

\section{Overview}

The work described in this main section resulted in the following publication \cite{campos2020quasinormal}. We were interested in the effect of turning the normal mode of the kink into a quasinormal mode (QNM). QNMs may appear in models where the stability equation has a potential with a ``volcano" shape. A few examples are described in \cite{gomes2012highly, bazeia2017sine, campos2021interaction} and in section \ref{chap_long}. This modification is important, taking into consideration the role of the normal mode of the kink on the resonant energy exchange mechanism. This mechanism states that the existence of resonance windows in kink-antikink collisions is caused by the energy exchange between the translational and vibrational modes of the kink. The vibrational mode is excited at the first bounce and, if it is turned into a QNM, its energy will start to leak, and only the remaining part will be converted back to translational energy at the successive collisions. Therefore, it is expected that in this case, the resonance structure will be gradually lost. 

The first work to investigate this effect was \cite{dorey2018resonant}. There, the authors constructed a modified $\phi^4$ model exhibiting a QNM. They showed, numerically, that when the decay rate of the QNM is increased, the resonance windows gradually disappear. This result was corroborated by a numerical analysis of the reduced Sugiyama model. As discussed before, this model has a typo, but the authors argued that it would be considered as a phenomenological model. 

Our contribution to this problem was to build a model that is similar to \cite{dorey2018resonant} but simple enough to allow an analytical solution of the kink profile and the QNM properties, such as shape, decay rate, and frequency. Hence, we were able to corroborate the results in \cite{dorey2018resonant}, explicitly showing how the QNM mode appears in an analytical model. In the next section, we describe the construction of the model and its kink solution.

It is important to mention that most models possess an infinite discrete set of QNMs. For the Schr\"{o}dinger and Klein-Gordon equation the set of QNMs can be found analytically for a few cases, as discussed in \cite{boonserm2011quasi}. However, according to the resonant energy exchange mechanism, the most important excitation mode in kink-antikink collisions is the shape mode. Therefore, we are actually interested in turning the shape mode into a QNM and this is the only QNM that we will focus on.

\section{Model}

\begin{figure}[tbp]
\centering
  \caption{Profile of the linearized potential in (a) the normal mode case and (b) the QNM case. Parameters are: (a) $\gamma=3.0$, $\lambda=1.0$ and $L=1.047$. (b) $\gamma=3.0$, $\Delta\gamma=2.0$, $\lambda=1.0$, $L=1.045$ and $\Delta L=1.365$.}
  \includegraphics[width=0.76\columnwidth]{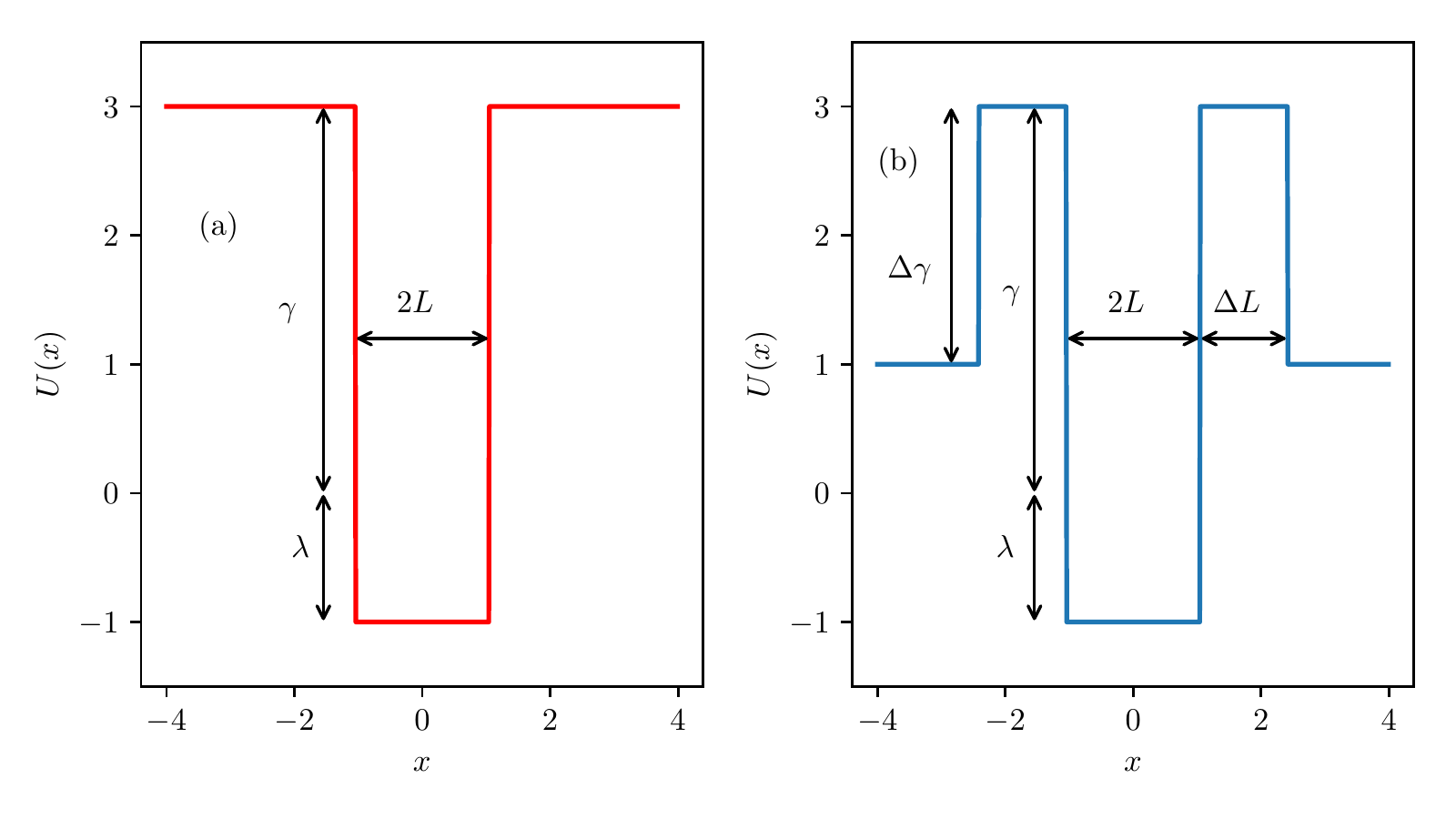}
        \caption*{Source: Results obtained by the author in \cite{campos2020quasinormal}.}
       \label{fig_linearized_potential}
\end{figure}

We start with the following scalar field Lagragian
\begin{equation}
\mathcal{L}=\frac{1}{2}\partial_\mu\phi\partial^\mu\phi-V(\phi),
\end{equation}
where we assume that the system has $Z_2$ symmetry with $V(-\phi)=V(\phi)$ and has two symmetric vacua at $\pm\phi_0$. Therefore, the system possesses a kink solution, $\phi_K$, which interpolates between the vacua. Then, considering perturbations around the kink solutions $\phi=\phi_K+\eta(x)e^{i\omega t}$, one arrives at the stability equation
\begin{equation}
\label{eq_stability}
H\eta\equiv\left[-\frac{d^2}{dx^2}+\left.\frac{d^2V}{d\phi^2}\right|_{\phi=\phi_K}\right]\eta=\omega^2\eta.
\end{equation}
This is a Schr\"{o}dinger-like equation with a linearized potential defined as $U(x)\equiv V^{\prime\prime}(\phi_K(x))$. The desired form of the linearized potential is in the form of a square well, which is one of the simplest quantum mechanics potentials with the correct qualitative behavior and analytical solutions. It reads
\begin{equation}
U(x)=\begin{cases}-\lambda,& 0\leq x<L,\\
\gamma, & x>L,
\end{cases}
\end{equation}
for positive $x$ and has even symmetry. In the definition above, $\lambda$, $\gamma$, and $L$ are positive constants. A typical profile of the linearized potential is shown in Fig.~\ref{fig_linearized_potential}(a). The linearized potential can be easily modified to exhibit QNMs as follows
\begin{equation}
\label{eq3_UQNM}
U(x)=\begin{cases}-\lambda,& 0\leq x<L,\\
\gamma, & L<x<L+\Delta L,\\
\gamma-\Delta\gamma, & x>L+\Delta L,
\end{cases}
\end{equation}
with new constants $\Delta L$ and $\Delta\gamma$. 
Again, we only write the potential for positive $x$ and consider even symmetry. This modified potential consists of a square well with two barriers and is shown in Fig.~\ref{fig_linearized_potential}(b). Clearly, if the square well potential has bound states with sufficiently high energy, it will become a QNM as $\Delta\gamma$ increases.

\begin{figure}[tbp]
\centering
  \caption{Profile of (a) the potential and (b) kink for the normal mode case ($\Delta\gamma=0.0$). (c), (d) the same for the QNM case with $\Delta\gamma=2.0$. The dashed lines mark the limiting points in the definition $V(\phi)$ and $\phi_K(x)$. Other parameters are: (a), (b) $\gamma=3.0$. (c), (d) $\gamma=3.0$ and $\epsilon=0.05$.}
  \includegraphics[width=0.76\columnwidth]{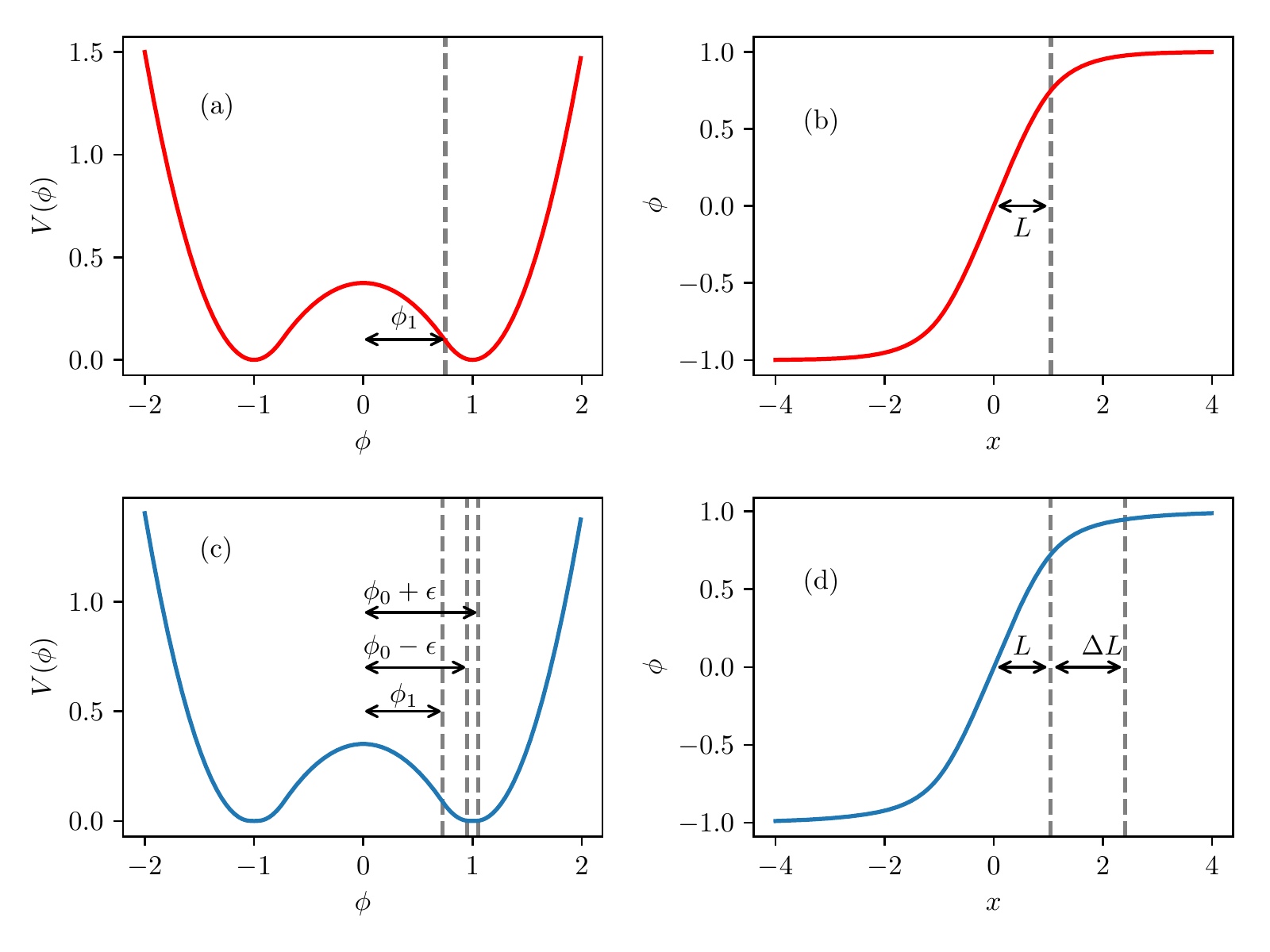}
       \caption*{Source: Results obtained by the author in \cite{campos2020quasinormal}.}       
       \label{fig_potential_kink}
\end{figure}

The task now is to construct a potential $V(\phi)$ that leads to the desired linearized one. This is accomplished by a piecewise-defined function with maximum powers of two in $\phi$, namely
\begin{equation}
\label{eq_Vsquarewell}
V(\phi)=\begin{cases}
\frac{\lambda}{2}(-\phi^2+A^2),& 0\leq \phi<\phi_1,\\
\frac{\gamma}{2}(\phi-\phi_0)^2, & \phi>\phi_1,
\end{cases}
\end{equation}
and
\begin{equation}
\label{eq_VQNM}
V(\phi)=\begin{cases}
\frac{\lambda}{2}(-\phi^2+A^2),& 0\leq \phi<\phi_1,\\
\frac{\gamma}{2}(\phi-\phi_0+\delta)^2+B^2, & \phi_1<\phi<\phi_0-\epsilon\\
\frac{\gamma-\Delta\gamma}{2}(\phi-\phi_0)^2, & \phi_0-\epsilon<\phi<\phi_0+\epsilon\\
\frac{\gamma}{2}(\phi-\phi_0-\delta)^2+B^2, & \phi>\phi_0+\epsilon
\end{cases}
\end{equation} 
for the normal modes and QNM cases, respectively. In the above definitions, all quantities are positive constants, except $\phi$. The expression for the potential above is written only for positive $\phi$, and we assume even symmetry. In total, there are five free parameters in the normal mode case and nine in the QNM case. To ensure that the system is well behaved, we choose the constants such that $V(\phi)$ and its derivative are continuous. This results in the following set of equations
\begin{align}
\lambda(-\phi_1^2+A^2)&=\gamma(\phi_1-\phi_0)^2,\\
-\lambda\phi_1&=\gamma(\phi_1-\phi_0),
\end{align}
for the normal mode case and
\begin{align}
\lambda(-\phi_1^2+A^2)&=\gamma(\phi_1-\phi_0+\delta)^2+B^2,\\
-\lambda\phi_1&=\gamma(\phi_1-\phi_0+\delta),\\
\gamma(\delta-\epsilon)^2+B^2&=(\gamma-\Delta\gamma)\epsilon^2\\
\gamma(\delta-\epsilon)&=(\Delta\gamma-\gamma)\epsilon,
\end{align}
for the QNM one. These equations reduce the number of free parameters in the normal mode case to three and the QNM one to five. We can also rescale the variables according to $x\to x/\sqrt{\lambda}$ and $\phi\to\phi_0\phi$. This is equivalent to set $\phi_0=1$ and $\lambda=1$, after redefining all constants. Therefore, in the end, the number of free parameters in eq.~(\ref{eq_Vsquarewell}) and (\ref{eq_VQNM}) are one and three, respectively. After fixing some constant in eq.~(\ref{eq_Vsquarewell}), the others can be found analytically using the continuity equations. After fixing three constants in eq.~(\ref{eq_VQNM}), the other can be found either numerically or analytically, depending on which constants one chooses to fix. Typical profiles of the potential are shown in Fig.~\ref{fig_potential_kink}.

From the definition of the potential it is easy to compute the kink profile using the Bogomol'nyi–Prasad–Sommerfield (BPS) equation. First, for the normal mode case, we find
\begin{equation}
\label{eq_phik_squarewell}
\phi_K(x)=\begin{cases}
A\sin(x),& 0\leq x<L,\\
1-(1-\phi_1)e^{-\sqrt{\gamma}(x-L)}, & x>L,
\end{cases}
\end{equation}
for positive $x$ with odd symmetry. The constants $\phi_1$ and $L$ are related by the expression $\phi_1=A\sin(L)$. For the QNM case, we find
\begin{equation}
\label{eq_phik_QNM}
\phi_K(x)=\begin{cases}
A\sin(x),& 0\leq x<L,\\
1-\delta+B\sinh(C+\sqrt{\gamma}(x-L)),& L<x<L+\Delta L,\\
1-\epsilon e^{-\sqrt{\gamma-\Delta\gamma}[x-(L+\Delta L)]}, & x>L+\Delta L,
\end{cases}
\end{equation}
for positive $x$ and also with odd symmetry. In this expression, we defined $C\equiv\ln(B/D)$, where
\begin{equation}
D\equiv \sqrt{B^2+(1-\delta-\phi_1)^2}+1-\delta-\phi_1.
\end{equation}
The constants $L$ and $L+\Delta L$ are the points where $\phi_K=\phi_1$ and $\phi_K=1-\epsilon$, respectively, and, thus, are not independent from the other constants. The first relation is the same obtained in the normal mode case and the second one reads
\begin{align}
\epsilon=\delta-B\text{sinh}(C+\sqrt{\gamma}\Delta L).
\end{align} 
The kinks' profiles are shown Fig.~\ref{fig_potential_kink} in the right panels.

This model is somewhat unusual, because it is piecewise defined. However, we found that it is very well-behaved with all the expected properties of a non-integrable model containing kinks. In the next section we will show how to solve the stability equation for the kink configuration.

\section{Stability equation}

The stability equation for the normal mode case is the same as the Schr\"{o}dinger equation with the square well potential, which has well-known analytical solutions \cite{griffiths2018introduction}. The eigenvalues of the even and odd eigenfunctions are the solutions of the following transcendental equations
\begin{align}
\label{eq_SW_even}
k=&p\tan(pL),\\
-k=&p\cot(pL),
\label{eq_SW_odd}
\end{align}
respectively, where $k\equiv\sqrt{\gamma-\omega^2}$ and $p\equiv\sqrt{\omega^2+1}$. The analytical expressions of the eigenfunctions are listed in section \ref{sec2_SW}. It can be shown that, due to the continuity relation of the potential, this system always possesses a zero-mode solution, where $\omega=0$. As $\gamma$ increases, a tower of alternating odd and even solutions starts to appear. In all simulations performed here, we choose the parameters such that the stability equation admits only the zero-mode and one normal mode. In this case, the system exhibits resonance windows, as expected by the resonant energy exchange mechanism (see below).

\begin{figure}[tbp]
\centering
  \caption{Relation between the parameters $\Delta L$ and $\Delta \gamma$ for several values of $L$. We fix $\gamma=4.0$.}
  \includegraphics[width=0.6\columnwidth]{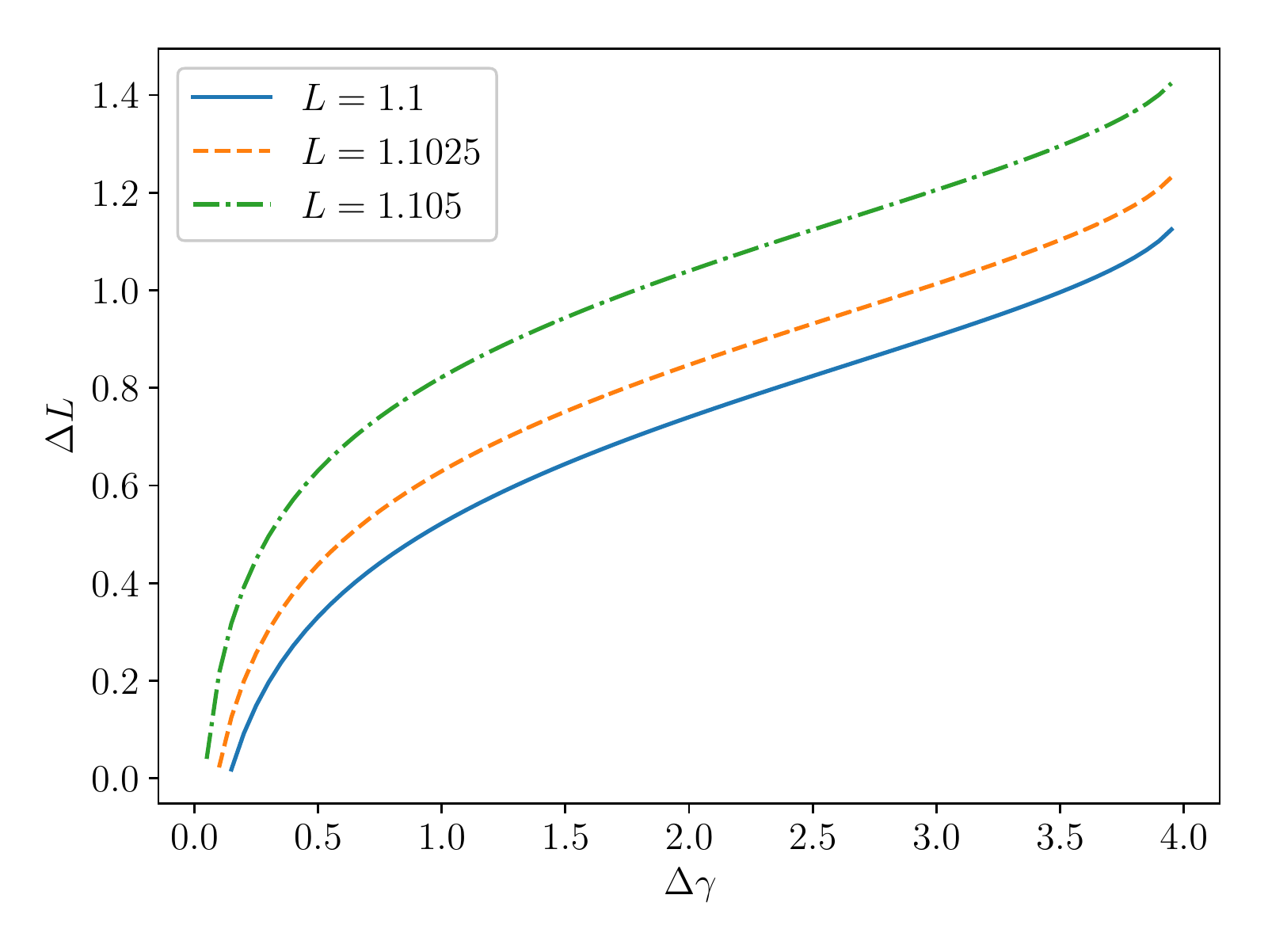}
       \caption*{Source: Results obtained by the author in \cite{campos2020quasinormal}.} 
       \label{fig_DgamDL}
\end{figure}

Now, let us focus on the QNM case. We are interested in comparing the stability potential in (\ref{eq_VQNM}) as we vary the parameters. In particular, we want to isolate the effect of $\Delta \gamma$ and $\Delta L$ variations while keeping the other parameters of the linearized potential fixed. Fixing the values of $L$ and $\gamma$, there remains only one degree of freedom in the potential due to the continuity conditions. In this case, $\Delta\gamma$ and $\Delta L$ are related as shown in Fig.~\ref{fig_DgamDL}. Notice that the point $\Delta\gamma=\Delta L=0$ is a solution to the continuity equations for the potential (\ref{eq_VQNM}) only if $L$ is precisely the solution of the continuity equations for the potential (\ref{eq_Vsquarewell}) with the same $\gamma$.

For the QNM case, we are interested in the scattering states and, naturally, in the QNMs. The scattering states of eq.~(\ref{eq_stability}) with the linearized potential described by eq.~(\ref{eq_VQNM}) can be written as
\begin{equation}
\label{eq_scattering}
\eta(x)=\begin{cases}
F_1e^{imx}+F_2e^{-imx},& x<-L-\Delta L,\\
G_1e^{-kx}+G_2e^{kx},& -L-\Delta L<x<-L,\\
H_1\sin(px)+H_2\cos(px),& -L<x<L,\\
I_1e^{-kx}+I_2e^{kx},& L<x<L+\Delta L,\\
Je^{imx},& x>L+\Delta L,
\end{cases}
\end{equation}
where $p$ and $k$ are defined as before and $m\equiv\sqrt{\omega^2+\Delta\gamma-\gamma}$. Imposing continuity of $\eta(x)$ and its derivative, we find a linear system, which can be solved straightforwardly. The solution of the linear system is given in section \ref{sec2_TR}. Using these expressions, we can obtain the transmission and reflection coefficients. They are defined, respectively, as $T=|\frac{J}{F_1}|^2$ and $R=|\frac{F_2}{F_1}|^2$ and are given by
\begin{align}
\label{eq_transmission}
\begin{split}
T=&4\bigg| e^{2k\Delta L}(1+i\alpha_{km}^-)[\cos(2pL)+\alpha_{kp}^-\sin(2pL)]-2i\alpha_{km}^+\alpha_{kp}^+\sin(2pL)\\
&+e^{-2k\Delta L}(1-i\alpha_{km}^-)[\cos(2pL)-\alpha_{kp}^-\sin(2pL)]\bigg|^{-2},
\end{split}\\
\begin{split}
R=&\frac{T}{4}\bigg\{ e^{2k\Delta L}\alpha_{km}^+[\cos(2pL)+\alpha_{kp}^-\sin(2pL)]-2\alpha_{km}^-\alpha_{kp}^+\sin(2pL)\\
&-e^{-2k\Delta L}\alpha_{km}^+[\cos(2pL)-\alpha_{kp}^-\sin(2pL)]\bigg\}^{2},
\end{split}
\end{align}
where
\begin{equation}
\label{eq2_alpha}
\alpha_{km}^{\pm}\equiv\frac{k^2\pm m^2}{2mk},\quad\alpha_{kp}^{\pm}\equiv\frac{k^2\pm p^2}{2pk}.
\end{equation}

From these analytical expressions, it is now easy to find the QNMs. They are defined as solutions with purely outgoing boundary conditions. This means that $F_1$ is set to zero in eq.~(\ref{eq_scattering}). In turn, this causes $T$ to diverge because the denominator vanishes. Similarly, it is possible to find the bound states by setting $im\to-q$ in eq.~(\ref{eq_scattering}), where $q\equiv\sqrt{\gamma-\Delta\gamma-\omega^2}$, and, again, setting $F_1$ to zero.

\begin{figure}[tbp]
\centering
  \caption{Transmission and reflection coefficients as a function of $\omega^2$ for several values of $L$ and $\Delta\gamma$. We fix $\gamma=4.0$. The grey vertical lines are the QNM or bound states frequencies.}
  \includegraphics[width=0.8\columnwidth]{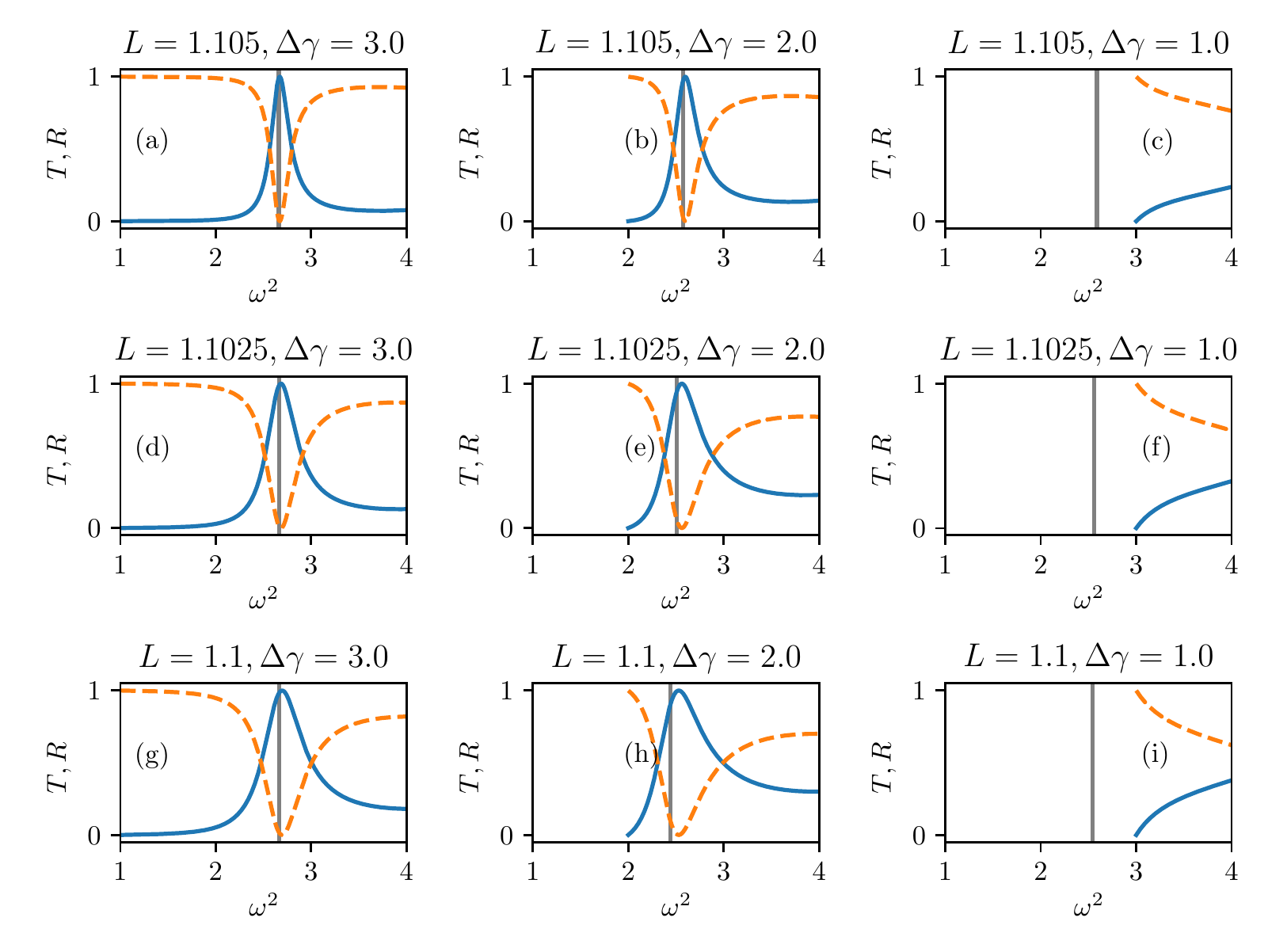}
       \caption*{Source: Results obtained by the author in \cite{campos2020quasinormal}.}        
       \label{fig_tun}
\end{figure}

In Fig.~\ref{fig_tun} we plot $T$ and $R$ as a function of the eigenvalue $\omega^2$. We clearly see that if $\Delta\gamma$ is large enough, the system exhibits transmission resonances, where the coefficient reaches the maximum value, one. These are linked to the presence of the QNMs and occur near the QNM frequency.

\begin{figure}[tbp]
\centering
  \caption{Analytical values of the frequency and decay rate of the QNMs and bound states as a function of $\Delta\gamma$. We consider several values of $L$ and fix $\gamma=4.0$.}
  \includegraphics[width=0.76\columnwidth]{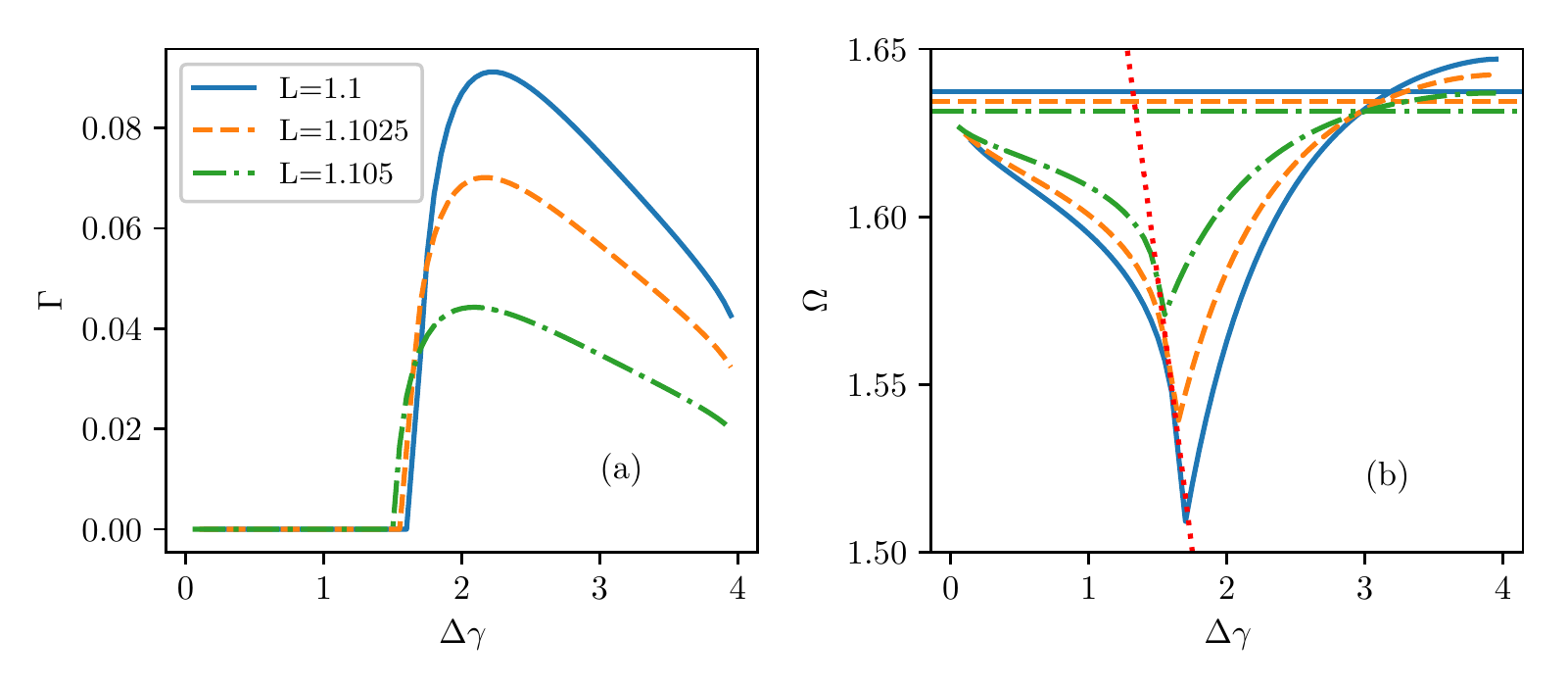}
       \caption*{Source: Results obtained by the author in \cite{campos2020quasinormal}.}
       \label{fig_decay}
\end{figure}

The transcendental equation obtained by setting the denominator of eq.~(\ref{eq_transmission}) equal to zero can be solved numerically for the normal mode and QNM. Writing the real part of $\omega$ as $\Omega$ and the imaginary part as $\Gamma$, we find the curves shown in Fig.~\ref{fig_decay}. For small $\Delta\gamma$ only the bound state equation has a solution and, of course, this solution possesses the decay rate $\Gamma=0$. At a critical value of $\Delta\gamma$, the bound state disappears, and the QNM appears, which is indicated by a nonzero decay rate $\Gamma$. These two regimes are separated in Fig.~\ref{fig_decay}(b) by the red dotted curve. At this curve, both bound state solution and quasinormal mode solutions are identical. This can only happen if $im=-q$, which implies that $m=q=0$ or $\omega=\sqrt{\gamma-\Delta\gamma}$. In the next section, we will see how the appearance of the QNM affects the kink-antikink interaction.

\section{Collision}

In this section, we will consider kink-antikink collision for our toy model. The collision is defined by the following initial conditions
\begin{align}
\phi(x,0)=&\phi_K(\gamma(x+x_0))-\phi_K(\gamma(x-x_0))-1,\\
\dot{\phi}(x,0)=&-\gamma v_i\left[\phi_K^\prime (\gamma(x+x_0))+\phi_K^\prime(\gamma(x-x_0))\right],
\end{align}
where $v_i$ is the initial velocity of both kinks and $\gamma\equiv 1/\sqrt{1-v_i^2}$. The kinks are initially separated symmetrically with respect to the origin by a distance equal to $2x_0$. After setting the initial conditions, the Euler-Lagrange equations of motion are integrated according to the numerical method described in section \ref{sec2_numeric}.

Considering first the case with $\epsilon=0$ or, equivalently, $\Delta \gamma=0$, we choose the free parameter $\gamma=2.667$. In this case, the system possesses only one vibrational mode in addition to the zero-mode. In agreement with the resonant energy exchange mechanism, the system exhibits a nested structure of resonance windows. 

\begin{figure}[tbp]
\centering
  \caption{Spacetime evolution of the scalar field for several initial velocities showing: (a) annihilation at $v=0.200$, (b) two-bounce resonance at $v=0.228$ and (c) reflection at $v=0.340$. Other parameters are $\gamma=2.667$ and $\epsilon=0$.}
  \begin{subfigure}[b]{0.3\textwidth}
    \centering
    \includegraphics[width=\textwidth]{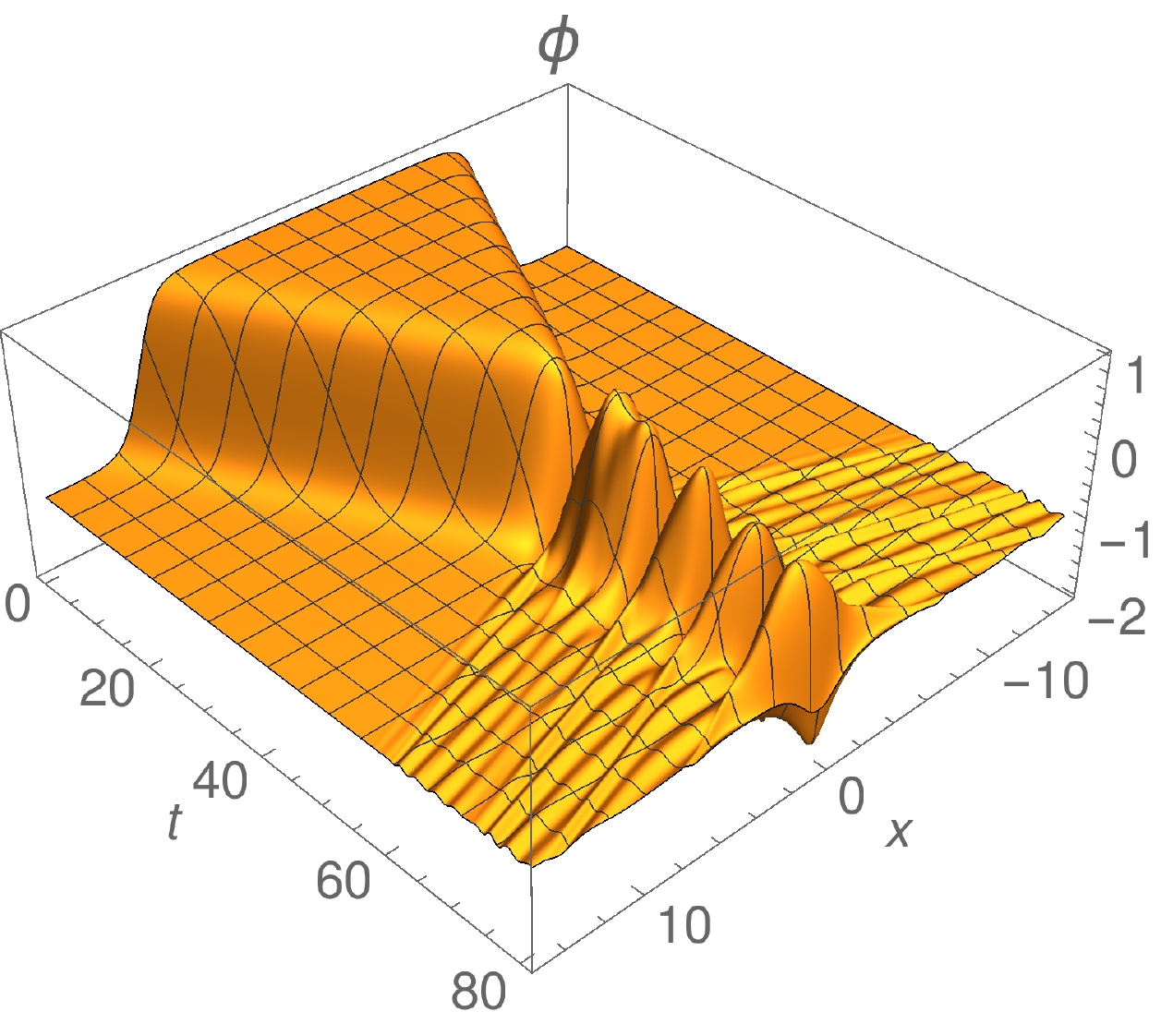}
    \caption{}
    \label{fig_v200}
  \end{subfigure}
  \begin{subfigure}[b]{0.3\textwidth}
    \centering
    \includegraphics[width=\textwidth]{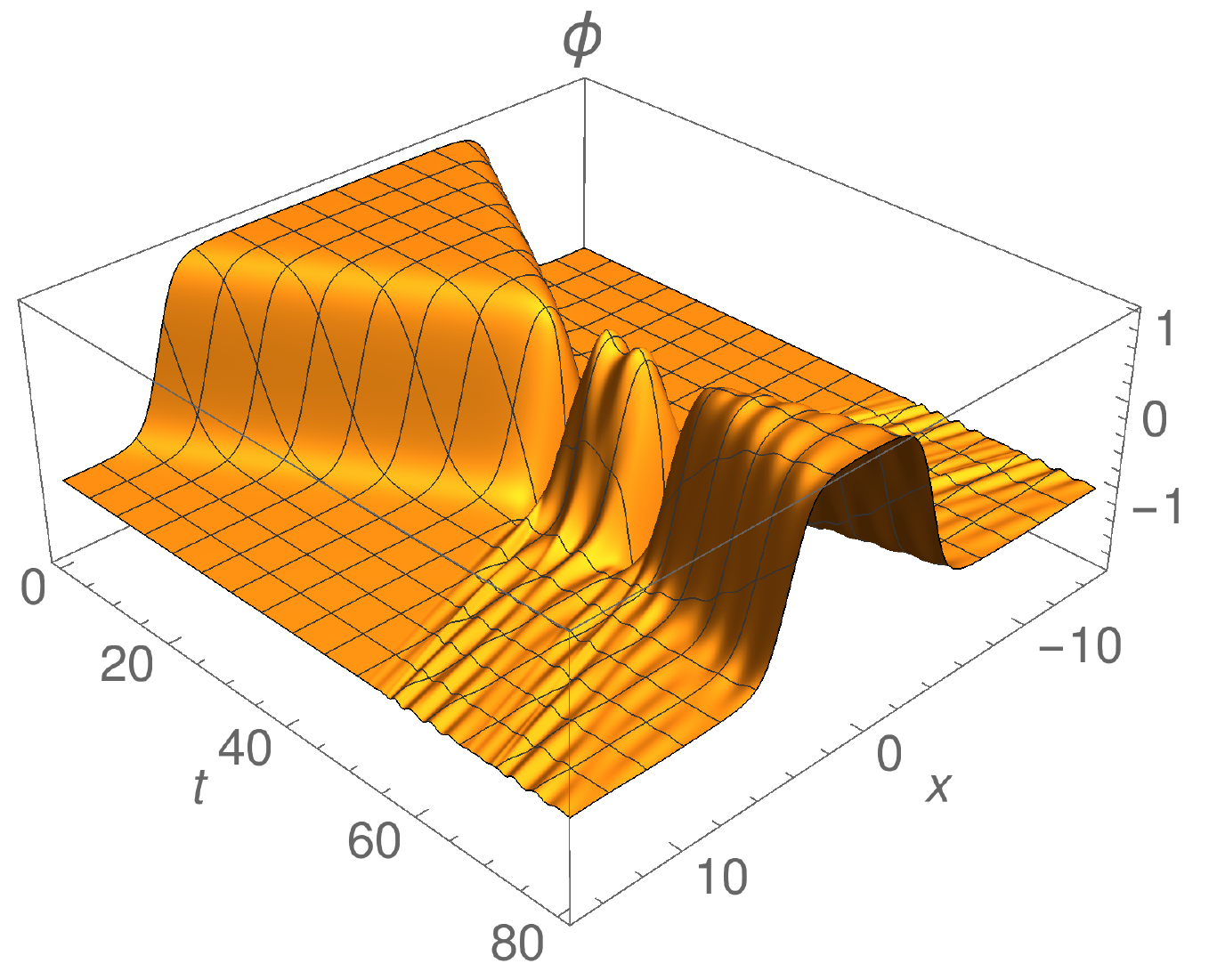}
    \caption{}
    \label{fig_v228}
  \end{subfigure}
  \begin{subfigure}[b]{0.3\textwidth}
    \centering
    \includegraphics[width=\textwidth]{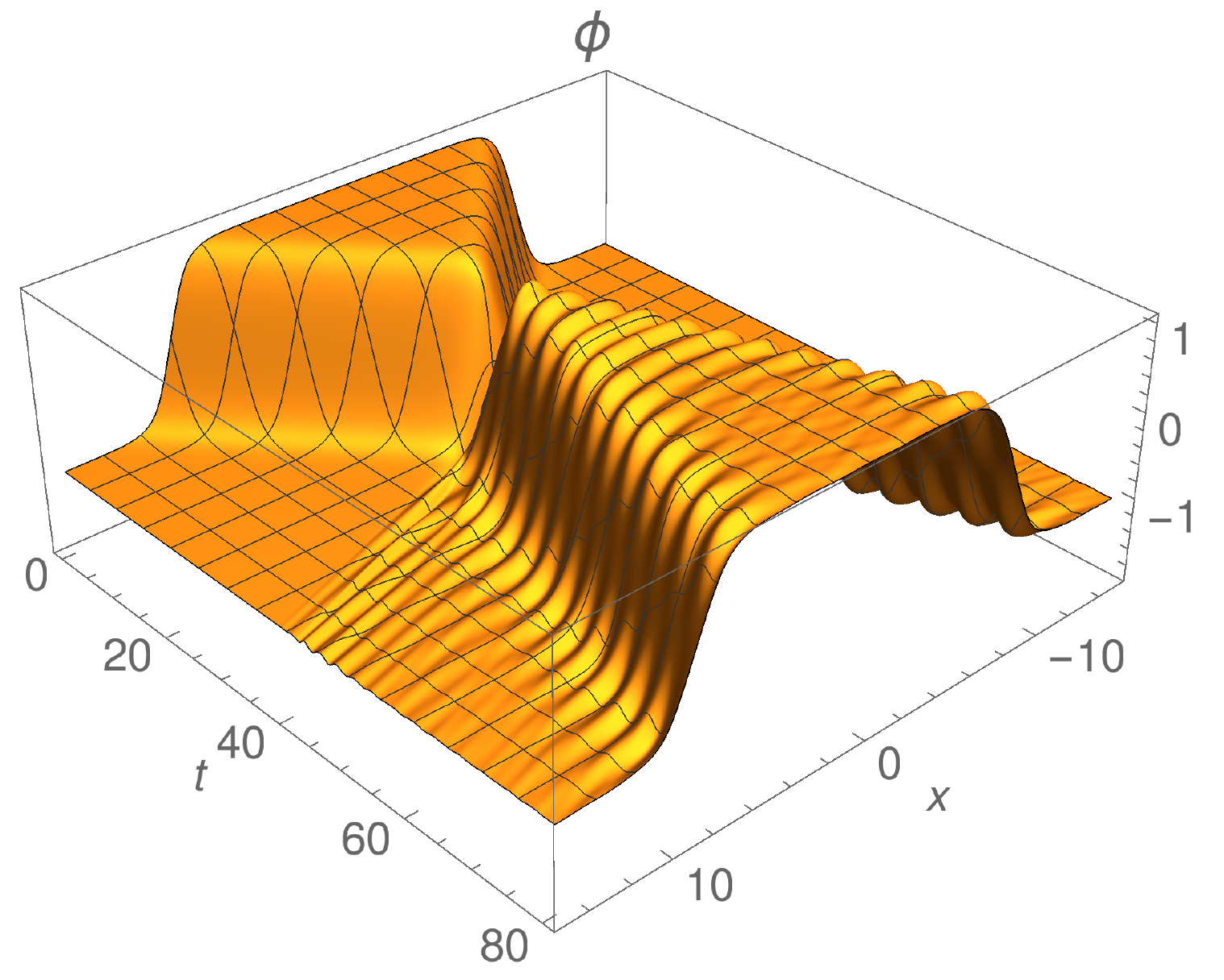}
    \caption{}
    \label{fig_v340}
  \end{subfigure}
       \caption*{Source: Results obtained by the author in \cite{campos2020quasinormal}.}
       \label{fig_3D}
\end{figure}

Typical spacetime evolutions of the field during collisions are shown in Fig.~\ref{fig_3D}. In all cases, the kinks start approaching each other, and in (a) they annihilate and form a bion that slowly radiates away the energy, in (b) the kinks bounce twice before separating, and in (c) the kinks separate after the first bounce. These are all typical scenarios in usual nonintegrable systems with kink solutions, such as the $\phi^4$ model.

\begin{figure}[tbp]
\centering
  \caption{(a) Final velocity as a function of the initial velocity. (b) Time between first and second bounce as a function of the two-bounce resonance windows order. (c) Field at the center of the collision as a function of time for the first eight two-bounce resonance windows. Parameters are $\gamma=2.667$ and $\epsilon=0$.}
  \includegraphics[width=0.94\columnwidth]{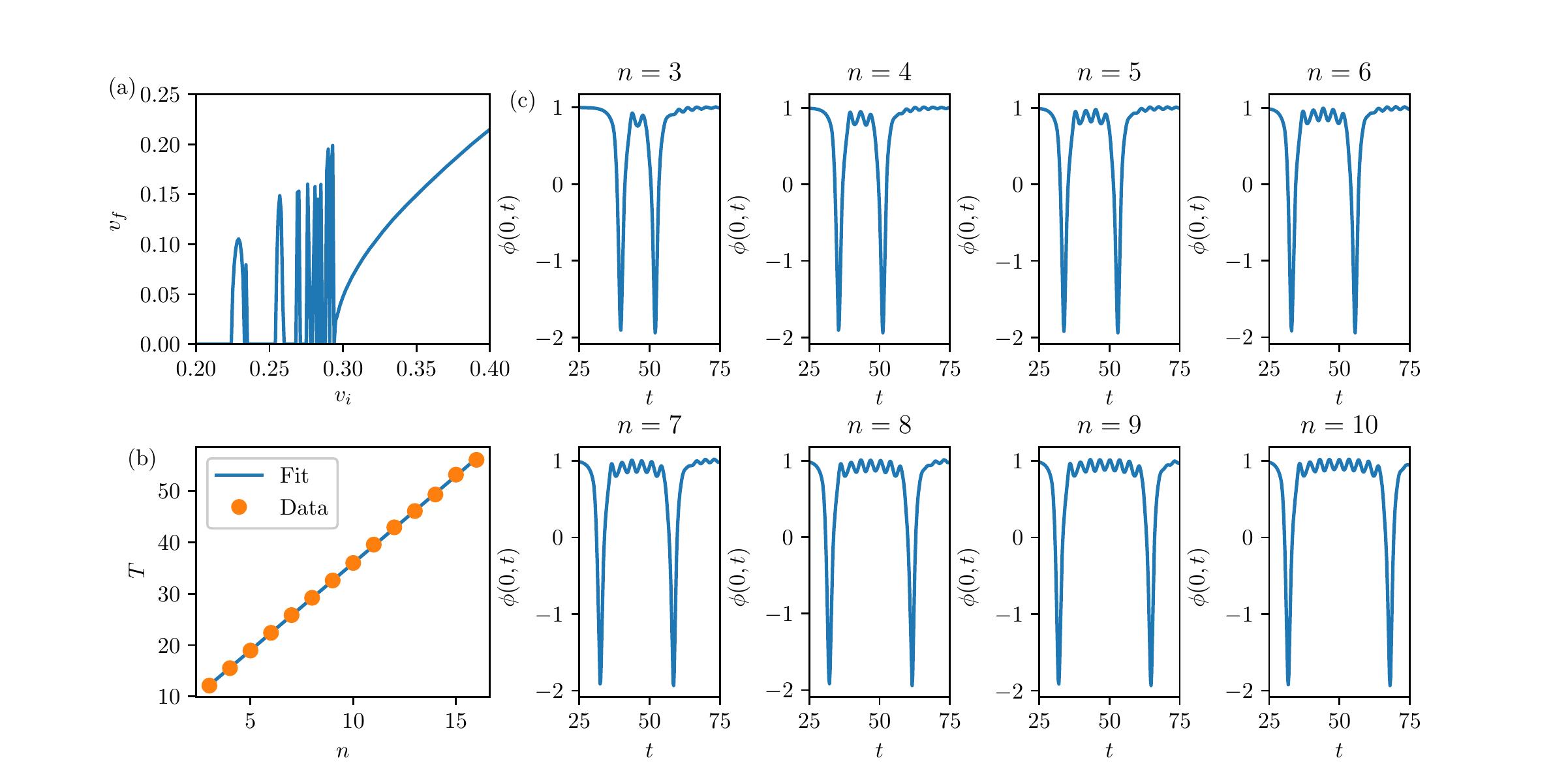}
       \caption*{Source: Results obtained by the author in \cite{campos2020quasinormal}.}
       \label{fig_T}
\end{figure}

We can simulate collisions in a range of initial velocities and measure the final velocity of the kinks. The result is shown in Fig.~\ref{fig_T}(a). There are many isolated peaks that correspond to resonance windows. After a critical velocity, the final velocity is always nonzero and monotonically increasing. This corresponds to the reflection curve.
Interestingly, we see that the resonance windows accumulate at the border of the reflection curve. Moreover, measuring the field at the center of the collision at the resonance windows, we obtain Fig.~\ref{fig_T}(c). The sharp peaks in the figure show the instants where the bounce occurs, and we find that the field has a small oscillation between bounces. We indexed the resonance windows in order with the integer $n$. The number of small oscillations increases by one as we move from one resonance window to the next, meaning that this oscillation has to be in phase for the separation to occur. According to \cite{campbell1983resonance} this implies in the following relation between the frequency of small oscillations $\omega_1$ and the time between collisions $T$
\begin{equation}
\label{eq_campbel}
\omega_1T=\delta+2\pi n.
\end{equation}
The time between bounces as a function of $n$ is shown in Fig.~\ref{fig_T}(b). To confirm the conjecture that the frequency of the small oscillations is due to the shape mode, we fit the curve according to eq.~(\ref{eq_campbel}) and obtain $\omega_1=1.510$. This result should be compared to the theoretical value $\omega=1.533$ obtained from eq.~(\ref{eq_SW_odd}). We chose the values of $n$ such that the constant $\delta$ is between $0$ and $2\pi$. The value of $\delta$ should be approximately equal to $\pi$, and the fit gives $\delta=3.719$, which is a reasonable result.

Let us turn now to the QNM case. In this case, there are three free parameters in the potential. We fix $\gamma=2.667$ and $\Delta\gamma=2.0$. This value of $\Delta\gamma$ is enough to guarantee that the normal mode has turned into a QNM. The last free parameter that we will vary is $\epsilon$ because we find it the most intuitive way to measure how far we are from the normal mode regime.

\begin{figure}[tbp]
\centering
  \caption{Final velocity as a function of the initial velocity for several values of $\epsilon$. Parameters are $\gamma=2.667$, $\Delta\gamma=2.0$ and (a) $\epsilon=0.0$, (b) $\epsilon=0.02$, (c) $\epsilon=0.04$, (d) $\epsilon=0.06$.}
  \includegraphics[width=0.8\columnwidth]{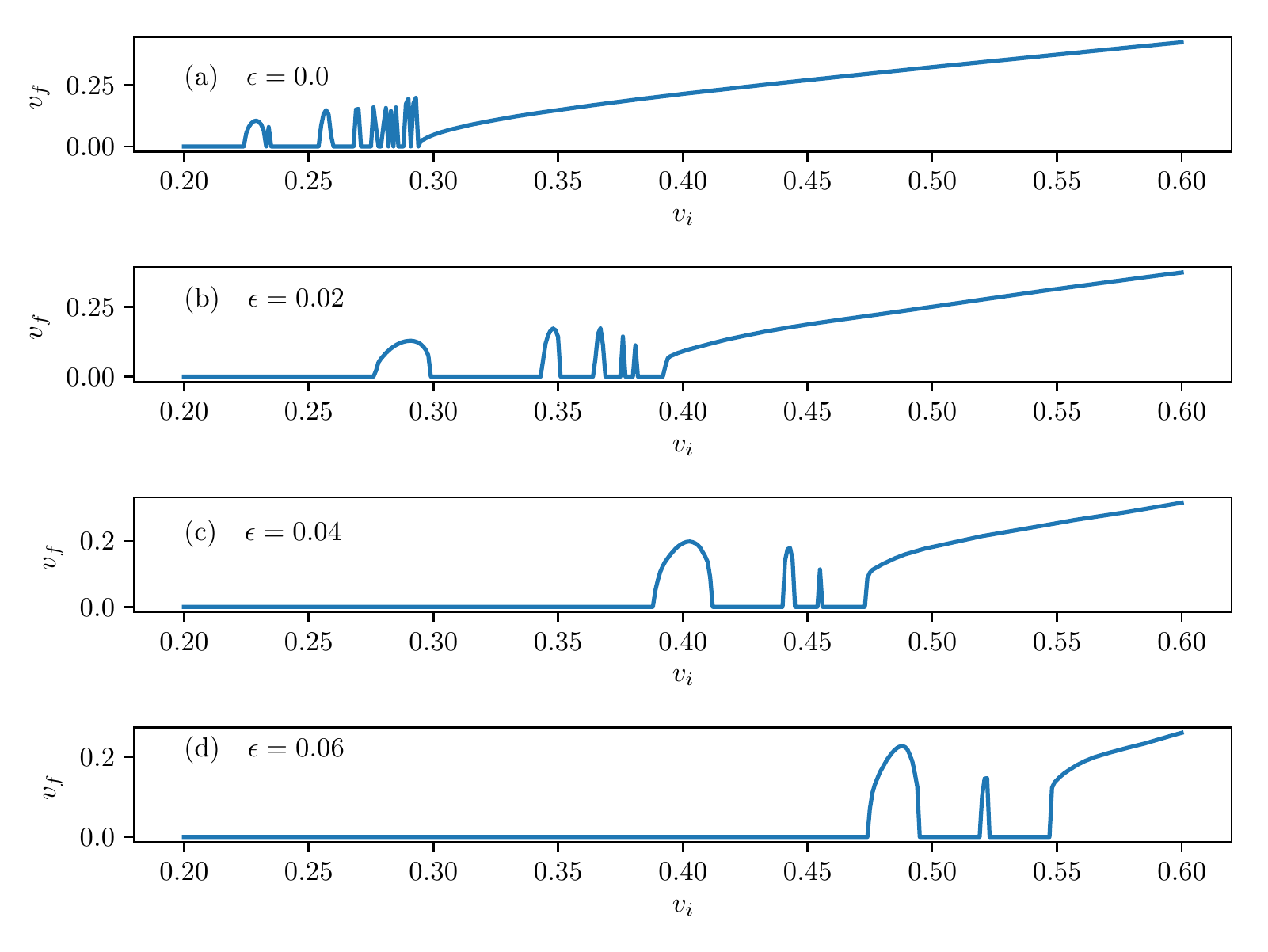}
       \caption*{Source: Results obtained by the author in \cite{campos2020quasinormal}.}
       \label{fig_vinvout}
\end{figure}

In Fig.~\ref{fig_vinvout}, we repeat the plot of the final velocity versus the initial one for several values of $\epsilon$. It is clear from the figure that, as $\epsilon$ increases, the resonance windows gradually start to disappear. This effect is linked to the following qualitative argument. The resonance windows occur because when translational energy is converted into vibrational energy at the bounces, it is stored and can be recovered. However, if the vibrational mode becomes a QNM, this energy will leak, and now only part of the initial energy can be recovered, making it increasingly harder for the kinks to separate. Notice that, accordingly, the critical velocity increases with $\epsilon$.  

\begin{figure}[tbp]
\centering
  \caption{Escape intervals of the initial velocity $v_i$. We zoom in the regions marked by the black box. (a) The blue region corresponds to $\epsilon=0.0$ and (b) the red region to $\epsilon=0.005$ and $\Delta\gamma=2.0$. We fixed $\gamma=2.667$ in both cases.}
  \includegraphics[width=0.8\columnwidth]{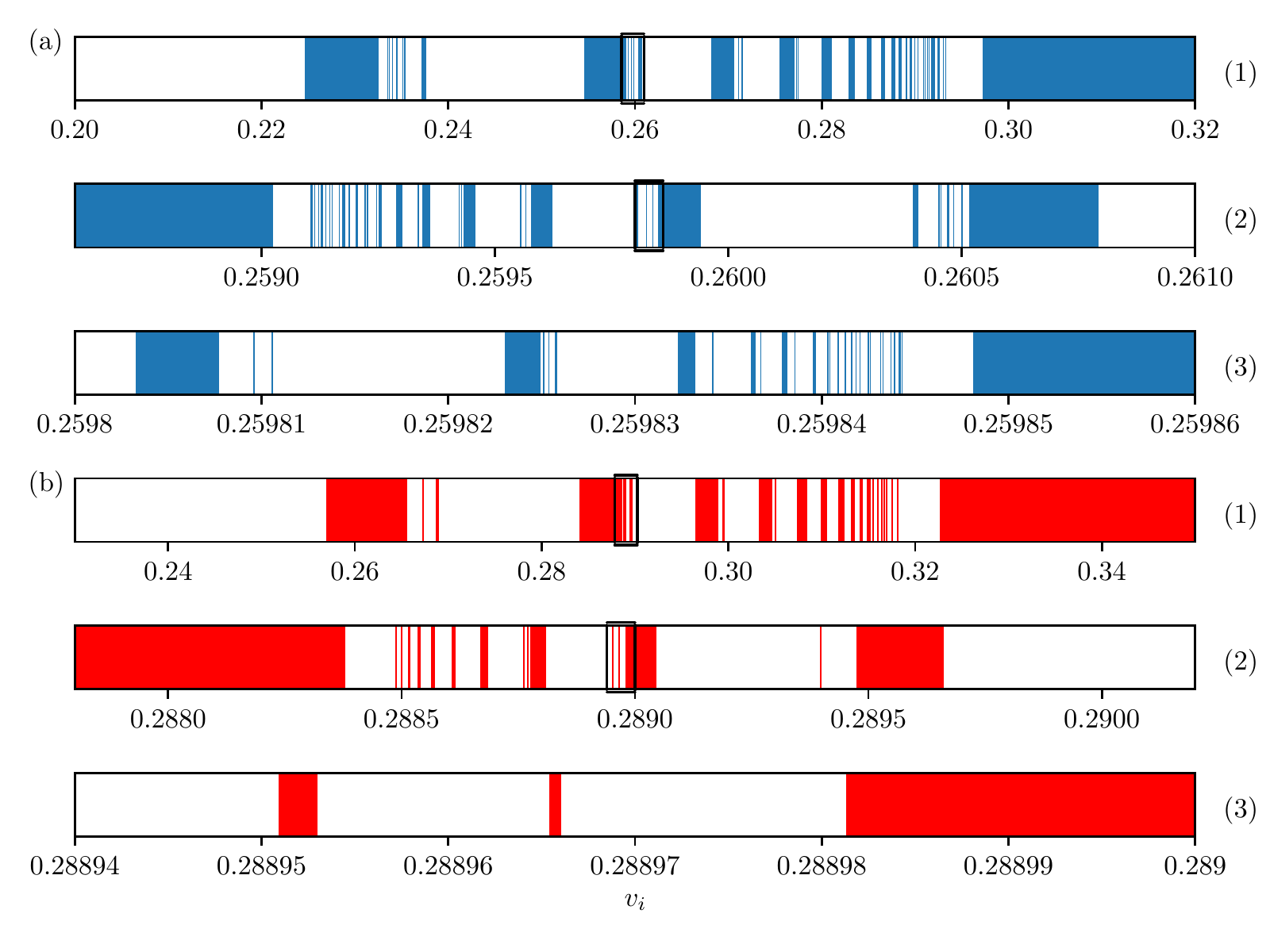}
       \caption*{Source: Results obtained by the author in \cite{campos2020quasinormal}.}
       \label{fig_fractal}
\end{figure}

The set of resonance intervals is known to exhibit fractal structure. To verify whether this is the case in our model, we depict the intervals where the kinks escape after the collision in Fig.~\ref{fig_fractal}. More specifically, the intervals are exhibited for three cases. The first one corresponds to the region near the border of the reflection curve. Then, we zoom in near the edge of an arbitrarily chosen two-bounce resonance window and, again, near the edge of an arbitrarily chosen three-bounce resonance window, as indicated in the figure. Accordingly, we vary $v_i$ between simulations in increasingly smaller steps as we zoom in. The resonance structure consists of a nested structure of three-bounce windows at the border of two-bounce windows, four-bounce windows at the border of the three-bounce windows, and so on. This is shown for $\epsilon=0.0$ in the upper panels.

Moving on to the QNM case, we set a small value $\epsilon=0.005$ to investigate the resonance structure's fate when the QNM mode is still close to the non-decaying normal mode. Again, we draw escape intervals for three cases. The first one is near the edge of the reflection curve, and we zoom twice near the edge of arbitrarily chosen resonance windows. We also vary $v_i$ with increasingly smaller steps as we zoom in. Clearly, a resonance structure is still visible near the critical velocity and even when we zoom in near the two-bounce window indicated in the figure. However, when we zoom in the three-bounce window indicated in the figure, we find a very poor structure of four-bounce windows. This effect is already visible despite the value of $\epsilon$ being small. On the other hand, for $\epsilon=0.0$, the structure goes to higher orders. This implies that the resonance structure is gradually disappearing, as expected by the leaking effect of the QNM.

\begin{figure}[tbp]
\centering
  \caption{Number of boxes to cover the escape regions in Fig.~\ref{fig_fractal} as a function of the box size. The slope is the box fractal dimension. Parameters are $\gamma=2.667$ and, in (b), $\Delta \gamma=2.0$.}  
  \begin{subfigure}[b]{0.46\textwidth}
    \centering
    \includegraphics[width=\textwidth]{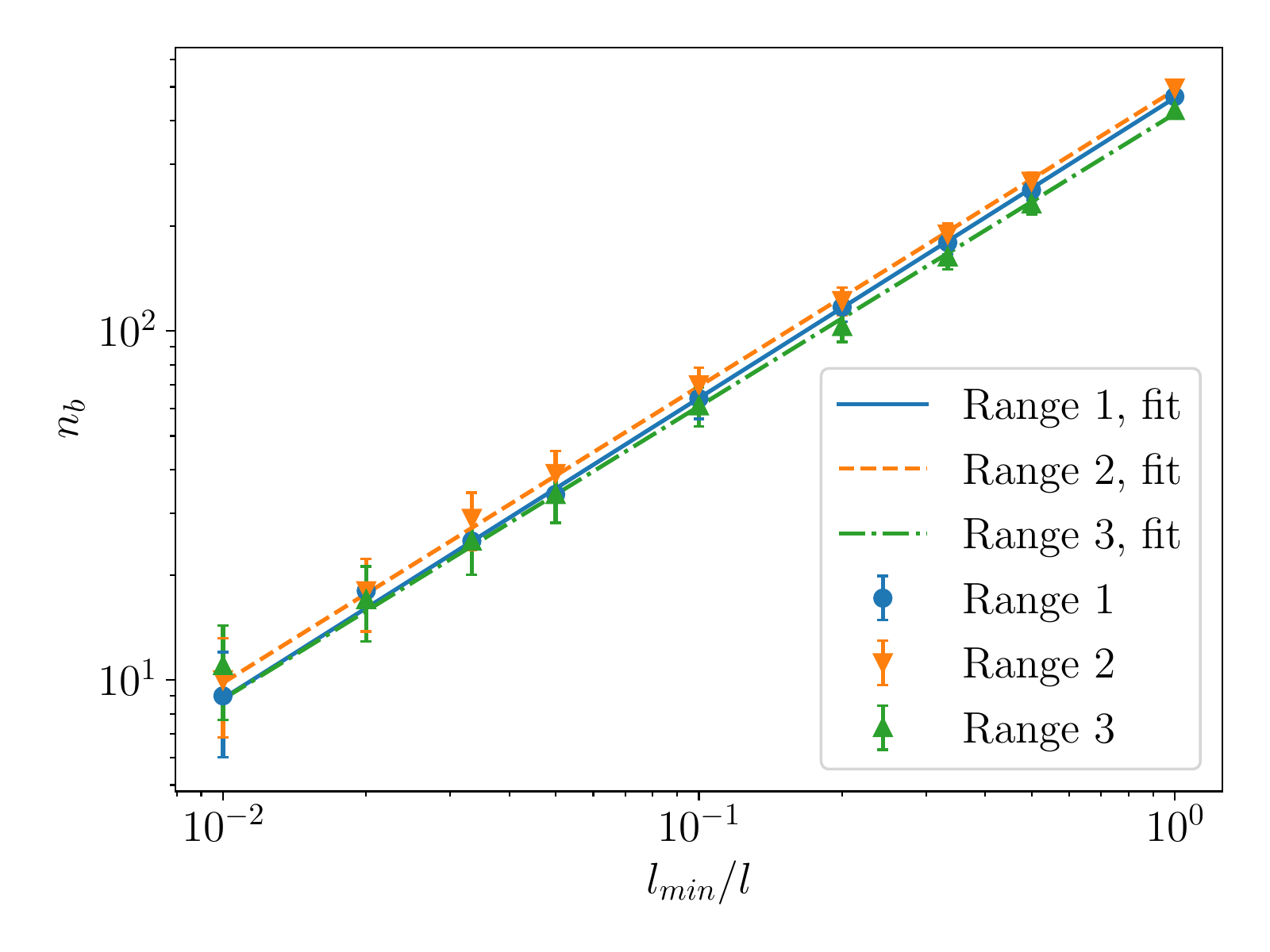}
    \caption{$\epsilon=0.0$}
    \label{fig_box}
  \end{subfigure}
  \begin{subfigure}[b]{0.46\textwidth}
    \centering
    \includegraphics[width=\textwidth]{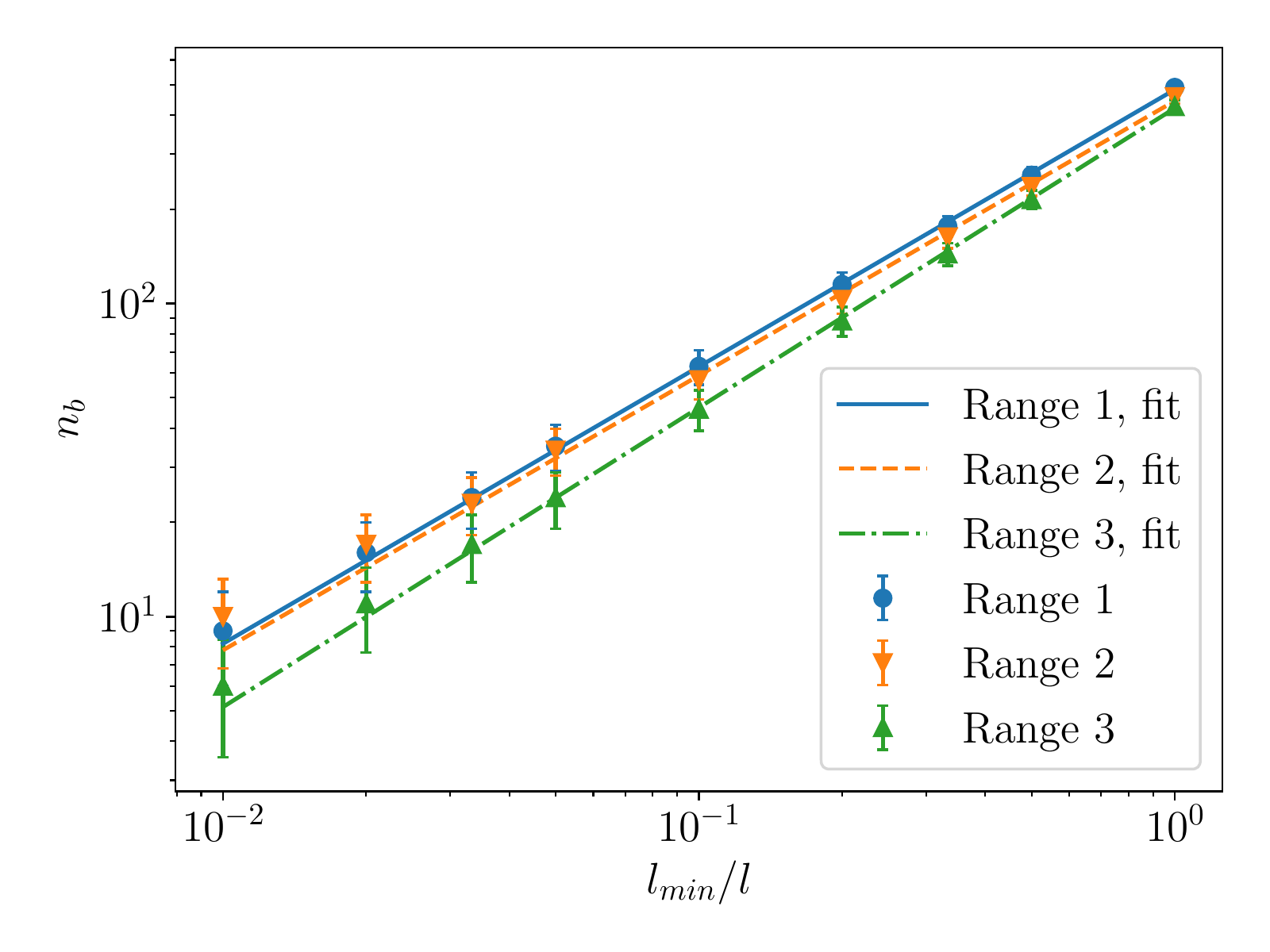}
    \caption{$\epsilon=0.005$}
    \label{fig_boxQNM}
  \end{subfigure}
  \caption*{Source: Results obtained by the author in \cite{campos2020quasinormal}.}
  \label{fig_boxes}
\end{figure}

We quantify the results described above by measuring the box fractal dimension of the intervals drawn in Fig.~\ref{fig_fractal}. We divide the interval into boxes and measure the number of boxes $n_b$ needed to cover the whole escape interval for several box lengths $l$. Then, the box fractal dimension is given by $D=-\lim_{l\to0}\log n_b/\log l$. The plot of $n_b$ versus $1/l$ is shown in Fig.~\ref{fig_boxes} and the box fractal dimension is estimated by the slope of a linear fit to the data. In the figure, $l$ is normalized by the smallest box size $l_{min}$. The values of the slopes are shown in Table \ref{dimension}. For the case with $\epsilon=0.0$, we obtain the box fractal dimension $D$ in the range $0.84-0.86$. On the other hand, for $\epsilon=0.005$, we obtain $D$ in the interval $0.88-0.885$ for the first two cases. However, a much larger dimension of $D=0.956$ is obtained for the last one, which is not very different from the non-fractal value $D=1$. This suggests that in the third interval, the fractal structure disappeared, as can be inferred from Fig.~\ref{fig_fractal}(b).

\begin{table}
\centering
\caption{Values of the slopes in Fig.~\ref{fig_boxes} as well as the range of $v_i$ and smallest box size used to compute the box fractal dimension. We consider $\gamma=2.667$ for both values of $\epsilon$ and $\Delta\gamma=2.0$ when $\epsilon=0.005$.}
\begin{tabular}{c|c|c|c|l}
Range number&$\epsilon$ & Range in $v_i$ & Smallest box size $l_{min}$ & Slope\\
\hline
1 &$0.0$ & $0.20-0.32$ & $1\times 10^{-4}$ & $0.860\pm 0.006$\\
2 &$0.0$ & $0.2586-0.2610$ & $2\times 10^{-6}$ & $0.849\pm 0.006$\\
3 &$0.0$ & $0.25980-0.25986$ & $5\times 10^{-8}$ & $0.84\pm 0.01$\\
\hline
1 &$0.005$ & $0.23-0.35$ & $1\times 10^{-4}$ & $0.885\pm 0.008$\\
2 &$0.005$ & $0.2878-0.2902$ & $2\times 10^{-6}$ & $0.88\pm 0.02$\\
3 &$0.005$ & $0.28894-0.28900$ & $5\times 10^{-8}$ & $0.956\pm 0.009$
\end{tabular}
\caption*{Source: Results obtained by the author in \cite{campos2020quasinormal}.}
\label{dimension}
\end{table}

\begin{figure}[tbp]
\centering
  \caption{Normalized width of resonance windows as a function of the window number for two bounce windows and some sets of three bounce windows. Parameters are $\gamma=2.667$, $\Delta\gamma=2.0$ and $\epsilon=0.0$, $0.005$.}
  \includegraphics[width=0.6\columnwidth]{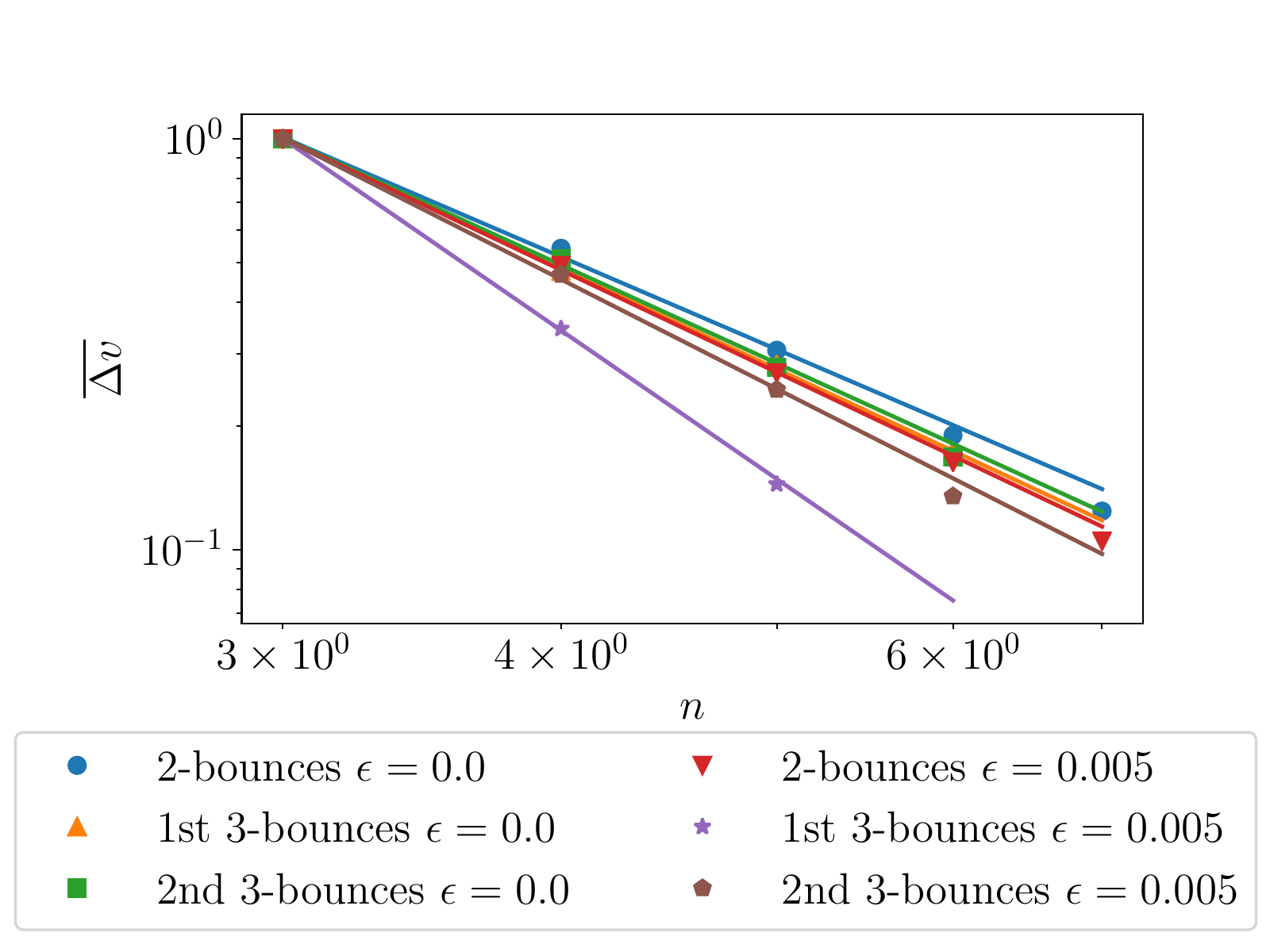}
       \caption*{Source: Results obtained by the author in \cite{campos2020quasinormal}.}
       \label{fig_colapse}
\end{figure}

To investigate further the disappearance of the fractal structure due to the change from normal mode to QNM, we measure the widths of the resonance windows $\overline{\Delta v}$. In a system exhibiting a true fractal structure, such as the $\phi^4$ model, the widths of the two-bounce windows should obey the same relation as the width of any other set of consecutive higher-bounce windows. In \cite{campbell1983resonance}, they found that the widths of resonance windows follow an approximate power-law behavior as a function of $n$. Repeating the analysis for our system, we also find approximate power-law decay for the widths. This can be seen in Fig.~\ref{fig_colapse}, where we plot the window's width as a function of the order of the windows for three different sets of resonance windows. These sets are the two-bounce windows and two chosen sets of three-bounce windows, which are the three-bounce windows near the first and second two-bounce ones. We plot the widths starting from the index $n=3$, and we normalize the value of all widths by the initial one. When $\epsilon=0.0$, the exponent of $n$ is approximately the same for the three sets. Repeating the calculation for the $\epsilon=0.005$ case, we did not find the same scaling behavior for different sets of resonance windows, meaning that the fractal structure is already lost to some degree for this value of $\epsilon$. Actually, we find that higher-order resonance windows decay faster with $n$, and this occurs because higher-bounce resonance windows take longer to occur and, therefore, there is more energy leak.

In the next section we will discuss and summarize the main findings of our work.

\section{Conclusion}

We started our work by considering the desired shape of the linearized potential in the stability equation. From the desired shape, we were able to construct the scalar field potential $V(\phi)$ systematically and find the kink profile. Our approach is novel and could be repeated to construct different scalar potential as was done later in \cite{basak2021new}, for instance, where the authors considered a linearized potential with a simple harmonic well in the center and constant elsewhere.

Another interesting feature of our work was that we were able to find the QNMs and normal modes analytically. Thus, we found explicitly that, at a critical value of $\Delta\gamma$, the bound state becomes a QNM. Moreover, the presence of the QNM mode is linked to resonances in the transmission coefficient.

We simulated kink-antikink collisions for our model, and we found annihilation, resonance, and reflection depending on the initial kink-antikink velocity, as expected. Furthermore, the resonances were shown to obey the resonant energy exchange mechanism. On the other hand, the resonant effect gradually disappears as the normal mode is gradually turned into a QNM due to the energy leakage of this mode. Hence, we were able to confirm the results in \cite{dorey2018resonant}. The disappearance of the self-similar structure for small values of $\epsilon$ was quantified by measuring the fractal structure via the box fractal dimension and comparing the scaling relation of the resonance windows width. We concluded that, even though resonance windows were still present in this case, higher-order windows are suppressed, and the structure loses self-similarity.

\section{Appendix: Numerical method}
\label{sec2_numeric}

This work employs the simplest method for integrating partial differential equations, which is a second-order finite difference method. We divide spacetime in a grid with spacings $\tau=0.002$ and $h=0.01$. We define the field at the gridpoints $(x_i,t_j)$ as $\phi_{i,j}$. Therefore, the partial derivatives are approximated by
\begin{align}
\frac{\partial \phi_{i,j}}{\partial t^2}=\frac{\phi_{i,j+1}-2\phi_{i,j}+\phi_{i,j-1}}{\tau^2}\\
\frac{\partial \phi_{i,j}}{\partial x^2}=\frac{\phi_{i+1,j}-2\phi_{i,j}+\phi_{i-1,j}}{h^2}.
\end{align}
After substituting these expressions in the equations of motion, one can solve for $\phi_{i,j+1}$ in terms of $\phi_{i,j}$ and $\phi_{i,j-1}$
\begin{equation}
\phi_{i,j+1}= 2(1-\lambda^2)\phi_{i,j}+ \lambda^2 (\phi_{i+1,j}+\phi_{i-1,j})-\phi_{i,j-1}-\tau^2V^\prime(\phi_{i,j}),
\end{equation}
where we defined $\lambda=\tau/h$. To perform the first iteration the method is modified and we need only the $\phi$ and $\phi_t$ at $t=0$ \cite{burden2015numerical}
\begin{equation}
\phi_{i,1}= (1-\lambda^2)\phi_{i,0}+ \frac{\lambda^2}{2} (\phi_{i+1,0}+\phi_{i-1,0})-\frac{\tau^2}{2}V^\prime(\phi_{i,j})+\tau\phi_t(x_i,0).
\end{equation} 
The final time in the simulations is fixed at $t=400.0$ and the boundaries are set at $x=\pm100.0$ with periodic boundary conditions. The box is large enough to guarantee that the boundary condition do not interfere with the bulk evolution. We tested the method's precision and observed that the energy is conserved up to a relative error of $10^{-4}$ during the full evolution of the system.

\section{Appendix: Transmission and reflection coefficients}
\label{sec2_TR}

The coefficients of eq.~(\ref{eq_scattering}) are given by
\begin{align}
I_1=&\frac{J}{2}e^{k(L+\Delta L)}e^{im(L+\Delta L)}\beta^-,\nonumber\\
I_2=&\frac{J}{2}e^{-k(L+\Delta L)}e^{im(L+\Delta L)}\beta^+,\nonumber\\
H_1=& \frac{Je^{im(L+\Delta L)}}{2}\big\{e^{k\Delta L}\beta^-[\sin(pL)-\frac{k}{p}\cos(pL)]+e^{-k\Delta L}\beta^+[\sin(pL)+\frac{k}{p}\cos(pL)]\big\},\nonumber\\
H_2=& \frac{Je^{im(L+\Delta L)}}{2}\big\{e^{k\Delta L}\beta^-[\cos(pL)+\frac{k}{p}\sin(pL)]+e^{-k\Delta L}\beta^+[\cos(pL)-\frac{k}{p}\sin(pL)]\big\},\nonumber\\
G_1=& \frac{Je^{im(L+\Delta L)}}{2}e^{-kL}\big\{e^{k\Delta L}\beta^-[\cos(2pL)+\alpha_{kp}^-\sin(2pL)]-e^{-k\Delta L}\beta^+\alpha_{kp}^+\sin(2pL)\big\},\nonumber\\
G_2=& \frac{Je^{im(L+\Delta L)}}{2}e^{kL}\big\{e^{k\Delta L}\beta^-\alpha_{kp}^+\sin(2pL)+e^{-k\Delta L}\beta^+[\cos(2pL)-\alpha_{kp}^-\sin(2pL)]\big\},\nonumber
\end{align}
\begin{align}
\begin{split}
F_1=& \frac{Je^{2im(L+\Delta L)}}{2}\big\{e^{2k\Delta L}[1+i\alpha_{km}^-][\cos(2pa)+\alpha_{kp}^-\sin(2pa)]-2i\alpha_{km}^+\alpha_{kp}^+\sin(2pL)\nonumber\\
  & +e^{-2k\Delta L}[1-i\alpha_{km}^-][\cos(2pL)-\alpha_{kp}^-\sin(2pL)]\big\},
\end{split}\nonumber\\	
\begin{split}
F_2=& \frac{J}{2}\big\{-ie^{2k\Delta L}\alpha_{km}^+[\cos(2pa)+\alpha_{kp}^-\sin(2pa)]+2i\alpha_{km}^-\alpha_{kp}^+\sin(2pL)\\
    &+ie^{-2k\Delta L}\alpha_{km}^+[\cos(2pL)-\alpha_{kp}^-\sin(2pL)]\big\},
\end{split}
\end{align}
where used the definition in eq.~(\ref{eq2_alpha}) and defined
\begin{equation}
\beta^\pm=1\pm\frac{im}{k}.
\end{equation}
The transmission coefficient is given by $T=|\frac{J}{F_1}|^2$ and the reflection coefficient by $R=|\frac{F_2}{F_1}|^2$.

\section{Appendix: Bound states of the square well potential}
\label{sec2_SW}

The even bound states of the square well potential have eigenvalues $\omega_{2n}$, which are solutions of the transcendental equation \ref{eq_SW_even}. Their respective eigenfunctions are given by
\begin{equation}
\eta_{2n}(x)= 
  \begin{cases} 
     C_{2n}e^{k_{2n}x}, & x<-L, \\
     C_{2n}e^{-k_{2n}L}\cos(p_{2n}x)/\cos(p_{2n}L), & -L<x<L, \\
     C_{2n}e^{-k_{2n}x}, & x>L,
  \end{cases}
\end{equation}
where $k_{2n}\equiv\sqrt{\gamma-\omega^2_{2n}}$, $p_{2n}\equiv\sqrt{\omega^2_{2n}+\lambda}$ and $C_{2n}$ are normalization constants.

The odd bound states of the square well potential have eigenvalues $\omega_{2n+1}$, which are solutions of the transcendental equation \ref{eq_SW_odd}. Their respective eigenfunctions are given by
\begin{equation}
\label{eigenodd}
\eta_{2n+1}(x)= 
  \begin{cases} 
     -C_{2n+1}e^{k_{2n+1}x}, & x<-L, \\
     C_{2n+1}e^{-k_{2n+1}L}\sin(p_{2n+1}x)/\sin(p_{2n+1}L), & -L<x<L, \\
     C_{2n+1}e^{-k_{2n+1}x}, & x>L,
  \end{cases}
\end{equation}
where $k_{2n+1}\equiv\sqrt{\gamma-\omega^2_{2n+1}}$, $p_{2n+1}\equiv\sqrt{\omega^2_{2n+1}+\lambda}$ and $C_{2n+1}$ are normalization constants.

\chapter{INTERACTIONS OF KINKS WITH DOUBLE LONG-RANGE TAILS}
\label{chap_long}

\section{Overview}

Recently, some attention has been given to kinks with long-range tails, which decay as a power-law. These kinks appear in scalar field models when the potentials are polynomials of higher-order \cite{khare2014successive, khare2019family}. They can be used to describe several physical systems, such as Rydberg atoms \cite{saffman2010quantum}, and quantum gases \cite{lahaye2009physics}. Other applications appear in cosmology \cite{valle2016relativistic, greenwood2009electroweak}, statistical mechanics \cite{campa2009statistical}, and supersymmetric quantum mechanics \cite{bazeia2017supersymmetric}.

In \cite{christov2019long}, the authors showed that usual initialization methods lead to wrong results in simulations of kinks with long-range tails, such as kink-antikink repulsion. To fix this issue, they developed a method that consists of a nonlinear least-square minimization of the initial condition such that the configuration of the system is as close as possible to a solution of the static field equation. This specialized method was proven to be very reliable and generated a very smooth scalar field evolution. However, the method could not generate initial conditions where the kink and the antikink are boosted.

Another difficulty due to the long-range tail of kinks is the computation of inter-kink force. The standard method used to estimate this force was developed by Manton and consists of integrating the momentum density around the kink position and computing the time derivative of the resulting expression \cite{manton2004topological}. However, this method fails to estimate the force between long-range kinks due to the large overlap between them. In \cite{manton2019forces}, the same author developed a more suitable method for long-range tails which consists of inserting an accelerating kink ansatz in the equations of motion and making some further approximations, which will be described below. This new method was shown to give very accurate results for a class of models with long-range tails \cite{christov2019kink}.

Later, in \cite{christov2021kink}, the authors were able to develop a specialized method to initialize kink-antikink as well as kink-kink configurations with a nonzero initial velocity. The method consists of starting the system at an initial guess and integrating the equations of motion for a short period. Then, the initial guess is modified in order to minimize the two-norm of the equations of motion in that period. However, this method was very computationally costly.

Here, we will consider a class of models where the kinks possess double long-range tails. This means that not only the tail that is facing the opposite kink is long-range, but also the one in the opposite direction. We will show that this property drastically alters the behavior of the system. Moreover, we will develop a method to initialize kinks with a nonzero velocity that is physically intuitive and more computationally efficient. This work resulted in the following publication \cite{campos2021interaction}.

The following section will describe the family of models containing kinks with double long-range tails.

\section{Model}

\begin{figure}[tbp]
\centering
  \caption{(a) Potential, (b) Kink and (c) Linearized potential of the stability equation for the model containing kinks with double long-range tails. We consider the first few values of $n$.}
  \includegraphics[width=1.0\columnwidth]{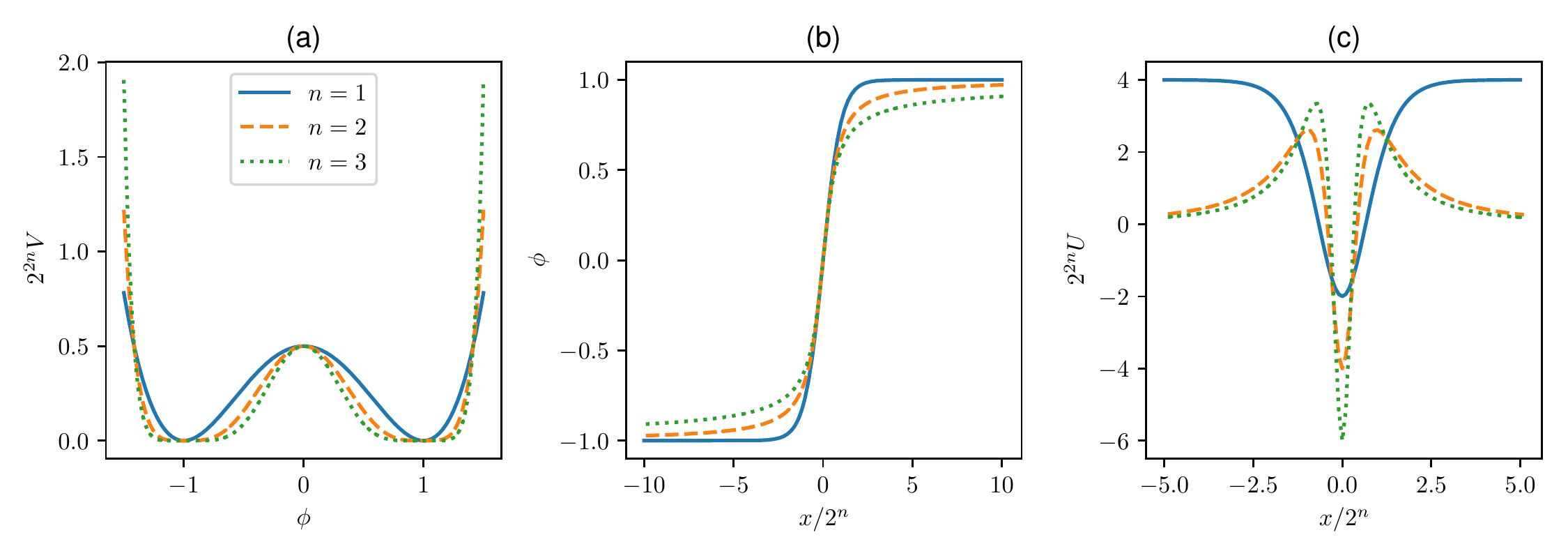}
       \caption*{Source: Results obtained by the author in \cite{campos2021interaction}.}
       \label{fig4_kink}
\end{figure}

We start with a scalar field Lagrangian in $(1+1)$ dimensions
\begin{equation}
\mathcal{L}=\frac{1}{2}\partial_\mu\phi\partial^\mu\phi-V_{4n}(\phi).
\end{equation}
The equations of motion are obviously
\begin{equation}
\frac{\partial^2\phi}{\partial t^2}-\frac{\partial^2\phi}{\partial x^2}+\frac{dV_{4n}}{d\phi}=0.
\end{equation}
Moreover, the family of potentials we consider are of the following form
\begin{equation}
V_{4n}(\phi)=\frac{1}{2^{2n+1}}(1-\phi^2)^{2n}.
\end{equation}
The constant $1/2^{2n+1}$ was chosen for later convenience without loss of generality. The first few potentials are shown in Fig.~\ref{fig4_kink}(a). It exhibits spontaneous symmetry breaking and $Z_2$ symmetry. The $n=1$ case corresponds to the well-known $\phi^4$ model, while for $n\geq 2$ the perturbations around the vacua are massless, which leads to the appearance of long-range tail. 

We can write the potentials in terms of a superpotential $W_{4n}$. The two are related as follows
\begin{equation}
V_{4n}(\phi)=\frac{1}{2}\left(\frac{dW_{4n}}{d\phi}\right)^2.
\end{equation}
Computing the superpotential explicitly for the first two relevant cases, we find
\begin{equation}
W_8=\frac{\phi}{4}-\frac{\phi^3}{6}+\frac{\phi^5}{20}+c_1,\quad W_{12}=\frac{\phi}{8}-\frac{\phi^3}{8}+\frac{3\phi^5}{40}-\frac{\phi^7}{56}+c_2.
\end{equation}
We can also set to zero the two integration constants $c_1$ and $c_2$ without loss of generality. 

The kink solutions can be found using the BPS equation
\begin{equation}
\frac{d\phi}{dx}=\frac{dW_{4n}}{d\phi}.
\end{equation}
Substituting the superpotentials in the expression above, we obtain respectively for the $\phi^8$ and $\phi^{12}$ models \cite{khare2014successive}
\begin{equation}
\frac{2\phi}{1-\phi^2}+\log\left(\frac{1+\phi}{1-\phi}\right)=x-A,\quad\frac{10\phi-6\phi^3}{(1-\phi^2)^2}+3\log\left(\frac{1+\phi}{1-\phi}\right)=2(x-A).
\end{equation}
These equations can be solved numerically for $\phi$, and the resulting kink profiles are shown in Fig.~\ref{fig4_kink}(b). Notice that both tails decay much more slowly than the solid curve, corresponding to the $\phi^4$ kink. The asymptotic form can be easily found as $1/x$ for $\phi^8$ and $1/\sqrt{2x}$ for the $\phi^{12}$. The masses of the kinks are given by the expression $M=W(1)-W(-1)$ leading to $4/15$ for $\phi^8$ and $4/35$ for $\phi^{12}$.

We can also study the behavior of perturbations around the kink solutions numerically. The linearized potential of the resulting stability equation is shown in Fig.~\ref{fig4_kink}(c). For $n\geq2$, the potentials have a volcano shape and tend to zero on both sides. This means that the only possible discrete solution is the zero mode, which is guaranteed to exist by translational symmetry. Therefore, the absence of a vibrational mode means we do not expect this system to exhibit resonance windows like other non-integrable models.

The next section will discuss how to compute the force between kinks with long-range tails.

\section{Inter-kink force}

In this section, we want to compute the inter-kink force for a kink-antikink configuration starting from rest. The following analysis was developed in \cite{manton2019forces} and generalized in \cite{christov2019kink}. We will start with a review of the previous works. The analysis starts by considering a symmetric configuration with an antikink at position $-A$ and a kink at position $A$. The solution for the kink at the right of the origin will be written as $\phi(x,t)=\eta(x-A(t))$. We denote the derivative with respect to $X\equiv x-A$ by a prime. Substituting this ansatz in the equation of motion, we find
\begin{equation}
\eta^{\prime\prime}-a\eta^\prime-\frac{dV(\eta)}{d\eta}=0,
\end{equation}
where we defined the acceleration $a\equiv\ddot{A}$. We have neglected the term proportional to $\dot{A}^2$ because we are considering that the kink starts from rest and, therefore, has a small velocity. Then, assuming that the accelerating kink solution approximately solves the BPS equation, we can substitute $\eta^\prime=dW/d\eta$. Then, integrating the resulting equation once this yields
\begin{equation}
\label{eq4_etap0}
\eta^\prime=\sqrt{2[V_{eff}(\eta)-aW(1)]},
\end{equation}
where $V_{eff}(\eta)\equiv V(\eta)+aW(\eta)$. 

So far, the approximations we made are nearly exact as long as the kinks' velocity is still small. The next simplification made in \cite{manton2019forces} was to expand $V_{eff}(\eta)$ until first order around $\eta=-1$, which is the approximate value of the scalar field in the overlapping region. We will see that it is necessary to include higher-order terms in some models. Ignoring this issue, for now, we obtain
\begin{equation}
\label{eq4_etap}
\eta^\prime=\sqrt{(1+\eta)^{2n}-2Ma}.
\end{equation}
Now that we found an equation for an accelerating kink near the overlapping region, we can ask the value of $a$ such that the solution best fits the left tail of the static kink solution centered at $A$. This can be achieved by integrating eq.~(\ref{eq4_etap}) with the appropriate boundary conditions. At $x=0$, or $X=-A$, $\eta^\prime=0$ due to the even symmetry of the kink-antikink configuration. According to eq.~(\ref{eq4_etap}), we have that $\eta^\prime=0$ when $(1+\eta)^{2n}=2Ma$. The second boundary condition at $x=A$, or $X=0$, is the point where the extrapolation of the static kink tail diverges, and we should have $\eta\to\infty$. Thus, we get
\begin{equation}
\int_{(2Ma)^{\frac{1}{2n}}}^\infty\frac{d\xi}{\sqrt{\xi^{2n}-2Ma}}=A,
\end{equation}
where $\xi=1+\eta$. After a change of variables, we find
\begin{equation}
A(2Ma)^{(n-1)/2n}=\int_1^\infty\frac{d\lambda}{\sqrt{\lambda^{2n}-1}}=\frac{-\sqrt{\pi}\Gamma\left(\frac{n-1}{2n}\right)}{\Gamma\left(-\frac{1}{2n}\right)},
\end{equation}
or equivalently
\begin{equation}
\label{eq4_accel}
a=\left[\frac{-\sqrt{\pi}\Gamma\left(\frac{n-1}{2n}\right)}{\Gamma\left(-\frac{1}{2n}\right)}\right]^{2n/(n-1)}\frac{1}{2M}A^{2n/(1-n)}.
\end{equation}
This result matches the expression found in \cite{christov2019kink} for a similar model. In that case, the analytical estimate was shown to have a very good agreement with numerical simulations. Observe that in this method, we match the tail of the accelerating kink facing the antikink to the same tail of the static kink. This means that the tail not facing the antikink, which we will call backtail, is not relevant for the interaction. This is a correct assumption. However, we will see that the behavior after the collision can be very different depending on the character of the backtail.  

We compared these analytical results to numerical simulations in our model, and we did not find a good agreement between the two. The reason for this discrepancy was that, while the second-order term in the tail expansion of the kink is absent in \cite{christov2019kink}, it is present in our model. Therefore, the approximation of $V_{eff}$ by the first-order term is much worse in our case. Here, the analytical result improves as the separation between the kink and the antikink increases because the second-order term becomes less relevant relative to the first-order one. However, the convergence of the numerical acceleration, and the analytical expression is very slow due to the long-range character of the tail.

In order to estimate the effect of the second-order term in the tail expansion, we make the following approximations. First, we write the expansion of the kink tail for the $\phi^8$ model up to second-order
\begin{equation}
\phi=-1+\frac{1}{A-x}+\frac{\log[2(A-x)]-\frac{1}{2}}{(A-x)^2}+\mathcal{O}\left((A-x)^{-3}\right).
\end{equation}
Then, we approximate $\log[2(A-x)]\simeq\log(2A)$ because the interaction between the kinks occurs in the overlapping region, where $x\ll A$. With this approximation, the second-order term can be interpreted as a shift in the point where the extrapolation diverges from $A\to A-\Delta A$. From the above expression, we find $\Delta A=\log(2A)-\frac{1}{2}$. Using this prescription in eq.~(\ref{eq4_accel}), we find better results for the $\phi^8$ model. For the $\phi^{12}$ model, however, the second-order term cannot be interpreted as a shift in the point where the extrapolated tail diverges. If we wish to include this term in the estimation of the force between the kinks, we need to integrate eq.~(\ref{eq4_etap0}) numerically. In this case, we chose to integrate this equation numerically, including all terms in $V_{eff}$ because the inclusion of more terms does not increase the difficulty in numerical integration.

In the next section, we will discuss the numerical method we used to simulate the kink-antikink collisions for the class of models where the kinks exhibit double long-range tails.

\section{Numerical method}

The first step in the simulation of the model is to construct a suitable initial condition for both the field and its time derivative, which we call the velocity field. We need both initial conditions because the equations of motion are second order in time. In order to understand the algorithm, it is important to investigate a simpler case of a single traveling kink. We will follow \cite{christov2021kink} in this discussion. We start by writing the field as $\phi(x,t)=u(x-vt)$, which is the functional form of the field of a traveling kink. Then, substituting in the equations of motion, we find that $u$ obeys the following equation
\begin{equation}
\label{eq4_u}
(1-v^2)u^{\prime\prime}(\xi)-V^\prime(u(\xi))=0,
\end{equation}
where $\xi\equiv x-vt$ and prime denotes derivative with respect to the argument. Moreover, the velocity field of a traveling kink is given by $\phi_t(x,t)\equiv f(\xi)=-vu^\prime(\xi)$. The derivative of the kink solution corresponds to the zero mode, which is the mode related to the translation invariance of the system. For a traveling kink, the velocity field is proportional to the zero mode of the kink. This is an important property that the authors did not give enough attention in \cite{christov2021kink}. This property means that the velocity field obeys the following equation
\begin{equation}
\label{eq4_zm}
(1-v^2)\frac{\partial^2f}{\partial\xi^2}-V^{\prime\prime}(u(\xi))f=0.
\end{equation}
In the above expression, the term $(1-v^2)$ is related to the Lorentz contraction of the traveling kink similar to the same factor in eq.~(\ref{eq4_u}). After this digression, we can now turn to the numerical method.

First, to initialize the field, we start with a guess, which will be given by the split-domain ansatz
\begin{equation}
\phi(x,t)=(1-\Theta(x))\phi_K(\gamma(x+A-vt))+\Theta(x)\phi_{\bar{K}}(\gamma(x-A+vt)).
\end{equation}
In words, this ansatz considers a kink to the left of the origin and an antikink to the right. Moreover, the kink is boosted to the right with velocity $v$, and the antikink is boosted to the left with velocity $-v$. Thus, this initial guess has a discontinuity at the center. This initial guess was suggested in \cite{christov2019long} for the static case and in \cite{christov2021kink} for the boosted one. There, the authors optimized this field configuration to a solution that obeys eq.~(\ref{eq4_u}) as close as possible. This is done via a nonlinear least-square minimization, which we also perform here. This minimization is supplemented by the constraint that the centers of the kinks should be fixed at $x=\pm A$, which in our case means that the field should vanish in those positions. After discretizing the field and using a Fourier spectral method to compute the derivatives, we perform the minimization via the least\_square function in the SciPy library in python. The program minimizes the following functional
\begin{equation}
\mathcal{I}[u]=\lVert(1-v^2)D_2u-V^\prime(u)\rVert_2^2+C|u(A)|^2+C|u(-A)|^2.
\end{equation}
The first term is a euclidean norm. The argument of the norm becomes a vector after discretizing the field, and the euclidean norm can be easily calculated. The second derivative is obtained by the Fourier Spectral matrix $D_2$ \cite{trefethen2000spectral}. The two last terms enforce the constraints for suitable values of the constant $C$. We found good results for $C=5.0$.

\begin{figure}[tbp]
\centering
  \caption{Initial guess (solid) and minimized one (dashed) for the scalar field (left) and the velocity field (right).}
  \includegraphics[width=0.9\columnwidth]{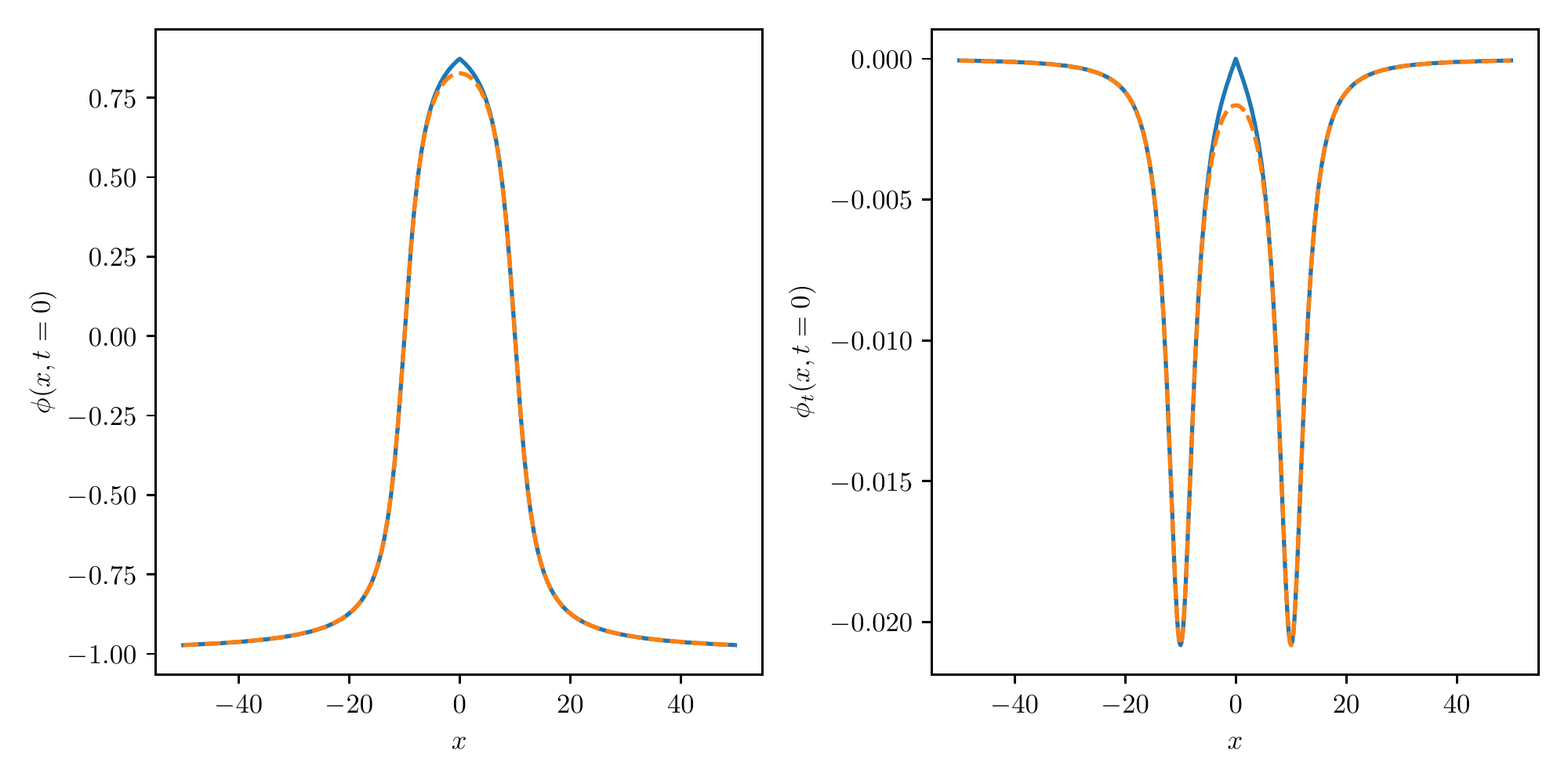}
       \caption*{Source: Results obtained by the author in \cite{campos2021interaction}.}
       \label{fig4_min}
\end{figure}

The next step is to find an initial condition for the velocity. We start with a guess given by
\begin{equation}
\phi_t(x,t=0)=v\text{sgn}(x)u^\prime(x).
\end{equation}
This guess is the analog of the split-domain ansatz to the velocity field because we are using the relation of the velocity field of a traveling kink with positive (negative) velocity to the left (right) of the origin. This guess is discontinuous at the center as well. In \cite{christov2021kink}, the authors suggested this initial guess and optimized it according to the equations of motion. They had to integrate the initial condition for a short time. This required a large computational time. Our approach is much more computationally efficient. We minimized the initial guess according to eq.~(\ref{eq4_zm}) via the same nonlinear least-square minimization. This means that we find the velocity field that obeys as close as possible the zero-mode equation for the field configuration $u(x)$. This was supplemented by the constraint that, near the centers of the kinks, the solution should be as close as possible to the initial guess. More specifically, the constraint was set in the intervals $-A-2<x<-A+2$ and $A-2<x<A+2$. The initial guesses and the minimized ones are shown in Fig.~\ref{fig4_min}. Observe that the minimization procedure smooths out the configuration at the center and preserves the configuration near the kinks.

The next step is to compute the actual evolution of the field. This is accomplished by discretizing the space in the interval $-200.0<x<200.0$ with spacings $h=0.2$ for $\phi^8$ and in the interval $-400.0<x<400.0$ with spacings $h=0.4$ for $\phi^{12}$. Then, we set periodic boundary conditions, but the integration is short enough to guarantee that the boundaries will not interfere with the bulk evolution. We approximate the second-order derivative with respect to $x$ by a Fourier spectral matrix. This consists in taking the Fourier transform of the field, performing the derivative in Fourier space, which is a trivial multiplication, and then transforming back to real space \cite{trefethen2000spectral}. After discretizing in space, the equations of motion become a set of first order ordinary diferential equations for a vector containing both the values field $\phi$ and its time derivative $\phi_t$ at the grid points. This procedure is known as the method of lines. The resulting ordinary differential equations are integrated in time using the solve\_ivp method from the SciPy library in python. This method implements an eighth order Runge-Kutta method with adaptive step size and error control.

\begin{figure}[tbp]
\centering
  \caption{Evolution of the velocity field in spacetime for (a) the reference method and (c) for an initial condition using our minimization methods. (b) Evolution of the velocity and (d) initial value of the velocity field. Dashed lines correspond to the reference method and solid ones for an initial condition using our minimization methods.}
  \includegraphics[width=0.9\columnwidth]{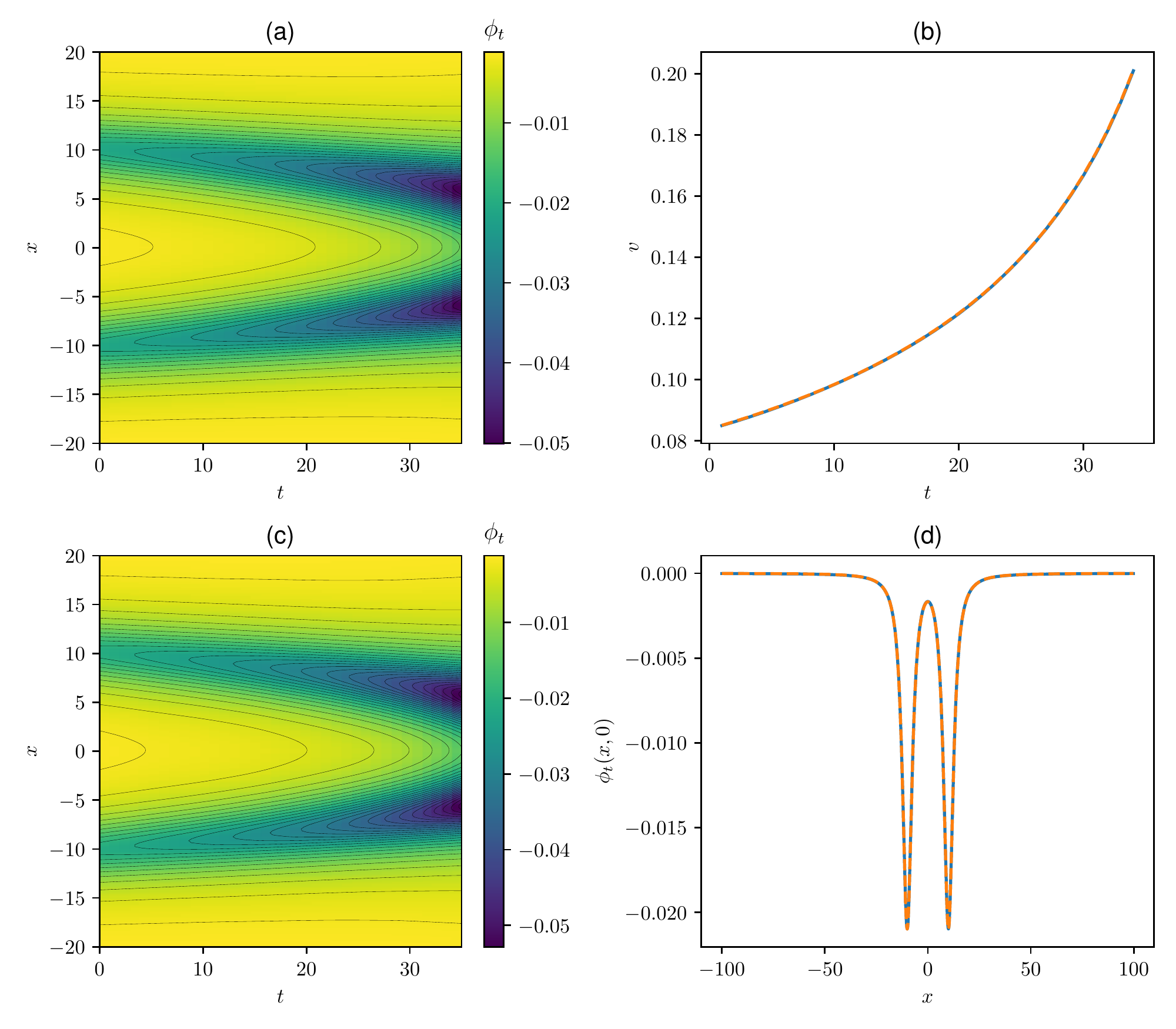}
       \caption*{Source: Results obtained by the author in \cite{campos2021interaction}.}
       \label{fig4_phit}
\end{figure}

We compare our initial conditions for a boosted kink-antikink configuration with a reference solution to test our minimization method. We obtain the reference solution by the following method. We start with a kink-antikink configuration where the elements are at rest with $A=30.0$. When the system starts at rest, the initialization method was shown to produce good results \cite{christov2019long}. If we let the kinks interact, the attraction makes them accelerate towards each other. After some time, the configuration becomes a boosted kink-antikink configuration. When the elements reach the position $A=10.0$, we measure the velocity $v\simeq0.0832$. Then, we initialize another simulation according to our minimization methods with the same $A$ and $v$ and let it evolve in time. Thus, we can compare the evolution of the simulation using our method and the continuation of the evolution of the reference method. The result is shown in Fig.~\ref{fig4_phit}.

In Fig.~\ref{fig4_phit}(a) and (c), we plot the evolution of the velocity field in both simulations. There are no visible differences between the two in the scale of the figure. This shows that our minimization procedure matches the natural evolution from a static configuration to a boosted one. In particular, we see in Fig.~\ref{fig4_phit}(d) that the initial condition of the velocity field matches perfectly the velocity field of the reference solution at $A=10.0$. Moreover, we can measure the position of the kinks as the point where the field crosses the value zero. Then, computing the derivative of the position with respect to time, we find the velocity of the kinks. The result is shown in Fig.~\ref{fig4_phit}(b). Both methods agree perfectly in the scale of the figure.

In the next section, we will summarize all the possible behaviors of kink-antikink collisions varying the initial velocity of the system.

\section{Collisions}

\begin{figure}[tbp]
\centering
  \caption{Evolution of the field in spacetime for kink-antikink interactions starting from a statitc configuration. We consider the $\phi^8$ model (left) and the $\phi^{12}$ model (right). Lines correspond to the evolution of a particle with acceleration given by (solid) eq.~(\ref{eq4_etap0}) without further approximations and (dashed) eq.~(\ref{eq4_accel}) after setting $A\to A-\Delta A$.}
  \includegraphics[width=0.9\columnwidth]{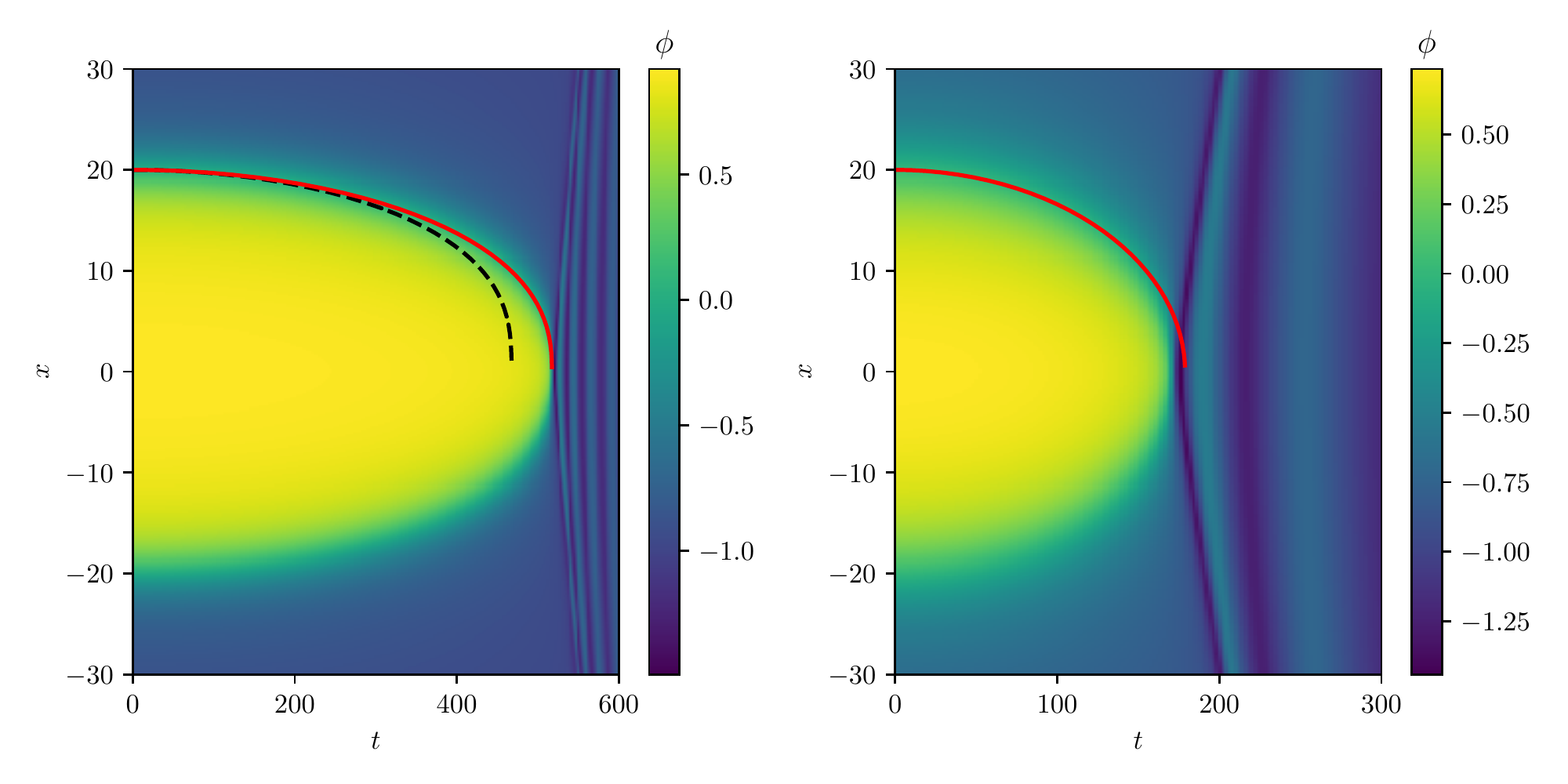}
       \caption*{Source: Results obtained by the author in \cite{campos2021interaction}.}
       \label{fig4_static}
\end{figure}

We start this analysis by the simulation of kink-antikink collisions that are initially static. The result is shown in Fig.~\ref{fig4_static}. First, we compare the evolution of the system before the collision to the analytical expression of the acceleration given by eq.~(\ref{eq4_accel}). The two do not match very well due to the presence of the second term in the tail expansion, as argued before. In the figure, we show for the $\phi^8$ the evolution of a particle accelerating according to that expression after prescribing $A\to A-\Delta A$. This solution has a reasonable agreement with simulations. On the other hand, we do not have a simple prescription for improving the acceleration estimate for the $\phi^{12}$ model. As a consistency check, we can compare the full evolution of the system with the acceleration obtained by integrating eq.~(\ref{eq4_etap0}) numerically without further approximations. More specifically, we find the acceleration numerically starting from an initial guess for $a$, and we integrate the equation starting with the value of $\eta$ such that $\eta^\prime=0$ from $x=0$ to $x=A$. Then, it becomes a root-finding problem. We search for a value of $a$ such that $\eta=0$ at $x=A$. This condition means that the kink is centered at $x=A$. As expected, the two results match very well, as shown in the solid lines in the figure.

\begin{figure}[tbp]
\centering
  \caption{Comparison of approximate calculations of the initial acceleration of an interacting kink to numerical simulations. Left plot corresponds to $\phi^8$ model and the right one, to the $\phi^{12}$ model.}
  \includegraphics[width=0.9\columnwidth]{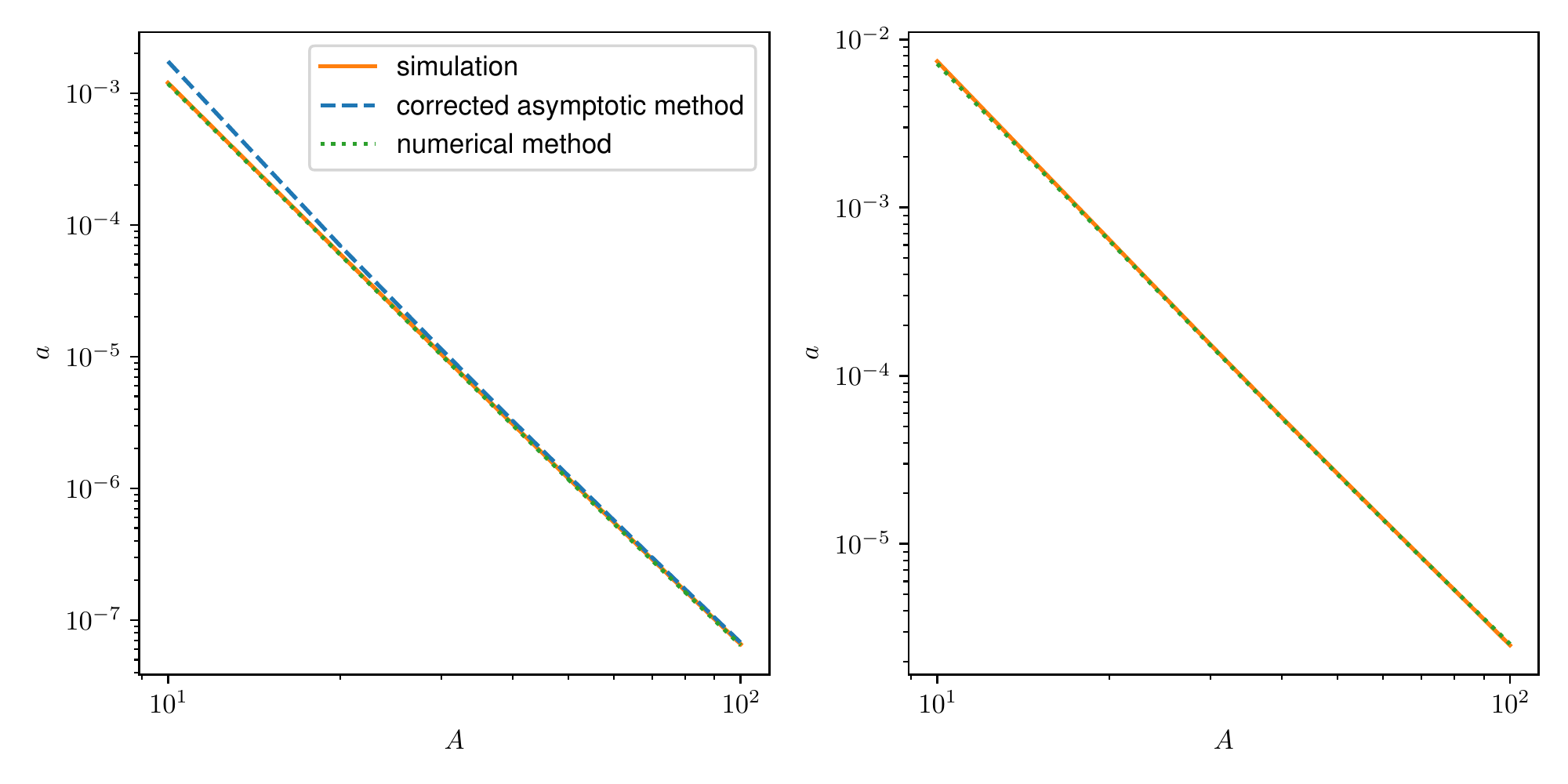}
       \caption*{Source: The author (2022).}
       \label{fig4_avsA}
\end{figure}

The comparison between the different methods to estimate the acceleration is shown in fig.~\ref{fig4_avsA}. We consider both the $\phi^8$ and $\phi^{12}$ models. To obtain the initial acceleration in the simulations, we evolve the system for a short time for several values of $A$ and obtain the acceleration using a stencil approximation for each simulation. The integration of eq.~(\ref{eq4_etap0}) without further approximations as described above is shown as the numerical method. It agrees perfectly with simulations and serves as a consistency check. For the $\phi^8$ model, we also show the result in eq.~(\ref{eq4_accel}) after including the prescription $A\to A-\Delta A$. In the figure, we call it the corrected asymptotic method. It has a reasonable agreement with the previous ones.

In Fig.~\ref{fig4_static}, it should be noticed that the system annihilates directly into radiation. This is in stark contrast to the behavior of kink-antikink collisions in most models, where a bion is formed. We argue that this occurs because the backtail is also long-range. The qualitative argument for this phenomenon will be described below.

First, let us discuss the behavior in the $\phi^4$ model or any similar models where both tails of the kinks decay exponentially. The interaction starts with the kink on the left and the antikink on the right. Then, the two elements temporarily annihilate, and the system reaches a state where the field is near the vacuum all over space. Due to the system's momentum, the field keeps decreasing, bounces back, and the kink-antikink pair reappears. The pair always separate above the critical velocity, while, for smaller velocities, it can either separate after multiple bounces or form a bion, which slowly decays to the trivial vacuum. During this process, the perturbations around the vacuum are massive, and it takes a long time to dissipate the energy. Therefore, the kink-antikink configuration can be easily recovered.

If the tail facing the opposing kink in a kink-antikink collision is long-range, the same picture applies. The kink and the antikink approach each other, temporarily annihilate and reach a vacuum value all over space. As the perturbations around this vacuum are also massive, the behavior remains the same. However, if the backtail is also long-range, perturbations around that vacuum are massless. Therefore, kink-antikink energy will dissipate much faster, meaning the kink-antikink configuration cannot be recovered, and the bion is not formed.

\begin{figure}[tbp]
\centering
  \caption{Field configuration in spacetime for kink-antikink collisions. We consider the $\phi^8$ model with several initial velocities.}
  \includegraphics[width=0.9\columnwidth]{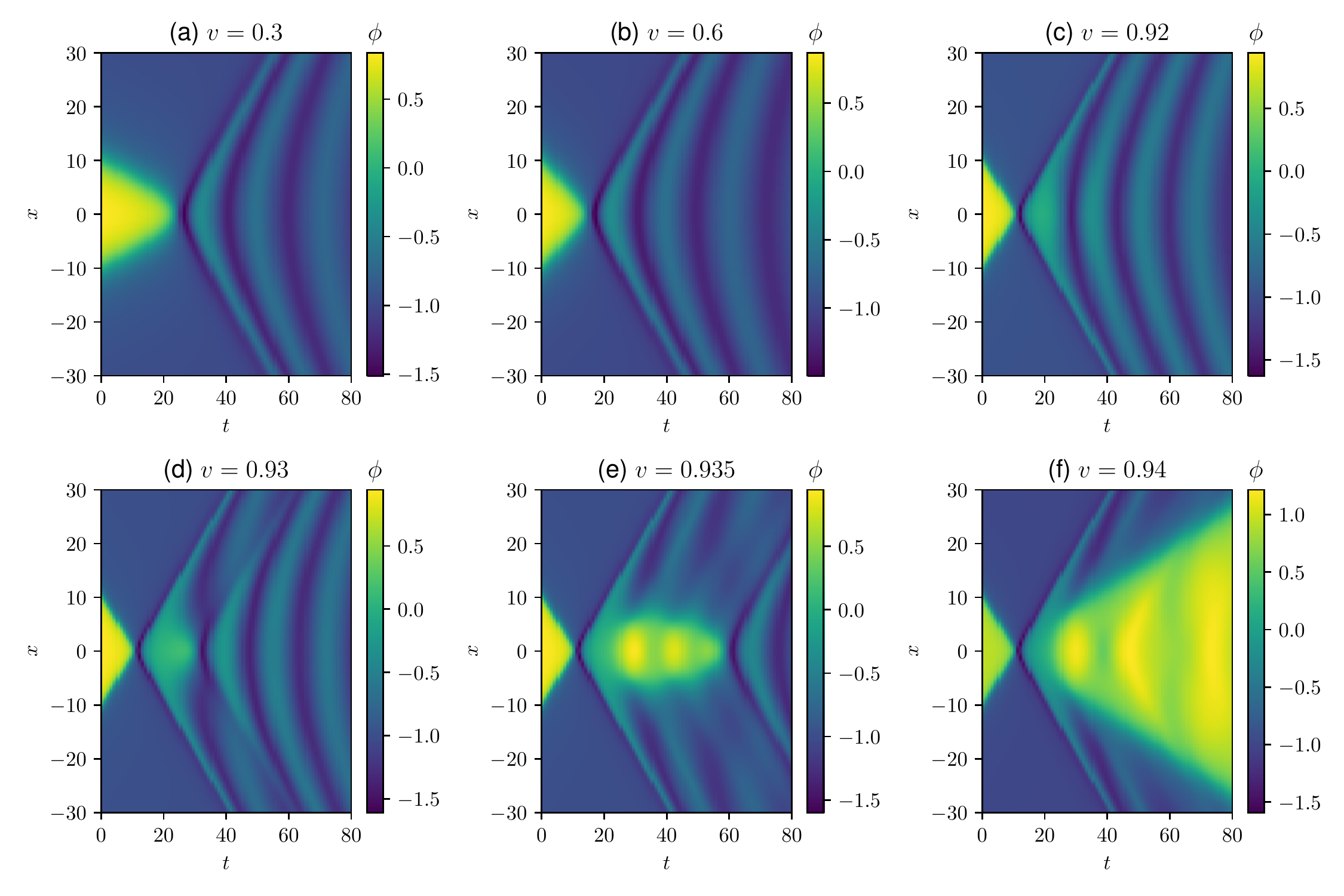}
       \caption*{Source: Results obtained by the author in \cite{campos2021interaction}.}
       \label{fig4_phi8}
\end{figure}

The next step is to consider kink-antikink collisions with a nonzero initial velocity. In this case, we need to use the full minimization method for the initial configuration of the field and the velocity field. The result is shown in Fig.~\ref{fig4_phi8} for the $\phi^8$ model. We see that the system annihilates directly into radiation until very large velocities. This behavior is quite similar to the static case. However, we find that the system can separate after a critical velocity that is ultra-relativistic. The critical velocity $v_c$ depends on the initial separation of the simulations. For $A=10.0$ we find $v_c\simeq 0.94$. We also observe that, in the reflection regime, the system still produces a large amount of radiation. Interestingly, at $v=0.935$, we observe that the system tries to separate, but the final velocity is not large enough to overcome the attraction between the kink and the antikink. Thus, they collide one more time and annihilate completely. This is the closest configuration we have to a bion, but they are not long-lived at all. This is because the velocity at the second collision is not large enough, and the system will always annihilate into radiation at the second bounce.

\begin{figure}[tbp]
\centering
  \caption{Field configuration in spacetime for kink-antikink collisions. We consider the $\phi^{12}$ model with several initial velocities.}
  \includegraphics[width=0.9\columnwidth]{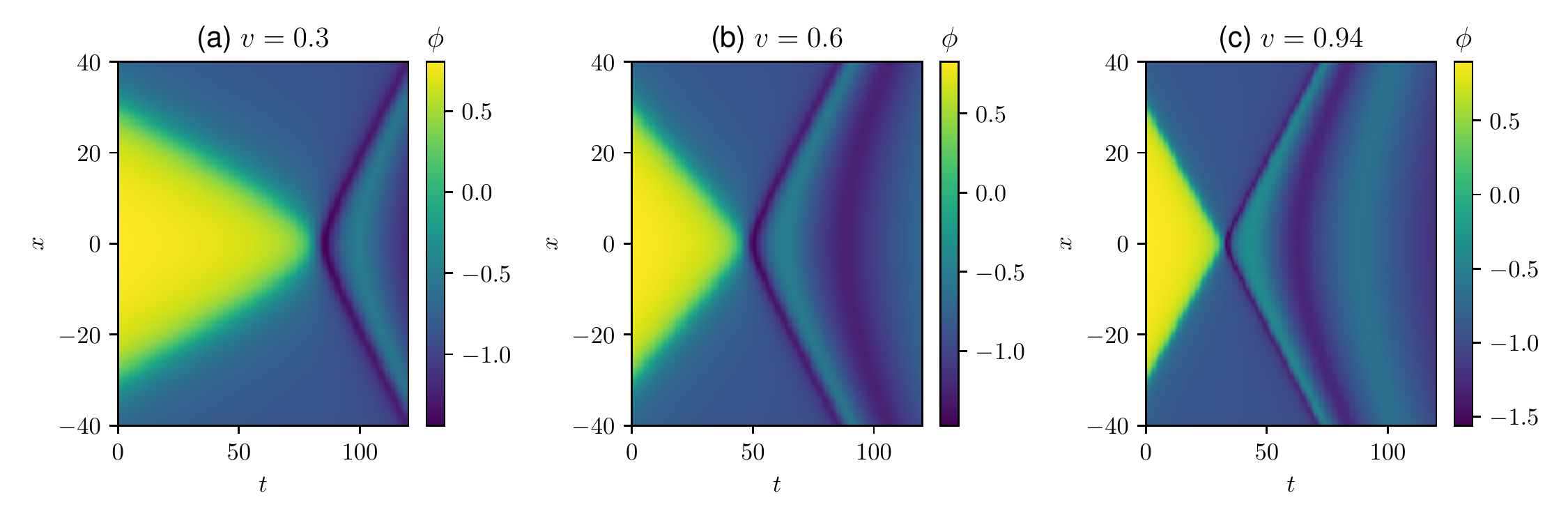}
       \caption*{Source: Results obtained by the author in \cite{campos2021interaction}.}
       \label{fig4_phi12}
\end{figure}

Finally, we repeat the process for the $\phi^{12}$ model. The result is shown in Fig.~\ref{fig4_phi12}. Again we find annihilation directly into radiation. Moreover, we expect that the critical velocity, in this case, is even larger than the previous one because the long-range character is even more pronounced. However, the minimization method does not converge to the correct solution if the velocity is too large. Hence, we were not able to find the critical velocity in this case.

It is clear from the results above that this model does not exhibit resonance windows. It is important to emphasize that this result was already expected due to the absence of a vibrational mode. Moreover, the long-range character of the backtail also suppresses bion formation and causes the system to annihilate directly into radiation. The next section will summarize our findings and discuss ideas for future works.

\section{Conclusion}

In this main section, we have discussed a class of scalar field models exhibiting kink solutions with long-range tails at both ends. We called them kinks with double long-range tails. We found the kink solutions explicitly and showed that the stability equation does not allow vibrational modes. Moreover, we showed that the method developed by Manton \cite{manton2019forces} can be applied in our case, but the approximations are considerably worse here because there is a second-order term in the tail expansion of the kinks.

In order to simulate kink-antikink collision, we developed a minimization method for the velocity field, building upon previous works on kinks with long-range tails \cite{christov2019long, christov2019kink, christov2021kink}. The idea behind the method is to find a configuration of the velocity field that obeys as close as possible the equation of the zero-mode of the field configuration. The field configuration, in turn, is obtained by the minimization of the equations of motion for a traveling wave. We checked the accuracy of our model against a reference configuration obtained from the evolution of a nearly static kink-antikink configuration.

We simulated kink-antikink collisions for several initial velocities, and we found that the system annihilates directly into radiation, except for ultra-relativistic velocities. We argued that this is related to the long-range character of the kink's backtail. After the ultra-relativistic critical velocity, the kinks separate, and near the critical velocity, a short-lived bound state appears. Furthermore, the system does not exhibit resonance windows as expected from the absence of a vibrational mode.

Ideas for future works include applying our initialization procedures for other classes of models as can be found, for instance, in \cite{khare2014successive, khare2019family}. Moreover, one could study interactions of long-range kinks with boundaries, which should be an exciting but challenging problem.

\chapter{WOBBLING DOUBLE SINE-GORDON KINKS}
\label{chap_wob}
\section{Overview}

Notably, the vibrational mode has an essential influence on the kink's behavior. First of all, in non-integrable theories, the vibrational mode is usually excited when the kink interacts with other elements. This mode can store energy and, therefore, if the translational energy is converted into vibrational during an interaction, it can be recovered in subsequent events. This property leads to the well-known resonance phenomenon. Furthermore, a vibrating kink will likely come up after the interaction. Therefore, it is indispensable to investigate the properties of vibrating kinks. Some of these properties are described in \cite{manton1997kinks, barashenkov2009wobbling, oxtoby2009resonantly}. For instance, it is now well known that the kink's vibration decays in time via the coupling between the first harmonic and radiation. 

These ideas lead to the question of how kinks interact if they are vibrating beforehand. This problem was investigated for the first time recently in \cite{izquierdo2021scattering} for the $\phi^4$ model. The authors argued that this analysis is important because the subsequent bounces in resonance windows can be interpreted as an iteration of wobbling kink collisions. Therefore, these collisions are important to understand the resonance phenomenon.

In \cite{izquierdo2021scattering} the authors found many exciting results, such as the appearance of one-bounce resonance windows. Moreover, they analyzed the final amplitude and frequency of the kinks after they separate and found that amplitude is usually large and that the frequency is Lorentz contracted.

Building upon the work mentioned above, we study collisions between wobbling kinks in the double sine-Gordon model. This model differs form the $\phi^4$ because the potential is periodic and has a free parameter $R$. Moreover, the model becomes integrable in both limits where $R$ is very small and very large. Interestingly, it gives an effective description of a wide range of physical systems, such as gold dislocations \cite{el1987double}, optical pulses and spin waves \cite{bullough1980double}, pseudo 1-D ferromagnets \cite{rettori1986double} and Josephson structures \cite{alfimov2014discrete}. 

There are quite a few works already on the double sine-Gordon model. Examples are \cite{kivshar1987radiative, malomed1989dynamics, kivshar1989dynamics}, where it is studied perturbatively using the inverse scattering method. Examples including analyses with a numerical focus are \cite{campbell1986kink, gani1999kink, gani2018scattering, gani2019multi, simas2020solitary}. However, the model is still not fully comprehended. For instance, the dependence of the critical velocity on the parameter $R$ was only found recently \cite{ gani2018scattering}. We will see that the study of colliding wobbling kinks in the double sine-Gordon leads to some novel results such as the appearance of spines in the region of one-bounce windows as well as of a structure of resonance windows near the integrable limits which is hidden in collisions between kinks that are not wobbling. The work described below resulted in the following publication \cite{campos2021wobbling}.

In the next section we describe the double sine-Gordon and its main properties.

\section{Model}

\begin{figure}[tbp]
\centering
  \caption{Double sine-Gordon (a) potential, (b) kink and (c) linearized potential for several values of the paramter $R$.}
  \includegraphics[width=0.9\columnwidth]{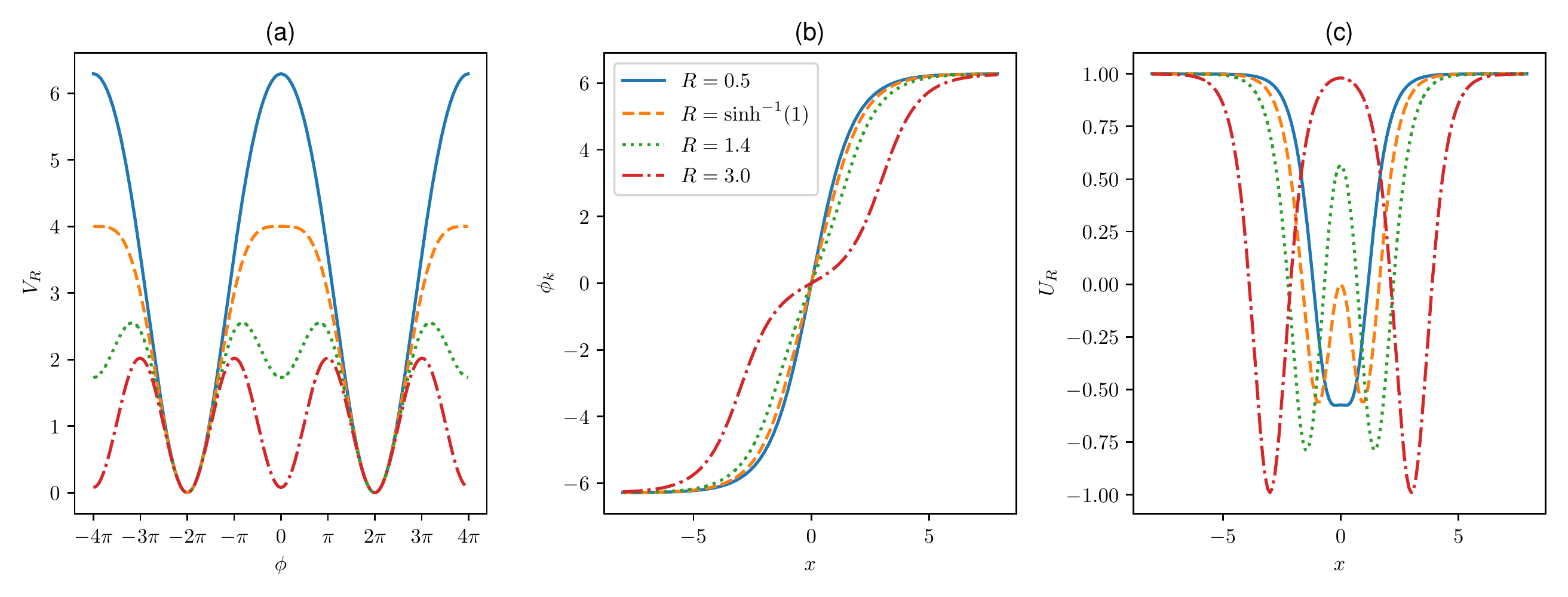}
       \caption*{Source: Results obtained by the author in \cite{campos2021wobbling}.}
       \label{fig5_kink}
\end{figure}

We will investigate the double sine-Gordon model described by a scalar Lagrangian
\begin{equation}
\mathcal{L}=\frac{1}{2}(\partial_\mu\phi)^2-V_\eta(\phi),
\end{equation}
with the following potential
\begin{equation}
V_\eta(\phi)=\frac{4}{1+4|\eta|}\left(\eta(1-\cos\phi)+1+\cos\frac{\phi}{2}\right).
\end{equation}
The case where $\eta<0$ was investigated in detail in \cite{simas2020solitary}. We are interested, however, in the regime with $\eta>0$. In this case, the constant $\eta$ can be written as $\eta=\frac{1}{4}\sinh^2R$ and the potential becomes
\begin{equation}
V_R(\phi)=\tanh^2R(1-\cos\phi)+\frac{4}{\cosh^2R}\left(1+\cos\frac{\phi}{2}\right).
\end{equation}
This potential is periodic with degenerate vacua at $\phi=2\pi(2n+1)$, with $n$ integer. The potential as a function of the field is shown in Fig.~\ref{fig5_kink} for some values of the parameter $R$. For $R=0$, the potential corresponds to the sine-Gordon one, which is integrable. At $R>\text{arcsinh}(1)\simeq0.881$, the potential possesses a local minima at $\phi=4\pi n$, for $n$ integer. These minima are responsible for the inner structure of the kink. For large values of $R$, the local minima of the potential gradually become global minima and the potential approaches again the sine-Gordon one, but with half the period.

Using the BPS equation (\ref{eq12_BPS2}), finding the kink solutions is a trivial task. They are given by
\begin{equation}
\phi_{k(\bar{k})}=4\pi n\pm4\tan^{-1}\left(\frac{\sinh(x)}{\cosh R}\right).
\end{equation}
This expression can be written in terms of the sine-Gordon kink expression, $\phi_{SG}=4\tan^{-1}\exp(x)$. One can express the double sine-Gordon kink as a linear combination of two sine-Gordon subkinks
\begin{equation}
\phi_{k(\bar{k})}=4\pi n\pm[\phi_{SG}(x+R)-\phi_{SG}(R-x)].
\end{equation}
When a kink is composed of more than one subkink, it possesses an inner structure. This general property occurs when the potential possesses a local minimum between its vacua. The kinks' profiles are shown in Fig.~\ref{fig5_kink} for some values of the parameter $R$. The inner structure can be clearly seen for $R=3.0$.

\begin{figure}
\centering
  \caption{Frequency of the shape mode as a function of the parameter $R$. The zero mode is indicated by a dashed line.}
  \includegraphics[width=0.5\columnwidth]{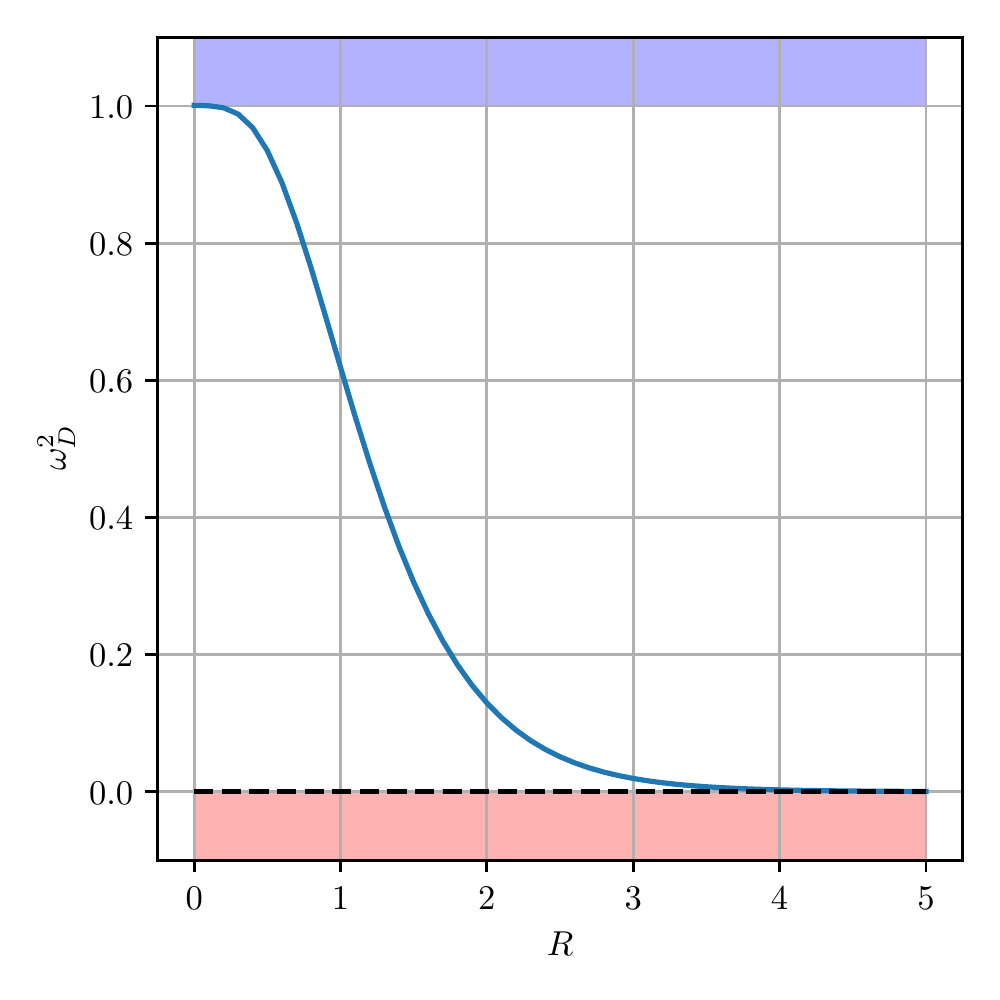}
       \caption*{Source: Results obtained by the author in \cite{campos2021wobbling}.}
       \label{fig5_stab}
\end{figure}

The resonance behavior depends on the spectrum of perturbations around the kink solution. Writing the field as
$\phi=\phi_k+\psi\cos(\omega t)$ and substituting in the equations of motion we arrive at the following stability equation
\begin{equation}
-\psi^{\prime\prime}+U_R\psi=\omega^2\psi,
\end{equation}
where $U_R$ is called the linearized potential and is defined as $U_R\equiv V_R^{\prime\prime}(\phi_k)$. More explicitly, one can arrive at \cite{gani2018scattering}
\begin{equation}
U_R=\frac{8\tanh^2R}{(1+\text{sech}^2R\sinh^2x)^2}+\frac{2(3-4\cosh^2R)}{\cosh^2R(1+\text{sech}^2R\sinh^2x)}+1.
\end{equation}
This stability equation possesses two discrete solutions. The first one is the zero mode, which is related to the translational invariance of the model, and the second one is the vibrational mode, also known as the shape mode. Finally, we have the continuum solutions for $\omega>1$. We denote the shape mode's frequency by $\omega_D$ and the normalized profile by $\psi_D$. Using the expression for the energy of the scalar field $\phi$, it is straightforward to show that the energy associated with the perturbation $A\psi_D\cos(\omega_Dt)$ is $E_D=\frac{1}{2}\omega_D^2A^2$. 

In \cite{campbell1986kink}, the authors were able to derive analytical approximations for the shape mode frequency both near $R=0$ and for large $R$. We will give a brief discussion of their analysis. For $R=0$, the model corresponds to the sine-Gordon, which does not possess a vibrational mode. However, the authors showed that as $R$ increases, it could be considered a good approximation that the threshold state becomes a vibrational mode. The threshold state is the lowest energy state in the continuum modes. For large $R$, the kink is composed of two widely separated sine-Gordon subkinks. As can be seen in Fig.~\ref{fig5_kink}, in this limit, the linearized potential $U_R$ becomes two widely separated sine-Gordon wells. The separate sine-Gordon potentials contains a zero mode each. Accordingly, the zero mode and shape mode of the double sine-Gordon kink correspond, respectively, to the symmetric and antisymmetric combinations of these two modes. The authors showed that this is a good approximation for the exact shape mode, even for intermediate values of the parameter $R$. 

\begin{figure}[tbp]
\centering
  \caption{(a) Amplitude and (b) frequency of the kink's vibration as a function of time for several values of $R$. (c)-(f) Power spectrum of the vibration near and far from the kink for large times with same values of $R$. The initial amplitude is $A=1.0$.}
  \includegraphics[width=\columnwidth]{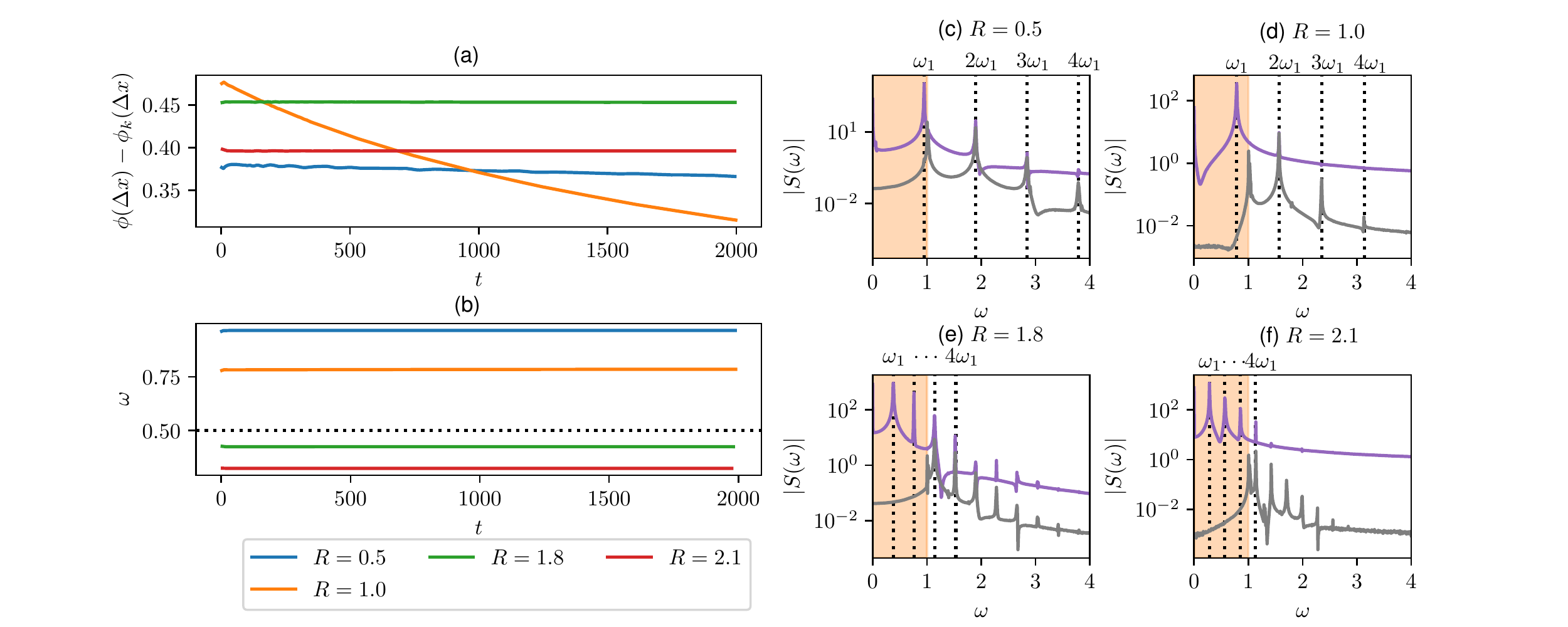}
       \caption*{Source: Results obtained by the author in \cite{campos2021wobbling}.}
       \label{fig5_deepmode}
\end{figure}

It is known that the amplitude of the kink's vibration slowly decays with time. This occurs due to the coupling of the first harmonic of the wobbling frequency with the radiating modes \cite{manton1997kinks}. This coupling appears when one considers discrete and continuous modes as a perturbation to the kink solution and expands the equations of motion up to higher orders. Moreover, if the shape mode's frequency is lower than half the threshold, the first harmonic is not large enough to couple with radiation, and the decay occurs via harmonics higher than the first one. It can be shown that this suppresses the decay \cite{romanczukiewicz2018oscillons}.  

The effects described above are also observed in the double sine-Gordon model as shown in Fig.~\ref{fig5_deepmode}. We simulate a vibrating kink at the origin. In the left panels, we plot the amplitude and frequency of the vibration at a point shifted by $\Delta x=1.465$ from the origin. This point is near the position where the amplitude of the shape mode is maximum. The vibration is measured by the difference $\phi(x)-\phi_k(x)$. From the figure, a few things can be noticed. First, as expected, for $R=0.5$ and $R=1.0$, we observe the expected decay for a frequency above half the threshold. However, for $R=0.5$, the decay is less pronounced. This effect becomes even more noticeable as we get closer to the integrable limit at $R=0$. Now, turning to large values of $R$, we see that the decay is suppressed for $R=1.8$ and $R=2.1$ because the frequency is below half the threshold. To corroborate these results, we plot the power spectrum of the time series of the vibration. Near the kink at $x=\Delta x$, we find that the largest peak is at the shape mode's frequency. Far from the kink at $x=25.0$, the largest peak is at harmonics higher than the threshold. These results are in agreement with the picture described above.

In the next section, we will discuss the behavior of kink-antikink collisions in the double sine-Gordon model. We will investigate two different scenarios, the usual collision, and collisions between wobbling kinks.

\section{Collision simulations}

\begin{figure}[tbp]
\centering
  \caption{(a) Final velocity as a function of initial velocity for usual kink-antikink collisions with $R=1.0$. (b) Bounce regions as a function of $R$ and $v_i$ for the same type of collisions.}
  \includegraphics[width=0.9\columnwidth]{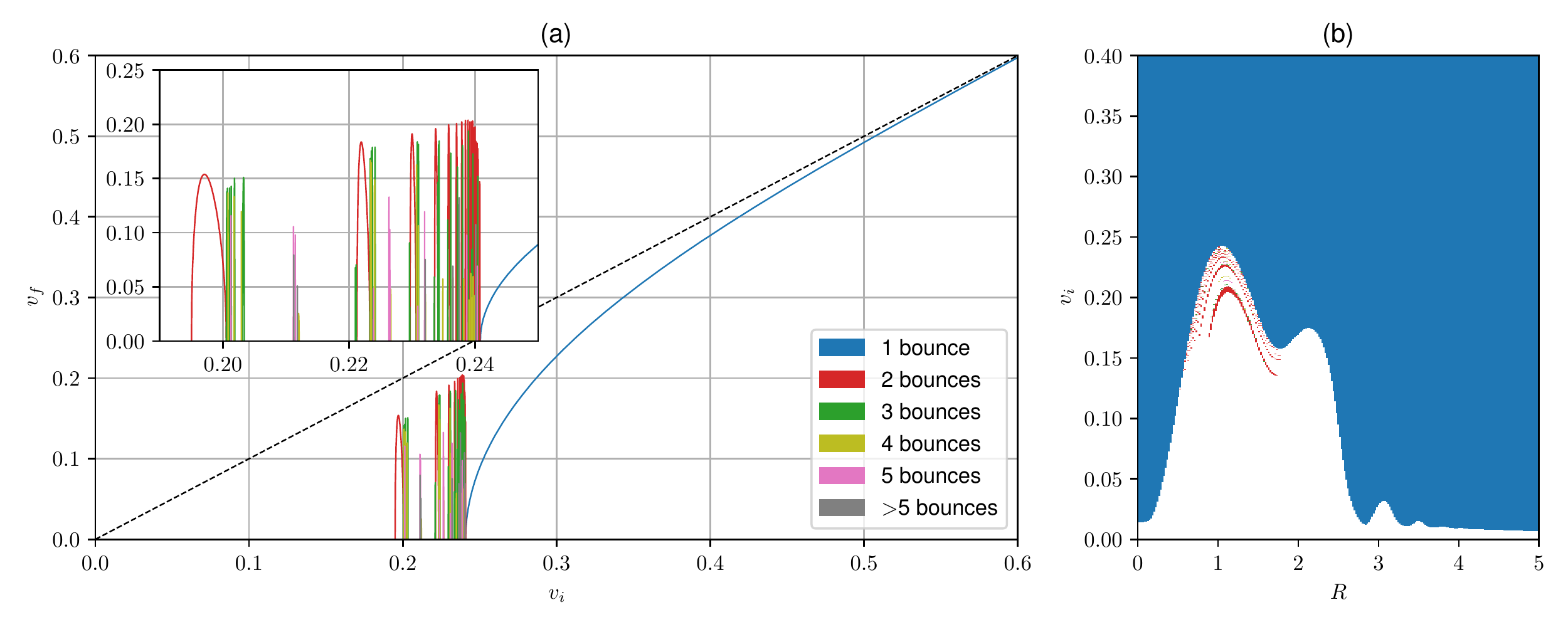}
       \caption*{Source: Results obtained by the author in \cite{campos2021wobbling}.}
       \label{fig5_res}
\end{figure}

Now, let us turn to collisions between wobbling double sine-Gordon kink and antikink. This corresponds to the initial conditions given by the following expression
\begin{equation}
\label{eq5_IC}
\phi=\phi_k(\xi_+)-\phi_k(\xi_-)-2\pi+A\sin(\omega_D\tau_+)\psi_D(\xi_+)-A\sin(\omega_D\tau_-)\psi_D(\xi_-),
\end{equation}
evaluated at $t=0$. We defined the Lorentz transformation of the coordinates as $\xi_{\pm}=\gamma(x\pm x_0\mp v_it)$ and $\tau_\pm=\gamma(t\mp v_ix)$, where $\gamma=1/\sqrt{1-v_i^2}$. These initial conditions correspond to a kink on the left vibrating with frequency $A$ and an antikink on the right vibrating with the same frequency. The two are approaching each other with velocities $\pm v_i$, respectively. The details of the simulations are described in section \ref{sec5_num}. It is clear from eq.~(\ref{eq5_IC}) that the wobbling frequency is Lorentz contracted in the center of mass frame. To see that, we measure the field at a position moving with the kink $x=v_it+\alpha$. In this case, we find that the wobbling term becomes
\begin{equation}
\label{eq5_lorentz}
A\sin\left(\frac{\omega_Dt}{\gamma}-\gamma\omega_Dv_i\alpha\right)\psi_D(\gamma(x_0+\alpha)),
\end{equation}
which oscillates with the Lorentz contracted frequency $\omega_D/\gamma$. It should be clear that, to measure this effect, we need to choose $\alpha\neq-x_0$, so that contribution from $\psi_D$ does not vanish at the point that we are measuring.

\begin{figure}[tbp]
\centering
  \caption{Value of the field at the center of the collision as a function of $t$ and $v_i$. We consider usual kink-antikink collisions with (a) a small value of $R=0.3$ and (b) a large value of $R=3.0$.}
  \includegraphics[width=0.9\columnwidth]{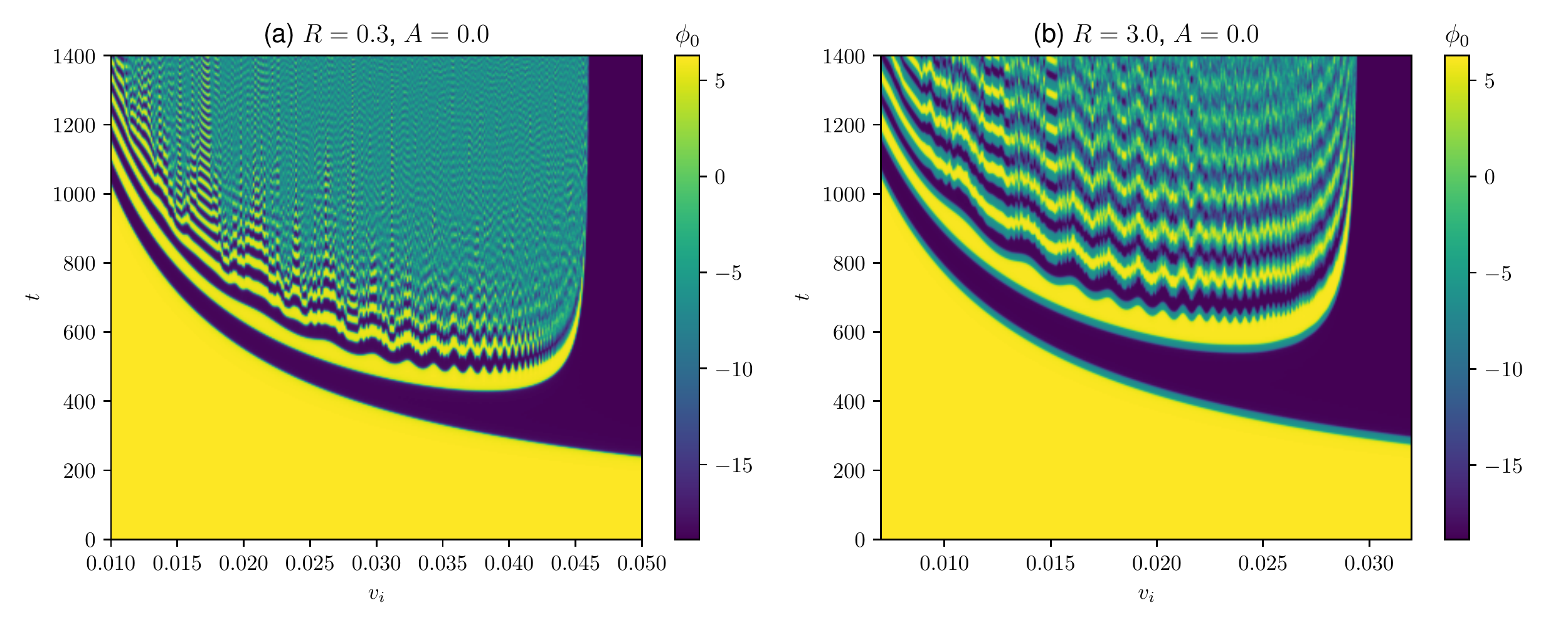}
       \caption*{Source: Results obtained by the author in \cite{campos2021wobbling}.}
       \label{fig5_center}
\end{figure}

We start the analysis by considering usual kink-antikink collisions with $A=0$. It is important to mention an important difference between the double sine-Gordon and the $\phi^4$ model. In the former, the potential is periodic, and a bounce occurs every time the kink and the antikink cross, while, in the latter, a bounce corresponds to a reflection between the two. The result considering $R=1.0$ is shown in Fig.~\ref{fig5_res} showing the usual behavior. For $v_i$ above a critical value, there is crossing after one bounce, while there are either annihilation or resonance windows for smaller velocities. We depict the number of bounces in the resonance window by a color code explained in the figure. In the right panel of the same figure, we summarize the regions where resonances occur using this color code. The blue region corresponds to crossing after one bounce and the white region to annihilation. The interface between the two corresponds to the critical velocity, which is in agreement with the non-monotonic behavior found in \cite{gani2018scattering}.
Moreover, the fractal structure appears in the red region. From the figure, we can estimate that fractal behavior occurs for $0.52<R<1.78$, which corresponds to wobbling frequencies $0.437<\omega_D<0.963$. Of course, the actual region should be slightly larger because the resonance windows get increasingly narrower and become harder to localize. These results indicate no resonance windows near-integrable regimes, presumably because the degrees of freedom decouple. This is in agreement with first-order perturbation theory, which predicts that there is no energy and momentum transfer between the kinks in this limit \cite{kivshar1989dynamics}. However, a more detailed analytical explanation for this phenomenon is still lacking.

We analyzed the system's behavior for large and small $R$ to explore the near-integrable limits and find out whether resonance windows exist or not. In Fig.~\ref{fig5_center}(a), we plot the value of the field at the center of the collision as a function of time and the initial velocity for $R=0.3$. This figure should be read as follows. Every vertical line corresponds to the field at the center of the collision for the corresponding initial velocity. Every time the field changes color in a vertical line, a bounce occurs. For large $v_i$, the field at the center starts yellow and then becomes purple. This means that the kink and the antikink bounced once and then separated. Below the critical velocity, the field oscillates between yellow and purple and never reaches a constant value. This means that a bion is formed and, therefore, annihilation occurs. Thus, we confirm that there are no resonance windows and, instead, we find false resonance windows. The false resonance windows are small peaks when the third bounce occurs. This means that the system is slightly closer to separating at a few points but returns and bounces again. The resonant energy exchange mechanism can also describe the position of false resonance windows. Hence, one can say that they are a weaker type of resonance. Likewise, in Fig.~\ref{fig5_center}(b), we do the same analysis for $A=3.0$, and we only observe false resonance windows. The absence of resonance windows in nearly integrable models has also been reported in \cite{simas2020solitary, dorey2021resonance}.

\begin{figure}[tbp]
\centering
  \caption{(a) Final velocity as a function of the initial velocity for $R=1.0$ and $A=0.1$. (b) Number of bounces as a function of $R$ and $v_i$ for $A=0.1$.}
  \includegraphics[width=0.9\columnwidth]{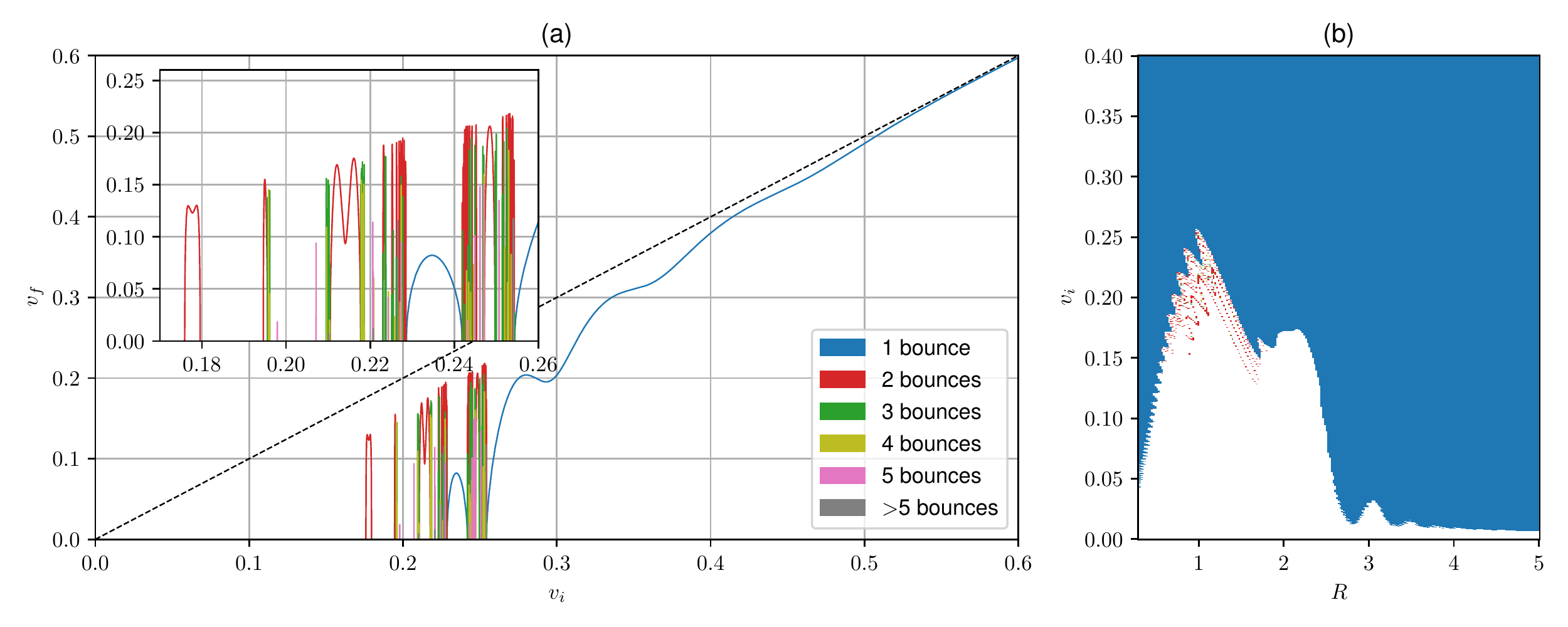}
       \caption*{Source: Results obtained by the author in \cite{campos2021wobbling}.}
       \label{fig5_res1}
\end{figure}

The next step is to turn on the wobbling. We initialize the system according to eq.~(\ref{eq5_IC}) with $A=0.1$. The result is shown in Fig.~\ref{fig5_res1}. We plot the final velocity as a function of the initial velocity on the left panel for $R=1.0$. The blue curves correspond to the regions where crossing occurs after one bounce. Due to the wobbling of the kinks, this curve oscillates. More specifically, this effect originates from the fact that the wobbling phase at the moment of a collision varies with $v_i$. The wobbling amplitude evolves approximately as
\begin{equation}
\label{eq5_St}
S(t)=A\exp\left[i\left(\frac{\omega_Dt}{\gamma}-\theta_0\right)\right],
\end{equation}
where $\theta_0$ is an initial constant. We consider it complex for simplicity, but the real part should be taken in the end. As the collision takes a time $t=x_0/v_i$ to happen, the phase at the moment of the collision is given by
\begin{equation}
\label{eq5_phase}
\theta=\sqrt{1-v_i^2}\frac{\omega_Dx_0}{v_i}+\theta_0.
\end{equation}
The variation of the phase with $v_i$ is the origin of the oscillation in the blue curve.

Another intriguing effect is that the blue curve splits, meaning that there are two separate blue regions. The small one is called a one-bounce resonance window, while the large is called the crossing curve. Interestingly, near the edges of one-bounce resonance windows, there appears a nested structure of higher-bounce resonance windows. As discussed in \cite{izquierdo2021scattering}, the appearance of one-bounce resonance windows can occur in two ways. The first one occurs as the oscillation of the crossing curve increases and eventually touches the $v_i$-axis. In this case, we say that the crossing curve splits. This occurs because the interaction between the vibrational and translational modes depends on the oscillation phase at the collision. Thus, if the oscillation is large and out of phase, it can lead to the annihilation of the kinks at some point in the crossing curve and split it. The second one occurs when a one-bounce window appears where previously it did not exist. This occurs when the wobbling is significant and in phase at the moment of the collision. In this case, the energy of the wobbling is transferred to the translational mode and allows the separation of the kinks. 

In Fig.~\ref{fig5_res1}(b), we show the number of bounces before separation as a function of $R$ and $v_i$. The color code is the same as before, and the white regions correspond to annihilation. One can see that the effect of wobbling is to create an intricate oscillating structure. These branches in the structure are called spines in \cite{dorey2021resonance}. In this work, the authors described how these structures appear in detail. In short, this structure can be understood if we follow how the resonance windows appear, move, and merge with the crossing curve as we vary the parameter $R$.   

\begin{figure}[tbp]
\centering
  \caption{(a) Final velocity as a function of the initial velocity for $R=1.0$ and $A=0.8$. (b) Number of bounces as a function of $R$ and $v_i$ for $A=0.8$.}
  \includegraphics[width=0.9\columnwidth]{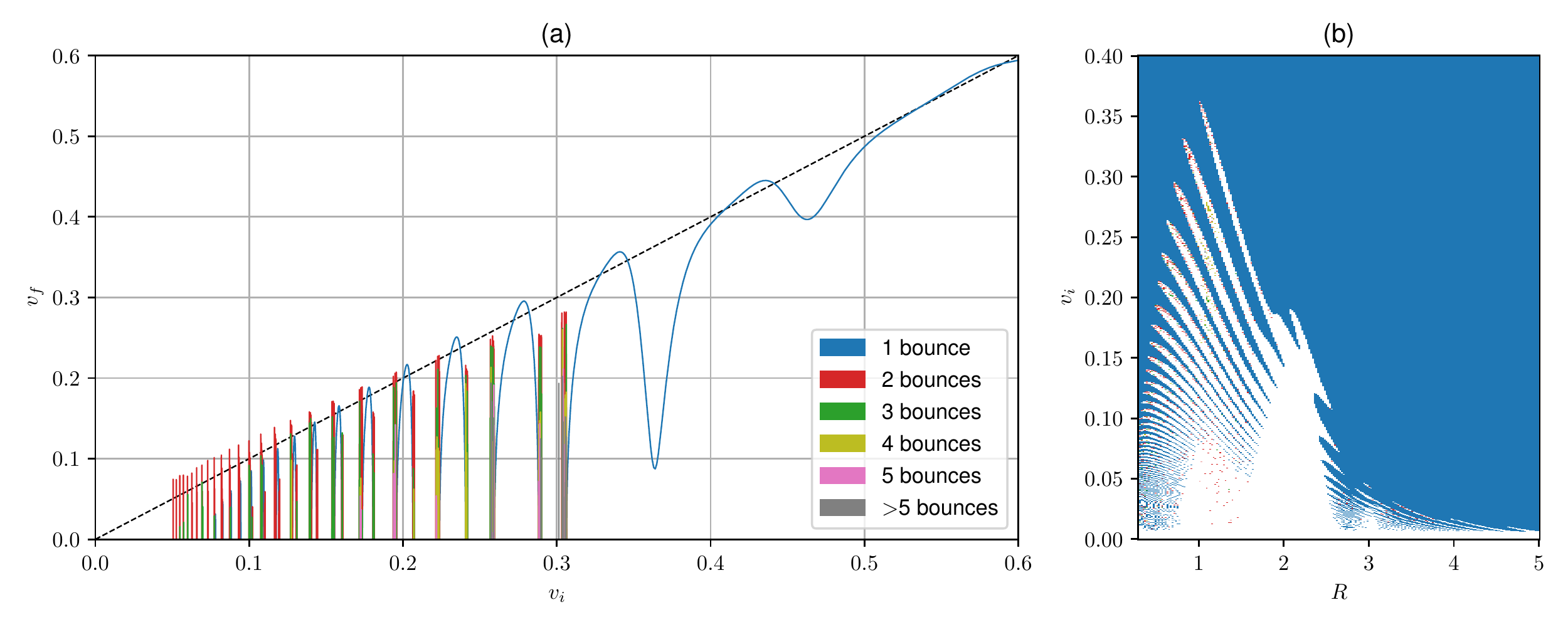}
       \caption*{Source: Results obtained by the author in \cite{campos2021wobbling}.}
       \label{fig5_res8}
\end{figure}

Increasing the wobbling amplitude to $A=0.8$, we arrive at Fig.~\ref{fig5_res8}. We plot the final velocity versus the initial velocity on the left panel. In this case, the wobbling energy is much larger, and its effect is much more noticeable. Accordingly, the oscillation in the crossing curve is more significant, and there is a larger number of one-bounce resonance windows. The one-bounce resonance windows appear in the two ways discussed above. Moreover, the final velocity can be above the identity curve due to the extra vibrational energy available. As the initial velocity decreases, we find many two-bounce resonance windows as well as higher bounce ones. They also appear due to the extra wobbling energy. This structure should continue all the way to $v_i=0$, but the curve is truncated because the computational time for the kinks to separate becomes increasingly larger. 

The number of bounces as a function of $v_i$ and $R$ is shown in Fig.~\ref{fig5_res8}(b). One can see that the spine structure is even more intricate in this case because there are many more one-bounce windows. Interestingly, it is possible to find annihilation for values of $v_i>0.35$. This occurs because if the wobbling is out of phase at the collision, it favors annihilation.

\begin{figure}
     \centering
     \caption{(a) Final velocity versus initial velocity for collisions between wobbling kinks with (a) $R=0.3$ and (b) $R=3.0$. (c) and (d) Value of the field at the center of the collision as a function of $t$ and $v_i$ for the same values of $R$. The amplitude is $A=0.8$.}
     \begin{subfigure}[b]{0.9\textwidth}
         \centering
         \includegraphics[width=\textwidth]{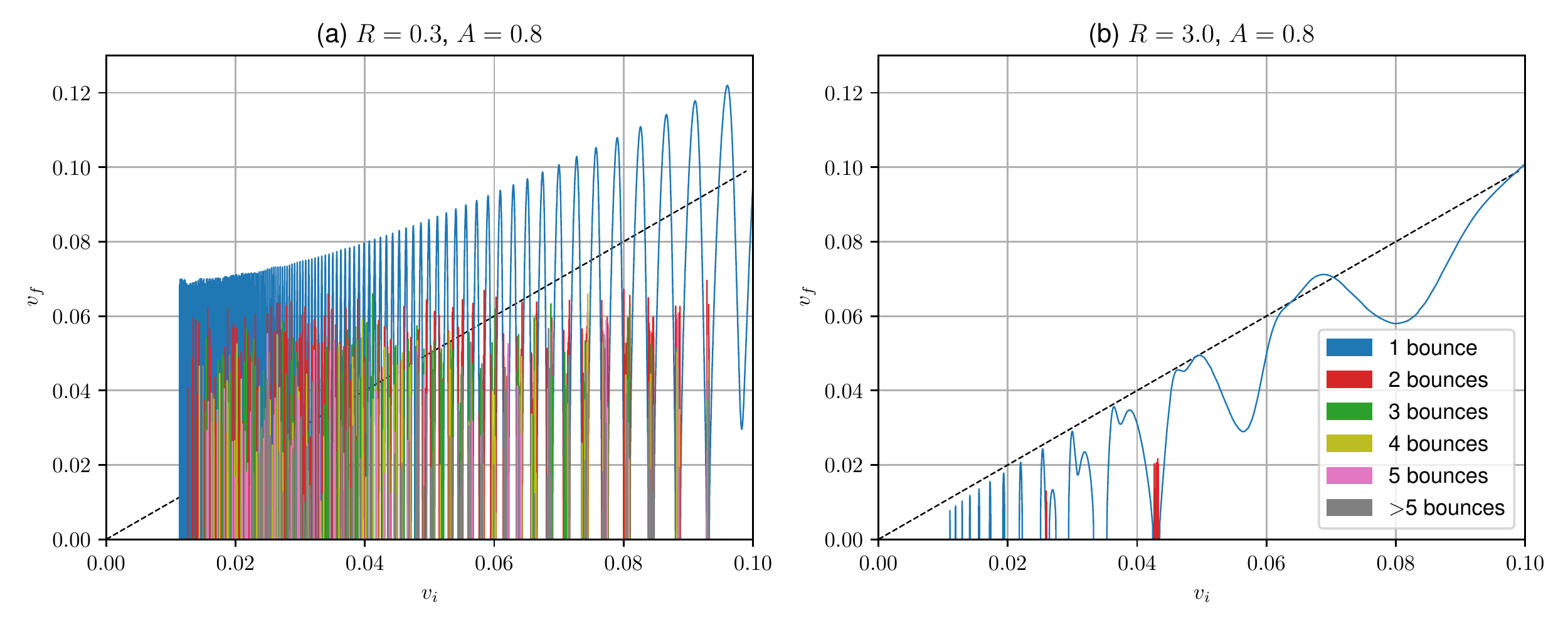}
     \end{subfigure}
     \hfill
     \begin{subfigure}[b]{0.9\textwidth}
         \centering
         \includegraphics[width=\textwidth]{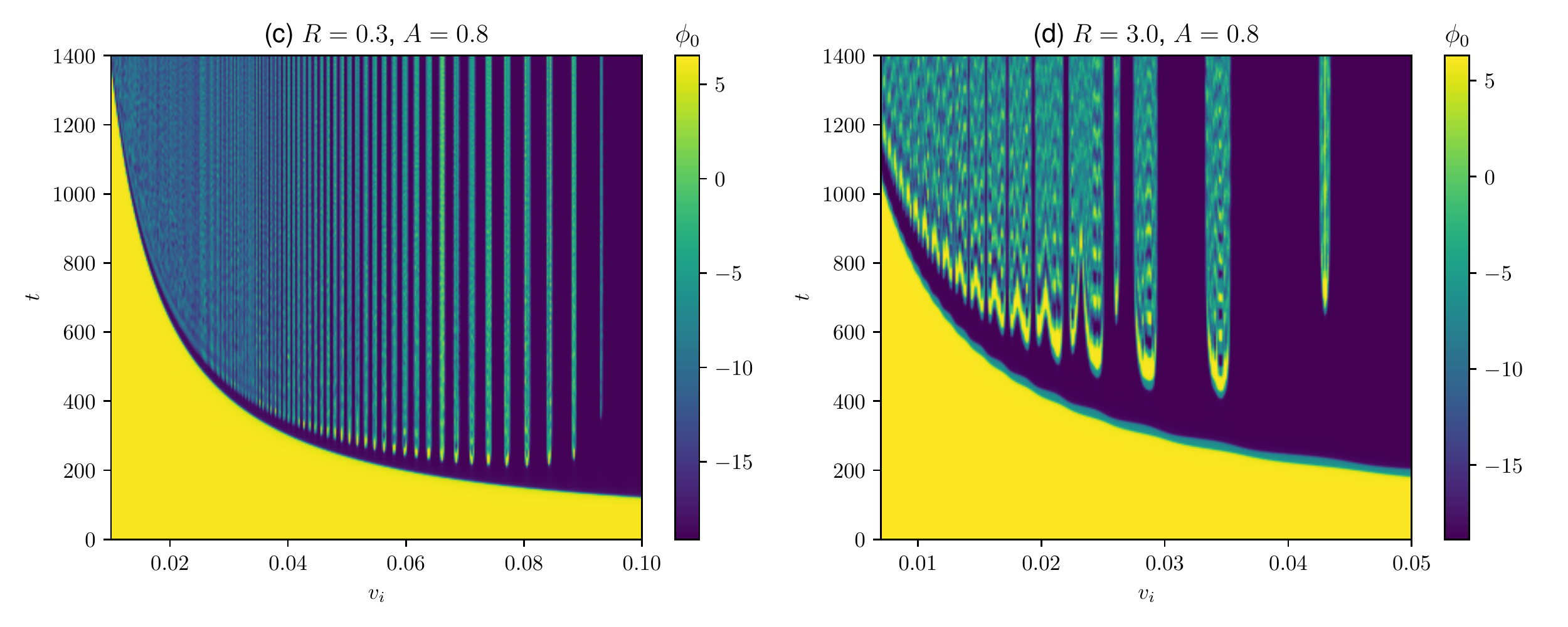}
     \end{subfigure}
        \caption*{Source: Results obtained by the author in \cite{campos2021wobbling}.}
        \label{fig5_rescent}
\end{figure}

If we take a look at Fig.~\ref{fig5_res1}(b) and Fig.~\ref{fig5_res8}(b), we will notice that there are two-bounce resonance windows, denoted as red dots, for small and large values of $R$. Previously, we found that these regions would not have resonance windows if the kinks were not wobbling initially. Therefore, we conclude that the resonant structure was hidden before turning the wobbling on. The plots in Fig.~\ref{fig5_rescent} can confirm it. First, we consider wobbling kink collisions with $A=0.8$ and compute the kinks' final velocity as a function of the initial one for both $R=0.3$ and $R=3.0$. In the former case, there are many higher-bounce windows at the border of the one-bounce ones, while, in the latter, there are only a few two-bounce windows. This confirms the hidden resonant structure. There are more higher-bounce resonance windows for $R=0.3$ than for $R=3.0$. This occurs because the wobbling energy $E_D$ is larger for larger wobbling frequencies, and $\omega_D$ is much larger for $R=0.3$ than for $R=3.0$ as shown in Fig.~\ref{fig5_stab}. Second, as a complement, we compute the field of the center of the collision as a function of time and $v_i$ for the same values of the parameters. In the scale of the figure, it is not possible to see higher-bounce resonance windows, but the structure of one-bounce resonance windows can be easily seen. In particular, for $R=0.3$, the one-bounce windows appear much more frequently. This is expected from eq.~(\ref{eq5_phase}) because, in this case, the wobbling frequency $\omega_D$ is much larger than the $R=3.0$ case.

\begin{figure}[tbp]
\centering
  \caption{(a) Final velocity as a function of the initial velocity for several values of $A$ and $R=1.0$. (b) Bounce regions as a function of $A$ and $v_i$ for $R=1.0$.}
  \includegraphics[width=0.9\columnwidth]{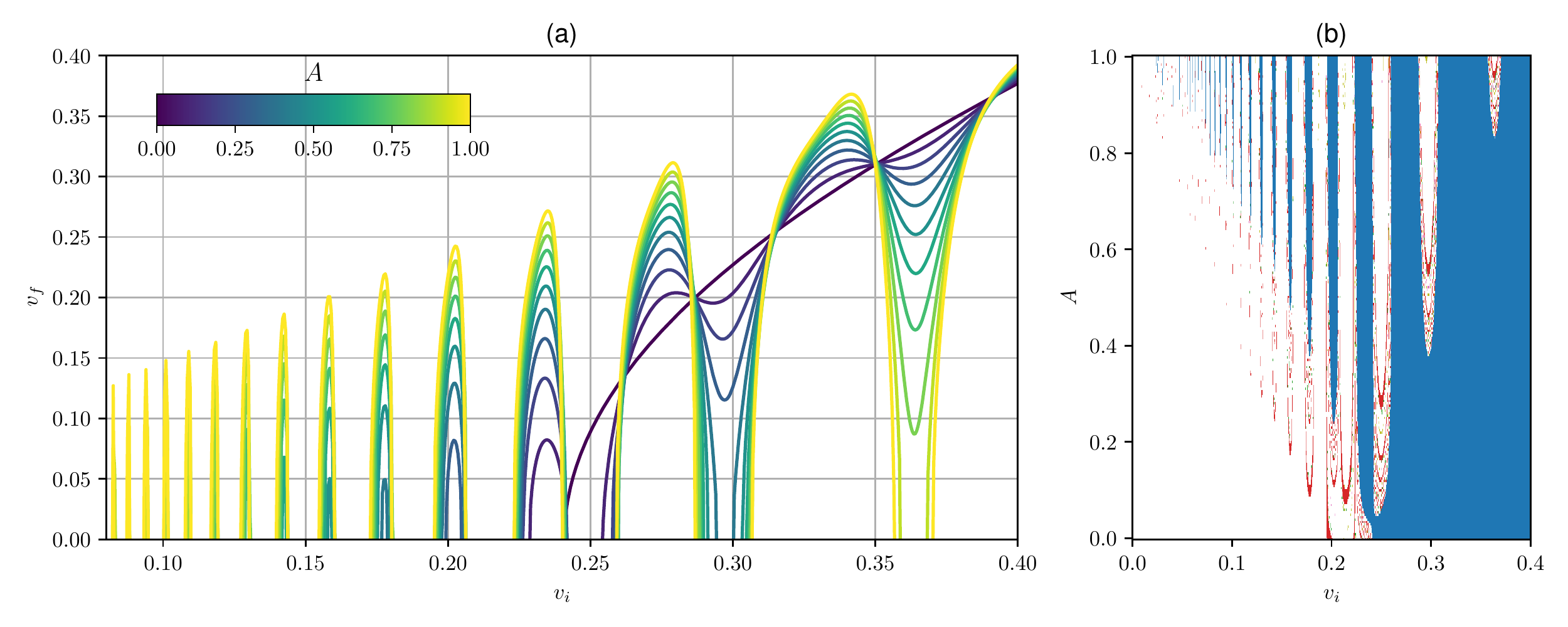}
       \caption*{Source: Results obtained by the author in \cite{campos2021wobbling}.}
       \label{fig5_resVA}
\end{figure}

In Fig.~\ref{fig5_resVA}(a), we plot the final velocity as a function of the initial one but only for one-bounce windows and the crossing curve. We fix $R=1.0$ and show the result for several values of the wobbling amplitude. This is the best way to visualize the appearance of one-bounce resonance windows. As the amplitude is increased, the crossing curve splits, giving rise to some one-bounce windows, whereas others appear as well in regions outside the crossing curve. The same behavior can be seen from a different angle in Fig.~\ref{fig5_resVA}(b). There, we show in blue the regions in the $v_iA$-plane where separation after one bounce occurs. The other colors indicate higher-bounce windows. Again, we are fixing $R=1.0$. As we move from the bottom, with $A=0$, to the top, we find that many one-bounce windows appear. The number of windows becomes larger as $A$ increases because the wobbling becomes more energetic. Again, the two ways that one-bounce windows appear can be distinguished in the figure. In the first one, the blue region is split as $A$ becomes larger due to the appearance of a white region in between. A blue region appears in a white region as $A$ increases in the second one.

\begin{figure}[tbp]
\centering
  \caption{Velocity at the center of a one-bounce resonance window as a function of an index. We fix $R=1.0$.}
  \includegraphics[width=0.56\columnwidth]{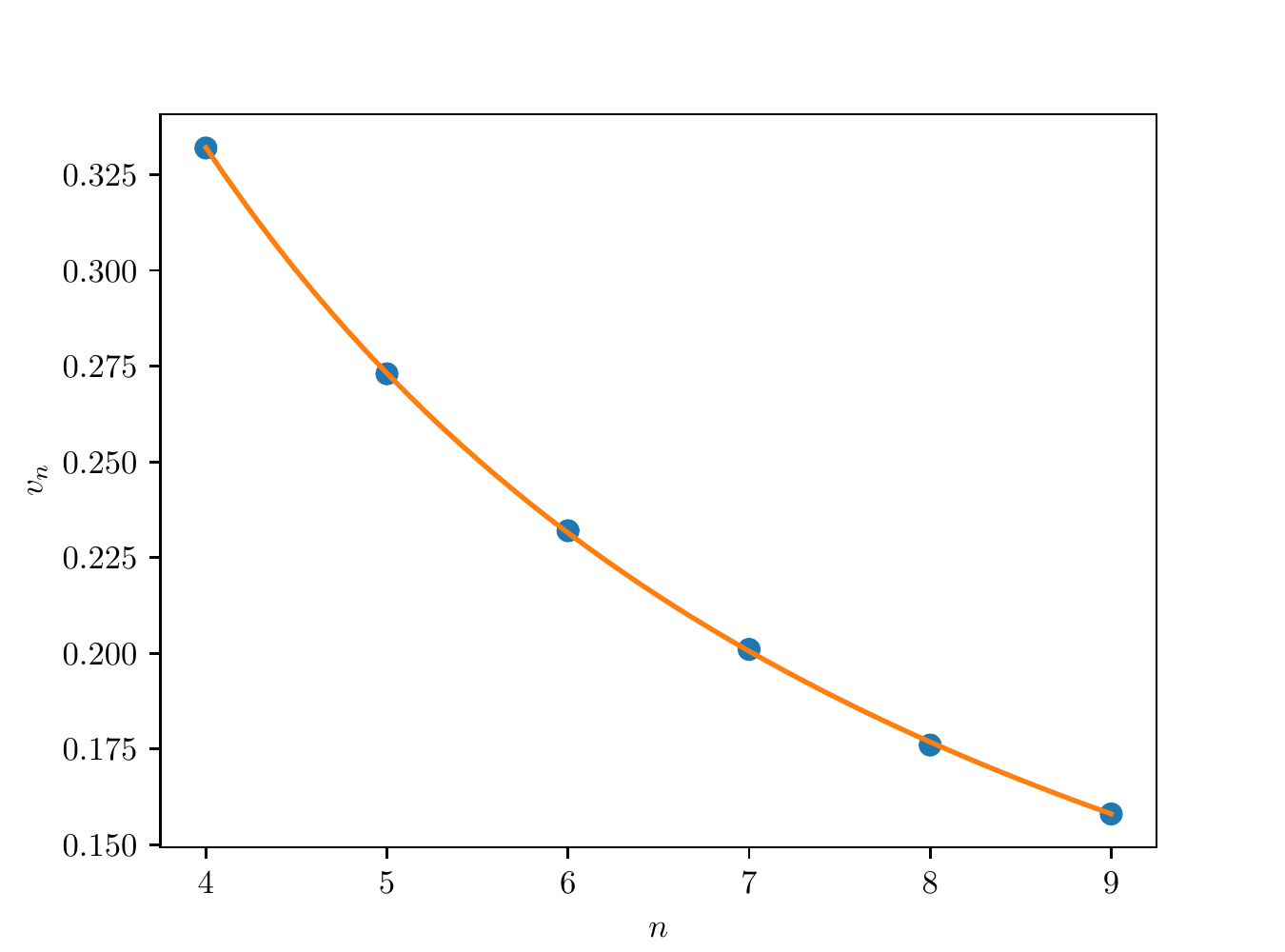}
       \caption*{Source: Results obtained by the author in \cite{campos2021wobbling}.}
       \label{fig5_vn}
\end{figure}

In Fig.~\ref{fig5_resVA}(a), one can notice that the position of the center of one-bounce windows is approximately the same for all values of $A$. We can find an approximate expression for the location of the windows using a simple model. In \cite{campbell1983resonance}, the authors assumed the wobbling amplitude is small and, therefore, the relation between the wobbling amplitude before and after the collision can be truncated at the first order. They also assumed that collisions possess time-reversal invariance. Under these assumptions, they were able to derive the following relation between the amplitude before and after the collision
\begin{equation}
S^\prime=-\frac{\rho}{\rho^*}(1+i\beta)S+i\beta S^*+\rho,
\end{equation}
where $\beta$ is a real constant and $\rho$ a complex one. Moreover, they found that $\beta\simeq 0$ for most systems and obtained the relation
\begin{equation}
S^\prime=-\frac{\rho}{\rho^*}S+\rho.
\end{equation}
Now, we can substitute eq.~(\ref{eq5_St}) in the relation above. It yields
\begin{equation}
S^\prime=-A\exp\left[i\left(\sqrt{1-v_i^2}\frac{\omega_Dx_0}{v_i}+\theta_0+2\theta_\rho\right)\right]+|\rho|\exp(i\theta_\rho),
\end{equation}
where $\theta_\rho$ is the phase of $\rho$. If the radiation is small, the sum of the vibrational and translational energy is approximately constant. Therefore, the maximum value of the translational energy occurs when the norm of the vibrational amplitude has the smallest value. The norm of $S^\prime$ will be a minimum when the phase of the first and second terms are aligned because, in this case, the final norm is the difference between the norms of the two separate terms. The alignment occurs when
\begin{equation}
\sqrt{1-v_i^2}\frac{\omega_Dx_0}{v_i}+\theta_0+2\theta_\rho=2\pi n+\theta_\rho,
\end{equation}
for any integer $n$. We can solve this equation for $v_i$ to obtain
\begin{equation}
\label{eq5_vn}
v_n=\frac{\omega_Dx_0}{\sqrt{(2\pi n-\delta)^2+\omega_D^2x_0^2}},
\end{equation}
where $\delta\equiv\theta_\rho+\theta_0$.

The center of two bounce windows as a function of $n$ is shown in Fig.~\ref{fig5_vn} also for $R=1.0$. To compare the numerical values of the position with our approximate expression above, we fit the points by a curve of the type 
\begin{equation}
v_n=\frac{a}{\sqrt{(2\pi n-b)^2+a^2}},
\end{equation}
for two constants $a$ and $b$. We find $a=9.22$, which should be compared with the theoretical value $\omega_Dx_0\simeq11.80$. The agreement is reasonable but not remarkable. This deviation is expected because we made many approximations to derive eq.~(\ref{eq5_vn}). Still, this analysis gives an approximate picture of how the wobbling phase affects the collision.

\begin{figure}[tbp]
\centering
  \caption{Maximum value of the energy density in spacetime (a) as a function of $v_i$ for $R=1.0$ and $A=0$ (b) as a function $R$ for $v_i=0.2$ and $A=0$. (c) and (d) Same as before with $A=0.5$. The grey areas correspond to resonance windows.}
  \includegraphics[width=0.9\columnwidth]{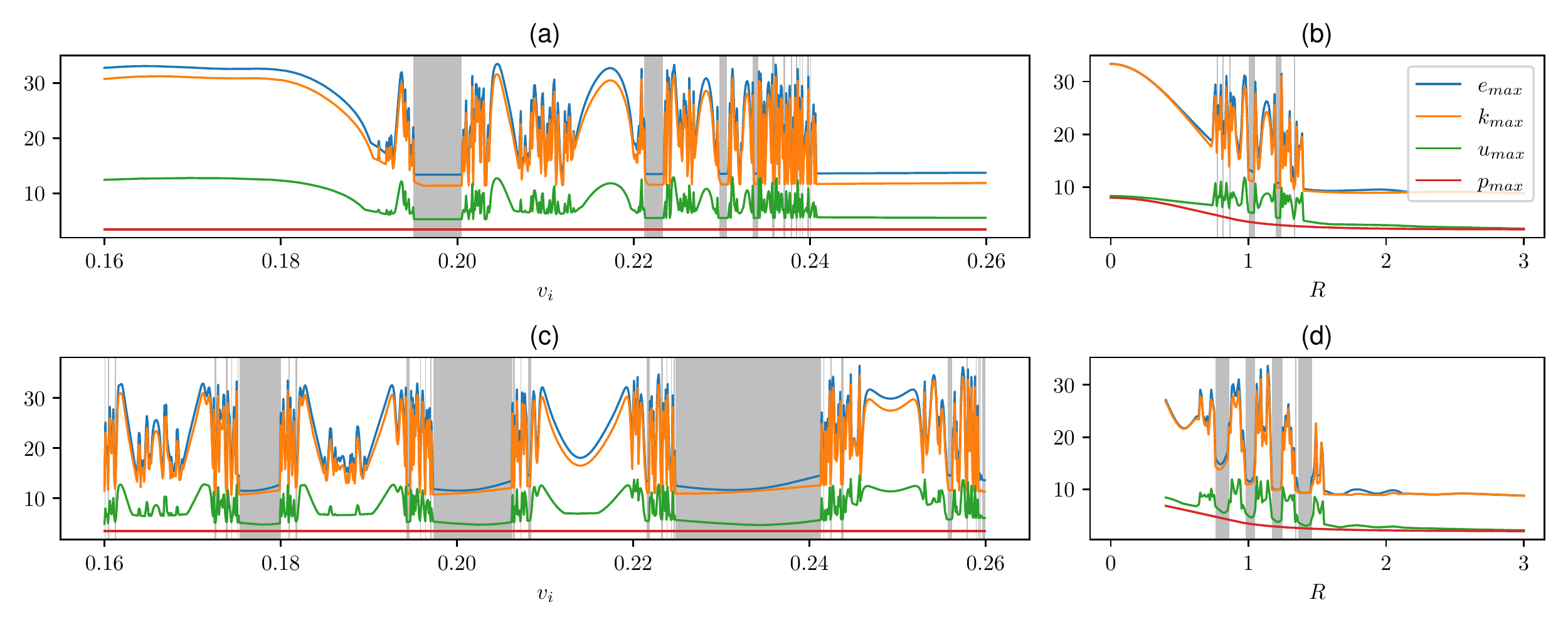}
       \caption*{Source: Results obtained by the author in \cite{campos2021wobbling}.}
       \label{fig5_maxdens}
\end{figure}

It is possible to study the evolution of the energy density in spacetime during a kink-antikink collision. The energy density is defined as $e=k+u+p$ where $k=\frac{1}{2}(\partial_t\phi)^2$ is the kinetic term, $u=\frac{1}{2}(\partial_x\phi)^2$ is the gradient term and $p=V(\phi)$ is the potential term. Interestingly, in some recent works, it was found that the fractal structure of kink-antikink collisions can be found by measuring the maximum value of the energy density for the whole spacetime evolution and plotting it as a function of the parameters \cite{yan2020kink,gani2019multi}. The surprising result was that the maximum energy density is continuous in the regions where there is separation and is chaotic in most regions where there is annihilation. 

We investigated if the same behavior could be found in kink-antikink collisions of the double sine-Gordon model. The result is shown in Fig.~\ref{fig5_maxdens}. In the figure, we plot the maximum values of the energy densities $e$, $k$, $u$ and $v$ as a function of $v_i$ for $R=1.0$ and $A=0.0$ and as a function $R$ for $v_i=0.2$ and $A=0.0$. Then, we repeat the procedure for $A=0.5$. In all cases, we mark the regions where there are resonance windows, including the one-bounce resonance windows for the case with wobbling. We find that the dependence of the maximum value of the energy density with the parameters is smooth in those regions. On the other hand, the behavior can be chaotic outside these regions. This result shows that there are other ways to identify resonant behavior besides the computation of the final velocity.

\begin{figure}[tbp]
\centering
  \caption{Wobbling frequency of the kinks after the collision in the frame where the kink is at rest. The dashed line corresponds to the shape mode frequency $\omega_D$ obtained from the stability equation. We fix $R=1.0$ and set $A=0.0$ in (a) and $A=0.1$ in (b). The colors indicate the number of bounces during the collision according to the scheme used in previous figures.}
  \includegraphics[width=0.8\columnwidth]{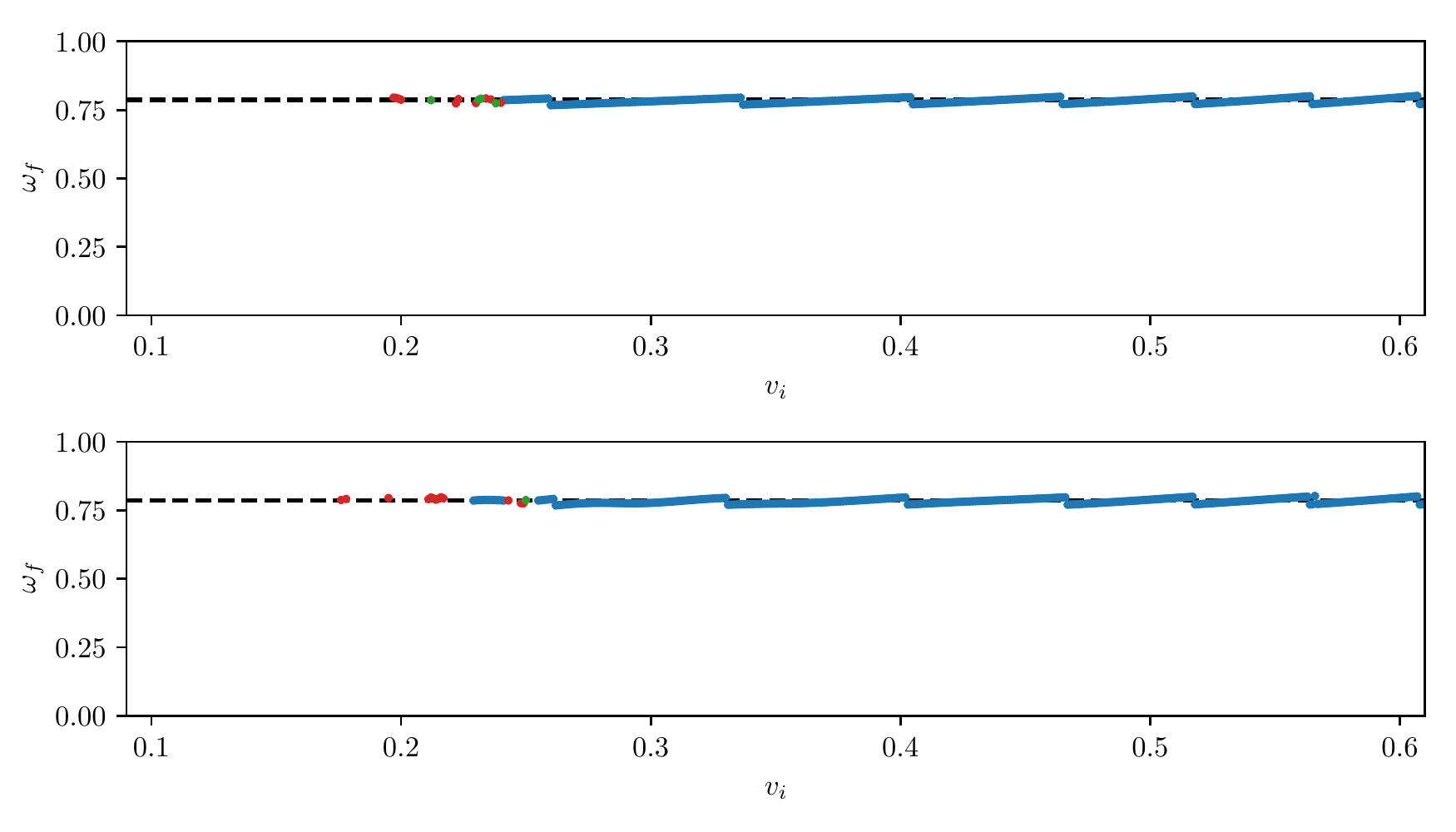}
       \caption*{Source: Results obtained by the author in \cite{campos2021wobbling}.}
       \label{fig5_freq}
\end{figure}

To complete our analysis, we compute the wobbling frequency of the kinks after the collision. We expect the kink to vibrate at the frequency $\omega_D$ obtained from the stability equation. However, the kinks are not static after the collision, they are actually moving. Therefore, the vibration should be Lorentz contracted according to eq.~(\ref{eq5_lorentz}). This is precisely what we found in our simulations. The result is shown in Fig.~\ref{fig5_freq}. We measure the vibration frequency by measuring the field's value at the position where the wobbling amplitude is expected to be a maximum from the shape mode's profile $\psi_D$. Then, we subtract the theoretical value of the kink's profile $\phi_k$ at this position from the field's value. We obtain the frequency of this time series as the one which gives the largest amplitude of the Fourier transform. After correcting for the Lorentz contraction, we obtain the excellent agreement shown in the figure.

In the next section, we will summarize our results and conclusions and also discuss ideas for future works.

\section{Conclusion}

We considered the double sine-Gordon model. This model is equivalent to the sine-Gordon model for $R=0$ and in the limit $R\to\infty$. It possesses a kink that can be described as a superposition of two sine-Gordon kinks. The kinks always have a vibrational mode. The frequency of this mode $\omega_D$ decreases from the threshold value of unity to zero as $R$ increases. For $\omega_D>0.5$, the vibration amplitude of a kink initially decays due to the coupling of the first harmonic of the wobbling frequency with continuum modes. This occurs when $R$ is not too large. As the wobbling frequency decreases, the decay is much slower and occurs via higher harmonics.

The usual kink-antikink collision for the double sine-Gordon model exhibits resonance windows for intermediate values of the parameter $R$ with the usual fractal structure. However, the windows are absent near the integrable limits, presumably due to the decoupling of the modes. Moreover, we were able to confirm the non-monotonic behavior of the critical velocity found in \cite{gani2018scattering}.

Considering collisions between wobbling kinks, we found that the collisions exhibit one-bounce windows similarly to the $\phi^4$ model. These windows appear in two ways. In the first one, they appear when the crossing curve splits. This occurs at a point where the wobbling is out of phase at the moment of the collision. In the second one, they appear at a point outside the crossing region. This occurs when the wobbling is in phase at the moment of the collision. The gradual appearance of the windows can be clearly seen in our figures. Mapping how the region of one-bounce resonance windows moves as the parameter $R$ is varied, we observe a spine structure. This structure is even more intricate for large values of the wobbling amplitude.

A significant result was that, when wobbling was included, we were able to find resonance windows much closer to the integrable limits than before, where there was no wobbling. This suggests that the system possesses a hidden resonant behavior when the wobbling is turned off.

We observed that the center of one-bounce windows did not depend on the wobbling amplitude. Moreover, we were able to find an approximate expression for the position of these windows. This was accomplished by expanding the wobbling amplitude after the collision up to linear order on the wobbling amplitude before the collision.

We showed that one could distinguish separation from annihilation by measuring how the maximum value of the energy densities varies with the system's parameters. Finally, we found consistently that, after the collision, the kink also vibrates at the frequency obtained from the stability analysis.  

Interesting continuations to this work include the reproduction of the wobbling kink effects using the effective collective coordinates system. Also, it is important to understand why the resonance windows disappear near the integrable regimes. Perhaps this could be achieved by a perturbative analysis.

\section{Appendix: Numerical method}
\label{sec5_num}

We simulate the system in a box with size $-100.0<x<100.0$. We apply the method of lines to integrate the equations of motion. The space is discretized with $N=2048$ points using periodic boundary conditions. Spatial derivatives are computed using a Fourier Spectral Method \cite{trefethen2000spectral}. The resulting set of ordinary differential equations is integrated using a fifth-order Runge-Kutta method with adaptive step size and error control \cite{dormand1980family}. It is implemented by the odeint library in C++ \cite{ahnert2011odeint}. The final time of the simulation is $t=1400.0$, except for frequency measurements, which need a more time-consuming simulation. The plots of the final velocity as a function of the initial one are computed with separation between points $\Delta v_i=10^{-5}$ and are truncated for small velocities. To avoid that radiation cross the boundaries and interact again with the system, we include damping for $x<-80$ and $x>80$ proportional to a bump function with maximum value $5$ and exactly zero outside this region \cite{christov2021kink}. The shape mode's profile and frequency are computed using the NDEigensystem method in Mathematica.

\chapter{FERMION-KINK SYSTEMS} 

\section{Definition and motivation}

The second half of this thesis consists of the study of fermion-kink systems, beginning in this section. First, let us define what a fermion-defect system is. It is a system that describes the interaction of a topological defect and a fermionic field. The two sets of fields are coupled via some interaction term. In two spacetime dimensions we have fermion-kink systems. The canonical example is the $\phi^4$ model coupled to a Dirac field via a Yukawa term
\begin{equation}
\label{eq6_phi4pf}
\mathcal{L}=\frac{1}{2}\partial^\mu\phi\partial_\mu\phi-\frac{1}{2}(\phi^2-1)^2+i\bar{\psi}\gamma^\mu\partial_\mu\psi-g\phi\bar{\psi}\psi.
\end{equation}
This is one of the most straightforward systems containing all these ingredients and will be studied in section \ref{sec6_analytic} and section \ref{chap7}. We will see that, in general, the kink will either bind to the fermion or scatter it.

So, why do we need to study these systems? The simplest explanation is that fermions are ubiquitous in nature and, therefore, it is likely that there exist topological defects that interact with them. The experimental realization of fermion-kink systems occurs in polymers, such as polyacetylene. A revision of these results can be found in \cite{niemi1986fermion}. People have also theoretically investigated interactions of fermions with vortices \cite{jackiw1981zero}, skyrmions \cite{hiller1986solutions, perapechka2018soliton, perapechka2019fermion}, solitons in a nonlinear sigma model \cite{kahana1984soliton}, monopoles \cite{yamagishi1983fermion, yamagishi1983fermionb, rubakov1982adler}, and planar defects \cite{bazeia2018dirac}. Thus, we imagine that experimental realization of those systems could be found in the future. Another theoretical application of fermion-kink systems appears in cosmology. In some higher dimensional cosmological models, domain walls, which are the extension of kinks to higher dimensions, present a plausible mechanism to localize fermions in the extra dimensions \cite{rubakov1983we, randjbar2000fermion, koley2005scalar, melfo2006fermion}.

\section{Literature revision}

One of the seminal works in the field is the reference \cite{jackiw1976solitons}. There, the authors observed that when fermions bind to topological defects, they possess a zero-energy solution with charge conjugation symmetry. They showed that this implies the defects have a fermion number of $\pm1/2$ when the system is quantized. This property is fascinating and is observed in polymers due to the fermion-kink interaction \cite{niemi1986fermion}. Later, it was shown that topological defects interacting with fermions could have any fractional fermionic charge \cite{goldstone1981fractional, alonso2019soliton}. Another attractive property of fermion-defect interaction is that it can lead to domain wall superconductivity \cite{vachaspati2006kinks} and string superconductivity \cite{vilenkin2000cosmic}.

Fermion-kink systems have been investigated for several types of potentials, and interaction terms \cite{shahkarami2011casimir, gousheh2013casimir, gousheh2014investigation, bazeia2017fermionic, bazeia2019fermion}. An important property of the interaction is that it is responsible for a nontrivial Casimir energy. In those works, the energy and eigenfunctions of the fermions are found numerically after the simplification that there is no back-reaction of the fermion to the kink. This is a good approximation when the coupling between the kink and fermion is weak. Recently, the interaction between kinks with inner structure and fermions has been investigated in \cite{bazeia2021fermions}. Interestingly, this study has applications in nanometric electronic devices.

The system given by eq.~(\ref{eq6_phi4pf}) is very important because it has analytical solutions when the back-reaction is ignored. These can be found, for instance, in \cite{chu2008fermions, charmchi2014complete, charmchi2014massive} for both massless and massive Dirac fields. This model will be studied in section \ref{chap7} considering a wobbling kink, which is a kink with the shape mode excited, instead of a static one. This main section is based in the following original study the we developed \cite{campos2021fermions}.

Instead of studying a single kink, it is possible to compute the energy of fermionic bound states considering a frozen kink-antikink configuration as the background. This was done numerically in \cite{chu2008fermions}. The authors found that the energy spectrum approaches two degenerate copies of an isolated kink configuration spectrum as the distance between the kinks increases. This result is expected for quantum mechanical systems. Similar results were found for the sine-Gordon potential in \cite{brihaye2008remarks}.

If one considers strong coupling between the fermion and the kink, it is necessary to include the back-reaction of the fermion to the kink. In this case, the problem becomes complicated and can only be solved numerically, even for a single kink. Self-consistent solutions have been found in \cite{shahkarami2011exact, amado2017coupled, klimashonok2019fermions, perapechka2020kinks} for kinks and in \cite{perapechka2018soliton, perapechka2019fermion} for baby-skyrmions. The back-reaction can have many interesting effects, such as creating kink-antikink pairs. 

Finally, one can consider again the model (\ref{eq6_phi4pf}) without back-reaction, but with a dynamical kink-antikink configuration. The dynamical configuration can be a kink-antikink collision as done in \cite{gibbons2007fermions, saffin2007particle}. The authors found that a fermion localized at the kink or the antikink can be transferred to the other defect due to the collision. We will consider this type of system for a slightly different model in section \ref{chap8}, which is based in the following original study that we developed \cite{campos2020fermion}. 

\section{Some theoretical aspects}
\label{sec6_analytic}

Our goal is to study fermion-kink systems. We will start considering  a Lagrangian slightly more general than eq.~(\ref{eq6_phi4pf}) in two spacetime dimensions. This is
\begin{equation}
\label{eq6_Vpf}
\mathcal{L}=\frac{1}{2}\partial^\mu\phi\partial_\mu\phi-V(\phi)+i\bar{\psi}\gamma^\mu\partial_\mu\psi-g\phi\bar{\psi}\psi.
\end{equation}
To solve the system, we start by writing the equations of motion\footnote{The classical limit of a Fermi field is a Grassmann number, which is anticommuting. However, it is justifiable to consider them as c-numbers in the equations of motion, as shown in \cite{rajaraman1982solitons}.}
\begin{align}
\label{chap6_eom}
&\partial_t^2\phi-\partial_x^2\phi+V^\prime(\phi)+g\bar{\psi}\psi=0,\\
&i\gamma^\mu\partial_\mu\psi-g\phi\psi=0.
\label{chap6_eomdirac}
\end{align}
We will use the approximation that there is no back-reaction of the kink to the fermion. This is done by setting the term $g\bar{\psi}\psi$ to zero in eq.~(\ref{chap6_eom}). This approximation is adequate when the coupling $g$ is small compared to kink's mass. The limit of a massive kink can be obtained by recovering the units in the potential and setting them to large values. Considering the approximation, the scalar field equation decouples and can be easily solved. Furthermore, we will consider that $V(\phi)$ has more than a single vacuum. In this case, the scalar field equation possesses a kink solution $\phi_k(x)$.

The next step is to solve the Dirac equation considering the kink solution as the scalar field configuration. First, we choose the representation of the gamma matrices as $\gamma^0=\sigma_1$ and $\gamma^1=i\sigma_3$. Then, we use the ansatz $\psi=e^{-iE_nt}\psi_n=e^{-iE_nt}(\psi_{n,+},\psi_{n,-})^T$. This leads to
\begin{align}
\label{eq6_psinp}
E_n\psi_{n,+}+\partial_x\psi_{n,-}-g\phi_k\psi_{n,-}=0,\\
E_n\psi_{n,-}-\partial_x\psi_{n,+}-g\phi_k\psi_{n,+}=0.
\label{eq6_psinm}
\end{align}
We can solve for each component, obtaining the following Schr\"{o}dinger-like equations which are uncoupled but have the same eigenvalues
\begin{equation}
\label{eq6_schrodingerlike}
-\partial_x^2\psi_{n,\pm}+V_\pm\psi_{n,\pm}=E_n^2\psi_{n,\pm}.
\end{equation}
The obtained effective potentials are $V_\pm=g(g\phi_k^2\mp\partial_x\phi_k)$. As the two Hamiltonians of the effective equations above are supersymmetric partners \cite{cooper1995supersymmetry}, the two equations are guaranteed to have the same set of eigenvalues, except for the zero mode, which has vanishing energy. Moreover, the corresponding eigenstates are guaranteed to solve the coupled first-order equations as well, after some normalization.

There are a few important properties of eq.~(\ref{eq6_schrodingerlike}) and the original ones (\ref{eq6_psinp}) and (\ref{eq6_psinm}). First, setting the fermion energy to zero ($E_0=0$), the original equations are easily integrated, and one obtains the zero mode $\psi_0$. However, due to the topological nature of the scalar field, one of the components is set to zero because it is not normalizable. Accordingly, the nonzero component of the zero mode is also a solution of the corresponding component of eq.~(\ref{eq6_schrodingerlike}). Depending on the depth of the potential $V_\pm(x)$, this set of equations can have more bound states. They appear in charge conjugated pairs with energies $\pm E_n$. Moreover, the threshold values of the effective potentials $V_\pm(x)$, which are the limits where both negative and positive values of $x$ are large, are finite. This implies that the equations also allow scattering states.

To make these statements more concrete, we consider the $\phi^4$ model as an example. In this case $V(\phi)=\frac{1}{2}(\phi^2-1)^2$ and $\phi_k=\tanh(x)$. This leads to effective potentials with P\"{o}schl-Teller form
\begin{equation}
V_\pm=g^2-\frac{g(g\pm1)}{\cosh^2(x)}.
\end{equation}
The corresponding effective Hamiltonian with these potentials have analytical solutions which can be found, for instance, in references \cite{chu2008fermions, charmchi2014complete}. The expressions for all the bound and scattering states are somewhat technical, but they will be listed in section \ref{sec6_bound} for completeness. They can be written in terms of the Hypergeometric function.

Let us analyze a few properties of this model. The threshold values, in this case, are equal to $g^2$ on both sides. This means that the scattering states start at the threshold energies $E_{\text{th}}^2=g^2$, or $E_{\text{th}}=\pm g$. It is instructive to calculate the zero mode for this model. Setting $E_0=0$ in eqs.~(\ref{eq6_psinp}) and (\ref{eq6_psinm}), we find
\begin{align}
\partial_x\psi_{0,-}-g\tanh(x)\psi_{0,-}=0,\\
-\partial_x\psi_{0,+}-g\tanh(x)\psi_{0,+}=0.
\end{align}
These have solutions $\psi_{0,\pm}=C_\pm\cosh^{\mp g}(x)$. Clearly, $\psi_{0,-}$ is not normalizable and, therefore, we set $C_-=0$. For $g>1$, there are other bound states. Considering the ones with non-negative energy, The total number of them is equal to the largest integer $n$ such that $g>n$.  

\begin{figure}[tbp]
\centering
  \caption{(upper left) Effective potentials $V_\pm(x)$ with $g=2.0$. (upper right) Fermion spectrum as a function of the Yukawa coupling $g$. (lower left) Fermion zero mode and (lower right) fermion first excited state wavefunctions considering $g=2.0$. We are considering a $\phi^4$ kink as the background scalar field.}
  \includegraphics[width=0.9\columnwidth]{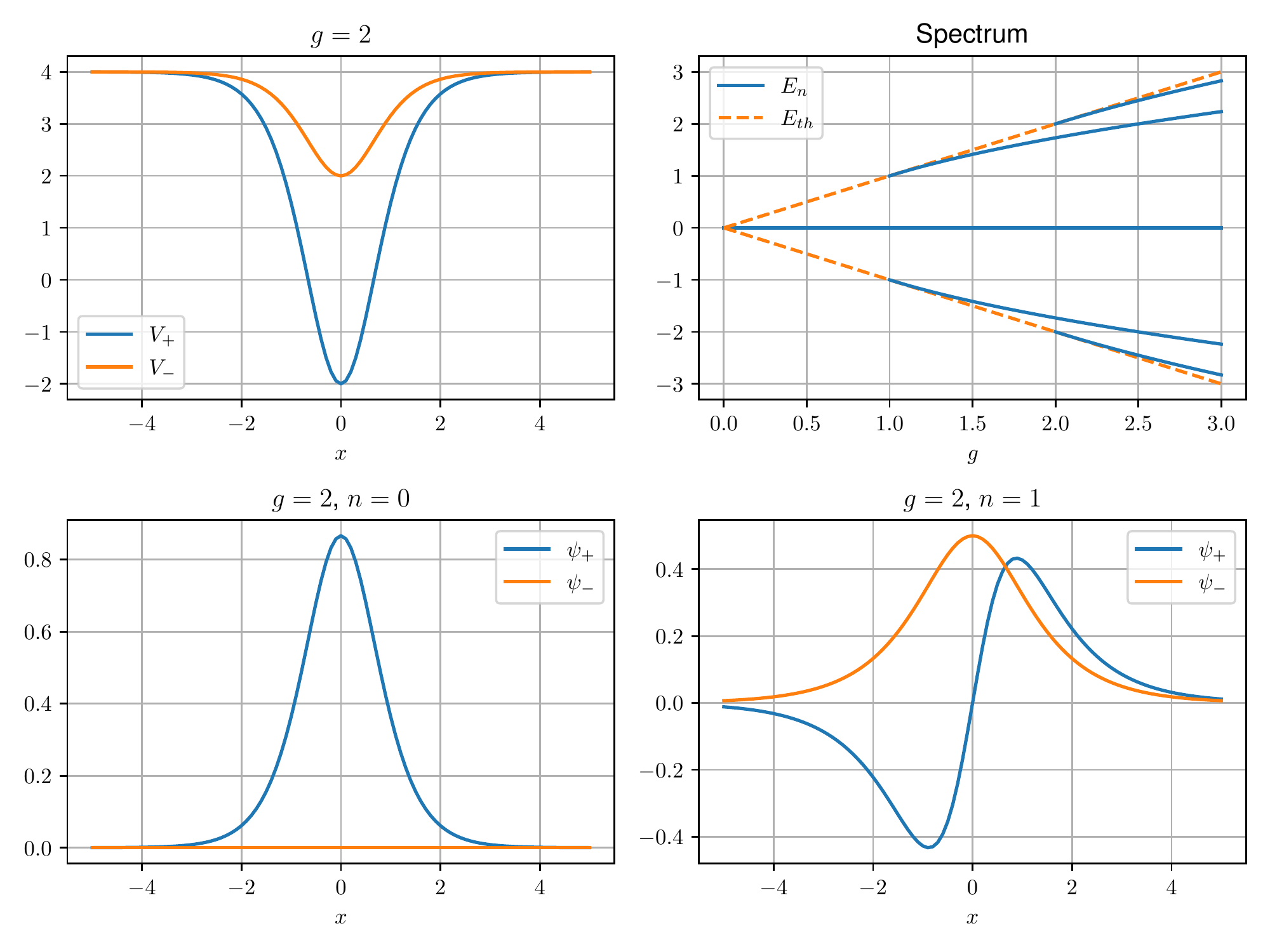}
  \caption*{Source: The author (2022).}
       \label{fig6_phi4pf}
\end{figure}

The effective potentials, energy spectrum, and a few bound state wavefunctions are shown in Fig.~\ref{fig6_phi4pf}. Looking at the spectrum, one can see how new bound states emerge from the threshold states as $g$ increases. Fixing $g=2.0$, we see that the effective potential $V_-$ is shallower than $V_+$. Hence, $V_+$ has one more energy eigenvalue than $V_-$, which corresponds to the zero mode. For these two potentials, the wavefunction of the zero mode is shown in the figure. We also plotted the wavefunction of the first excited state with positive energy. One intriguing property that can be observed in the figure is that the wavefunctions $\psi_{n,+}$ and $\psi_{n,-}$ have definite and opposite parity. Moreover, the parity $\psi_{n,\pm}$ is inverted when $n$ increases by one because the effective potentials are even.

Before finishing this main section, we would like to discuss a condition under which the background approximation is exact. Writing $\psi=(u_1+iu_2,v_1+iv_2)^T$, eq.~(\ref{chap6_eomdirac}) becomes
\begin{align}
\dot{u}_1&=-\partial_xv_2+g\phi v_2,\\
\dot{u}_2&=\partial_xv_1-g\phi v_1,\\
\dot{v}_1&=\partial_xu_2+g\phi u_2,\\
\dot{v}_2&=-\partial_xu_1-g\phi u_1.
\end{align}
Clearly, $u_1$ and $v_2$ are decoupled from $u_2$ and $v_1$ and vice-versa. Moreover, if we choose, for instance, an initial condition such that $u_2(t=0)=v_1(t=0)=0$, we will have $u_2(t)=v_1(t)=0$ for all times. As the backreaction is proportional to
\begin{equation}
g\bar{\psi}\psi=2g(u_1v_1+u_2v_2),
\end{equation}
we will have that the backreaction vanishes for all times. Interestingly, the condition $u_2(t=0)=v_1(t=0)=0$ is not rare. For instance, it is respected when the fermion is initialized at zero mode.

In the next main section, we will study the effect of the kink's vibration on the fermion in the same model. In section \ref{chap8}, we will study in a different model what happens to a fermion localized on a kink after a collision.

\section{Appendix: Fermion bound states of the $\phi^4$ model}
\label{sec6_bound}

The fermion bound states in the presence of a $\phi^4$ kink are listed in references \cite{chu2008fermions, charmchi2014complete}. For every integer $n<g$, the bound state solutions with positive energy are given by the following expressions
\begin{align}
\psi_{n,+}&=N_+^n\text{sech}^{g-n}(x)F\left(-n,2g-n+1,g-n+1;\frac{1}{2}(1-\tanh(x))\right),\\
\psi_{n,-}&=N_-^n\text{sech}^{g-n}(x)F\left(-n+1,2g-n,g-n+1;\frac{1}{2}(1-\tanh(x))\right),
\end{align}
with energy $E_n=\sqrt{2gn-n^2}$. The function $F$ corresponds to the hypergeometric function. For $g<1$, there is only the zero mode with $n=0$ and $N_-^0=0$. A new positive energy bound state appear every time $g$ crosses an integer value. For these states, the normalization constants obey the relation $N_-^n/N_+^n=n/E_n$.

For every positive momentum $k$, the scattering states are given by the expressions below
\begin{align}
\psi_{k,L}^\pm&=N_\pm^{k,L}\text{cosh}^{ik}(x)F\left(\frac{1}{2}-ik-\zeta_\pm,\frac{1}{2}-ik+\zeta_\pm,1-ik;\frac{1}{1+e^{2x}}\right),\\
\psi_{k,R}^\pm&=N_\pm^{k,R}\text{cosh}^{ik}(x)F\left(\frac{1}{2}-ik-\zeta_\pm,\frac{1}{2}-ik+\zeta_\pm,1-ik;\frac{1}{1+e^{-2x}}\right),
\end{align}
with energy $E_k=\sqrt{k^2+g^2}$. Moreover, we define $\zeta_\pm\equiv g\pm\frac{1}{2}$. The normalization constants obey the relation $N_-^{k,L}/N_+^{k,L}=-N_-^{k,R}/N_+^{k,R}=(ik+g)/E_k$.

\chapter{FERMIONS ON WOBBLING KINKS}
\label{chap7}
\section{Overview}

In this main section, we will discuss a model with a fermion coupled to a wobbling kink background. This work resulted in the following publication \cite{campos2021fermions}. We will consider the simple realizations of a fermion-kink system discussed in the previous main section. It consists of a fermion in the background of a $\phi^4$ kink interacting via a Yukawa coupling in two spacetime dimensions. As said before, this system has an analytical solution for a static kink background. This can be found, for instance, in \cite{chu2008fermions, charmchi2014complete, charmchi2014massive}.  

This work aims to analyze the consequences of including vibration in the kink background, considering the abovementioned model. We studied the fermion's behavior coupled to a wobbling kink, which is a kink with the vibrational mode excited. This idea was inspired by the resonant energy exchange mechanism, which states that kinks may lose translational energy and start wobbling during an interaction. More recently, in \cite{izquierdo2021scattering}, the authors considered collisions between wobbling kinks and showed that, in general, after a kink-antikink collision, they emerge with the vibrational mode excited. Therefore, it is essential to consider the effect of the vibration on any fermions that may be interacting with the kink. Moreover, if one considers the wobbling kink as a background, it is possible to find analytical solutions for the system's evolution using perturbation theory. We will show that wobbling has exciting effects.

We will consider the wobbling effect on the fermion due to both normal and quasinormal modes. As shown in \cite{dorey2018resonant, campos2020quasinormal}, it is possible to modify the kink potential in order to turn the shape mode into a quasinormal mode. This has interesting consequences for the resonant behavior of the system. Hence, we will also investigate how this modification can affect the fermion.

The next section will discuss the fermion-kink model for a general potential. Moreover, we will discuss how to perform perturbation theory for this system. Later, we will specialize in the $\phi^4$ model and a toy model containing either a normal or quasinormal mode.

\section{Model}

We will consider the Lagrangian in two spacetime dimensions containing a scalar field and a fermion field interacting via a Yukawa coupling
\begin{equation}
\mathcal{L}=\frac{1}{2}\partial_\mu\phi\partial^\mu\phi-V(\phi)+i\bar{\psi}\gamma^\mu\partial_\mu\psi-g\phi\bar{\psi}\psi.
\end{equation}
The potential $V(\phi)$ is chosen such that it has degenerate vacua with vanishing energy. Therefore, the scalar field possesses a kink solution. The Euler-Lagrange equation for the fermion field
\begin{equation}
\label{eq6_dirac}
i\gamma^\mu\partial_\mu\psi-g\phi_k\psi=0,
\end{equation}
where we considered that the scalar field is fixed as a background and given by the kink profile. We choose the following representation for the gamma matrices, $\gamma^0=\sigma_1$ and $\gamma^1=i\sigma_3$. Multiplying the Dirac equation above by $\gamma^0$, we obtain
\begin{equation}
i\partial_t\psi=H_0\psi,
\end{equation}
where the Hamiltonian $H_0$ is defined as
\begin{equation}
H_0=-i\sigma_2\partial_x+g\phi_k\sigma_1.
\end{equation}
Let us diagonalize the Hamiltonian by solving the following time-independent equation
\begin{equation}
\label{eq6_TI}
H_0\psi_n=E_n\psi_n.
\end{equation}
Writing the eigenfunctions in term of its components as $\psi_n=(\psi_{n,+},\psi_{n,-})^T$, we obtain eqs.~(\ref{eq6_psinp}) and (\ref{eq6_psinm}) again. This is expected because we are studying the same system considered in section \ref{sec6_analytic}. Thus, this set of equations has analytical solutions, which are listed in section \ref{sec6_bound}.

Now, we consider wobbling kinks as a background. This can be achieved by including a perturbation in the scalar field as follows $\phi=\phi_k+A\eta_S\cos(\omega_St)$. In this expression, $A$ is the amplitude, and $\eta_S$ and $\omega_S$ are the shape mode's profile and frequency, respectively. Thus, the Dirac equation becomes
\begin{equation}
i\partial_t\psi=(H_0+AH_1(t))\psi,
\end{equation}
where the perturbation Hamiltonian is given by $H_1(t)=g\eta_S\cos(\omega_St)\sigma_1$. The solution of the perturbed equation can be written as a linear combination of the eigenfunctions of the non-perturbed Hamiltonian with time-dependent coefficients
\begin{equation}
\psi=\sum_nc_n(t)e^{-iE_nt}\psi_n.
\end{equation}
Moreover, we can expand the coefficients in powers of the perturbation amplitude
\begin{equation}
c_n(t)=c_n^{(0)}(t)+Ac_n^{(1)}(t)+\cdots.
\end{equation}
We choose the initial condition such that the system starts at state $i$. This can be written as $c_n(0)=\delta_{ni}$. Solving in zero order, we find $c_n^{(0)}(t)=\delta_{ni}$. Then, the first-order coefficients are given by time-dependent perturbation theory as \cite{Sakurai:1341875}
\begin{equation}
\label{eq6_cn1}
c_n^{(1)}(t)=-i\int_0^te^{i\omega_{ni}t}\langle n|H_1(t^\prime)|i\rangle dt^\prime,
\end{equation}
where we defined $\omega_{ni}=E_n-E_i$ and the matrix elements between two states is defined as
\begin{equation}
\langle n|\mathcal{O}|i\rangle=\int\psi_n^T\mathcal{O}\psi_idx.
\end{equation}
The transition probability is quantified by coefficients $c_n(t)$, obtained by projecting our system in the state $|n\rangle$. They are also known as Bogoliubov coefficients. For $n\neq i$, the first-order transition probability from the state $i$ to a state $n$ is given by
\begin{equation}
P_{i\to n}(t)=A^2|c_n^{(1)}(t)|^2.
\end{equation}
This expression is only valid as long as $A^2|c_n^{(1)}(t)|^2\ll 1$. The time integration in eq.~(\ref{eq6_cn1}) leads to
\begin{equation}
\label{eq6_cn1integrated}
c_n^{(1)}(t)=\frac{g}{2}\langle n|\eta_S\sigma_1|i\rangle\left(\frac{1-e^{i(\omega_{ni}+\omega_S)t}}{\omega_{ni}+\omega_S}+\frac{1-e^{i(\omega_{ni}-\omega_S)t}}{\omega_{ni}-\omega_S}\right).
\end{equation}
We can define the transition rate from state $i$ to $n$ as the ratio between the probability and the total time for long times. It is easy to show that, in our case, it is given by Fermi's golden rule
\begin{equation}
R_{i\to n}=A^2g^2\frac{\pi}{2}|\langle n|\eta_S\sigma_1|i\rangle|^2[\delta(\omega_{ni}+\omega_S)+\delta(\omega_{ni}-\omega_S)],
\end{equation}
for $n\neq i$. This expression must be integrated over a set of final states to make sense of the delta functions.

In the next section, we apply this formalism to the $\phi^4$ model and compare the analytical results to numerical analysis.

\section{$\phi^4$ model}

In this section, we specialize to the the case where the scalar field is described by $\phi^4$ model. For this model, we set $V(\phi)=\frac{1}{2}(\phi^2-1)^2$, $\phi_k=\tanh(x)$, $\eta_S=\tanh(x)\text{sech}(x)$ and $\omega_S=\sqrt{3}$. The effective fermionic potential in this case is given by
\begin{equation}
V_\pm=g^2-\frac{g(g\pm1)}{\cosh^2x},
\end{equation}
which has analytical solutions as shown, for instance, in \cite{chu2008fermions,charmchi2014complete}. For completeness, the solutions as listed in section \ref{sec6_bound}. One important property of these solutions is that for $g<1$, there is only one discrete state, the zero mode, and, as $g$ increases, a positive and a negative discrete solutions appear every time $g$ crosses an integer value. Moreover, there is a continuum of scattering solutions with fermion energy obeying $|E|>g$. 

\begin{figure}
     \centering
     \caption{Matrix elements of the operator $\eta_S\sigma_1$ (a) between discrete states and (b) between discrete states and the continuum. We consider the $\phi^4$ model and in (b) $g=1.6$.}
     \begin{subfigure}[b]{0.42\textwidth}
         \centering
         \includegraphics[width=\textwidth]{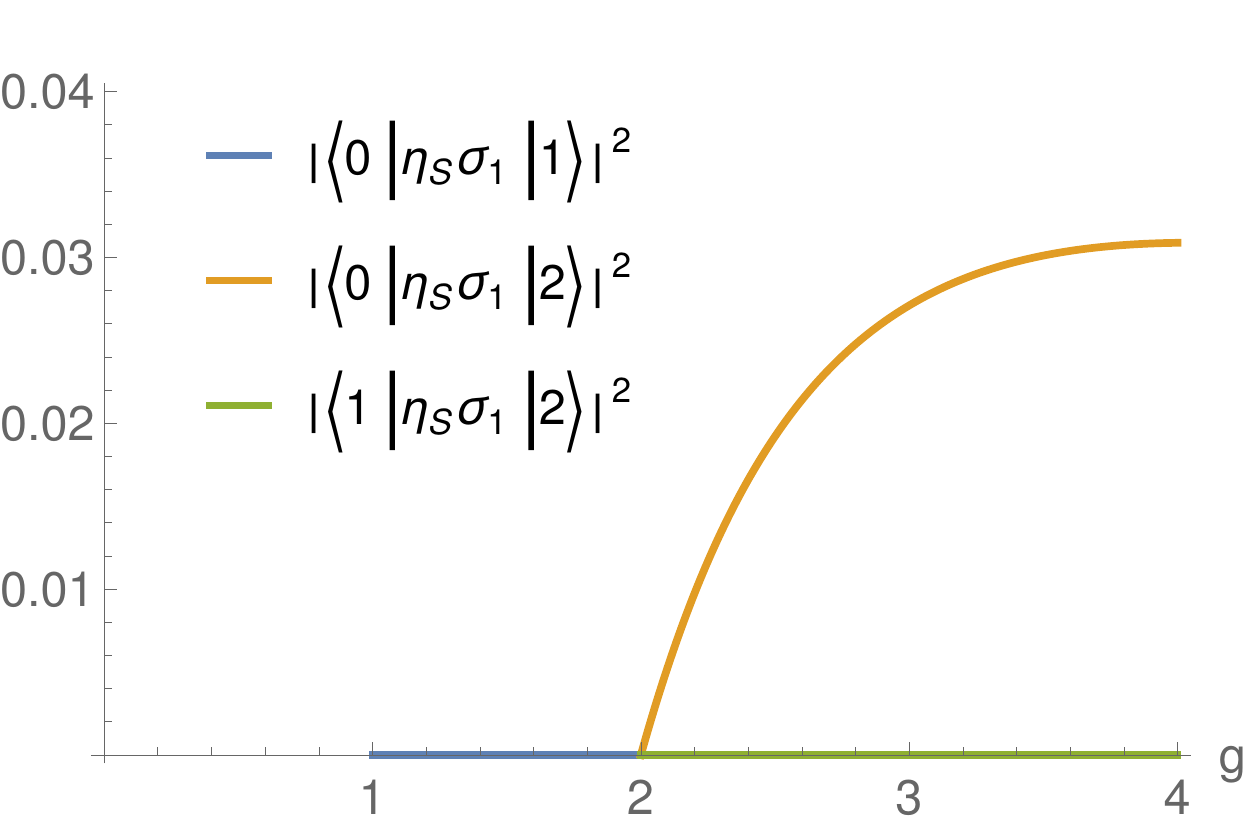}
         \caption{}
     \end{subfigure}
     \hfill
     \begin{subfigure}[b]{0.42\textwidth}
         \centering
         \includegraphics[width=\textwidth]{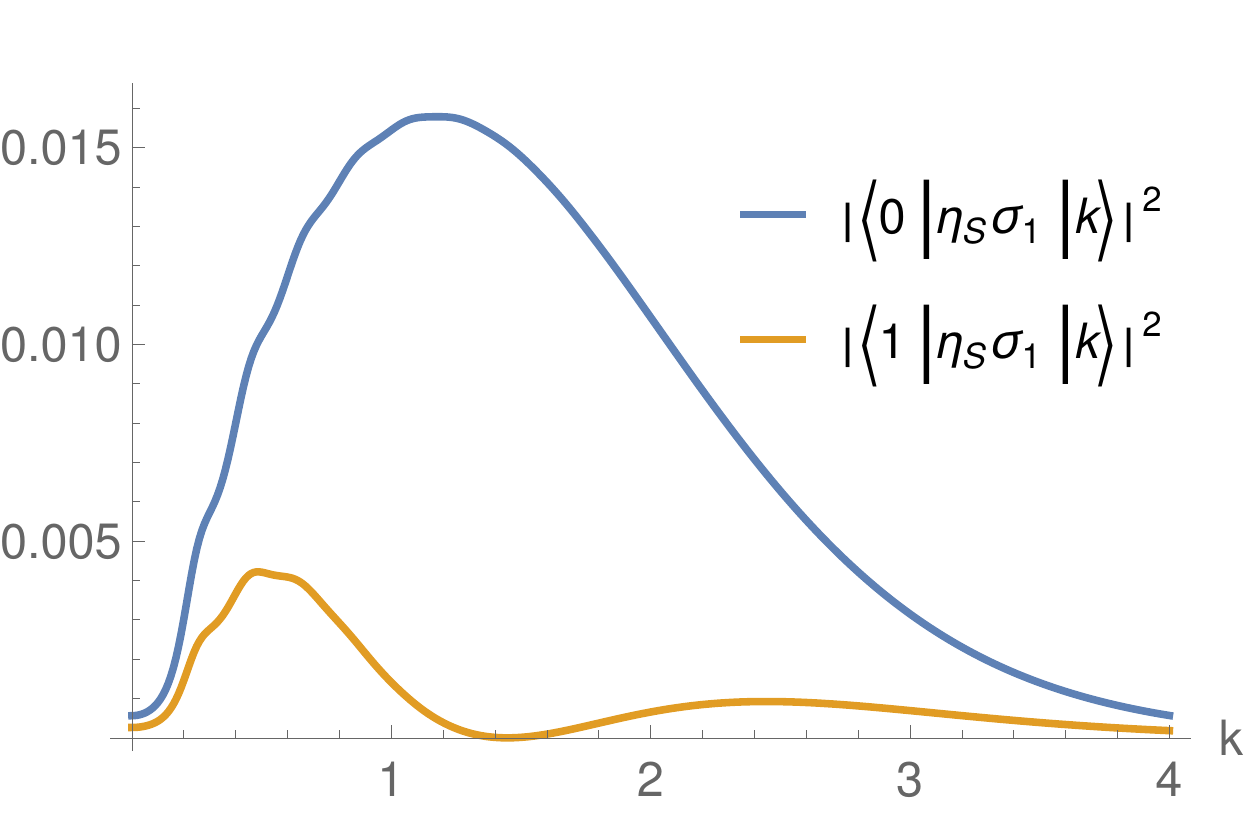}
         \caption{}
     \end{subfigure}
        \caption*{Source: Results obtained by the author in \cite{campos2021fermions}.}
        \label{fig6_eta}
\end{figure}

The first step in the analysis is to compute the matrix elements of the operator $\eta_S\sigma_1$. They are shown in Fig.~\ref{fig6_eta}. We see the matrix elements between the discrete states in the left panel. Transitions are only allowed when the $n$ and $i$ are both odd or even, due to the parity of the respective eigenfunctions and the function $\eta_S$. For instance, the matrix element between the zero mode and the first excited state is given by
\begin{equation}
\langle1|\eta_S\sigma_1|0\rangle=\int(\psi_{1,-}\eta_S\psi_{0,+}+\psi_{1,+}\eta_S\psi_{0,-})dx.
\end{equation}
The first term vanishes because $\psi_{1,-}$ is even, $\psi_{0,+}$ is even, and $\eta_S$ is odd. The second term also vanishes because $\psi_{0,-}=0$. The first allowed transition between positive energy states appears for $g>2$ as shown in the figure. Moreover, we can compute the transition between discrete states and the continuum using the analytical fermionic solutions of the $\phi^4$ model. In general, they are allowed, as shown in the right panel.

\begin{figure}
     \centering
     \caption{Evolution of the transition probabilities in time for the $\phi^4$ model. We compute transitions from the initial states to the continuum and to other discrete states. In (a) the transition rate to the continuum is the slope of the dotted line. We fix the wobbling amplitude $A=0.1$.}
     \begin{subfigure}[b]{0.48\textwidth}
         \centering
         \includegraphics[width=\textwidth]{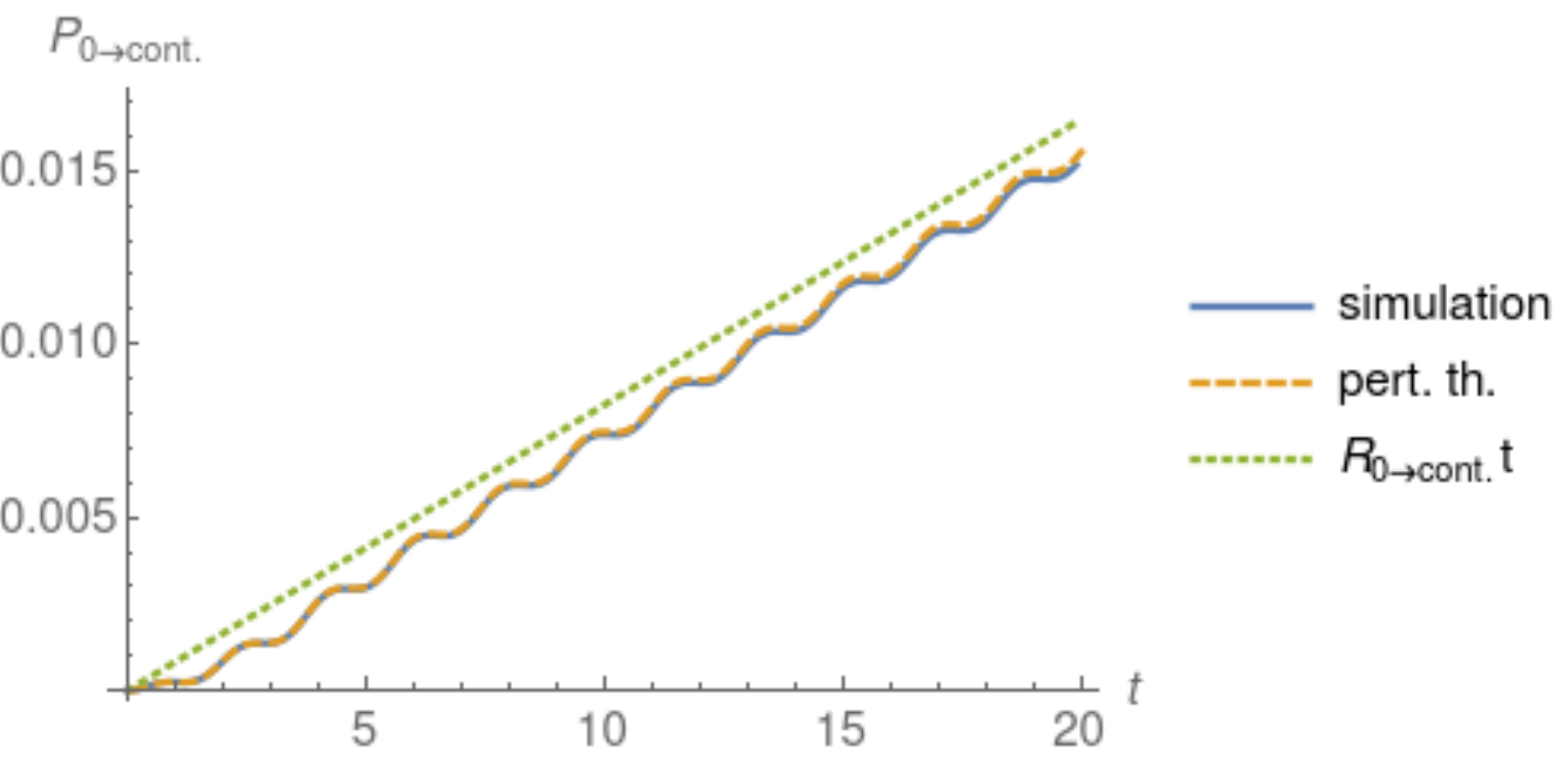}
         \caption{$g=0.8$}
     \end{subfigure}
     \hfill
     \begin{subfigure}[b]{0.48\textwidth}
         \centering
         \includegraphics[width=\textwidth]{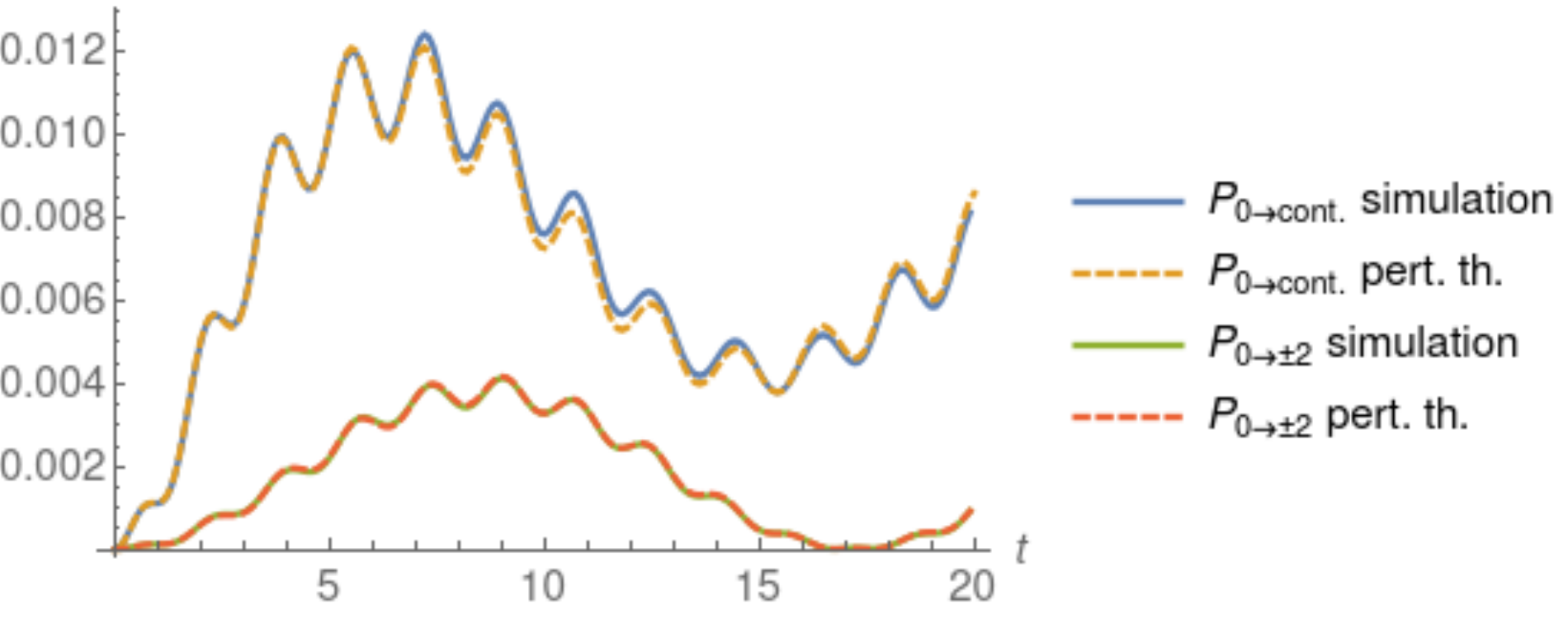}
         \caption{$g=2.1$}
     \end{subfigure}
        \caption*{Source: Results obtained by the author in \cite{campos2021fermions}.}
        \label{fig6_coef}
\end{figure}

We will first consider the fermion's evolution in the presence of a wobbling kink starting from the zero mode. Most importantly, when we specialize to this case, the back-reaction to the kink vanishes, and the background approximation is exact. However, we are still making approximations. We ignore the back-reaction to obtain the form of the perturbation due to the kink's vibration that we are considering.

We turn on the perturbation with amplitude $A=0.1$. This value is interesting because it is a typical vibration amplitude of a kink after colliding with an antikink. When the perturbation starts, the fermionic density, which is initially localized at the kink, starts to oscillate and may radiate. The radiation is significant when the transition to the continuum is allowed by Fermi's golden rule, as explained below. In Fig.~\ref{fig6_coef}, we plot the evolution of the transition probabilities in this case. The probabilities are computed in two ways. First, we use the perturbation theory equations derived above. Second, we compute the evolution of the fermionic field via numerical integration of the equations of motion. The details of the numerical integration are given in section \ref{sec6_num}. 

For $g=0.8$, we only have the zero mode and the continuum states, and we compute the transition rate to the continuum. This is the transition to the set of all continuum states. The results obtained from the two methods have a good agreement. The transition rate can be computed by integrating Fermi's golden rule. Due to the delta functions, the integration is only nonzero if $g<\omega_S$. In this range, it yields
\begin{equation}
R_{i\to\text{cont.}}=2A^2g^2\pi\left\vert\left\langle k=\sqrt{\omega_S^2-g^2}\left\vert\eta_S\sigma_1\right\vert i\right\rangle\right\vert^2\frac{\omega_S}{\sqrt{\omega_S^2-g^2}}.
\end{equation}
In the left panel of the figure, the transition rate is shown as the slope of the dotted line. Accordingly, the evolution of the transition probability has the same slope on average. In the right panel, we consider $g=2.1$. the transition probability to the second excited state has complex oscillatory behavior, as described by eq.~(\ref{eq6_cn1integrated}). Moreover, the transition to the continuum is not allowed by Fermi's golden rule in this case, and the transition probability is bounded.

The results in this section are very relevant because they predict that fermions will escape the kink at some rate if the kink is wobbling and if the Yukawa coupling $g$ is not too large. Based on the fermion escape rate in those systems, this result could be used to measure, for instance, the wobbling amplitude or the Yukawa coupling $g$. In the next section, we will repeat the analysis for a toy model that possesses a normal mode that can be turned into a quasinormal one.

\section{Toy model}

The model we will consider in this section was described in detail in section \ref{chap2}. This model describes a kink with vibrational modes that can be turned into quasinormal modes when the model is extended. The potential is given by eq.~(\ref{eq_Vsquarewell}) for the normal mode case and by eq.~(\ref{eq_VQNM}) for the quasinormal mode case. The corresponding kinks are given by eq.~(\ref{eq_phik_squarewell}) and eq.~(\ref{eq_phik_QNM}), respectively. The toy model has one free parameter for the normal mode case that we fix at $\gamma=3.0$. For this value of the parameter, the kink has a single shape mode that can be found analytically as shown in section \ref{sec2_SW}. For the quasinormal case, the toy model has three free parameters that we fix at $\gamma=3.0$, $\Delta\gamma=2.0$, and $\epsilon=0.05$. For these values of the parameters, the shape mode of the previous case becomes a quasinormal mode.

\begin{figure}
     \centering
     \caption{Energy of fermion bound states on a kink background as a function of the Yukawa coupling $g$. The two plots corresponds to the toy model containing (a) a normal mode and (b) a quasinormal mode.}
     \begin{subfigure}[b]{0.45\textwidth}
         \centering
         \includegraphics[width=\textwidth]{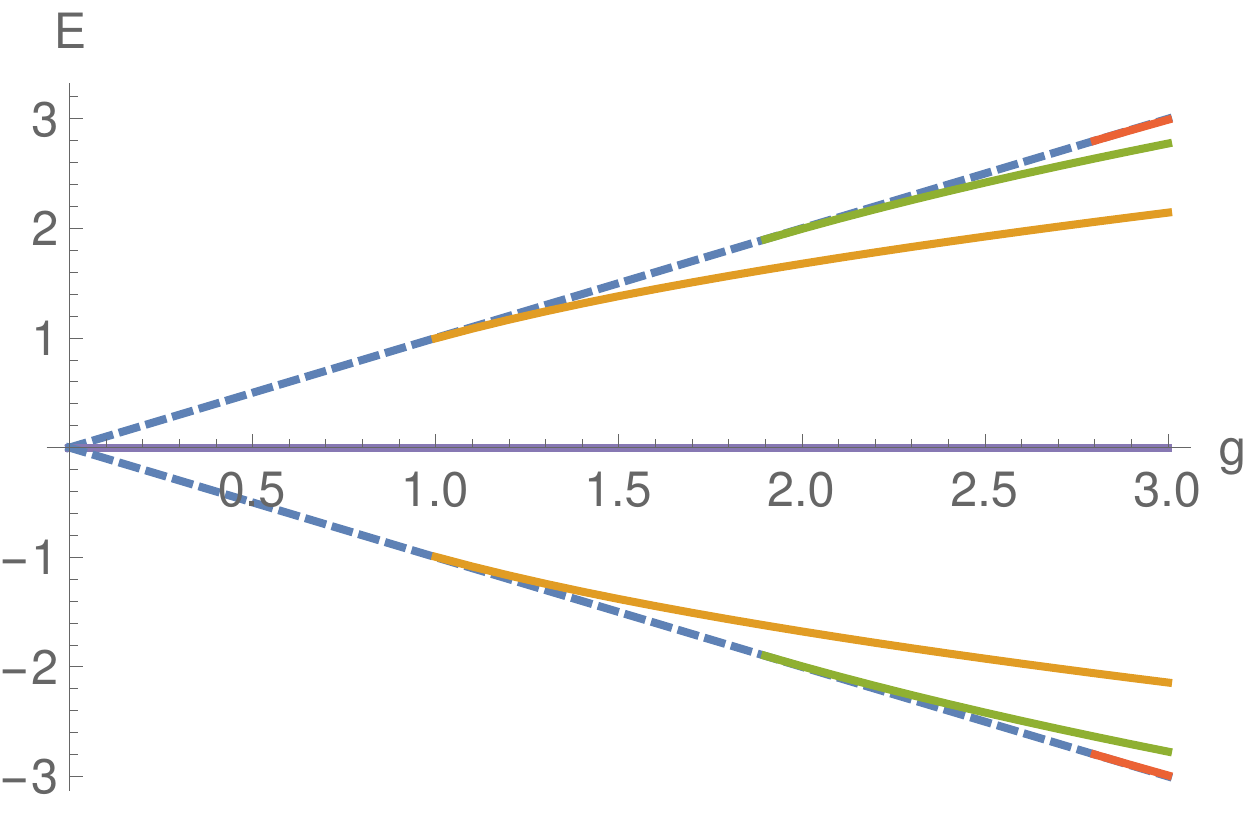}
         \caption{$\epsilon=0.0$}
     \end{subfigure}
     \hfill
     \begin{subfigure}[b]{0.45\textwidth}
         \centering
         \includegraphics[width=\textwidth]{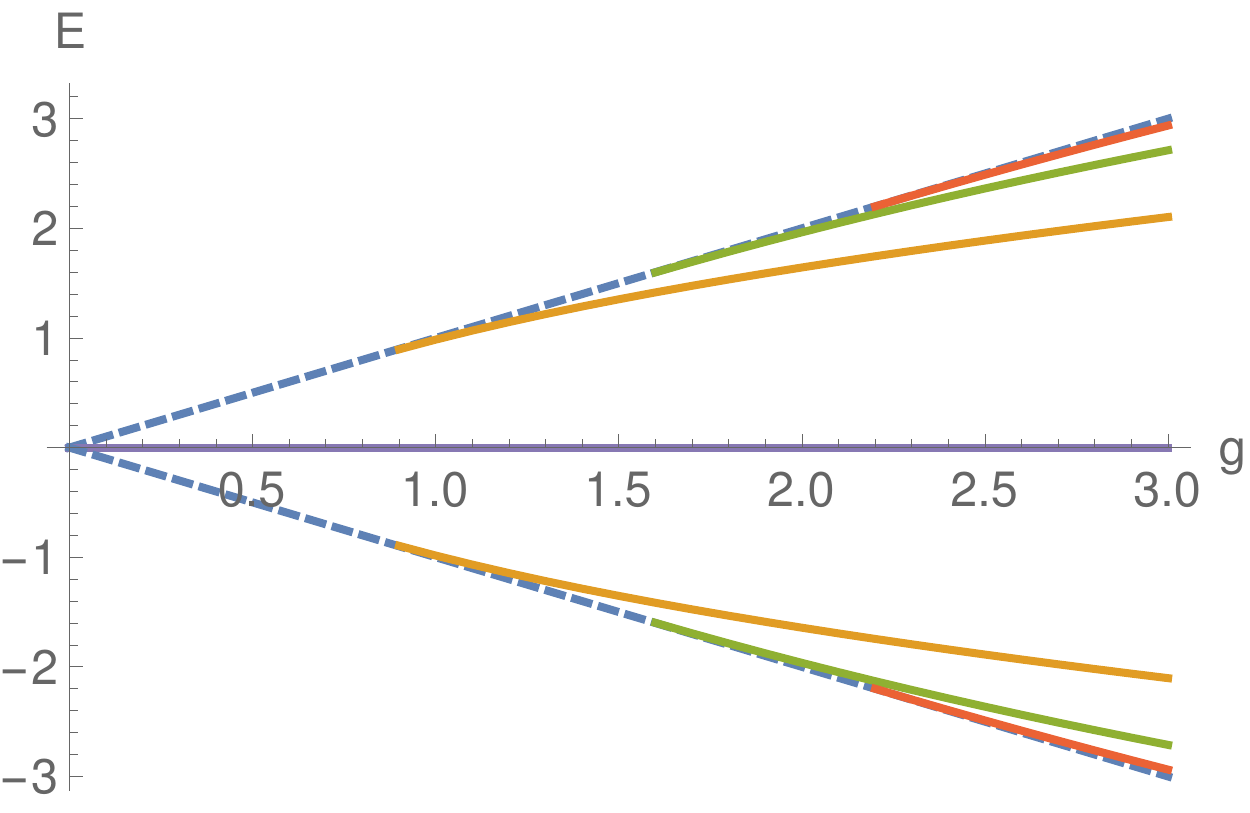}
         \caption{$\epsilon=0.05$}
     \end{subfigure}
        \caption*{Source: Results obtained by the author in \cite{campos2021fermions}.}
        \label{fig6_Evsg}
\end{figure}

The fermionic spectra in the presence of the kinks are shown in Fig.~\ref{fig6_Evsg} for both normal and quasinormal mode cases. This can be easily computed numerically after substituting both kink solutions in the effective fermionic potential given $V_\pm$. This is in contrast with the $\phi^4$ model, where the fermionic bound states are found analytically. Similar to the $\phi^4$ model, both positive and negative energy states appear as the Yukawa coupling $g$ is increased. In the quasinormal mode case, the kink's tail decays more slowly to the vacuum, and the effective potential is wider than in the normal mode case. Therefore, the bound states appear at smaller values of $g$.

\begin{figure}
     \centering
     \caption{Evolution of the fermion density in spacetime for the toy model with several values of the parameters and $A=0.1$.}
     \begin{subfigure}[b]{0.45\textwidth}
         \centering
         \includegraphics[width=\textwidth]{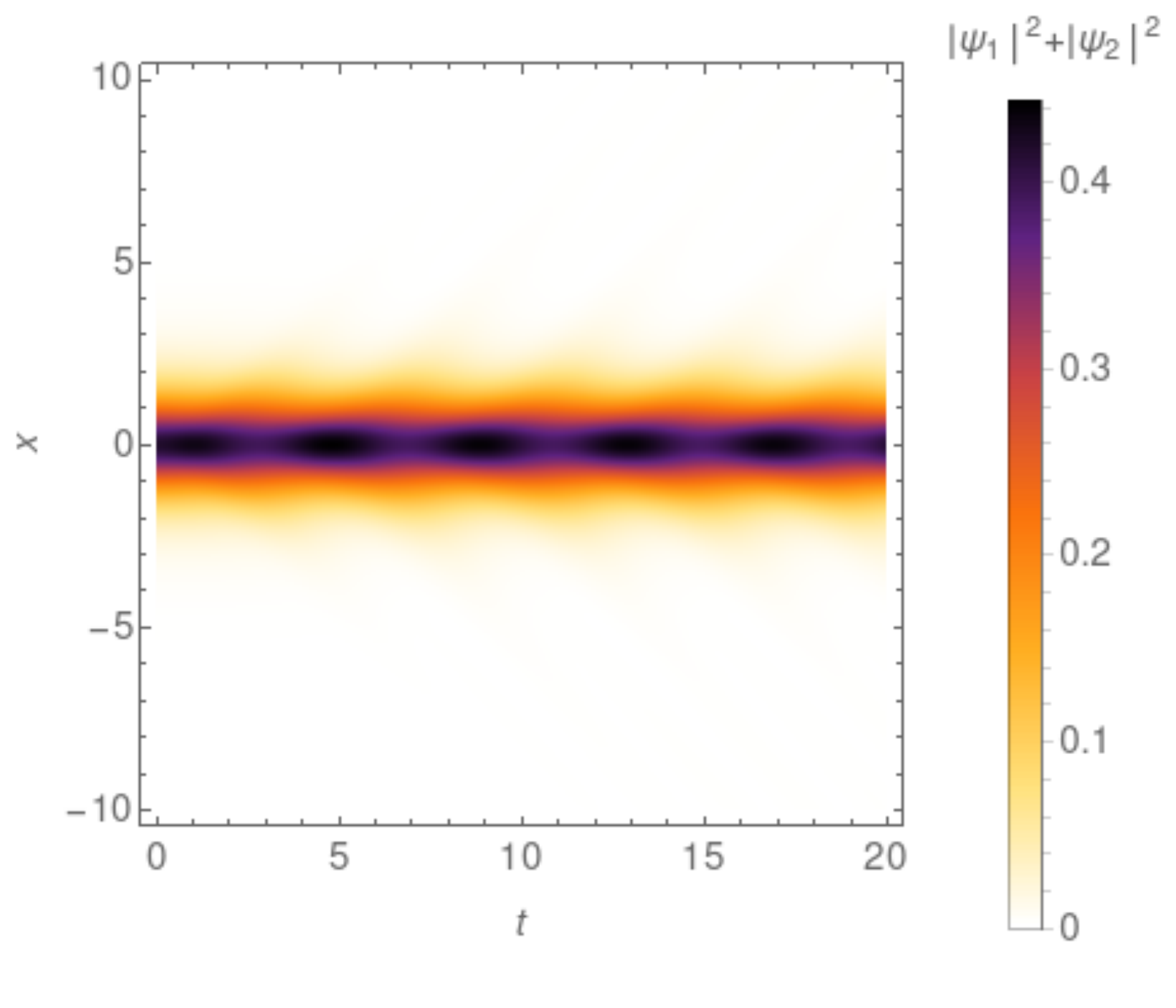}
         \caption{$\epsilon=0.0$, $g=0.8$}
     \end{subfigure}
     \hfill
     \begin{subfigure}[b]{0.45\textwidth}
         \centering
         \includegraphics[width=\textwidth]{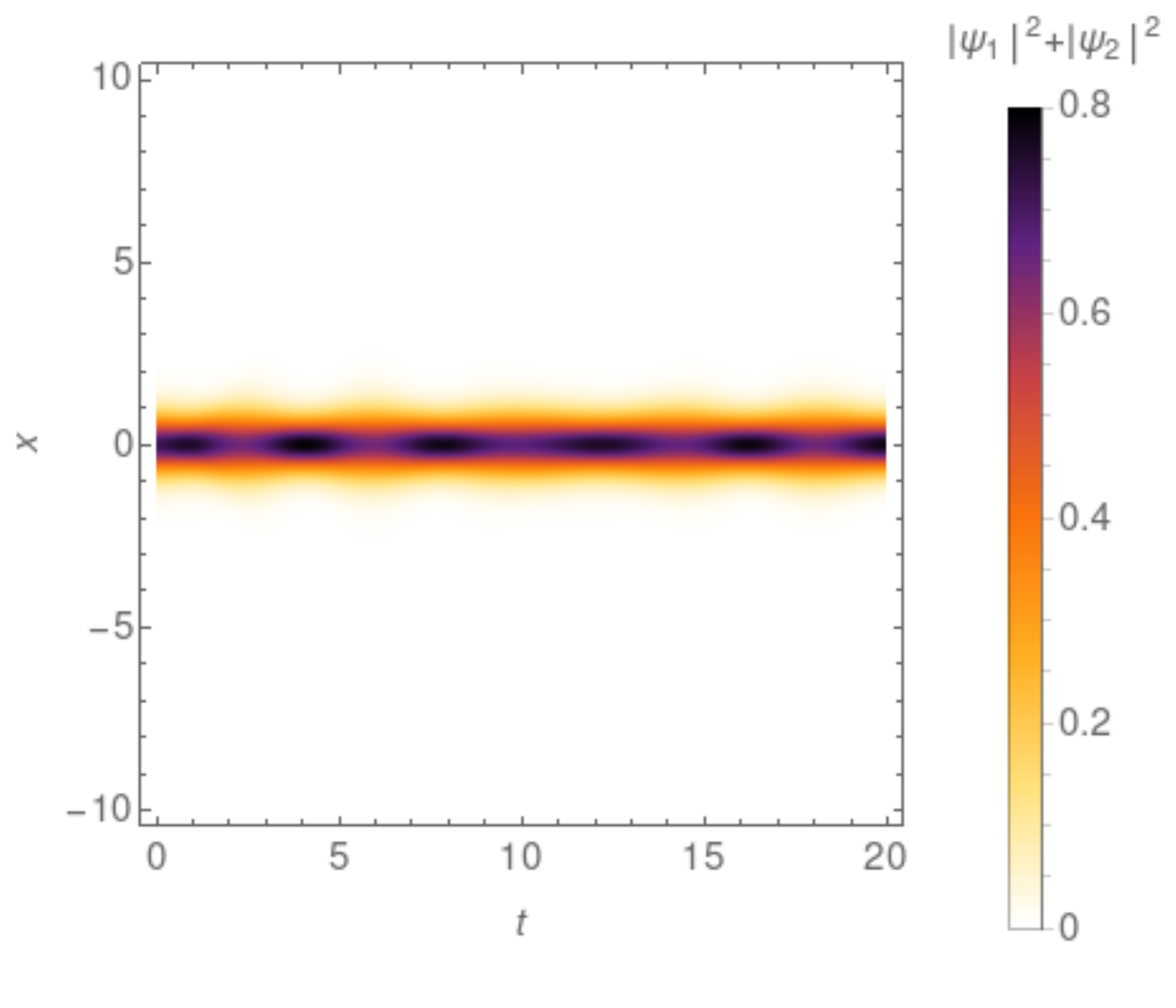}
         \caption{$\epsilon=0.0$, $g=2.1$}
     \end{subfigure}
     \hfill     
     \begin{subfigure}[b]{0.45\textwidth}
         \centering
         \includegraphics[width=\textwidth]{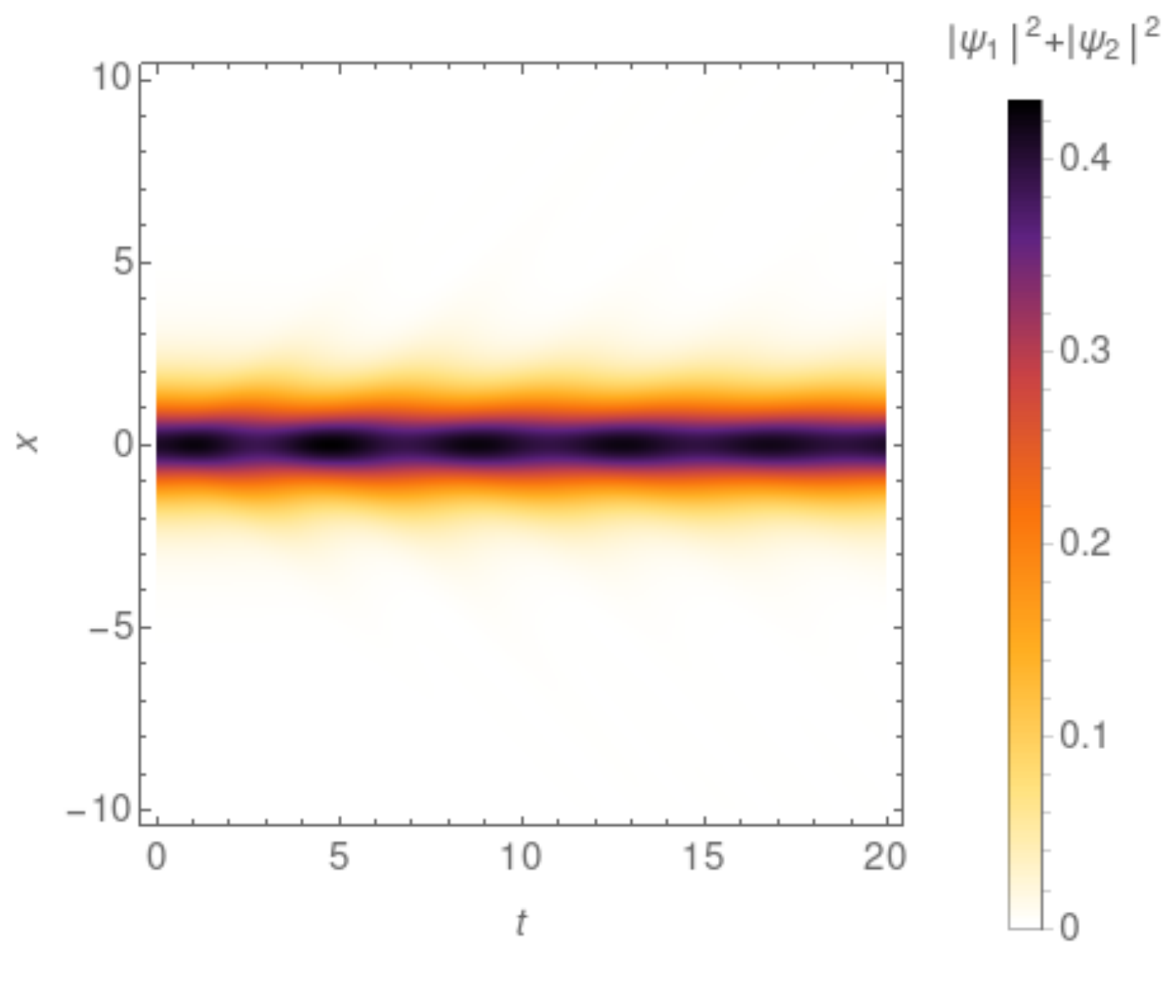}
         \caption{$\epsilon=0.05$, $g=0.8$}
     \end{subfigure}
     \hfill     
     \begin{subfigure}[b]{0.45\textwidth}
         \centering
         \includegraphics[width=\textwidth]{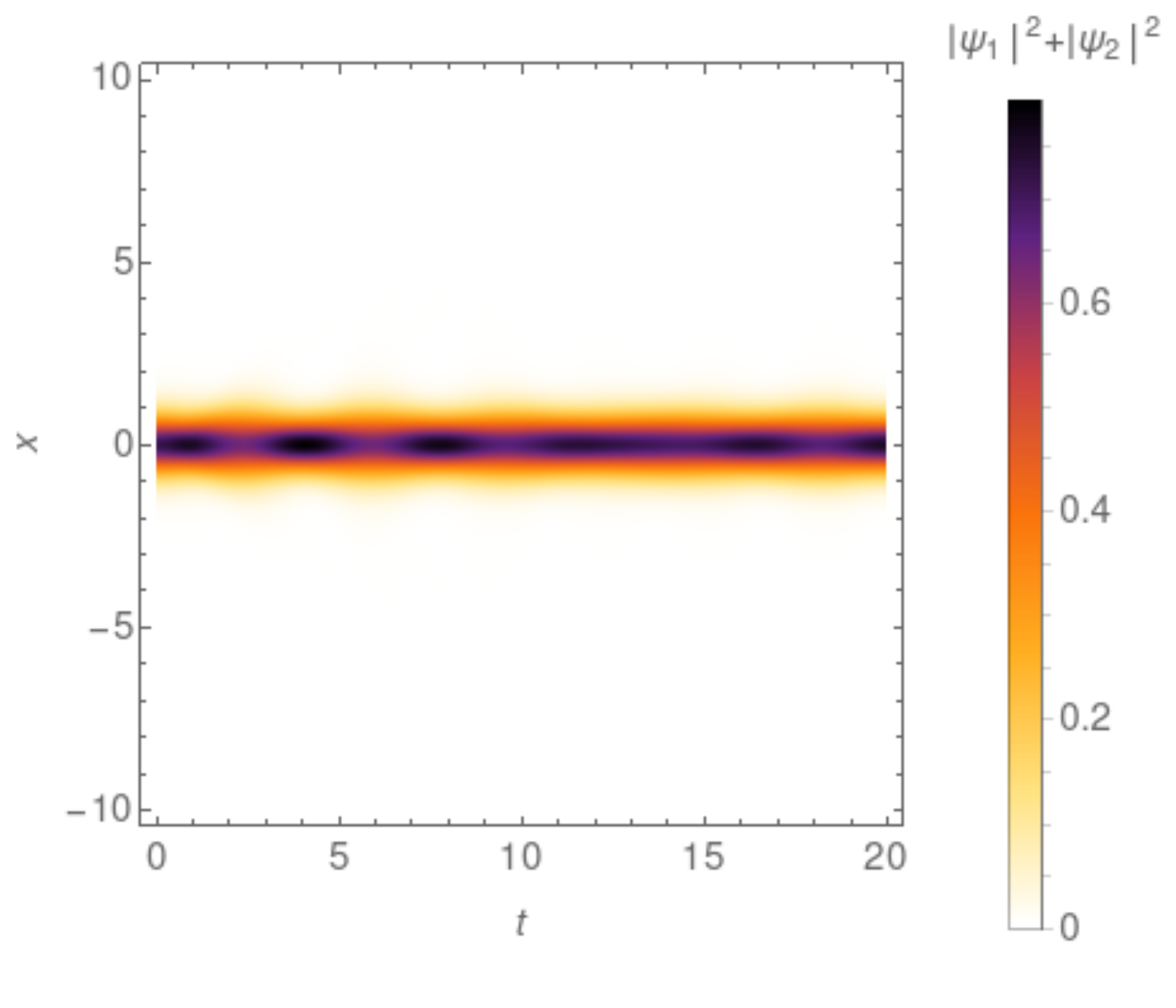}
         \caption{$\epsilon=0.05$, $g=2.1$}
     \end{subfigure}
        \caption*{Source: Results obtained by the author in \cite{campos2021fermions}.}
        \label{fig6_dens}
\end{figure}

In Fig.~\ref{fig6_dens}, we show the evolution of the fermionic density obtained from numerical integration of the equations of motion with wobbling amplitude $A=0.1$. Similar to the $\phi^4$ case, we choose to initialize the fermion at the zero mode. Hence, it starts confined at the kink, located at the origin. Then, due to the perturbation, the system oscillates, and it is possible to see some radiation for smaller values of $g$. In the right panels, $g$ is larger, and the fermion density starts more localized at the origin than the fermion density shown in the left panels. This occurs because the unperturbed Hamiltonian becomes more confining as $g$ increases. However, the perturbation effect as $g$ varies is more complicated, as shown in the $\phi^4$ case. In fact, it can favor the escape of fermions initially, as $g$ increases from zero. If we continue increasing $g$, it does not favor the escape anymore, after the energy gap to the continuum becomes too large.

\subsection{Normal mode}

\begin{figure}
     \centering
     \caption{Matrix elements of the operator $\eta_S\sigma_1$ (a) between discrete states and (b) between discrete states and the continuum. We consider the toy model with $\gamma=3.0$ and $\epsilon=0.0$. In (b), we fix $g=1.6$.}
     \begin{subfigure}[b]{0.42\textwidth}
         \centering
         \includegraphics[width=\textwidth]{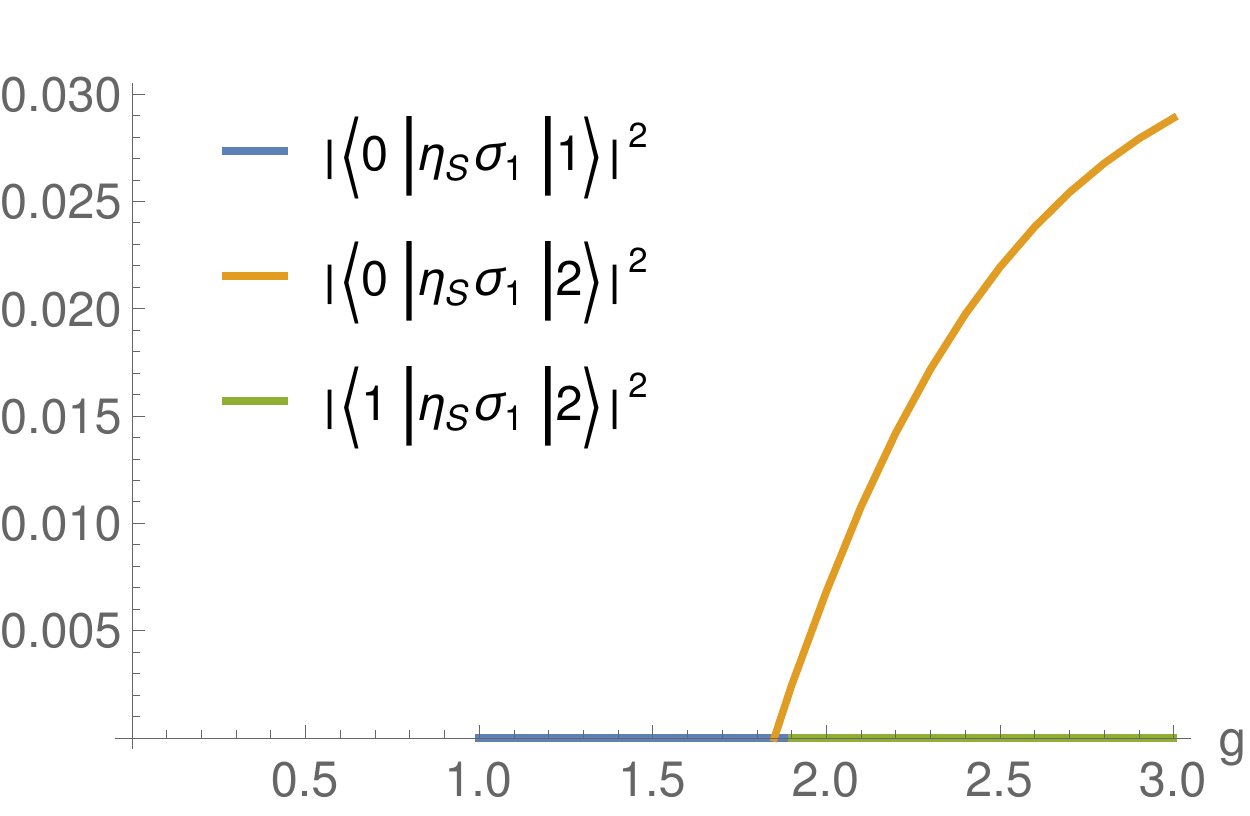}
         \caption{}
     \end{subfigure}
     \hfill
     \begin{subfigure}[b]{0.42\textwidth}
         \centering
         \includegraphics[width=\textwidth]{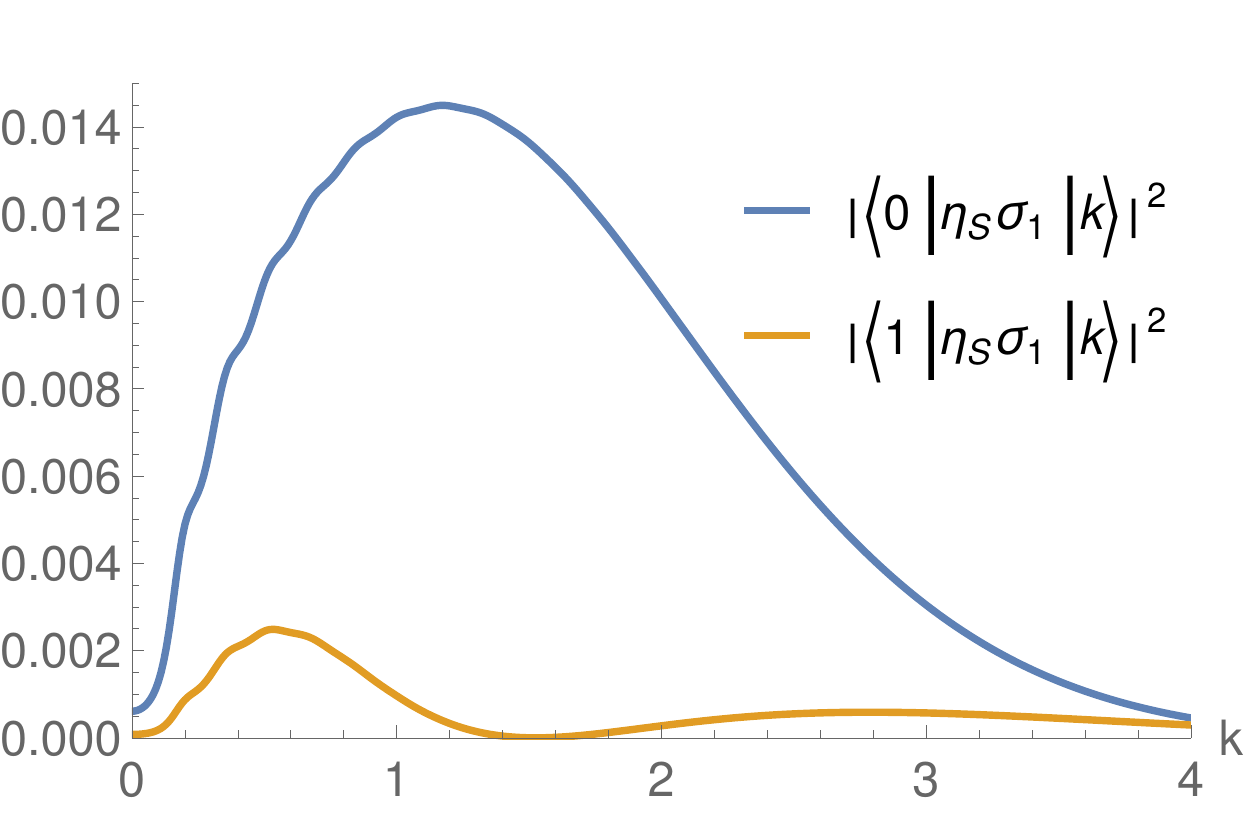}
         \caption{}
     \end{subfigure}
        \caption*{Source: Results obtained by the author in \cite{campos2021fermions}.}
        \label{fig6_SWeta}
\end{figure}

Now, we focus on the case where the kink has only normal modes. This means that $\epsilon=0.0$ and, as mentioned before, $\gamma=3.0$. We start by computing the matrix elements of the operator $\eta_S\sigma_1$, which in this case must be done using the numerical values of the fermion bound states. The result is shown in Fig.~\ref{fig6_SWeta}. The results are completely analogous to the $\phi^4$ case. We also find the parity selection rule and the same qualitative behavior of the matrix elements from discrete states to the continuum. From this analysis, we expect that these elements should have the same dependence with the parameters for other scalar potentials with $Z_2$ symmetry and containing kinks.

\begin{figure}
     \centering
     \caption{Evolution of the transition probabilities in time. We compute transitions from the initial states to the continuum and to other discrete states. In (a) the transition rate to the continuum is the slope of the dotted line. We consider the toy model with $\gamma=3.0$, $\epsilon=0.0$ and $A=0.1$.}
     \begin{subfigure}[b]{0.48\textwidth}
         \centering
         \includegraphics[width=\textwidth]{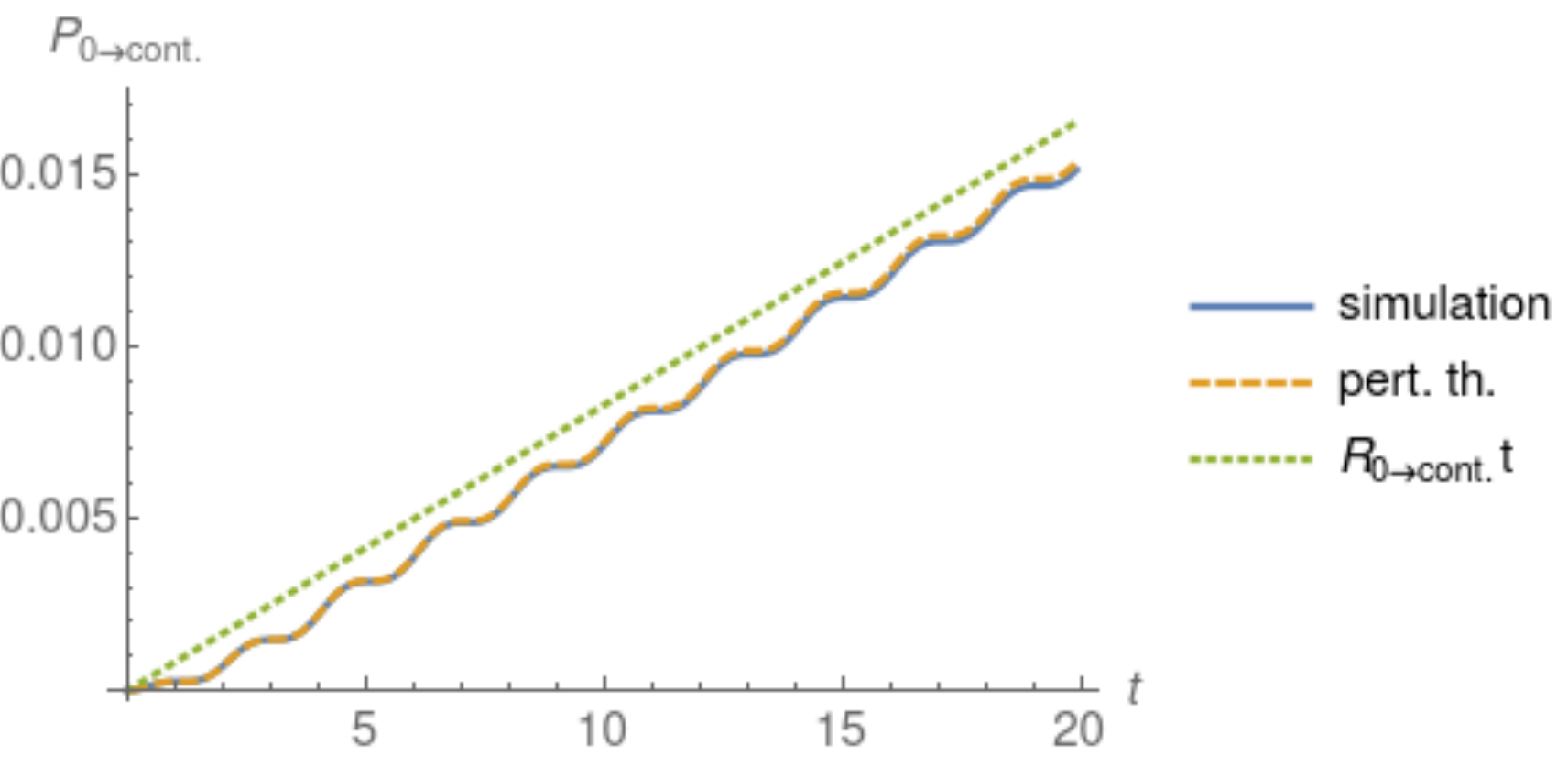}
         \caption{$g=0.8$}
     \end{subfigure}
     \hfill
     \begin{subfigure}[b]{0.48\textwidth}
         \centering
         \includegraphics[width=\textwidth]{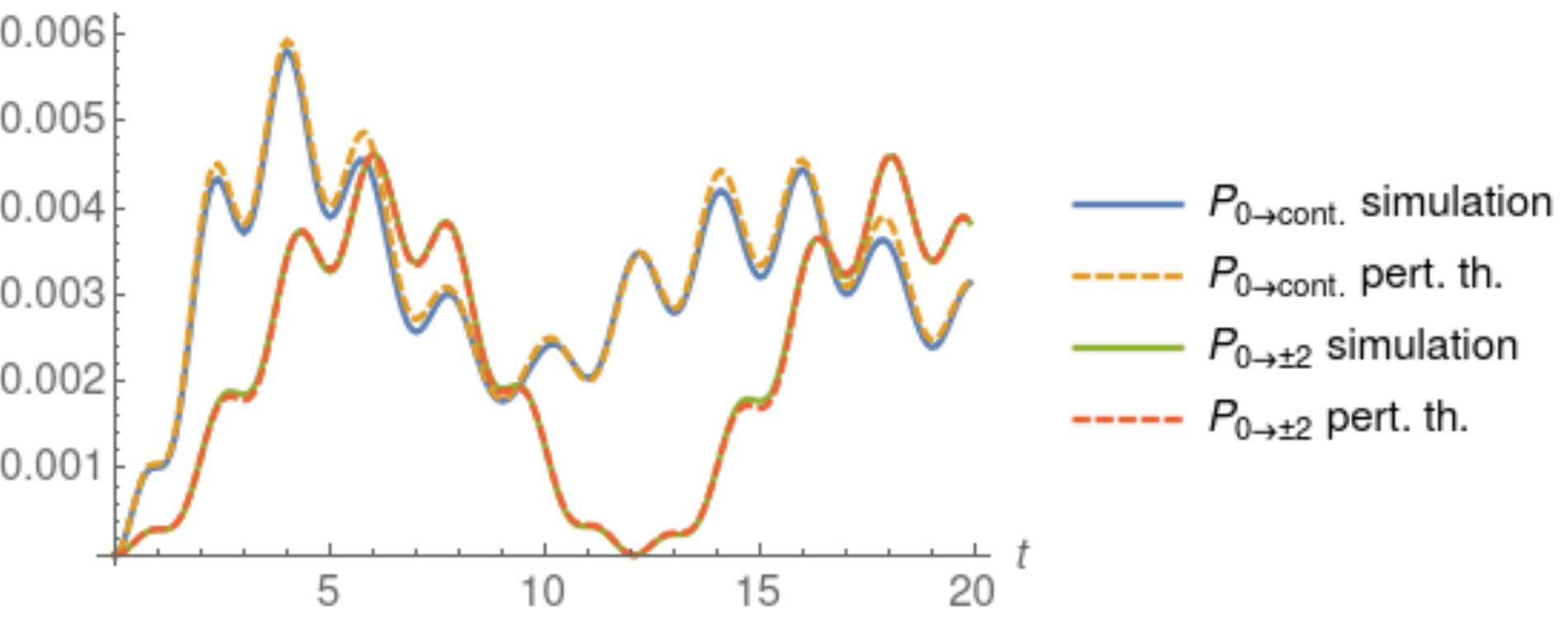}
         \caption{$g=2.1$}
     \end{subfigure}
        \caption*{Source: Results obtained by the author in \cite{campos2021fermions}.}
        \label{fig6_SWcoef}
\end{figure}

Next, we compute the evolution of the transition probability in time for a fermion in the presence of a wobbling kink with amplitude $A=0.1$. This amplitude is measured with respect to the following shape mode profile
\begin{equation}
\label{eq6_shapemode}
\eta_S(x)=\begin{cases}-e^{k_Sx},&x<-L,\\
e^{-k_SL}\sin(p_Sx)/\sin(p_SL),&-L<x<L,\\
e^{-k_Sx},&x>L.\end{cases}
\end{equation}
This is the first odd solution listed in section \ref{sec2_SW} and the normalization was fixed arbitrarily. In the expression above, we defined $k_S\equiv\sqrt{\gamma-\omega_S^2}$ and $p_S\equiv\sqrt{\omega_S^2+1}$ with $\omega_S$ being the solution of the transcendental equation $k_S=-p_S\cot(p_SL)$. Again, we initialize the systems with the fermion at zero mode. The result for the transition to all continuum states and other discrete states is shown in Fig.~\ref{fig6_SWcoef}. The figure also compares the evolution computed with the perturbative analysis to numerical simulations.
Moreover, the value of the transition rate to the continuum obtained with Fermi's golden rule is shown as the slope of the dotted line. Similarly to the $\phi^4$ case, we observe a transition to the continuum for $g=0.8$, which agrees with Fermi's golden rule when averaged over large times. Considering the case with $g=2.1$, the transition to the continuum does not exhibit a steady increase in the transition probability and has a complicated oscillatory behavior because the energy gap between the continuum and the zero mode is too large. Moreover, the transition probability to the second excited state also oscillates in a complicated manner as expected from eq.~(\ref{eq6_cn1integrated}). Again, this result is very similar to what was obtained in the $\phi^4$.

\subsection{Quasinormal mode}

Now, we modify the toy model, turning the normal mode into a quasinormal mode. The main goal of this analysis is to see how the previous results are altered when this change is made. Again we start by computing the fermion bound states numerically and using the result to compute the matrix elements of the operator $\eta_S\sigma_1$. We also find the parity selection rules and the same qualitative behavior of the matrix elements from discrete states to the continuum. Hence, we choose not to show the results graphically.

Now, we would like to study how the fermions evolve when subject to a quasinormal mode as a perturbation. Hence, we need to define precisely the perturbation Hamiltonian. One possibility is to use the analytical solution of the quasinormal mode for $\eta_S$ and $\omega_S$. This solution was computed in section \ref{chap2}. It diverges for large $x$, but it describes the behavior of the quasinormal mode for large times if carefully truncated. However, it cannot be used to describe the transient behavior quasinormal mode as it leaks through the potential barrier. In situations where a normal mode turns into a quasinormal mode, the evolution of the perturbation starting with the normal mode configuration is a good approximation of the transient behavior that we are looking for. We will adopt this approach because we find it more sensible. In other words, we will compute the evolution of scalar field perturbation with the normal mode profile of the corresponding model with the same value of $\gamma$ but with $\epsilon=0.0$ as an initial condition and use it as the perturbation Hamiltonian. The initial profile is given by eq.~(\ref{eq6_shapemode}). Its evolution is described by the time-dependent equation of perturbations around the kink solution given below \begin{equation}
\label{eq7_lineta}
\frac{\partial^2\eta(x,t)}{\partial t^2}-\frac{\partial^2\eta(x,t)}{\partial x^2}+U(x)\eta(x,t)=0,
\end{equation}
where the linearized potential $U(x)$ is given by eq.~(\ref{eq3_UQNM}). Then, the evolution of the coefficients $c_n^{(1)}(t)$ can be computed using eq.~(\ref{eq6_cn1}) with $H_1(t)=g\eta(x,t)\sigma_1$.

We can find an approximation for the evolution of the quasinormal mode. It will approximately oscillate with a frequency $\Omega$ but also decay with the rate $\Gamma$, which were found analytically in section \ref{chap2}. In this case, the perturbed scalar field is approximated by $\phi=\phi_k+A\eta_S\cos(\Omega t)e^{-\Gamma t}$, where $\eta_S$ is given by eq.~(\ref{eq6_shapemode}). Then it is easy to show that the coefficient $c_n^{(1)}(t)$ is given by 
\begin{equation}
\label{eq6_cn1QNM}
c_n^{(1)}(t)=\frac{g}{2}\langle n|\eta_S\sigma_1|i\rangle\left(\frac{1-e^{i(\omega_{ni}+\Omega+i\Gamma)t}}{\omega_{ni}+\Omega+i\Gamma}+\frac{1-e^{i(\omega_{ni}-\Omega+i\Gamma)t}}{\omega_{ni}-\Omega+i\Gamma}\right).
\end{equation}
This description is incomplete because the quasinormal mode will also generate outgoing radiation as it decays, but it allows a closed expression for the transition coefficients.

\begin{figure}
     \centering
     \caption{Evolution of the transition probabilities in time. We compute transitions from the initial states to the continuum and to other discrete states. The numerical evolution is compared to two results obtained from perturbation theory. We consider the toy model with $\gamma=3.0$, $\Delta\gamma=2.0$, $\epsilon=0.05$ and $A=0.1$.}
     \begin{subfigure}[b]{0.48\textwidth}
         \centering
         \includegraphics[width=\textwidth]{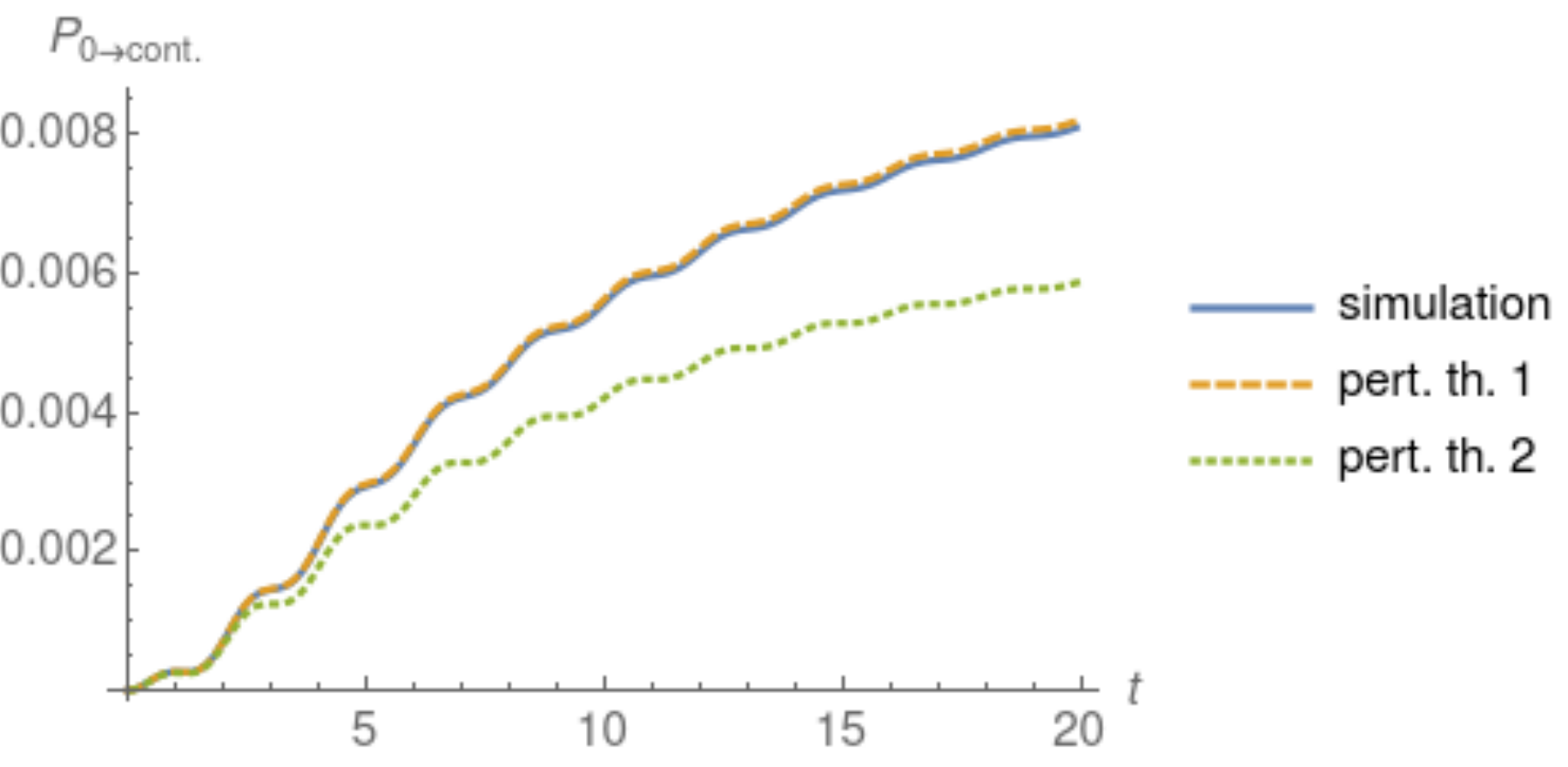}
         \caption{$g=0.8$}
     \end{subfigure}
     \hfill
     \begin{subfigure}[b]{0.48\textwidth}
         \centering
         \includegraphics[width=\textwidth]{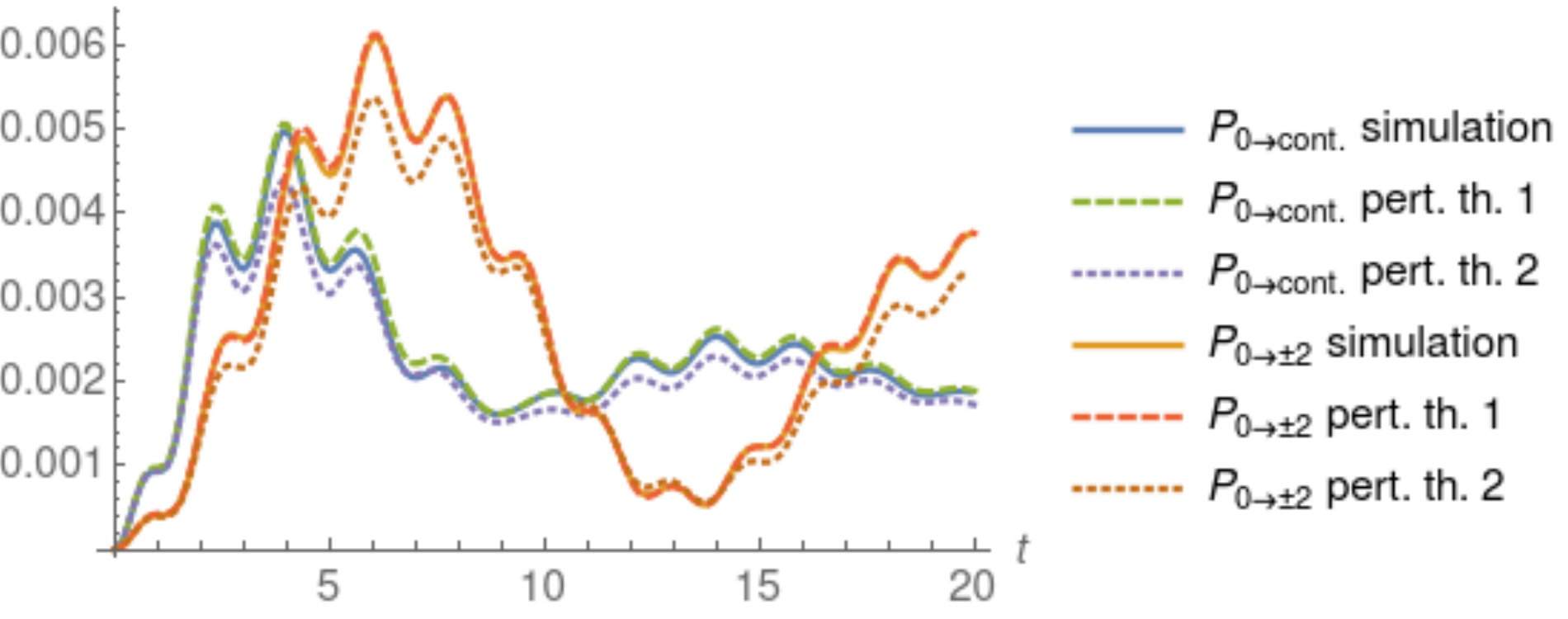}
         \caption{$g=2.1$}
     \end{subfigure}
        \caption*{Source: Results obtained by the author in \cite{campos2021fermions}.}
        \label{fig6_QNMcoef}
\end{figure}

After defining the perturbation, we can compute the evolution of the transition probability with time. First, we compute the evolution numerically using the full transient evolution of the quasinormal mode. Second, we compute the evolution according to eq.~(\ref{eq6_cn1}) also using the full transient evolution of the quasinormal mode. In the figure, this method is called perturbation theory 1. Finally, we compute the evolution according to eq.~(\ref{eq6_cn1QNM}). In the figure, this method is called perturbation theory 2. The results are shown in Fig.~\ref{fig6_QNMcoef} considering the perturbation amplitude $A=0.1$. The first two methods have a good agreement.
On the other hand, the third one gives the correct behavior but is less accurate. In the left panel, we show the evolution for $g=0.8$. This result is similar to the cases with normal modes, but the transition probability increases at slower rates as time evolves. This occurs because the perturbation gets weaker as it leaks through the potential barrier of the kink. In the right panel, we show the results for $g=2.1$. The transition probabilities oscillate similarly to the previous cases, but the oscillations also get weaker with time due to decreased perturbation strength.

\begin{figure}[tbp]
\centering
  \caption{Asymptotic value of the transition probability as a function of $g$. We consider the toy model with parameters $\gamma=3.0$, $\Delta\gamma=2.0$, $\epsilon=0.05$ and $A=0.1$.}
  \includegraphics[width=0.6\columnwidth]{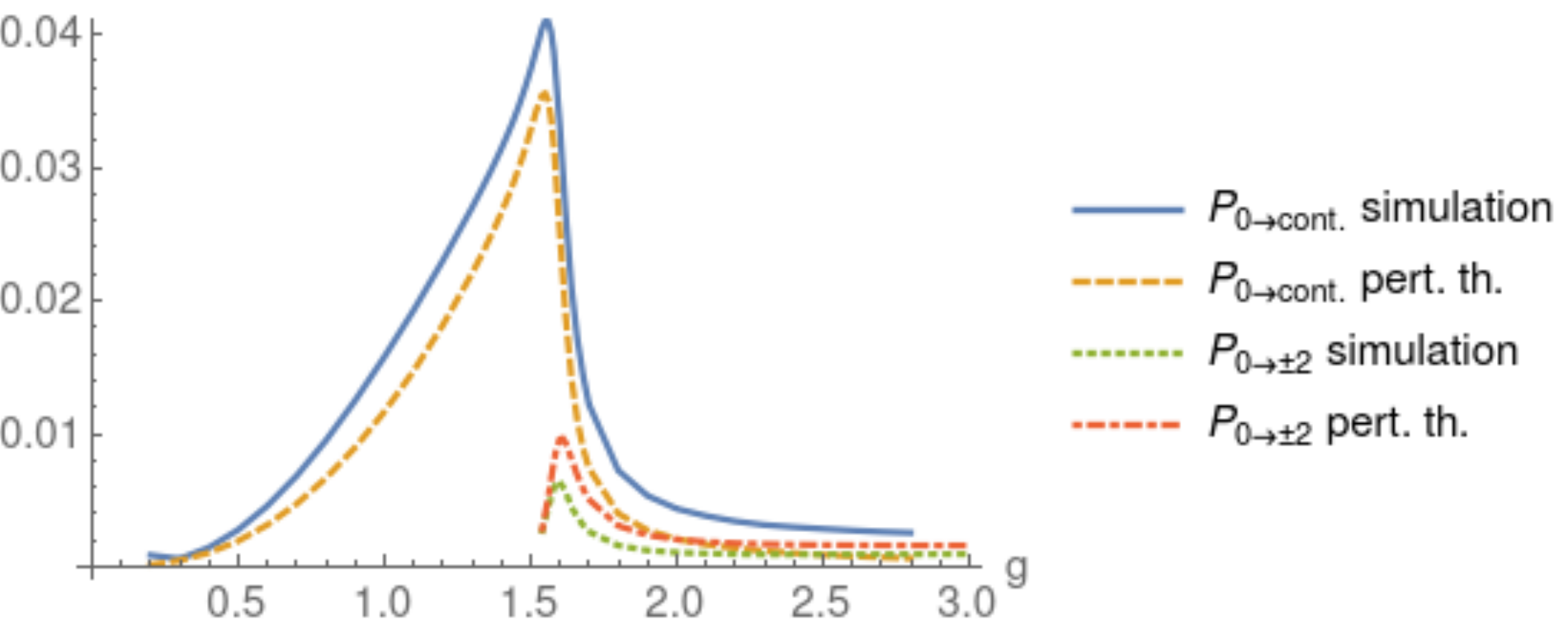}
       \caption*{Source: Results obtained by the author in \cite{campos2021fermions}.}
       \label{fig5_final}
\end{figure}

As the perturbation gets weaker with time, the transition probability will reach an asymptotic state if we wait long enough. We can use this value to summarize the system's behavior as a function of the Yukawa coupling $g$. The result is shown in Fig.~\ref{fig5_final}. We compare the results from numerical simulations with the value of eq.~(\ref{eq6_cn1QNM}) in the limit of very large $t$. The two results have very similar behaviors. Observe that the asymptotic value of the transition probability to the continuum initially increases with $g$ and then suddenly decays to a very small value for $g\simeq\Omega$. This behavior is reminiscent of Fermi's golden rule, which states that the transition is forbidden if the energy gap to the continuum is too large. Near $g=1.54$, the second excited state appears. The asymptotic value of the transition probability to this state quickly reaches a peak and then decays as the energy gap between the two states gets larger.

The following section will summarize the results described above, discuss our conclusions, and list some ideas for future works.

\section{Conclusion}

In this work, we investigated the effect of an excited vibrational mode on a fermion that is bound to a kink. We started by considering the $\phi^4$ model and a toy model. Both have fermionic bound states in the presence of a non-vibrating kink, but the former can be solved analytically, while the latter can only be solved numerically. Then we include the vibration as a perturbation and compute the transition between these states perturbatively. This analysis leads to an expression for the transition probability and the transition rate, which obeys Fermi's golden rule.

The fermionic spectrum for the toy model is similar to the $\phi^4$ case. Interestingly, the bound states in the quasinormal case appear earlier than in the normal mode one because the effective potential is wider in this case. The full evolution fermionic field can be computed numerically. We observe that the fermion density is mainly located on the origin, where the kink is. The density oscillates and can also radiate when the transition to the continuum is allowed.

The perturbative expressions for the transition probabilities depend on matrix elements that can be easily computed. We specialize to the case where the fermion starts at zero mode. The perturbative analysis is compared to numerical simulations, and the two have good agreement. We find that for small values of the Yukawa coupling $g$, the transition to the continuum increases with an average increasing rate given by Fermi's golden rule, while for large values of $g$, the probabilities are bounded. Thus, the only effect of the perturbation is a complex oscillation. This result is significant because it predicts that a fermion bound to a kink will have a fixed escape rate if the kink is wobbling and the Yukawa coupling constant is not too large. This could possibly be measured in some experimental setting of a fermion-kink system.

Finally, when considering the quasinormal mode case, we can describe the transient behavior starting with the corresponding normal mode with $\epsilon=0.0$ and computing the evolution of this perturbation. The perturbation evolves according to a time-dependent equation subject to the linearized potential $U(x)$ of the quasinormal mode. We find that the probability reaches an asymptotic value due to the leakage of the perturbation. When plotted as a function of the Yukawa coupling constant $g$, this value initially increases and then sharply decreases when the gap between the zero mode and the continuum becomes too large. This behavior is the analogue of Fermi's golden rule for quasinormal modes. One interesting continuation of this work is considering excited states as the initial condition. One could also try to formulate the problem considering the fermion back-reaction at every step and see how the results are altered. However, this problem is much more challenging.

\section{Appendix: Numerical method}
\label{sec6_num}

We simulate the system in a box with range $-100.0<x<100.0$ with spacing $h=0.1$. Again, we apply the method of lines to integrate the equations of motion. The partial derivatives with respect to $x$ are given by a five-point stencil approximation
\begin{equation}
\frac{\partial\psi}{\partial x}=\frac{-\psi(x+2h)+8\psi(x+h)-8\psi(x-h)+\psi(x-2h)}{12h},
\end{equation}
where the dependence on $t$ is suppressed for brevity. Moreover, at the boundaries, we set $\psi(x=\pm L,t)=0$. These conditions reflect the radiation, but simulations are short enough to avoid that the boundary conditions have any effect on the bulk evolution. At $x=-L+h$ we modify the partial derivative in order to consider only points inside the box
\begin{equation}
\frac{\partial\psi}{\partial x}=\frac{\psi(-L+4h)-6\psi(-L+3h)+18\psi(-L+2h)-10\psi(-L+h)-3\psi(-L)}{12h}.
\end{equation}
We use a similar expression at $x=L-h$. This results in a set of first-order ordinary differential equations for both the field $\phi$ and its time derivative $\phi_t$. The equations are integrated using a fifth-order Runge-Kutta algorithm with adaptive step size and error control \cite{dormand1980family}. The algorithm implemented in the solve\_ivp method of the SciPy library in Python. For the $\phi^4$ model and the toy model with $\epsilon=0.0$, we ignore the slow decay of the shape mode\footnote{The amplitude of the shape mode is not constant because it slowly emits radiation \cite{manton1997kinks}.} and we write the scalar field as $\phi=\phi_k+A\eta_S\cos(\omega_S t)$. When considering quasinormal modes, we must numerically integrate the linearized equation for the perturbation, eq.~(\ref{eq7_lineta}). This is done using the method of lines in a similar manner. The partial derivatives with respect to $x$ are computed using a five-point stencil approximation, and the resulting set of ordinary differential equations are integreted using the same Runge-Kutta method. At the boundaries, we also set $\eta(x=\pm L,t)=0$.

For the toy model, the fermion bound states are found numerically using the NDEigensystem method in Mathematica and the solve\_bvp method of the SciPy library in Python. The continuum states are found as described in \cite{gousheh2013casimir}. For instance, the scattering state for a wave coming from the left with wavenumber $k$ and energy $E_k=\sqrt{k^2+g^2}$ can be found factoring out the oscillatory behavior of the transmitted wave contribution as follows
\begin{equation}
\psi_k=e^{ikx}\chi_k(x).
\end{equation}
Then, we find the asymptotic value $\chi_k(\infty)$ by solving eqs.~(\ref{eq6_psinp}) and (\ref{eq6_psinm}) in the limit $x\to\infty$. Using $\phi_k(\infty)=1.0$ and $\chi_k^\prime(\infty)=0$, one can show that
\begin{equation}
\chi_k(\infty)=\begin{pmatrix}
1\\
\frac{g+ik}{\sqrt{g^2+k^2}}
\end{pmatrix}.
\end{equation}
After that, it is a simple matter to integrate eqs.~(\ref{eq6_psinp}) and (\ref{eq6_psinm}) backwards and obtain $\chi_k(x)$. Numerically, we start the integration from $L=20.0$, as we cannot reach infinity. Finally, the scattering functions are properly normalized.

\chapter{FERMION TRANSFER IN THE $\phi^4$ MODEL WITH AN IMPURITY}
\label{chap8}
\section{Overview}

The work in this main section resulted in the following publication \cite{campos2020fermion}. We will discuss a model describing a fermion coupled to a scalar field and compute the effect of a kink-antikink collision on the fermion. 

One of the earliest works to consider fermions in this scenario was \cite{gibbons2007fermions}. The authors considered a five-dimensional spacetime where a brane, described by a $\phi^4$ kink, localizes the fermion in the fifth dimension. Under a few assumptions, they show that the motion of the fermion in the fifth dimension decouples from the other spatial dimensions and can be effectively described by motion in $(1+1)$ dimensional spacetime. They considered a fermion bound to the kink in the zero mode as an initial condition. After colliding with an antikink, most fermions either stay bound to the kink or are transferred to the antikink, and only a few are lost as radiation. However, this is only true in the reflecting regime of kink-antikink collisions because, when the kink and the antikink annihilate, they form a bion. As the bion oscillates, it makes the fermions radiate at each oscillation, and all fermions will end up radiating to infinity. 

In \cite{saffin2007particle}, the authors made a very similar study, but with a broader range of parameters and higher numerical precision, and found similar results. They also included collisions of the kink with boundaries, which can be analyzed with the same formalism. The authors did not realize in these works that the background approximation is exact when the fermion starts at the zero mode of the kink or antikink.

We decided to investigate fermion transfer in a different setting to derive something new. However, finding relevant models is not an easy task because when the kink is not centered around the origin, the effective fermionic potential for a kink background may not allow bound states. This occurs, for instance, for the sine-Gordon and the $\phi^6$ models. 

In light of some recent works where the authors considered a family of models with a half-BPS preserving impurity \cite{adam2019phi, adam2019spectral, adam2019bps, adam2019solvable, manton2019iterated, adam2020kink}, we decided to study the fermion's behavior in background collisions in one of these models. In general, for topological models, Bogomol'nyi–Prasad–Sommerfield (BPS) solutions are static and saturate the energy in a given topological sector. Saturation means that it is the solution with the lowest energy value among all solutions obeying the boundary conditions that define the topological sector. Typical scalar field systems in $(1+1)$ dimensions, such as the $\phi^4$ and sine-Gordon, possess BPS solutions consisting of kinks and antikinks. When a half-BPS preserving impurity is included, one-half of the BPS solutions disappear, and the others acquire a nontrivial structure as opposed to the family of BPS solutions without an impurity, which is related by a trivial translation. This allows the nontrivial evolution in the configuration space of BPS solutions to be viewed as a multi-defect BPS interaction. In general, multi-defect BPS interaction only appear in models in higher spacetime dimensions, such as the abelian Higgs model and Yang-Mills-Higgs theory, and the one-dimensional models serve as a guide to these more complicated systems. 

In the following section, we will describe the model that we investigated. It consists of a fermion interacting with a background scalar field model with a half-BPS preserving impurity. This scalar field model possesses BPS and non-BPS defect solutions. In particular, we will consider, as a background, collisions between these defects. However, the background approximation is exact in the cases we will consider because the fermion will start at the zero mode.

\section{Model}

\subsection{Lagrangian and field equations}
We considered the following Lagrangian containing three types of terms
\begin{equation}
\mathcal{L}=\mathcal{L}_{\text{scalar}}+\mathcal{L}_{\text{fermion}}+\mathcal{L}_{\text{int}}.
\end{equation}
The scalar field term consists of the $\phi^4$ model with a half-BPS preserving impurity
\begin{equation}
\mathcal{L}_{\text{scalar}}=\frac{1}{2}\phi_t^2-\frac{1}{2}\phi_x^2-U(\phi)-2\sigma\sqrt{U(\phi)}-\sqrt{2}\sigma\phi_x-\sigma^2,
\end{equation}
where the potential is given by
\begin{equation}
U(\phi)=\frac{1}{2}(1-\phi^2)^2,
\end{equation}
and the $\sigma$ term describes the impurity, which will be defined below. The fermion Lagrangian has the standard massless form
\begin{equation}
\mathcal{L}_{\text{fermion}}=i\bar{\psi}\gamma^\mu\partial_\mu\psi. 
\end{equation}
The two fields interact via a Yukawa term
\begin{equation}
\mathcal{L}_{\text{int}}=-g\phi\bar{\psi}\psi.
\end{equation}

The scalar field Lagrangian was first considered in \cite{adam2019phi} without the fermion field. Here, we will briefly summarize some of its properties. The sigma term was added such that the system still possesses one BPS equation, instead of two. If we write the energy of the scalar field for a static configuration, we find
\begin{align}
E&=\int\left(\frac{1}{2}\phi_x^2+U(\phi)+2\sigma\sqrt{U(\phi)}+\sqrt{2}\sigma\phi_x+\sigma^2\right)dx,\\
&=\int\left(\frac{1}{\sqrt{2}}\phi_x+\sigma+\sqrt{U}\right)^2dx-\sqrt{2}\int\phi_x\sqrt{U}dx\geq-\sqrt{2}\int_{\phi(-\infty)}^{\phi(\infty)}\sqrt{U}d\phi.
\end{align}
The inequality above is clearly saturated by the BPS condition
\begin{equation}
\phi_x+\sqrt{2}\sigma+(1-\phi^2)=0.
\end{equation}
This solution corresponds to the antikink solution of the $\phi^4$ theory without impurity. On the other hand, the kink solution can only be found solving the full second-order field equations. To fix the impurity term, we require that the kink solution $\phi_{K_0}(x)=\tanh(x)$ is still a solution of the field equations. In this case, one finds
\begin{equation}
\sigma=\frac{\lambda}{\cosh^2(x)},
\end{equation}
where $\lambda$ is a constant that can be anywhere in the range $\lambda>-\sqrt{2}$. However, if we translate the kink solution, it is not a solution anymore.

The field equations can be found via the Euler-Lagrange equations. They give
\begin{align}
\label{eq3_ELphi}
\phi_{tt}-\phi_{xx}+2(\phi^2-1)\phi&=\frac{2\sqrt{2}\lambda}{\cosh^2(x)}(\phi-\tanh(x)),\\
i\gamma^\mu\partial_\mu\psi-g\phi\psi&=0.
\label{eq3_ELpsi}
\end{align}
In eq.~(\ref{eq3_ELphi}) we ignored the term proportional to $g\bar{\psi}\psi$ describing fermion back-reaction on the scalar field. Moreover, we can see from the same equation that the impurity term appearing on the right-hand side vanishes for $\phi(x)=\tanh(x)$, as desired. In this case, we are left with the equation of motion for the $\phi^4$ model, which is also solved by $\phi(x)=\tanh(x)$. 

For the Dirac matrices, we consider the following representation $\gamma^0=-\sigma^2$ and $\gamma^1=i\sigma^3$. In this case, the fermion field splits into two decoupled Majorana fields
\begin{equation}
\psi=\psi_1^M+i\psi_2^M.
\end{equation}
We will consider the field equations only for $\psi_1^M$ because they are identical to the ones for $\psi_2^M$. Defining the components as $\psi_1^M=(\psi_1,\psi_2)^T$, we find
\begin{align}
\label{eq3_dirac1}
\partial_t\psi_1=-\partial_x\psi_2+g\phi\psi_2,\\
\partial_t\psi_2=-\partial_x\psi_1-g\phi\psi_1.
\label{eq3_dirac2}
\end{align}
The representation of the gamma matrices and the notation of the Fermi field are not the same as before, but the analysis in this main section is very similar to what we did in section \ref{sec6_analytic}. Moreover, if we choose a field configuration where $\psi_2^M=0$ as an initial condition, the back-reaction term $g\bar{\psi}\psi$ vanishes. For this set of configurations, the background approximation is justified.

\subsection{Scalar field solutions}

Now let us review the solutions to the scalar field equations that were found in \cite{adam2019phi}. The first one we will discuss is the kink-on-impurity $\phi_{K_0}=\tanh(x)$. The impurity was chosen such that this is still a solution to the field equations, as can be seen in eq.~(\ref{eq3_ELphi}). The subscript $0$ denotes that it is bound to the impurity at the origin. 

\begin{figure}[tbp]
\centering
  \caption{Profile of BPS solutions. (Solid) Isolated lump, (Dashed) antikink and lump, (Dotted) antikink-on-impurity. We fix $\lambda=-1.0$.}
  \includegraphics[width=0.6\columnwidth]{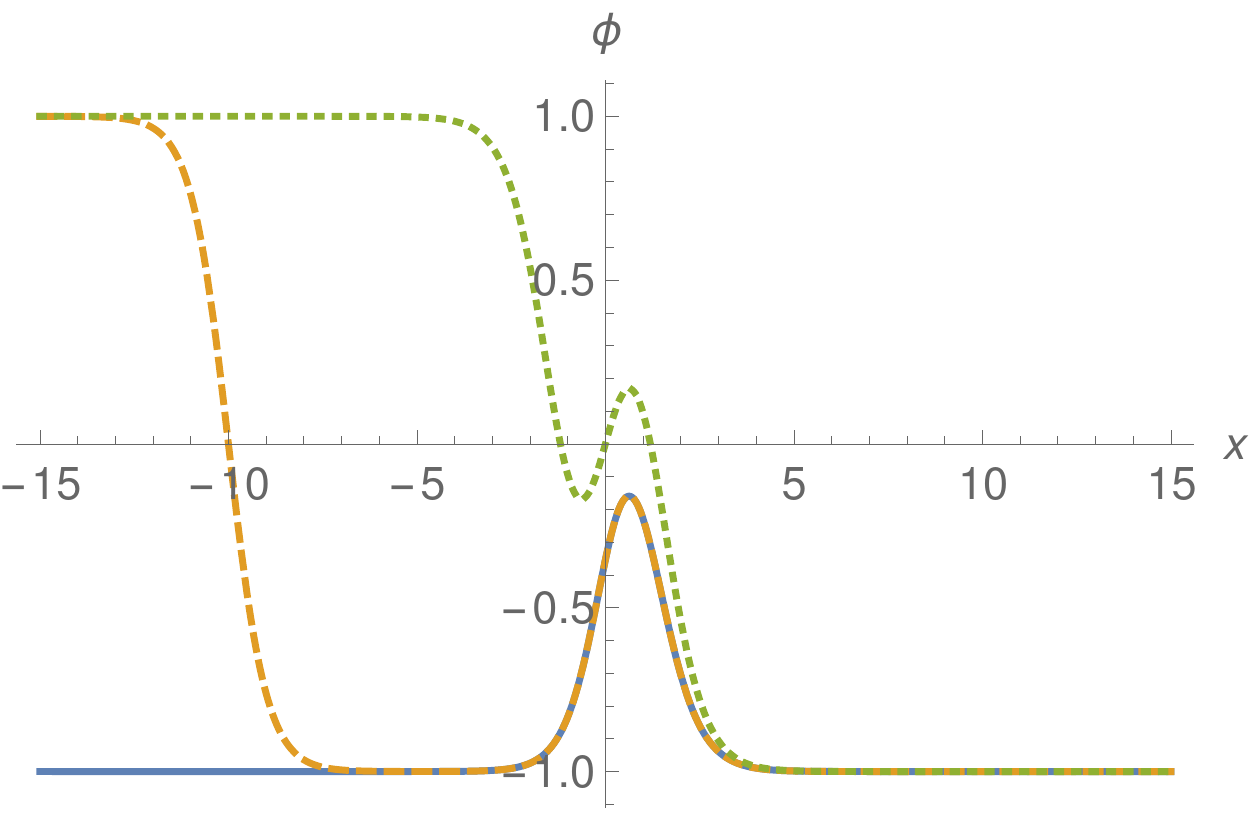}
       \caption*{Source: Results obtained by the author in \cite{campos2020fermion}.}
       \label{fig3_antikink-lump}
\end{figure}

Then, we turn to the BPS solutions $\phi_{LA}(x;x_0)$. The system does not possess translation symmetry anymore due to impurity. However, as half of the BPS property is preserved, there is still a generalized translational symmetry between BPS solutions. Defining the generalized coordinate, or modulus, of the BPS solution as the point $x_0$ where $\phi_{LA}(x_0;x_0)=0$, one can find a family of BPS solutions. The profile is obtained integrating the BPS equation numerically from this initial condition and, as we move in moduli space, the solutions vary non-trivially. Due to the presence of the impurity near the origin, the $\phi^4$ antikink is deformed and, in general, consists of an antikink $A$ and a lump $L$ as shown in the dashed curve in Fig.~\ref{fig3_antikink-lump}. When the modulus $x_0=0$, we have the antikink-on-impurity solution $A_0$, which is shown in the dotted curve in Fig.~\ref{fig3_antikink-lump}. In the limits where $x_0\to\pm\infty$, we find the lump solutions $L^\pm$, which are the deformed solution corresponding to the trivial vacuum solutions $\phi=\pm1$ in the absence of impurities. The negative solution $L^-$ is shown in the solid curve in Fig.~\ref{fig3_antikink-lump}. The aforementioned solutions vary with the parameter $\lambda$ and only exist for $\lambda>-\sqrt{2}$. We fixed $\lambda=-1$ for reasons that will appear when we discuss the fermion bound states.

Besides the solutions discussed above, we can also find approximate solutions using the additive ansatz. One prescription to construct such solutions is to add a solution of the $\phi^4$ model far from the origin and a solution of the full model with the impurity. For instance, we have the approximate solution
\begin{equation}
\label{eq3_phi1}
\phi(x)=\phi_{K_0}(x)+\phi_A(x;x_0)\pm1,
\end{equation}
where $\phi_A(x;x_0)=-\tanh(x-x_0)$ is the antikink solution of the $\phi^4$ model. This is a good solution for $|x_0|\gg 1$ and the sign on the last term depends on the sign of $x_0$. Of course, these solutions are not static because there is a force between the kink-on-impurity and the antikink. We could also consider boosted antikink solutions $\phi_A(x,t;x_0,v)=-\tanh(\gamma(x-vt-x_0))$ instead of the $\phi_A(x;x_0)$ in the ansatz.

Using the additive ansatz, it is possible to write an approximation to the solution $\phi_{LA}(x;x_0)$ for $|x_0|\gg 1$
\begin{equation}
\label{eq3_phi2}
\phi(x)=\phi_{L^\pm}(x)+\phi_A(x;x_0)\mp1.
\end{equation}
Again, the sign in the equation depends on the sign of $x_0$. Similarly, in the expression above, we could include a boost in the $\phi^4$ antikink. In this case, and if the $\phi^4$ antikink velocity is small, the solution approximates the system's dynamics near the BPS regime. In the next section, we will consider the evolution of the scalar field, starting with the aforementioned approximate solutions.

\subsection{Fermion bound states}

Now, let us investigate the behavior of the fermion in the presence of some solutions of the scalar field. As mentioned before, we ignore the fermion's back-reaction to the kink. We start by considering the ansatz $\psi_1=\eta_+\cos(\omega t-\theta)$ and $\psi_2=\eta_-\sin(\omega t-\theta)$. Substituting in eqs.~(\ref{eq3_dirac1}) and (\ref{eq3_dirac2}), we find two decoupled equations for $\eta_\pm$
\begin{equation}
-\partial_x^2\eta_\pm+g(g\phi^2\mp\partial_x\phi)\eta_\pm=\omega^2\eta_\pm.
\end{equation}
These are Scr\"{o}dinger-like equations with potentials $V_\pm=g(g\phi^2\pm\partial_x\phi)$. If the scalar field consists of the $\phi^4$ kink, these equations have well-known analytical solutions, which can be found, for instance, in \cite{chu2008fermions, charmchi2014complete}. They consist of a bound state for every integer in the range $0<n<g$ and the zero mode. Moreover, if one considers boosted $\phi^4$ kinks, it is necessary to boost the fermion field as well. This is achieved in the usual way, which we will revise following \cite{tong}.  

A vector is defined as being transformed under a Lorentz transformation by the matrix 
\begin{equation}
\Lambda=\exp\left(\frac{1}{2}\Omega_{\rho\sigma}\mathcal{M}^{\rho\sigma}\right),
\end{equation}
where $\Omega_{\rho\sigma}$ is an antisymmetric constant matrix and $\mathcal{M}^{\rho\sigma}$ are the Lorentz group generators in the vector representation. They are given by
\begin{equation}
(\mathcal{M}^{\rho\sigma})^{\mu\nu}=\eta^{\rho\mu}\eta^{\sigma\nu}-\eta^{\sigma\mu}\eta^{\rho\nu}.
\end{equation}
For a boost in the positive $z$ direction, we set $\Omega_{01}=-\Omega_{10}=\chi$. After lowering one index in $\mathcal{M}^{\rho\sigma}$, we get
\begin{equation}
{\Lambda^\mu}_\nu=\exp\begin{pmatrix}
0&\chi\\
\chi&0
\end{pmatrix}=
\begin{pmatrix}
\cosh\chi&\sinh\chi\\
\sinh\chi&\cosh\chi
\end{pmatrix}.
\end{equation}
The parameter $\chi$ is known as rapidity, defined as $\chi=\tanh^{-1} v$. Using this definition, the matrix assumes the usual form. 

Now, we consider the spinor representation of the Lorentz group. In this case, a Lorentz transformation is performed by the operator
\begin{equation}
S[\Lambda]=\exp\left(\frac{1}{2}\Omega_{\rho\sigma}S^{\rho\sigma}\right),
\end{equation}
where $S^{\rho\sigma}$ are the generetors of the Lorentz group in the spinor representation. They are given by
\begin{equation}
S^{\rho\sigma}=\frac{1}{4}[\gamma^\rho,\gamma^\sigma].
\end{equation}
In particular,
\begin{equation}
S^{01}=-S^{10}=\frac{1}{4}[\gamma^0,\gamma^1]= \begin{pmatrix}
0&\frac{1}{2}\\
\frac{1}{2}&0
\end{pmatrix}.
\end{equation}
Using the same values for $\Omega^{\rho\sigma}$, the spinor transformation matrix becomes
\begin{equation}
S[\Lambda]=\exp\begin{pmatrix}
0&\chi/2\\
\chi/2&0
\end{pmatrix}=
\begin{pmatrix}
\cosh(\chi/2)&\sinh(\chi/2)\\
\sinh(\chi/2)&\cosh(\chi/2)
\end{pmatrix}.
\end{equation}
Finally, the full transformation will be given by 
\begin{equation}
\psi(x)\to S[\Lambda]\psi(\Lambda^{-1}x).
\end{equation}
We define the transformed coordinates $x^\prime=\gamma(x-vt)$ and $t^\prime=\gamma(t-vx)$ and the full Lorentz transformation for the spinor system is
\begin{align}
\psi_1^\prime=\cosh(\chi/2)\psi_1(x^\prime,t^\prime)+\sinh(\chi/2)\psi_2(x^\prime,t^\prime),\\
\psi_2^\prime=\sinh(\chi/2)\psi_1(x^\prime,t^\prime)+\cosh(\chi/2)\psi_2(x^\prime,t^\prime).
\end{align}

\begin{figure}[tbp]
\centering
  \caption{Spectrum of fermion coupled to (a) the lump and (b) to the BPS antikink solution with $\lambda=-1$. We set $g=2.0$, which corresponds to the supersymmetric case.}
  \includegraphics[width=0.9\columnwidth]{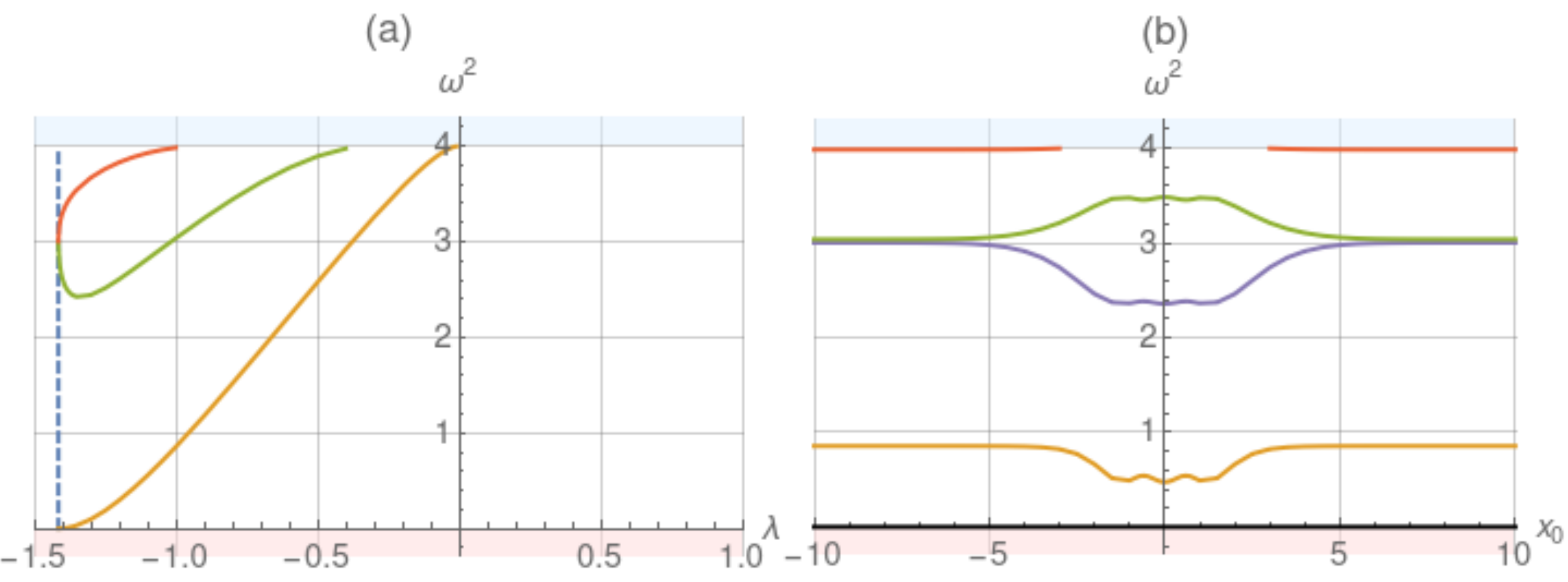}
       \caption*{Source: Results obtained by the author in \cite{campos2020fermion}.}
       \label{fig3_spectrum}
\end{figure}

Moving on to other configurations of the scalar field, we calculated the fermion spectrum in the presence of the lump, which can only be done numerically. The result is shown in Fig.~\ref{fig3_spectrum}. In the figure, we show all parameter $\lambda$ values, but we fix $g=2.0$, which is the supersymmetric value. This value is interesting because, in this case, the equation for perturbations around the kink solutions is identical to the equation for fermionic bound states. The spectrum of perturbations around the kink solution was computed in \cite{adam2019phi}, and indeed it coincides with the fermion spectrum that we found. For positive values of $\lambda$, it is easy to show graphically that the potential $V_\pm$ has no minimum and, therefore, has no bound states. For $\lambda<0$, a bound state appears. The difference between the two cases is that the peak at the center of the lump $L^\pm$ is inverted with respect to the value $\phi=\pm1$ when $\lambda$ changes sign. As $\lambda$ approaches the value $-\sqrt{2}$, the lump consists of infinitely separated $\phi^4$ kink and antikink. In this case, the spectrum consists of a zero mode and two-degenerate discrete modes with $\omega=\sqrt{3}$. In the following analysis, we fixed $\lambda=-1$ because this is an intermediate value in the range where the lump possesses a bound state.

The next configuration of the scalar field that we analyzed was the BPS antikink $\phi_{LA}(x)$. In this case, the spectrum as a function of the modulus $x_0$ is shown in Fig.~\ref{fig3_spectrum}(b) for $\lambda=-1$. For large $|x_0|$ the spectrum consists of the ones of a separated $\phi^4$ antikink and the lump. This configuration has the zero mode and the shape mode with $\omega=\sqrt{3}$ from the antikink and three discrete modes from the lump. For smaller values of $|x_0|$, the spectrum is slightly deformed, and the highest-energy solution disappears in the continuum.

In the next section, we will simulate the behavior of fermions during collisions of kinks and lumps, and we will show how to use these solutions to measure the asymptotic states of the system.

\section{Results}

\subsection{Scalar field collisions}

\begin{figure}[tb]
\centering
  \caption{Spacetime evolution of the scalar field in (a) near BPS regime and (b) non-BPS regime. (c) and (d) Same as before for the fermion density. Parameters are $g=2.0$, $v=0.3$ and $\lambda=-1.0$.}
  \begin{subfigure}[b]{0.85\textwidth}
    \centering
    \includegraphics[width=\textwidth]{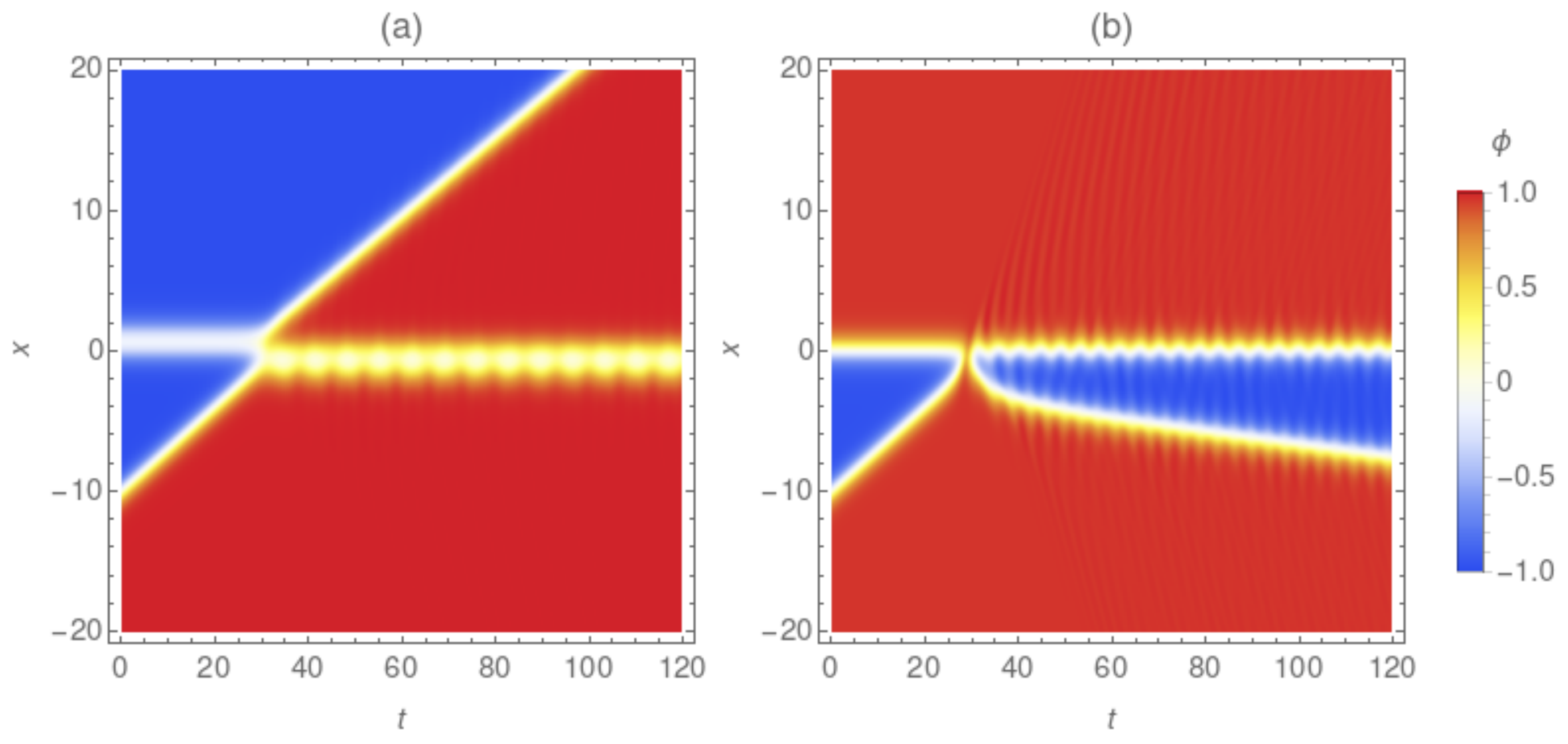}
    \label{fig3_phi}
  \end{subfigure}
  \begin{subfigure}[b]{0.85\textwidth}
    \centering
    \includegraphics[width=\textwidth]{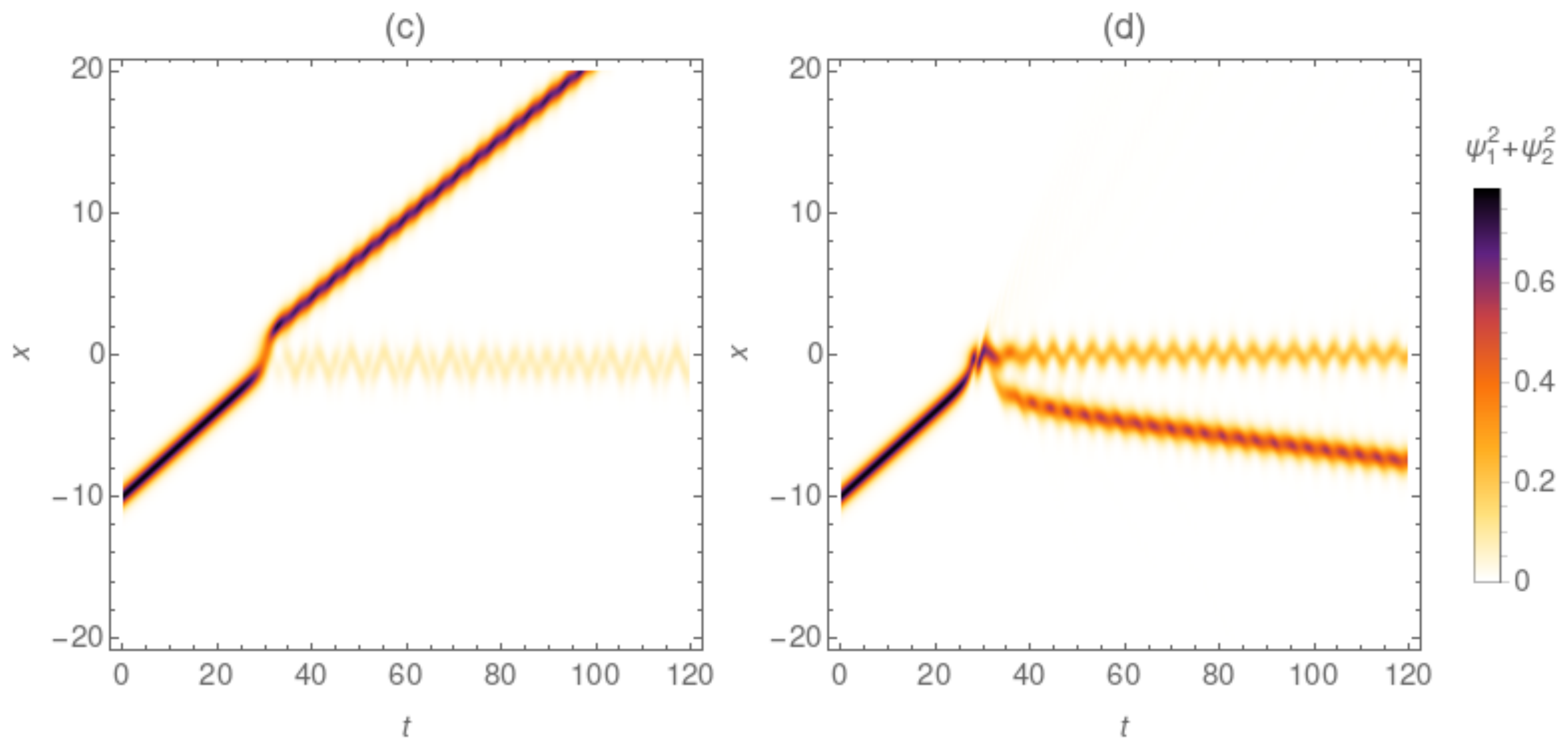}
    \label{fig3_density}
  \end{subfigure}
  \caption*{Source: (a) and (b) Reproduction of the results in \cite{adam2019phi}. (c) and (d) Results obtained by the author in \cite{campos2020fermion}.}
       \label{fig3_colormap}
\end{figure}

We will consider two distinct types of collisions in the scalar field, one near the BPS regimes and one that is not. The first one is the evolution of eq.~(\ref{eq3_phi1}) and the second one of eq.~(\ref{eq3_phi2}). In both cases, we will consider that the $\phi^4$ antikink is boosted instead. These solutions will be used as the initial condition, and the equations of motion will be integrated numerically according to the numerical method described in section \ref{sec3_num}. Both cases were analyzed in \cite{adam2019phi}, and here we reproduce the results for completeness.

Considering eq.~(\ref{eq3_phi1}), it consists of $\phi^4$ antikink and a lump. After boosting the antikink with $v=0.3$, we obtain the result shown in Fig.~\ref{fig3_colormap}(a). The antikink moves in the direction of the lump and crosses, leaving behind a lump in the opposite vacuum. This can be written schematically as
\begin{equation}
A+L^-\to L^+ + A.
\end{equation}
As the collision is close to the BPS regime, we expect the antikink to move smoothly through the lump. However, we considered $v=0.3$, which is quite far from the small velocity regime. Therefore, we can see in the figure that the lump is excited in the process and oscillates.

Considering eq.~(\ref{eq3_phi1}), it consists of an antikink and the kink-on-impurity. After boosting the antikink with $v=0.3$, we obtain the result shown in Fig.~\ref{fig3_colormap}(b). The antikink moves in the direction of the kink-on-impurity and is reflected after a single bounce. The kink-on-impurity is tightly bound at the impurity and remains at the origin. This can be written schematically as
\begin{equation}
\label{eq3_process2}
A+K_0\to A+K_0.
\end{equation}
This collision is far from the BPS regime, and it generates much more radiation than the previous case, as is clearly seen in the figure. Moreover, the kink-on-impurity and the antikink are excited in the process. Similar to the pure $\phi^4$ model, for smaller velocity values, the antikink, and the kink-on-impurity annihilate and, for other velocity values, the system can even exhibit resonance windows. However, we will be focusing on the reflection regime because, otherwise, the system does not have a well-defined final state. In the following subsections, we will consider the evolution of the fermion field in the presence of these scalar field collisions.

\subsection{Bogoliubov coefficients}

Now, turning to the fermion field we are interested in the transition probability between the allowed fermion states. After initializing the scalar field, we set the initial state of the fermion field at one of the fermion states bounded to one of the defects present on the scalar field at $t=0$. This construction is possible because the defects are well separeted in the initial condition. We call this state $j$ and its wavefunction $\psi_{\text{in}}^j(t=0)$. Then, we integrate the evolution of the fermion field according to the dirac equation subject to the time-dependent scalar field. This solution is denoted $\psi_{\text{in}}^j(t)$. At a later time, the scalar field will reach an asymptotic state composed by a different set of defects, which possesses another set of fermion bound states. We call this new basis set $\psi_{\text{out}}^k$, where $k$ can be any allowed fermion bound state of the final configuration. We can expanded our evolved solution in terms of the new basis set as follows
\begin{equation}
\label{eq3_bog}
\psi_{\text{in}}^j(t)=\sum_k B_{j\to k}(t)\psi_{\text{out}}^k(t)+\text{continuum}.
\end{equation}
The expansion coefficients $B_{j\to k}$ are called Bogoliubov coefficients and are precisely the transition amplitude between the initial state $j$ and the final state $k$. They can be found by projecting the evolved state into the new basis set
\begin{equation}
\label{eq3_dotprod}
B_{j\to k}=(\psi_{\text{out}}^k(t),\psi_{\text{in}}^j(t))\equiv\int (\psi_{\text{out}}^k(t))^T\psi_{\text{in}}^j(t) dx.
\end{equation}
A detailed derivation of the quantum interpretation of the Bogoliubov coefficients can be done by expanding the fermion field in terms of particle operators. This can be found, for instance, in \cite{saffin2007particle}.

\subsection{Adiabatic evolution}

\begin{figure}[tbp]
\centering
  \caption{Snapshots of the evolution of the (a) scalar field and (b) fermion field for several instants during a near BPS evolution of the system. Parameters are $v=0.02$ and $g=2.0$. Dotted curves are the static BPS solutions and the fermion densities of the fermion zero mode of the static BPS solutions at the same positions. The two curves are indistinguishable in this scale.}
  \includegraphics[width=0.9\columnwidth]{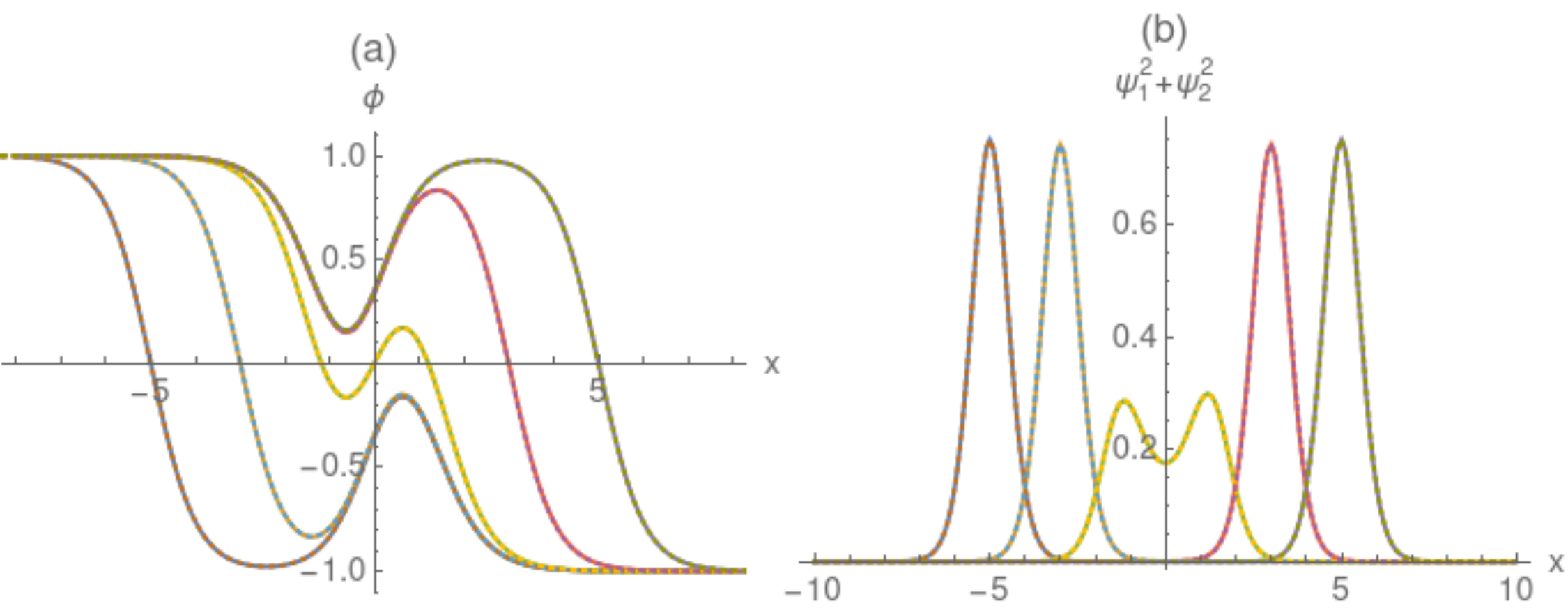}
       \caption*{Source: Results obtained by the author in \cite{campos2020fermion}.}
       \label{fig3_adiabatic}
\end{figure}

Now, we will study the system's evolution in the near BPS regime. The initial condition, in this case, is given by eq.~(\ref{eq3_phi2}) considering a boost in the antikink with a small velocity $v$. Of course, we are considering that the antikink and the lump start well separated. For small velocities, the configuration of the scalar field is very close to the true BPS state, and a BPS configuration can well approximate the evolution at any time. In this case, the system is said to be evolving adiabatically. This was shown in \cite{adam2019phi}. To confirm these results we plot the scalar field evolution for $v=0.02$ in Fig.~\ref{fig3_adiabatic}(a). The solid lines are the solution of the equations of motion at the instants where the zero of the scalar field is at $x_0\simeq-5.0$, $-3.0$, $0.0$, $3.0$ and $5.0$. The dotted lines are the BPS solution with a modulus at the same values of $x_0$. As expected, both curves are indistinguishable in the scale of the figure. This confirms that the evolution is adiabatic.

This analysis raises the question of whether the evolution of the fermion field also evolves adiabatically near the BPS regime. To answer this question, we start the fermion field at the fermion zero mode bound to the antikink of the initial scalar field configuration. Then, we let it evolve according to the Dirac equation with the time-dependent scalar field as a background. The numerical method employed to integrate the Dirac equation is described in section \ref{sec3_num}. We define the fermion density as
\begin{equation}
n=\psi_1^2+\psi_2^2.
\end{equation} 
Its configuration at the same instants of time considered above is shown in the solid lines of Fig.~\ref{fig3_adiabatic}(b). We compare these solutions to the fermion zero mode bound of the BPS antikink solutions with a modulus at the same values of $x_0$. The fermion density of these solutions is shown in dotted lines in the figure. In this case, the solid and dotted lines are indistinguishable in the scale of the figure. This confirms that the fermion field also evolves adiabatically near the BPS regime.

\subsection{Relativistic evolution}

\begin{figure}[tbp]
\centering
  \caption{Bogoliubov coefficients as a function of time for a collision between (a) antikink and lump, and (b) antikink and kink-on-impurity. Parameters are $v=0.3$, $g=2.0$ and $\lambda=-1.0$.}
  \includegraphics[width=0.9\columnwidth]{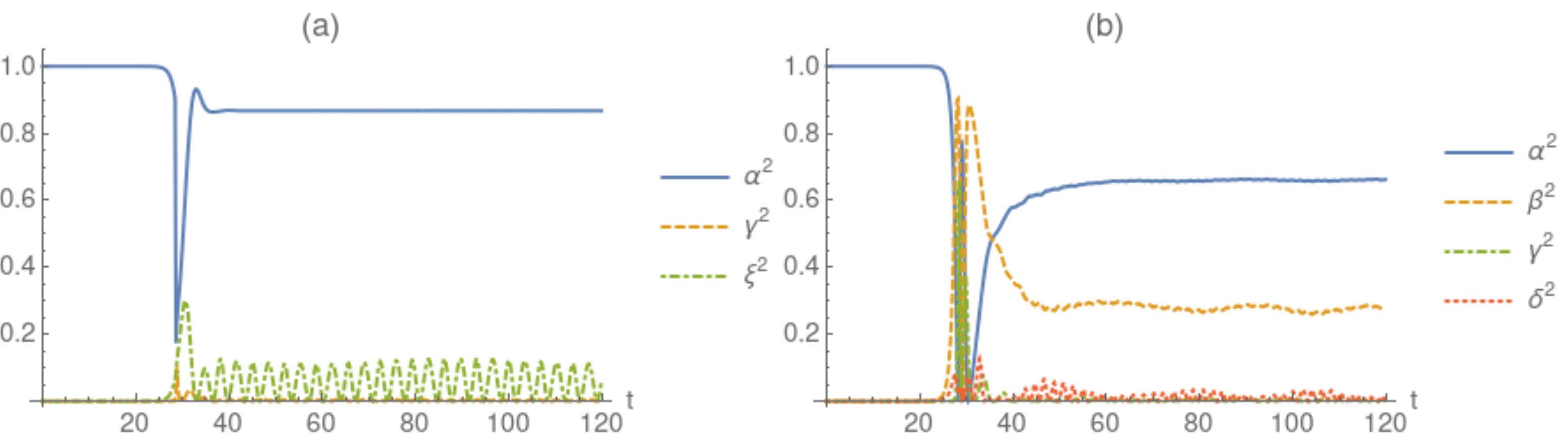}
       \caption*{Source: Results obtained by the author in \cite{campos2020fermion}.}
       \label{fig3_bogvst}
\end{figure}

For the two types of collision considered for the scalar field, we will study the evolution of the fermion field. These are the near BPS collision, which occurs between an antikink and a lump, and the non-BPS one, which occurs between an antikink and the kink-on-impurity. To simplify the analysis, we will fix the initial condition of the fermion field at the fermion zero mode, bound to the antikink, which we will denote as $A$. Other possible allowed states\footnote{The notation for the fermionic bound states is very similar for the notation of defects present in the scalar field. However, it is usually clear from the context which of the two we are referring to.} are the first excited state bound to the antikink $AE$, the zero mode bound to the kink-on-impurity $K$ as well as the first excited state $KE$ and the lowest fermion state bound to the lump $L$. The Bogoliubov coefficients between our initial state and the others are defined as follows
\begin{equation}
\alpha\equiv B_{A\to A}(t),\quad\beta\equiv B_{A\to K},\quad\gamma\equiv B_{A\to AE}(t),\quad\delta\equiv B_{A\to KE}(t),\quad\xi\equiv B_{A\to L}(t).
\end{equation}
Higher excited states also exist for large enough $g$, but we will not measure the transition to these states.

The evolution of the fermion density for this initial condition is shown in Fig.~\ref{fig3_colormap}(c) and (d) for the near BPS and non-BPS regimes, respectively. In both cases, the fermion density is initially localized at the antikink, which is approaching the origin. After colliding with the lump, in the near BPS case, and with the kink-on-impurity, in the non-BPS case, part of density stays bound at the antikink, and part is transferred to the other defect. Moreover, part of the fermion density is lost in the form of radiation. This effect is more pronounced in the non-BPS collision. Interestingly, the radiation of the fermion field is not very significant, even in the non-BPS case, and most fermions stay bound to one of the two defects.

We can compute the evolution of the Bogoliubov coefficients with time by calculating the inner product at each time instant according to eq.~(\ref{eq3_dotprod}). The result is shown in Fig.~\ref{fig3_bogvst}. Before the collision, we have $\alpha=1.0$ due to our choice of initial conditions. During the collision, we cannot separate the different defects, and the coefficients' value is not reliable. In particular, the coefficient $\xi$ refers to the transition to the lowest fermion state bound to $L^-$ before the near-BPS collision, and at some point, we have to change the definition to compute the transition to the lowest fermion state bound to $L^+$. This is necessary because the lump adjusts to the boundaries as the antikink passes through the origin. Finally, when the defects are farther apart and can be distinguished again, the Bogoliubov coefficients reach an asymptotic value that tells the transition amplitude between states due to the collision.

An important technical detail is that the Bogoliubov coefficients oscillate with time except for the zero mode. This can be understood by looking at our ansatz and noticing that the fermion field oscillates with phase $\omega t+\theta$. However, in the simulations, the value of $\theta$ is not known. Thus, we project the fermion field onto an excited state with a fixed phase at an arbitrary value $\theta_0$. Due to the field's oscillation, the phases will coincide once every period and, when they coincide, the coefficient will reach a peak, which is precisely the value of the Bogoliubov coefficient. 

An intriguing effect that should be noticed in Fig.~\ref{fig3_bogvst}(b) is that $\beta^2$ does not reach a well-defined asymptotic value because there is a small oscillation in its value. This oscillation is accompanied by an oscillation in the amplitude of $\delta^2$. Moreover, the two are negatively correlated. This can be understood by noting that the states we are using when computing the projection in eq.~(\ref{eq3_dotprod}) are fermion bound states to static or boosted kink and antikink as well as bound states to the lump. However, the final configuration of the process in eq.~(\ref{eq3_process2}) is not any of those because the kink-on-impurity is excited and oscillates around the origin. This oscillation can be interpreted as a perturbation around the kink fixed at the origin. Therefore, if we project our final state on the fermion states bound to the kink at the origin, there will be a transition between the states due to the perturbation.

\begin{figure}[tbp]
\centering
  \caption{Final value of the Bogoliubov coefficient as a function of $g$ for several values of $v$. We consider the near BPS collision between antikink and lump and fix $\lambda=-1.0$.}
  \includegraphics[width=0.9\columnwidth]{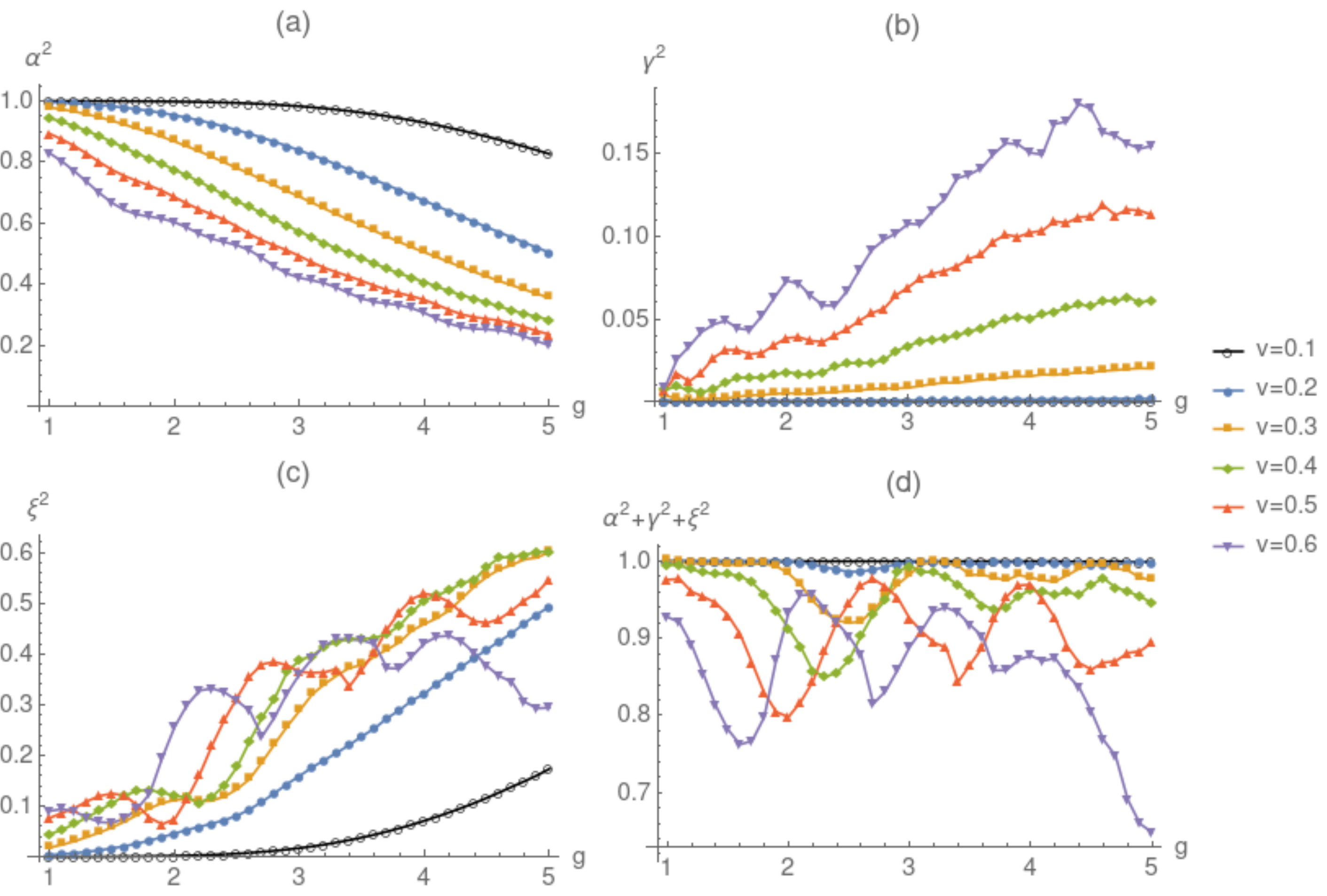}
       \caption*{Source: Results obtained by the author in \cite{campos2020fermion}.}
       \label{fig3_bog1}
\end{figure}

Having said that, we will use the asymptotic value of the Bogoliubov coefficients as a measurement of the amount of fermions that are transferred from one defect to the other due to the collision. First, considering the near BPS collision between the $\phi^4$ antikink and the lump, we find the result shown in Fig.~\ref{fig3_bog1}. These values are shown for relativistic velocities in the range $v\geq 0.1$. For $v=0.1$, the value of $\alpha^2$ is close to $1.0$, which means that the fermion stays at the fermion zero mode bound to the antikink. This is expected because we are approaching the adiabatic limit discussed previously. Moreover, for this value of $v$, a small fraction of fermions are transferred to the lump. This fraction gets more significant as $g$ increases. In general, as the velocity or the coupling constant increases, less fermions stay at the zero mode bound to the antikink. In Fig.~\ref{fig3_bog1}, we see the $\alpha^2$ decreases as $v$ and $g$ increases. Accordingly, $\gamma^2$ has the opposite behavior, while $\xi^2$ has a similar tendency but has a more complicated behavior. The amount of fermions transferred to higher excited states or lost as radiation is quantified by the difference of the sum $\alpha^2+\gamma^2+\xi^2$ from unity. In Fig.~\ref{fig3_bog1}(d), we see that there is less tendency for the fermions to stay at the lowest excited states as $v$ increases, presumably because the collisions are more energetic. However, we do not see the same tendency as $g$ increases.

\begin{figure}[tbp]
\centering
  \caption{Final value of the Bogoliubov coefficient as a function of $g$ for several values of $v$. We consider the non-BPS collision between antikink and kink-on-impurity and fix $\lambda=-1.0$.}
  \includegraphics[width=0.9\columnwidth]{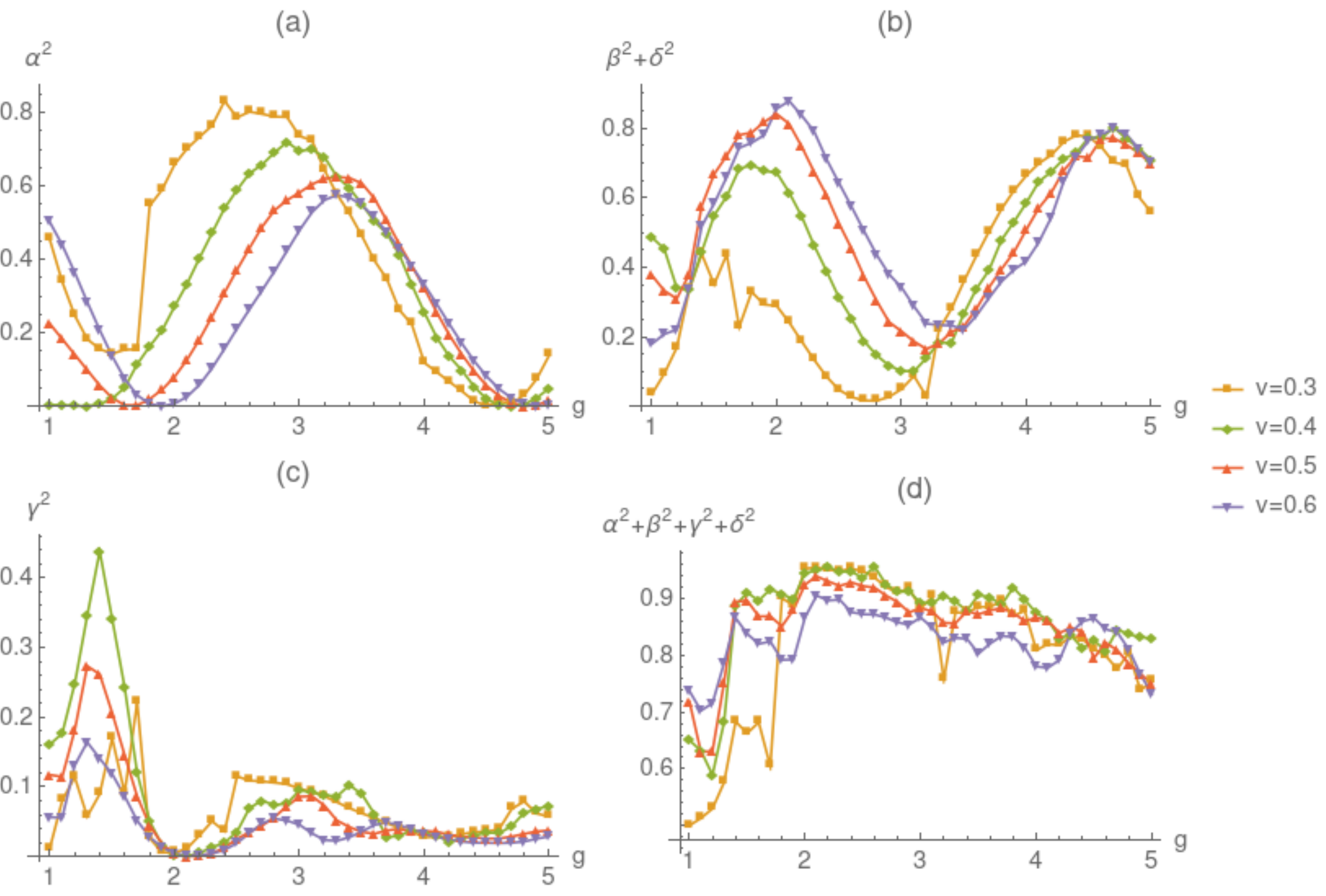}
       \caption*{Source: Results obtained by the author in \cite{campos2020fermion}.}
       \label{fig3_bog2}
\end{figure}

Then, we consider the non-BPS collision between an antikink and the kink-on-impurity. The values of $v$ we used are also relativistic, and we set $v\geq 0.3$ to guarantee that the antikink and the kink-on-impurity will not annihilate. The behavior of fermions when the kink and antikink annihilate is to escape as radiation in a chaotic manner slowly. We will not discuss this case further. The result for larger velocities is shown in Fig.~\ref{fig3_bog2}. We show in the plot the final values of $\alpha^2$, $\beta^2+\delta^2$, $\gamma^2$, and the sum of the three quantities. We do not plot the terms $\beta^2$ and $\delta^2$ separately because they oscillate, as discussed previously, while the sum has a well-defined value. It is clear that in the non-BPS case, the system's behavior is more complicated and very sensitive to the parameters. The behavior of $\alpha^2$ and $\beta^2+\delta^2$ are approximately sinusoidal and negatively correlated. This approximate behavior was justified in \cite{gibbons2007fermions} by solving the Dirac equation with an ansatz for the fermion field symmetric with respect to $x$, with time-dependent amplitude and phase, and interacting with a constant scalar field at its maximum value during the collision. 

In Fig.~\ref{fig3_bog2}(c), we show the values of $\gamma^2$. They have a complicated dependence on $g$. The fraction of fermions that are transferred to this state is small in general but can be significant in a few cases. Again, we quantify the amount of fermions that is not transferred to higher excited states and not lost as radiation by the sum $\alpha^2+\beta^2+\gamma^2+\delta^2$. It is plotted in Fig.~\ref{fig3_bog2}(d). We can also see that the sum tends to decrease for larger values of $v$, where the collision is more energetic.

We will discuss and summarize our results and conclusions in the following sections and mention some ideas for future projects.

\section{Conclusion}

In this work, we considered a scalar field with a half-BPS preserving defect coupled to a fermion field via a Yukawa coupling. The scalar field possesses BPS solutions as well as non-BPS solutions. The BPS ones can be interpreted as being composed of two defects: a $\phi^4$ antikink and a lump. When the fermion couples to the scalar field, it gains states bound to the defect configurations of the system considered.

Among the many possible collisions of the scalar field, we considered a near-BPS one, which occurs between the antikink and the lump, and a non-BPS one, which occurs between the antikink and the kink-on-impurity. Then, we coupled the fermion to the scalar field, starting with the fermion bound to the antikink at the zero mode. For the near-BPS collision, the scalar field configuration and the fermion evolve adiabatically in moduli space for non-relativistic velocities. 

We found that after the collision, most fermions stay bound to one of the two defects at the final state. Only a tiny fraction is lost as radiation, especially for the near-BPS regime. Then, we quantified these transitions via the Bogoliubov coefficients, which reach an asymptotic value for late times. This asymptotic value measures the transition probabilities between defects due to the whole collision process. They have complicated behavior, but it is possible to see a few tendencies. For the near-BPS case, the collision is almost adiabatic for small $v$ and $g$. As these quantities increase, the probability of being transferred to the other defect or to move to an excited state at the same defect also increases. Moreover, we computed the difference of the sum of the probability to stay at the lowest excited states from unity. We found that the amount of fermions transferred to higher excited states or lost as radiation increases as the collision becomes more energetic. For the non-BPS, the behavior is much less regular, but the amount of fermions that are transferred to higher excited states or lost as radiation also tends to increase as the collision becomes more energetic.

This work is most relevant because, if we view the BPS antikink and a multi-defect configuration composed of a $\phi^4$ antikink and a lump, we can use our analysis to guide the study of higher dimensional systems, where multi-defect BPS configurations can exist as well. A natural continuation of this work is to consider situations where there is a back-reaction of the fermion to the scalar field and investigate how the inclusion of this term affects kink-antikink collisions. Moreover, it would be interesting to analyze whether it is possible to define Bogoliubov coefficients in this case as well.

\section{Appendix: Numerical method}
\label{sec3_num}

To integrate the equations of motions (\ref{eq3_ELphi}) and (\ref{eq3_ELpsi}), we employ the method of lines. It consists of discretizing space with spacings $h=0.01$. The scalar field at the gridpoints $x_i$ is denoted by $\phi_i(t)$, where $i=0,1,\ldots,N$. We make similar definitions for the fermion field. The spatial derivatives are discretized according to a second-order finite difference method. For the fermion field, we need a first derivative, which reads
\begin{equation}
\frac{\partial\psi_i}{\partial x}=\frac{\psi_{i+1}-\psi_{i-1}}{2h}.
\end{equation}
For the scalar field, we need a second derivative, which reads
\begin{equation}
\frac{\partial^2\phi_{i}}{\partial x^2}=\frac{\phi_{i+1}-2\phi_i+\phi_{i-1}}{h^2}.
\end{equation}
In the expressions above, we omitted the time dependence for simplicity. This results in a set of first-order ordinary differential equaitons for a vector containing the values at the grid points of the components of the fermion field, the scalar field $\phi$ and its derivative $\phi_t$. Note, that we need to define an auxiliary field equal to $\phi_t$ to transform the equation of motion for $\phi_i(t)$, which is second-order, into two first-order equations. The equations are integrated using a fourth-order Runge-Kutta method with step size $\tau=0.01$. We fix the boundaries at $x=\pm100.0$, which gives $N=20000$. The boundary conditions are $\psi_0(t)=\psi_N(t)=0.0$, $\phi_0(t)=1.0$ and, depending on the case considered, we can have either $\phi_N(t)=\pm 1.0$. The final time of the integration is in the range $100.0<t_f<400.0$. These values are short enough, and therefore, the boundary does not interfere with the bulk evolution of the system. The fermion field is normalized to one, and the time integration is approximately unitary and conserves the normalization during the whole evolution of the system up to a relative error of order $10^{-5}$.

\chapter{FINAL REMARKS}

We started this thesis by defining solitons, solitary waves and topological defects, and giving a brief historical introduction to these concepts. Then, we discussed important theoretical and applied works in the literature. In section 2, we discussed some general properties of kinks and made a careful revision of kink-antikink collisions and the resonant energy exchange mechanism.

In section 3, we described a model that we constructed such that the vibrational mode of the kink becomes a quasinormal mode. Our toy model made it possible to compute the frequency of the quasinormal mode and the decay rate analytically because the stability potential was constant in different domains. In most models, it is impossible to perform this task, and numerical evaluation of the quasinormal mode can be tricky. Moreover, we were able to confirm previous results in the literature, such as the gradual disappearance of the resonant structure. The toy model potential was constructed using a continuous and piecewise-defined function with a continuous derivative. This is an exciting possibility that has been little explored. This prescription could be used to construct new topological models as done, for instance, in \cite{basak2021new}. One interesting idea for future work is to study the collective coordinates effective model of this toy model because the kink and vibrational modes are simple. The collective coordinates method could also be applied for newly constructed toy models.

In section 4, we considered a scalar field with a family of polynomial potentials of higher order. The potentials in this family are analogous to the $\phi^4$ one with two minima. The crucial difference is that, for some values of the parameters, the first contribution when we expand the potential around the vacua is not quadratic. It can be quartic, sextic, and so on. In this case, we obtain a kink with long-range tails on both sides of the kink, which decay as a power-law. The long-range character of the tails makes it impossible to use the usual ansatz to initialize the system. Therefore, it is necessary to introduce a minimization procedure to find a suitable initial condition. The procedure consists of finding a configuration that obeys the equations of motion of a traveling wave as close as possible. This is done starting with an initial guess and using a least-square minimization algorithm. Similarly, the velocity field must obey the zero-mode equation of a traveling wave as close as possible. 

After correctly initializing the system, we found the kink-antikink annihilated directly into radiation for velocities below an ultra-relativistic critical value. One interesting continuation of this work is to apply the minimization procedure to other families of models containing kinks with long-range tails. In particular, one could construct a model that starts from the $\phi^4$ model and gradually turns into a model with long-range tails. This way, it should be possible to understand how the critical velocity becomes ultra-relativistic, and the bion formation is suppressed. Finally, it should be interesting to study the interaction of kinks with long-range tails in the presence of boundaries obeying Neumann boundary conditions or any other types of boundary conditions. Interactions of kinks with boundaries have very rich physics and have not been fully explored yet.

In section 5, we investigated collisions of the double sine-Gordon model. We simulated kink-antikink collisions in this model for a wide range of model parameters and indicated the regions where annihilation and separation occur after one or more bounces. Surprisingly, the multi-bounce region disappears as we approach the integrable limits of the model, despite the presence of a vibrational mode. Then, we simulated collisions where the kink and the antikink have their vibrational mode symmetrically excited beforehand. This excitation led to the appearance of one-bounce resonance windows and a more complex fractal pattern of higher-resonance windows. After characterizing the behavior of each region in parameter space, we found that a spine structure appears in the region of separation after one bounce. Interestingly, the excitation increased the range in parameter space where higher-bounce resonance windows occur, showing that the resonance structure was only hidden near the integrable limits. Finally, we were able to find an approximate model to the position of one-bounce resonance windows and to characterize the separation regions using the maximum value of the energy densities.

One fascinating continuation to the work in section 5 is to study other modifications of the sine-Gordon model. Also, one could try to understand the behavior near the integrable limits analytically using the perturbation theory. Finally, it should also be possible to compare the simulations results to an effective model using collective coordinates. However, the effective Lagrangian can only be found numerically in this case.

In section 6, we introduced the concept of kinks coupled with fermions. Then, we discussed important results in the literature of fermion-kink systems and listed a few general analytical properties relevant to those systems.

In section 7, we analyzed the effect of the shape mode excitation on a fermion that is bound to a kink. This system is very important because the kink is expected to be excited after an interaction. In this case, the equations of motion can be solved analytically using perturbation theory. However, it is necessary to neglect the fermion back-reaction on the scalar field in order to do so. We considered both the $\phi^4$ model and the toy model used in section 3, which possesses quasinormal modes. The fermionic bound states can be found analytically in the former and can only be found numerically in the latter. According to our results, the fermion should escape the kink at a constant rate whenever the transition to the continuum is allowed by Fermi's golden rule. This is an interesting prediction that could be verified or used to constraint the wobbling amplitude of a kink or the value of the Yukawa coupling constant. The transition rate diminishes with time when the shape mode is turned into a quasinormal mode, and the transition probabilities reach an asymptotic value. Interestingly, the behavior of the asymptotic value can be understood as being reminiscent of Fermi's golden rule. 

An exciting continuation of this work is to study fermions interacting with kinks with an inner structure, such as the double sine-Gordon. This could be done with and without wobbling. Moreover, one could consistently include the fermion back-reaction and see how the results are modified. Of course, when fermion back-reaction is considered, it opens many possibilities for future works as very few works in the literature include this effect.

Finally, in section 8, we considered defect collisions of a scalar field model with a half-BPS preserving impurity. The scalar field was used as a background for a fermionic field which evolves as the defects collide. We considered two types of collisions. The first one was a collision between the antikink and the lump, which occurs near the BPS sector of the system. The second one was between the antikink and the kink-on-impurity, which occurs far from the BPS sector of the system. We started the fermion at the zero mode of the antikink and measured the fermion transfer between the defects participating in the collision process.
Interestingly, the background approximation is exact for this initial condition. One of the main results of this work was to conclude that near the BPS regime, the evolution becomes adiabatic, and there is no fermion transfer. For relativistic collisions, the behavior is more complicated, but, in general, the fermion is more likely to be transferred to the other defect, move to a higher excited state at the same defect, or be lost as radiation.  

An interesting continuation of this work is to consider cases where the fermion back-reaction is non-vanishing and compute the evolution consistently. For strong coupling constants, the effect of this term can be considerable. Thus, the fermion back-reaction to the scalar field should be considered whenever possible. However, the problem becomes considerably more complicated and more numerically involved. For this reason, the study of fermions without back-reaction can be used to give some intuition on the expected behavior of the system. An interesting question in light of the work \cite{adam2019spectral} is whether a fermion excitation could also lead to the appearance of spectral walls when the back-reaction is included. We plan to study this and other questions in future works.

\renewcommand\bibname{REFERENCES}
\phantomsection
\titleformat{\chapter}{\normalfont\bfseries\centering}{\thechapter}{1em}{}
\addcontentsline{toc}{chapter}{\protect\numberline{}\sffamily REFERENCES}
\bibliography{joao}

\phantompart

\end{document}